\documentclass[aps,pra,amsmath,amssymb,reprint,floatfix,superscriptaddress,showpacs]{revtex4-1}
\usepackage{graphicx}
\usepackage[bookmarks=false,colorlinks=true,linkcolor=blue,filecolor=blue,citecolor=blue,urlcolor=blue]{hyperref}
\usepackage{bm}
\usepackage{soul}

\newcommand{\bfa}{\mathbf{a}}
\newcommand{\bfb}{\mathbf{b}}
\newcommand{\bff}{\mathbf{f}}
\newcommand{\bfg}{\mathbf{g}}
\newcommand{\bfl}{\mathbf{l}}
\newcommand{\bfp}{\mathbf{p}}
\newcommand{\bfq}{\mathbf{q}}
\newcommand{\bft}{\mathbf{t}}
\newcommand{\bfw}{\mathbf{w}}
\newcommand{\bfA}{\mathbf{A}}
\newcommand{\bfB}{\mathbf{B}}
\newcommand{\bfC}{\mathbf{C}}
\newcommand{\bfG}{\mathbf{G}}
\newcommand{\bfJ}{\mathbf{J}}
\newcommand{\bfK}{\mathbf{K}}
\newcommand{\bfM}{\mathbf{M}}
\newcommand{\bfN}{\mathbf{N}}

\newcommand{\bfU}{\mathbf{U}}
\newcommand{\bfV}{\mathbf{V}}
\newcommand{\bfW}{\mathbf{W}}
\newcommand{\bfX}{\mathbf{X}}
\newcommand{\bfY}{\mathbf{Y}}

\newcommand{\bfalpha}{{\bm \alpha}}
\newcommand{\bfbeta}{{\bm \beta}}
\newcommand{\bfgamma}{{\bm \gamma}}
\newcommand{\bfupsilon}{{\bm \upsilon}}
\newcommand{\bfomega}{{\bm \omega}}

\newcommand{\bfOmega}{{\bm \Omega}}
\newcommand{\bfchi}{{\bm \chi}}
\newcommand{\bfvarphi}{{\bm \varphi}}
\newcommand{\bfvartheta}{{\bm \vartheta}}

\newcommand{\bJ}{\bar{J}}
\newcommand{\bcalJ}{\bar{\mathcal{J}}}

\newcommand{\bE}{\bar{E}}
\newcommand{\bcalE}{\bar{\mathcal{E}}}

\newcommand{\bH}{\bar{H}}
\newcommand{\bcalH}{\bar{\mathcal{H}}}
\newcommand{\bT}{\bar{T}}
\newcommand{\bz}{\bar{z}}
\newcommand{\bw}{\bar{w}}
\newcommand{\bbfJ}{\bar{\mathbf{J}}}

\newcommand{\bPsi}{\bar{\Psi}}
\newcommand{\brho}{\bar{\rho}}
\newcommand{\bXi}{\bar{\Xi}}

\newcommand{\calE}{\mathcal{E}}
\newcommand{\calL}{\mathcal{L}}
\newcommand{\calH}{\mathcal{H}}
\newcommand{\calJ}{\mathcal{J}}
\newcommand{\calN}{\mathcal{N}}
\newcommand{\calO}{\mathcal{O}}
\newcommand{\calS}{\mathcal{S}}
\newcommand{\calV}{\mathcal{V}}

\newcommand{\tb}{\tilde{b}}
\newcommand{\tn}{\tilde{n}}
\newcommand{\tK}{\tilde{K}}
\newcommand{\tE}{\widetilde{E}}
\newcommand{\tX}{\widetilde{X}}
\newcommand{\tY}{\widetilde{Y}}
\newcommand{\talpha}{\tilde{\alpha}}
\newcommand{\tepsilon}{\tilde{\epsilon}}

\newcommand{\tchi}{\widetilde{\chi}}
\newcommand{\tphi}{\widetilde{\phi}}
\newcommand{\ttheta}{\widetilde{\theta}}
\newcommand{\tPsi}{\widetilde{\Psi}}
\newcommand{\tPhi}{\widetilde{\Phi}}
\newcommand{\tTheta}{\widetilde{\Theta}}
\newcommand{\tvarphi}{\widetilde{\varphi}}
\newcommand{\tbfalpha}{\tilde{{\bm \alpha}}}
\newcommand{\tbfomega}{\tilde{{\bm \omega}}}
\newcommand{\tbfPhi}{\widetilde{{\bm \Phi}}}
\newcommand{\tbfTheta}{\widetilde{{\bm \Theta}}}
\newcommand{\tbfchi}{\widetilde{{\bm \chi}}}
\newcommand{\tbfW}{\widetilde{\mathbf{W}}}
\newcommand{\tbfX}{\widetilde{\mathbf{X}}}
\newcommand{\tbfY}{\widetilde{\mathbf{Y}}}

\newcommand{\ua}{\uparrow}
\newcommand{\da}{\downarrow}
\newcommand{\sgn}{\textrm{sgn}}

\begin{document}

\title{Non-Abelian $SU(N-1)$-singlet fractional quantum Hall states from coupled wires}

\author{Y. Fuji}
\affiliation{Max-Planck-Institut f\"ur Physik komplexer Systeme, D-01187 Dresden, Germany}

\author{P. Lecheminant}
\affiliation{Laboratoire de Physique Th\'eorique et Mod\'elisation, CNRS UMR 8089, Universit\'e de Cergy-Pontoise, Site de Saint-Martin, F-95300 Cergy-Pontoise Cedex, France}

\date{\today}

\begin{abstract}
The construction of fractional quantum Hall (FQH) states from the two-dimensional array of quantum wires provides a useful way to control strong interactions in microscopic models and has been successfully applied to the Laughlin, Moore-Read, and Read-Rezayi states. 
We extend this construction to the Abelian and non-Abelian $SU(N-1)$-singlet FQH states at filling fraction $\nu=k(N-1)/[N+k(N-1)m]$ labeled by integers $k$ and $m$, which are potentially realized in multi-component quantum Hall systems or $SU(N)$ spin systems. 
Utilizing the bosonization approach and conformal field theory (CFT), we show that their bulk quasiparticles and gapless edge excitations are both described by an $(N-1)$-component free-boson CFT and the $SU(N)_k/[U(1)]^{N-1}$ CFT known as the Gepner parafermion. 
Their generalization to different filling fractions is also proposed. 
In addition, we argue possible applications of these results to two kinds of lattice systems: bosons interacting via occupation-dependent correlated hoppings and an $SU(N)$ Heisenberg model.
\end{abstract}

\maketitle

\tableofcontents

\section{Introduction}

Topologically ordered phases have attracted considerable interests in recent decades due to their robustness against local perturbations and their possible applications to quantum computation \cite{Nayak08,BZeng15}. 
A prominent example is the fractional quantum Hall (FQH) state, which hosts chiral gapless modes at the boundary and quasiparticle excitations obeying nontrivial statistics, called the Abelian or non-Abelian anyons, in the bulk. 
One of major subjects in this research area is to provide an effective description of the topologically ordered phases and to classify them on the basis of it. 
In this respect, several mathematical frameworks are developed to classify topologically ordered phases in two spatial dimensions (2D) (see Refs.~\cite{Wen16,Schoutens16,TLan16} and references therein). 
Another important subject is to investigate microscopic realizations of these phases. 
However, beyond exactly solvable models which are often not quite realistic, the strong interaction, which is essential to stabilize the topological orders, becomes a main obstacle on pursuing this problem. 

A useful tool to tackle this problem is the so-called coupled-wire construction, which was originally applied to Abelian FQH states \cite{Kane02}. 
An advantage of this approach lies in its broad application to interacting systems, even for lattice systems where the construction of trial wave functions designed for the lowest Landau level cannot be applied. 
In this approach, one starts from the array of one-dimensional (1D) fermionic or bosonic wires and then adds interactions among them. 
If the interactions are appropriately chosen in such a way that they open a bulk gap but leave gapless excitations localized along the outermost wires, the coupled wires provide an effective description of the 2D topologically ordered phase with gapless boundaries. 
The latter edge excitations described by a conformal field theory (CFT) in (1+1)-dimensions is a fingerprint of the underlying 2D topological order.
In the last few years, the coupled-wire construction has been extensively developed for a variety of topological phases in 2D interacting systems: Abelian and non-Abelian FQH states \cite{Teo14,Klinovaja14a,Meng14a,Meng14b,Cano15,Sagi15a}, chiral spin liquids \cite{Meng15a,Gorohovsky15,PHHuang16,Lecheminant16}, fractional topological insulators and superconductors \cite{Mong14,Klinovaja14b,Neupert14,Oreg14,Sagi14,Seroussi14,Vaezi14a,Vaezi14b,Santos15}, and symmetry-protected topological phases \cite{YMLu12,Fuji16}. 
There are also interesting applications to 3D topological phases \cite{Sagi15b,Meng15b,Meng16,Iadecola16} and their surface states \cite{Mross15,Mross16,Sahoo16}. 

The coupled-wire construction allows us to strictly treat the interactions for Abelian topological orders with the help of the Luttinger liquid theory \cite{Gogolin,Giamarchi}. 
For non-Abelian topological orders, the precise control of the interactions is tricky since the underlying CFT is by itself an interacting system. 
Nevertheless, the precise control is achieved especially when the corresponding CFT, or precisely the simple current algebra, admits a free-field representation in terms of bosonic and/or fermionic fields. 
This is a key idea of the seminal work by Teo and Kane \cite{Teo14}, in which they succeeded in describing the Moore-Read \cite{Moore91} and Read-Rezayi states \cite{Read99} based on the $\mathbb{Z}_k$ or $SU(2)_k/U(1)$ parafermion CFT \cite{Zamolodchikov85}. 
Although the variety of non-Abelian topological orders is undoubtedly rich, it appears that the microscopic understanding of them is mostly limited to those associated with the $\mathbb{Z}_k$ parafermion CFT. 
An aim of this paper is to provide a theoretical tool to microscopically deal with non-Abelian topologically orders beyond the $\mathbb{Z}_k$ parafermion CFT. 

In this paper, we thus investigate the generalization of the coupled-wire construction developed by Teo and Kane for multi-component Abelian and non-Abelian FQH states with a general internal $SU(N)$-algebraic structure. 
Multi-component FQH states have been studied in various contexts by including the spin degeneracy of electrons \cite{Halperin83}, an isospin index for bilayer FQH states \cite{[{See for instance, }]DasSarma}, or valley degrees of freedom of the graphene \cite{Arovas99,Nomura06,Goerbig07,Dean11}. 
Our starting point is the coupled-wire construction of $(N-1)$-component Abelian FQH states that correspond to generalized Halperin wave functions at filling $\nu=(N-1)/N$ and are $SU(N-1)$ singlet. 
The simplest cases for $N=2$ and $3$ respectively correspond to the well-known bosonic Laughlin state at $\nu =1/2$ and the Halperin (221) state at $\nu =2/3$ \cite{Halperin83}, whose coupled-wire constructions have been achieved previously \cite{Kane02,Teo14}. 
In this work, we further reveal a hidden $SU(N)$ symmetry of these $SU(N-1)$-singlet FQH states, which most prominently appears in gapless edge states. 
These edge states are indeed described by the chiral $SU(N)_1$ Wess-Zumino-Witten (WZW) CFT \cite{dFMS} with central charge $c=N-1$ and thus correspond to $N-1$ massless chiral bosons. 

A non-Abelian extension of these $SU(N-1)$-singlet states can be obtained by symmetrizing $k$ copies of the generalized Halperin wave functions. 
The resulting state for bosons at filling $\nu= k(N-1)/N$ is a generalization of the non-Abelian spin-singlet (NASS) FQH states introduced by Ardonne and Schoutens for $N=3$ \cite{Ardonne99,Ardonne01a}. 
The quasiparticle excitations of the non-Abelian $SU(N-1)$-singlet states carry the same fractional charges as those of the Abelian ones but obey non-Abelian statistics. 
The physical realizations of these non-Abelian states for $N=4$ have been proposed for rotating spin-1 cold bosons \cite{Reijnders02,Reijnders04}. 
Those for general $N$ have been considered as natural candidates for non-Abelian topological phases in fractional Chern insulators with higher Chern numbers \cite{Sterdyniak13}.

We device a coupled-wire system for the bosonic non-Abelian $SU(N-1)$-singlet states at $\nu= k(N-1)/N$ in terms of $k$ channels of $(N-1)$-component bosonic wires. 
In this setup, the construction is achieved by first taking $k$ copies of the Abelian $SU(N-1)$-singlet state and then introducing suitable interactions made from their excitations. 
Those interactions are the tunnelings of unit-charge particle excitations between adjacent wires and interactions among quasiparticle excitations within the same wire. 
We then discuss that the resulting state has a spectral gap in the bulk while has gapless edge states described by the chiral $SU(N)_k$ WZW CFT. 
Our approach employs free-bosonic (vertex) representations of the $SU(N)_k$ WZW CFT and the $SU(N)_k/[U(1)]^{N-1}$ CFT. 
The latter CFT is a generalization of the $\mathbb{Z}_k$ parafermion and is known as the Gepner parafermion \cite{Gepner87}. 
The vertex representations allow us to identify the neutral sector of the edge states as those parafermions and to rigorously prove the existence of the bulk gap for some special case. 
We also discuss that the neutral parts of bulk quasiparticles are described by spin fields of the parafermion CFT, which are associated with a $\mathbb{Z}_k^{N-1}$ symmetry breaking. 

This construction is further extended to non-Abelian FQH states generically described by $[U(1)]^{N-1} \times SU(N)_k/[U(1)]^{N-1}$ CFTs at different filling fractions, which include fermionic FQH states. 
The construction proceeds in a similar way by starting from certain parent Abelian states that share the same filling fraction and the same quasiparticle lattice structure with the non-Abelian states. 
In fact, those parent Abelian states are associated with the $K$ matrices proposed in Refs.~\cite{Ardonne00,Ardonne01b}. 
Our approach reveals an intimate relation between certain Abelian and non-Abelian FQH states at the microscopic level.

The coupled-wire approach also has a particular advantage for the applications to spatially anisotropic lattice systems. 
Interactions required to stabilize the multi-component FQH states take the form of tunnelings dressed by particle fluctuations. 
Such tunnelings are naturally realized by occupation-dependent correlated hoppings on the lattice, giving rise to the effect like a mutual flux attachment \cite{Senthil13}. 
Interestingly, they can be engineered for ultracold atoms in optical lattices with periodically modulated interactions \cite{Rapp12,Meinert16}. 
Another, perhaps most natural, way to realize the multi-component FQH states is to consider $SU(N)$ spins. 
They will give an effective description of a Mott-insulating phase of alkaline-earth or ytterbium ultracold gases loaded into 2D optical lattices \cite{Fukuhara07,DeSalvo10,Taie12,Hofrichter16}. 
An interesting issue about this model is that several theoretical studies show a tendency to stabilize the $SU(N)$ Abelian chiral spin states for large $N$ \cite{Khveshchenko89,Khveshchenko90,Hermele09,Hermele11,GChen16}. 
We briefly argued the applications to these lattice systems.

\subsection*{Outline of the paper}

Considering the length and technical complexity of this paper, we here provide a short summary for each section along with some tips for readers. 
It will allow the readers to skip around to find sections of their interest, depending on their knowledge about the coupled-wire construction, $SU(N)$ algebra, and CFT. 
As a technical remark, all the analyses given in this paper are basically understood within free bosonic theory.

Section~\ref{sec:Preliminaries} presents several preliminaries of our approach. 
We start with a conventional way of understanding the Abelian $SU(N-1)$-singlet FQH states in terms of the trial wave function and Chern-Simons theory. 
We then introduce the non-Abelian $SU(N-1)$-singlet FQH states by symmetrizing the Abelian states.
The most general setup of our coupled-wire system is also presented in the basis of Luttinger liquid theory.

In Sec.~\ref{sec:221State}, we construct the Halperin (221) state, which corresponds to the bosonic Abelian $SU(2)$-singlet FQH states. 
The readers who are not familiar with the coupled-wire construction will find its basic idea from Sec.~\ref{sec:Int221State}. 
While a main purpose of this section is to reveal a hidden link with the $SU(3)_1$ WZW CFT (Sec.~\ref{sec:SU31Currents}), the analysis proceeds in a heuristic way and the specific knowledge about the CFT is not assumed. 
The quasiparticles of the Halperin (221) state is also obtained in the manner of Teo and Kane (Sec.~\ref{sec:SU31Quasiparticles}).

In Sec.~\ref{sec:NASSState}, we extend the construction to the bosonic NASS state at $\nu=2k/3$, particularly focusing on the $k=2$ case. 
As the NASS state is described by the $SU(3)_k$ WZW CFT, the analysis in the large extent relies on the underlying $SU(3)$-algebraic structure of the Halperin (221) state discussed in the preceding section. 
We present the coupled-wire system in terms of the $SU(3)_k$ CFT (Sec.~\ref{sec:SU3kCurrents}) and also the $[U(1)]^2 \times SU(3)_2/[U(1)]^2$ CFT (Sec.~\ref{sec:SU32Parafermion}). 
The latter description involving the parafermionic CFT turns out to be convenient for examining quasiparticles (Sec.~\ref{sec:SU32Quasiparticles}) as well as for the extension to general filling fractions in Sec.~\ref{sec:GeneralFilling}.
It also reveals the level-rank duality between $SU(3)_2$ and $SU(2)_3$, which exchanges the interwire and intrawire interactions in the neutral sector between the $k=2$ NASS and $k=3$ Read-Rezayi states.

In Sec.~\ref{sec:SUN}, the construction is generalized to the bosonic $SU(N-1)$-singlet FQH states with Abelian (Sec.~\ref{sec:AbelianSUN}) and non-Abelian statistics (Sec.~\ref{sec:NonAbelianSUN}), which are described by the $SU(N)_k$ WZW CFT.
The construction is a straightforward generalization of the previous two sections and is therefore presented in a more systematic and abstract manner by highlighting the $SU(N)$-algebraic structure. 
Quasiparticles of the non-Abelian $SU(N-1)$-singlet state are identified by the relation with a $\mathbb{Z}_k^{N-1}$ statistical mechanical model in Sec.~\ref{sec:SUNkQuasiparticles}. 
A somewhat ad hoc argument in Sec.~\ref{sec:SU32Quasiparticles} is complemented here. 
This analysis may be interesting by its own for those familiar with similar statistical mechanical models.

Section~\ref{sec:GeneralFilling} presents extensions of the non-Abelian $SU(N-1)$-singlet states constructed in Sec.~\ref{sec:SUN} to different filling fractions. 
The construction is achieved by turning on interactions mixing different channels within the same component in certain Abelian $(N-1)k$-layer FQH states. 
The resulting non-Abelian states are generically described by the $[U(1)]^{N-1} \times SU(N)_k/[U(1)]^{N-1}$ CFT. 
While the discussion starts with the general $N$ case, two specific examples are given for $N=3$ and $k=2$ and thus one may jump from Sec.~\ref{sec:NASSState}. 
These examples include the NASS states at $\nu=4/(4m+3)$ by Ardonne and Schoutens \cite{Ardonne99,Ardonne01a} and a bilayer non-Abelian state at $\nu=4/(4m+1)$ by Barkeshli and Wen \cite{Barkeshli10}. 

Section~\ref{sec:LatticeSystems} provides the application of coupled-wire approach to two kinds of lattice system: lattice bosons with occupation-dependent correlated hoppings and an $SU(N)$ Heisenberg model. 
This section is written in a different taste from previous sections and only focuses on the Abelian FQH states for the sake of simplicity. 
Those who are interested in concrete microscopic models to realize the FQH states may directly come here after short glances at Sec.~\ref{sec:221State} and the first part of Sec.~\ref{sec:SUN}.

Section~\ref{sec:Conclusion} concludes this paper with several outlooks. 
Six appendices are devoted to complete technical details of the analysis in the main text. 
Appendix~\ref{app:RootWeightSUN} summarizes our conventions of roots and weights of the $SU(N)$ algebra.
The rests provide explicit proofs of the vertex representations of the WZW and parafermion CFTs.

\section{Preliminaries}
\label{sec:Preliminaries}

In this section, we first introduce Abelian and non-Abelian FQH $SU(N-1)$-singlet states that are studied in this paper. 
The former is described by a generalized Halperin trial wave function, while the latter is obtained by a symmetrization of the Abelian states. 
We then describe our coupled-wire systems where those FQH states are constructed. 

\subsection{Abelian $SU(N-1)$-singlet FQH states and $K$ matrix}

We start from multi-component Abelian FQH states whose trial wave functions on a disk geometry are given by 
\begin{align} \label{eq:MultiFQHS}
\Psi_\bfK (\{ z_i^\sigma \}) &= \prod_{\sigma=1}^{N-1} \prod_{i<j} (z_i^\sigma -z_j^\sigma)^{K_{\sigma \sigma}} \nonumber \\ 
&\ \ \ \times \prod_{\sigma' < \sigma''} \prod_{k,l} (z_k^{\sigma'} -z_l^{\sigma''})^{K_{\sigma' \sigma''}} e^{-\frac{1}{4} \sum_{i,\sigma} |z^\sigma_i|^2}, 
\end{align}
where $\{ z^\sigma_i \} = z^\sigma_1, \cdots, z^\sigma_{\calN_\sigma}$ with $\calN_\sigma$ being the number of particles with the $\sigma$-th component ($\sigma= 1, \ldots N-1$), $z^\sigma_i \equiv x^\sigma_i +iy^\sigma_i$ denotes the 2D complex coordinate of the $i$-th particle with the $\sigma$-th component, and the magnetic length has been normalized to one. 
$\bfK$ is the so-called \emph{K matrix} \cite{Wen92,Wen95}, an $(N-1) \times (N-1)$ integer symmetric matrix that fully determines the topological property of Eq.~\eqref{eq:MultiFQHS} in the absence of extra symmetry. 
For example, the ground-state degeneracy on a torus is given by $\left| \det \bfK \right|$. 
This matrix also appears in the effective low-energy description of the state \eqref{eq:MultiFQHS}, that is the (2+1)-dimensional Chern-Simons theory. 
If we assign the $U(1)$ charge $q$ to each component of particles, the corresponding Lagrangian is given by 
\begin{align} \label{eq:ChernSimons}
\calL = \frac{1}{4\pi} \epsilon_{\mu \nu \lambda} K_{\sigma \sigma'} a^\sigma_\mu \partial_\nu a^{\sigma'}_\lambda -\frac{q}{2\pi} \epsilon_{\mu \nu \lambda} t_\sigma A_\mu \partial_\nu a^\sigma_\lambda, 
\end{align}
where we have assumed summation over repeated indices. 
Here, $a^\sigma_\mu$ are internal gauge fields, $A_\mu$ is an external $U(1)$ gauge field, and $t_\sigma$ is an $(N-1)$-dimensional vector called the charge vector and given by $\bft = (1,\cdots,1)$. 
This choice of charge vector is called the symmetric or multi-layer basis \cite{Wen92,Wen95}. 

In the following, we consider the $(N-1)$-component Abelian FQH states \eqref{eq:MultiFQHS} with the $K$ matrix whose diagonal entries are $2$ and off-diagonal entries are $1$:  
\begin{align} \label{eq:KmatSUN}
\bfK_{SU(N)} = 
\begin{pmatrix} 
2 & 1 & 1 & \cdots & 1 & 1 \\ 
1 & 2 & 1 & \cdots & 1 & 1 \\ 
1 & 1 & 2 & & & 1 \\ 
\vdots & \vdots & & \ddots & & \vdots \\ 
1 & 1 & & & 2 & 1 \\ 
1 & 1 & 1 & \cdots & 1 & 2 
\end{pmatrix}. 
\end{align}
Since all the diagonal entries are even integer, the corresponding FQH state is bosonic. 
Such a FQH state is realized at the filling fraction,
\begin{align}
\nu=\frac{N-1}{N}. 
\end{align}
The state is not just $SU(N-1)$ symmetric but actually $SU(N-1)$ singlet. 
For the single-component case, $\bfK_{SU(2)}=2$ and Eq.~\eqref{eq:MultiFQHS} represents the $\nu=\frac{1}{2}$ bosonic Laughlin FQH state. 
It has been pointed out that this state possesses a hidden $SU(2)$ symmetry \cite{Fradkin99}. 
Indeed, its edge state and the underlying Chern-Simons theory are both described by the $SU(2)_1$ WZW CFT. 
For the two-component case, the $K$ matrix \eqref{eq:KmatSUN} gives the spin-singlet Halperin (221) state \cite{Halperin83}, 
\begin{align} \label{eq:Halperin221}
\Psi_\textrm{221} (\{ z^\uparrow_i, z^\downarrow_i\}) =& \ \prod_{i>j} (z^\uparrow_i-z^\uparrow_j)^2 (z^\downarrow_i-z^\downarrow_j)^2 \nonumber \\ 
& \times \prod_{k,l} (z^\uparrow_k-z^\downarrow_l) e^{-\frac{1}{4} \sum_i (|z^\uparrow_i|^2 +|z^\downarrow_i|^2)}. 
\end{align}
This state exhibits a hidden $SU(3)$ symmetry as its edge states are described by the $SU(3)_1$ WZW CFT \cite{Ardonne99}. 
The case for $N=4$ and its relation to the $SU(4)_1$ WZW CFT have been pointed out in Ref.~\cite{Reijnders04}. 

The hidden $SU(N)$ symmetry of the above Abelian FQH states is revealed by considering the underlying lattice structure \cite{Read90}. 
In fact, the $K$ matrix \eqref{eq:KmatSUN} is regarded as a Gram matrix that is formed by the scalar products of primitive vectors of the $SU(N)$ root lattice. 
The change of primitive vectors with preserving the lattice structure is given by the $(N-1) \times (N-1)$ integer matrix $\bfG \in GL(N-1,\mathbb{Z})$ with determinant $\pm 1$. 
Then we can find a matrix $\bfG$ that maps the $K$ matrix \eqref{eq:KmatSUN} to the familiar Cartan matrix of the $SU(N)$ algebra (see Appendix~\ref{app:RootWeightSUN}), 
\begin{align} \label{eq:CartanSUN}
\bfA_{SU(N)} = \begin{pmatrix}
2 & -1 & 0 & \cdots & 0 & 0 \\ 
-1 & 2 & -1 & & 0 & 0 \\
0 & -1 & 2 & \ddots & & \vdots \\
\vdots & & \ddots & \ddots & -1 & 0 \\
0 & 0 & & -1& 2 & -1 \\
0 & 0 & \cdots & 0 & -1 & 2 
\end{pmatrix}, 
\end{align}
where
\begin{align}
\bfG \bfA_{SU(N)} \bfG^T = \bfK_{SU(N)}. 
\end{align}
If we do not consider additional symmetries, such as charge conservation, two Abelian FQH states given by $K$ matrices that transform each other by the $GL(N-1,\mathbb{Z})$ transformation share the same topological properties \cite{Read90,Wen92}. 
For example, both $\bfK_{SU(N)}$ and $\bfA_{SU(N)}$ give the same ground-state degeneracy $N$ on a torus. 

We remark that the transformation $\bfG$ does not preserve the form of the charge vector $\bft$. 
Indeed, it maps the Chern-Simons Lagrangian in the multilayer basis Eq.~\eqref{eq:ChernSimons} to that in the hierarchical basis \cite{Wen92,Wen95}, 
\begin{align}
\calL = \frac{1}{4\pi} \epsilon_{\mu \nu \lambda} (\bfA_{SU(N)})_{\sigma \sigma'} b^\sigma_\mu \partial_\nu b^{\sigma'}_\lambda -\frac{q}{2\pi} \epsilon_{\mu \nu \lambda} t^h_\sigma A_\mu \partial_\nu b^\sigma_\lambda, 
\end{align}
where $\bft^h = (0,\cdots,0,1)$ and $b^\sigma_\mu=\sum_{\sigma'} G^T_{\sigma \sigma'} a^{\sigma'}_\mu$. 
Thus it gives an $(N-1)$-th-level Haldane-Halperin hierarchical state at $\nu=(N-1)/N$ \cite{Haldane83,Halperin84}. 

Throughout this paper, we use the multi-layer basis, where each component of particles carries charge $q$. 
The hidden $SU(N)$ structure of the $K$ matrix is a most important key observation on constructing the non-Abelian extension of $SU(N-1)$-singlet states. 

\subsection{From Abelian to non-Abelian $SU(N-1)$-singlet FQH states}

From the Abelian FQH states given by $\bfK_{SU(N)}$, we can construct the trial wave functions for non-Abelian FQH states in a way similar to Ref.~\cite{Cappelli01}. 
By dropping the Gaussian factor in Eq.~\eqref{eq:MultiFQHS}, we define the reduced wave function for the Abelian $SU(N-1)$-singlet FQH state, 
\begin{align}
\tPsi_{SU(N)}(\{ z^\sigma_i \}) = \prod_{\sigma=1}^{N-1} \prod_{i<j} (z^\sigma_i-z^\sigma_j)^2 \prod_{\sigma' < \sigma''} \prod_{k,l} (z^{\sigma'}_k-z^{\sigma''}_l). 
\end{align}
The trial wave functions for non-Abelian FQH states, which we will focus on, can be represented by 
\begin{align} \label{eq:SUNkWaveFunc}
\tPsi_{SU(N)}^k(\{ z^\sigma_i \}) = \calS_\textrm{$k$-groups} \prod_\textrm{$k$ groups} \tPsi_{SU(N)}. 
\end{align}
This wave function is constructed as follows. 
We first partition the total $\calN = (N-1)pk$ bosons into $k$ groups, each of which contains $p$ bosons with each component index $\sigma$. 
For each group of $(N-1)p$ bosons, we then write the Abelian $SU(N-1)$-singlet FQH wave function $\tPsi_{SU(N)}$. 
After these wave functions from the $k$ groups are multiplied together, we finally apply the symmetrization operation $\calS_{k\textrm{-groups}}$ over all possible partitions into groups to obtain $\tPsi_{SU(N)}^k$. 
The resulting non-Abelian FQH states are realized at filling factor
\begin{align}
\nu=\frac{k(N-1)}{N}, 
\end{align}
and they are still $SU(N-1)$ singlet. 

For $N=2$, the non-Abelian FQH states constructed in this way correspond to the bosonic Moore-Read state at $\nu=1$ \cite{Moore91} and the bosonic Read-Rezayi states at $\nu=k/2$ \cite{Read99}. 
For $N=3$, they correspond to the bosonic NASS states at $\nu=2k/3$ \cite{Ardonne99,Ardonne01a}. 
Those for $N=4$ are proposed in the context of rotating spin-1 bosons \cite{Reijnders02,Reijnders04} and the general $N$ case is considered in Ref.~\cite{Sterdyniak13}. 
For $N=2$ and $3$, the trial wave function \eqref{eq:SUNkWaveFunc} and associated quasi-hole wave functions have been constructed from conformal blocks of the chiral $SU(N)_k$ WZW CFT \cite{Moore91,Read99,Ardonne99,Ardonne01a}. 
We expect that this holds for general $N$ and hence the bulk quasiparticle statistics and edge states of these non-Abelian states are both described by the chiral $SU(N)_k$ WZW CFT. 
From the bulk-edge correspondence, the ground-state degeneracy $D$ on a torus should coincide with the number of sectors of the edge states \cite{Moore91}. 
For the above non-Abelian FQH state, the latter is given by the number of primary fields of the chiral $SU(N)_k$ current algebra, 
\begin{align}
D = \frac{(N+k-1)!}{(N-1)!k!}. 
\end{align}

We can further generalize the non-Abelian $SU(N-1)$-singlet FQH state to other filling factors, 
\begin{align}
\nu=\frac{k(N-1)}{N+k(N-1)m},
\end{align}
where $m$ is integer. 
Even $m$ corresponds to a bosonic FQH state, including the $m=0$ state constructed above, while odd $m$ corresponds to a fermionic FQH state. 
Although we can explicitly write down the trial wave functions for these states, it is not important for our purpose. 
Instead, we just mention several properties of them. 
Since we are considering $(N-1)$-component particles, there are $N-1$ conserved charges. 
The charge part of the underlying CFT is modified with $m$, while the neutral part is unchanged. 
Therefore, the full chiral CFT is not $SU(N)_k$ but rather $[U(1)]^{N-1} \times SU(N)_k/[U(1)]^{N-1}$ for $m>0$. 
Here, the CFT in the neutral part, $SU(N)_k/[U(1)]^{k-1}$, is a generalization of the $\mathbb{Z}_k$ parafermion CFT \cite{Zamolodchikov85}, known as the Gepner parafermion \cite{Gepner87}. 
Owing to the fact that the ground-state degeneracy on a torus must be divisible by the denominator of the filling factor \cite{Haldane85}, we expect that the degeneracy is given by 
\begin{align}
D= \frac{(k+N-1)! [N+(N-1)km]}{N!k!}. 
\end{align}
The main purpose of this paper is to construct these non-Abelian $SU(N-1)$-singlet FQH states from coupled wires. 

\subsection{Array of Luttinger liquids} \label{sec:ArrayLL}

\begin{figure}
\includegraphics[clip,width=0.45\textwidth]{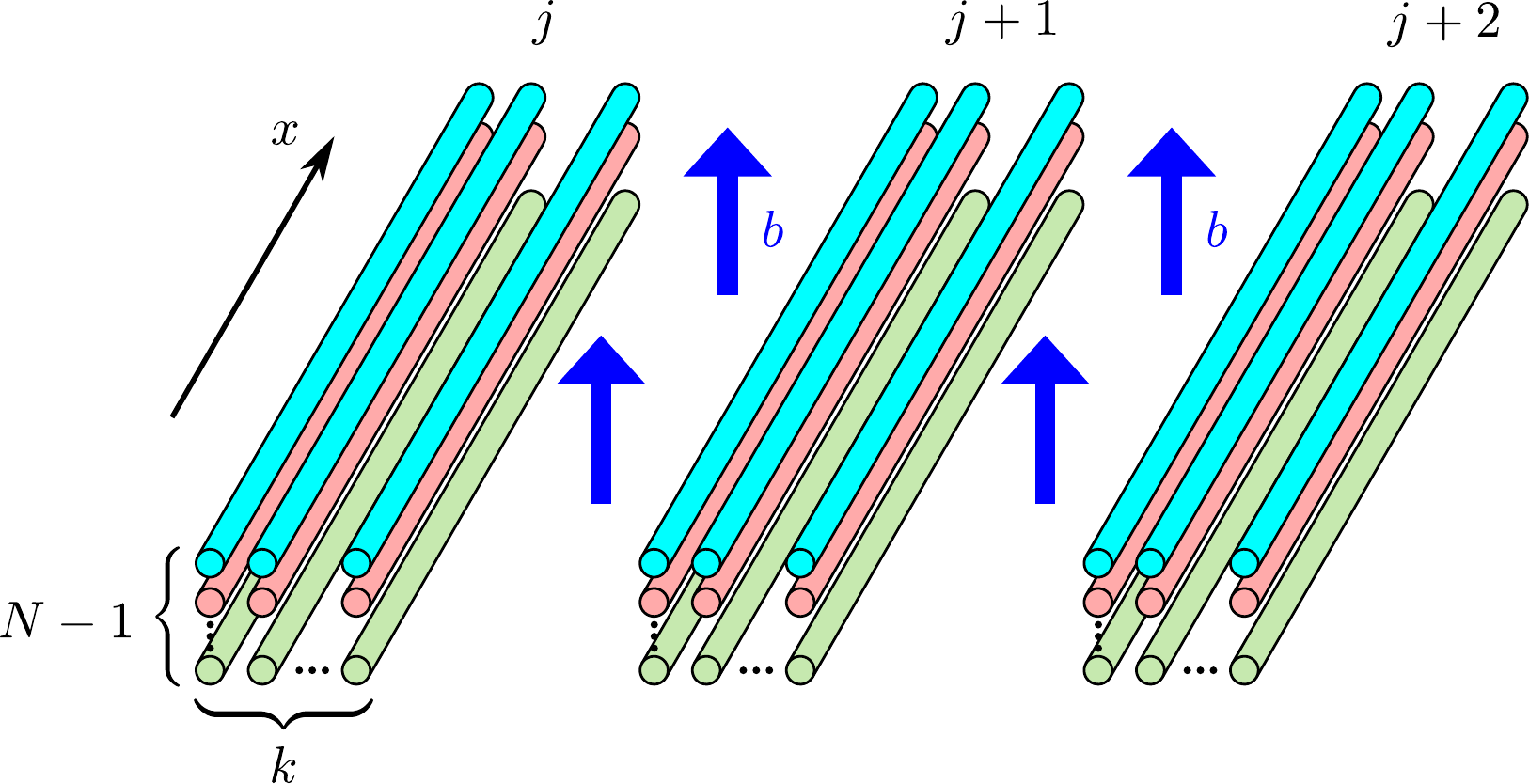}
\caption{The array of Luttinger liquids.
The wire labeled by $j$ consists of $k$ copies of $(N-1)$-component bosons. 
They are put on a magnetic field with the magnetic flux $b$.}
\label{fig:CoupledWires}
\end{figure}

Following Refs.~\cite{Kane02,Teo14}, we apply the coupled-wire construction to the Abelian and non-Abelian $SU(N-1)$-singlet FQH states. 
As a building block, we first introduce an array of Luttinger liquids \cite{Gogolin,Giamarchi},
\begin{align} \label{eq:DecoupledWires}
\calH_0 = \sum_{j=1}^{N_w} \sum_{\sigma=1}^{N-1} \sum_{a=1}^k \frac{v_F}{2\pi} \int dx \left[ \frac{1}{g} (\partial_x \theta_{j,\sigma,a})^2 +g (\partial_x \varphi_{j,\sigma,a})^2 \right],
\end{align}
where $N_w$ is the number of wires, and $v_F$ and $g$ are the velocity and stiffness of the Luttinger liquids. 
Here $j$, $\sigma$, and $a$ respectively stand for the indices of wire, component of boson, and channel. 
Thus each wire has $N-1$ components and each component has $k$ channels. 
This setup is schematically drawn in Fig.~\ref{fig:CoupledWires}. 
The bosonic fields $\theta_{j,\sigma,a}(x)$ and $\varphi_{j,\sigma,a}(x)$ are dual to each other and satisfy the commutation relations,
\begin{align} \label{eq:BosonComm}
[\theta_{j,\sigma,a}(x), \varphi_{j',\sigma',a'}(x')] = i\pi \delta_{jj'} \delta_{\sigma \sigma'} \delta_{aa'} \Theta(x-x'), 
\end{align}
and therefore, 
\begin{align} \label{eq:ChainFieldComm}
[\partial_x \theta_{j,\sigma,a}(x), \varphi_{j',\sigma',a'}(x')] = i\pi \delta_{jj'} \delta_{\sigma \sigma'} \delta_{aa'} \delta(x-x'). 
\end{align}
Here $\Theta(x)$ is a step function that takes $1$ for $x>0$ while $0$ for $x<0$. 
The bosonic field $\theta_{j,\sigma,a}(x)$ corresponds to the density fluctuation and is related to the boson density operator by 
\begin{align}
\rho_{j,\sigma,a}(x) &= \bar{\rho} +\sum_{n \in \mathbb{Z}} \rho^{(n)}_{j,\sigma,a}(x), \\
\rho^{(0)}_{j,\sigma,a}(x) &= \frac{1}{\pi} \partial_x \theta_{j,\sigma,a}(x), \\
\rho^{(n)}_{j,\sigma,a}(x) &\propto e^{in[2\pi \brho x+2\theta_{j,\sigma,a}(x)]} \hspace{10pt} (n \neq 0), 
\end{align}
where $\brho$ is the average density of bosons. 
Throughout this paper, we assume that each species of boson in each wire takes the same average density. 
We also introduce the ``Fermi'' momentum $k_F=\pi \brho$ from the correspondence with the Dirac fermions. 
The other bosonic field $\varphi_{j,\sigma,a}(x)$ represents the current fluctuation and is related to the operator creating a boson with charge $q$, 
\begin{align}
\Phi^\dagger_{j,\sigma,a}(x) \propto e^{i\varphi_{j,\sigma,a}(x)}. 
\end{align}

Next we consider the interactions between wires. 
The forward-scattering interactions will be summed up with the form, 
\begin{align}
\calH_\textrm{forward} =& \ \sum_{j,j'=1}^{N_w} \sum_{\sigma,\sigma'=1}^{N-1} \sum_{a,a'=1}^k \int dx \left( \partial_x \varphi_{j,\sigma,a}, \ \partial_x \theta_{j,\sigma,a} \right) \nonumber \\
& \times \bfU_{j \sigma a; j' \sigma' a'} \begin{pmatrix} \partial_x \varphi_{j',\sigma',a'} \\ \partial_x \theta_{j',\sigma',a'} \end{pmatrix}, 
\end{align}
where $\bfU_{j \sigma a; j' \sigma' a'}$ are $2 \times 2$ matrices. 
Combining with the decoupled wires \eqref{eq:DecoupledWires}, we obtain the sliding-Luttinger-liquid (SLL) Hamiltonian \cite{Emery00,Vishwanath01,Sondhi01,Mukhopadhyay01}, 
\begin{align} \label{eq:SLLHam}
\calH_\textrm{SLL} = \calH_0 +\calH_\textrm{forward}. 
\end{align}
We also consider backscattering interactions of the form, 
\begin{align} \label{eq:GeneralInt}
& \calV_j^{\{ n_{p \sigma a}, m_{p \sigma a} \}}(x) \nonumber \\
& = \prod_{p \geq 0} \prod_{\sigma=1}^{N-1} \prod_{a=1}^k \left[ \Phi^\dagger_{j+p,\sigma,a}(x) e^{b(j+p)x} \right]^{n_{p \sigma a}} \rho^{(\frac{1}{2} m_{p \sigma a})}_{j+p,\sigma,a}(x), 
\end{align}
where $n_{p \sigma a} \in \mathbb{Z}$, $m_{p \sigma a} \in 2 \mathbb{Z}$, and $b$ is the magnetic flux related to the filling factor by $\nu=2k_F/b$. 
If $n_{p \sigma a}<0$, these interactions are interpreted as $|n_{p \sigma a}|$ times applications of $\Phi_{j+p,\sigma,a}$, instead of $\Phi^\dagger_{j+p,\sigma,a}$. 
In general, $p$ runs over some nonnegative integers, but in this paper we only consider interactions within the same wire ($p=0$) and those between adjacent wires ($p=1$). 
Equation \eqref{eq:GeneralInt} is expressed in terms of the bosonic fields as 
\begin{align}
\calV_j^{\{ n_{p \sigma a}, m_{p \sigma a} \}}(x) \propto R_j^{\{ n_{p \sigma a}, m_{p \sigma a} \}}(x) \calO_j^{\{ n_{p \sigma a}, m_{p \sigma a} \}}(x), 
\end{align}
where 
\begin{align}
\calO_j^{\{ n_{p \sigma a}, m_{p \sigma a} \}}(x) &= e^{ i\sum_{p,\sigma,a} \left( n_{p \sigma a} \varphi_{j+p,\sigma,a} +m_{p \sigma a} \theta_{j+p,\sigma,a} \right)}, \\
\label{eq:GenIntOsc}
R_j^{\{ n_{p \sigma a}, m_{p \sigma a} \}}(x) &= e^{i \sum_{p,\sigma,a} \left[ b(j+p)n_{p \sigma a} +k_F m_{p \sigma a} \right] x}. 
\end{align}
Assuming spatial homogeneity of the coupling constants, the interaction Hamiltonian takes the form, 
\begin{align} \label{eq:IntHam}
\calH_\textrm{int} =& \ \sum_{j=1}^{N_w} \sum_{\{ n_{p \sigma a}, m_{p \sigma a} \}} \int dx \nonumber \\
& \times  \left[ v^{\{ n_{p \sigma a}, m_{p \sigma a} \}} \calV_j^{\{ n_{p \sigma a}, m_{p \sigma a} \}}(x) +\textrm{H.c.} \right]. 
\end{align}
Since the factor $R_j^{\{ n_{p \sigma a}, m_{p \sigma a} \}}$ rapidly oscillates with the spatial coordinate $x$ parallel to the wires, the interaction \eqref{eq:IntHam} vanishes after the integration unless $R_j^{\{ n_{p \sigma a}, m_{p \sigma a} \}}=1$. 
Hence this condition determines the possible forms of interactions in low energy, which are specified by the set of integers $\{ n_{p \sigma a}, m_{p \sigma a} \}$. 

To further restrict the allowed forms of the interactions, we impose several physical constraints: particle and momentum conservations. 
Although these constraints are not necessary in the following argument, as topological order is free from symmetries, it will be reasonable to assume them for actual setups of the FQH states. 
The particle conservation implies that 
\begin{align}
\sum_{p \geq 0} \sum_{\sigma=1}^{N-1} \sum_{a=1}^k n_{p \sigma a} =0. 
\end{align}
This slightly changes Eq.~\eqref{eq:GenIntOsc} to 
\begin{align}
R_j^{\{ n_{p \sigma a}, m_{p \sigma a} \}}(x) &= e^{i \sum_{p,\sigma,a} (bpn_{p \sigma a} +k_F m_{p \sigma a}) x}. 
\end{align}
Moreover, we impose the \emph{separate} conservation of each component of boson, 
\begin{align} \label{eq:ParticleConsv}
\sum_{p \geq 0} \sum_{a=1}^k n_{p \sigma a} =0. 
\end{align}
Finally the momentum conservation requires that
\begin{align} \label{eq:MomConsv}
\sum_{p \geq 0} \sum_{\sigma=1}^{N-1} \sum_{a=1}^k \left( bpn_{p \sigma a} +k_F m_{p \sigma a} \right) =0. 
\end{align}
If $\sum_{p,\sigma,a} m_{p \sigma a} \neq 0$, this gives the direct relation to the filling factor, 
\begin{align}
\nu = \frac{-2\sum_{p,\sigma,a} pn_{p \sigma a}}{\sum_{p,\sigma,a} m_{p \sigma a}}. 
\end{align}
We note that $\nu$ represents the filling factor averaged over all the components and channels in each wire. 
Thus the \emph{total} filling factor in each wire is given by $\nu_\textrm{tot}=(N-1)k\nu$. 

\section{Halperin $(221)$ state} \label{sec:221State}

We first show the construction of the Halperin $(221)$ state \eqref{eq:Halperin221} in an array of the Luttinger liquids with two-component bosons. 
This is achieved in a parallel way to the construction of Abelian FQH states at the second level hierarchy \cite{Teo14}. 
However, for the later purpose, we further elucidate the underlying $SU(3)$-algebraic structure of the Halperin $(221)$ state from the coupled-wire construction. 

\subsection{Interactions at $\nu_\textrm{tot}=2/3$}
\label{sec:Int221State}

We assign up and down spins to each component of the boson and write $\sigma = \ua, \da$. 
Since we here only consider a single channel $k=1$, we simply drop the channel index $a$. 
Then the commutation relations of the bosonic fields \eqref{eq:ChainFieldComm} are reduced to 
\begin{align}
[\partial_x \theta_{j,\sigma}(x), \varphi_{j',\sigma'}(x')] = i\pi \delta_{jj'} \delta_{\sigma \sigma'} \delta(x-x'). 
\end{align}
At filling factor $\nu=1/3$ for each component, correlated hoppings between neighboring wires, which are allowed by the particle conservation \eqref{eq:ParticleConsv} and momentum conservation \eqref{eq:MomConsv}, are listed in Table~\ref{tab:Int221}. 
\begin{table*}
\caption{Possible correlated hoppings for two-component bosons at $\nu_\textrm{tot}=1/3+1/3$.
The notation follows Sec.~\ref{sec:ArrayLL}.}
\label{tab:Int221}
\begin{ruledtabular}
\begin{tabular}{l|llllllll|l}
$I$ & $n_{0\ua}^{(I)}$ & $n_{0\da}^{(I)}$ & $n_{1\ua}^{(I)}$ & $n_{1\da}^{(I)}$ & $m_{0\ua}^{(I)}$ & $m_{0\da}^{(I)}$ & $m_{1\ua}^{(I)}$ & $m_{1\da}^{(I)}$ & $\calO_j^{\{ n_{p \sigma}^{(I)}, m_{p \sigma}^{(I)} \}}$ \\ \hline
1 & 1 & 0 & -1 & 0 & 2 & 2 & 2 & 0 & $\exp i[\varphi_{j,\ua}-\varphi_{j+1,\ua}+2(\theta_{j,\ua}+\theta_{j,\da}+\theta_{j+1,\ua})]$ \\
2 & 0 & 1 & 0 & -1 & 0 & 2 & 2 & 2 & $\exp i[\varphi_{j,\da}-\varphi_{j+1,\da}+2(\theta_{j,\da}+\theta_{j+1,\da}+\theta_{j+1,\ua})]$ \\
3 & 1 & 0 & -1 & 0 & 2 & 0 & 2 & 2 & $\exp i[\varphi_{j,\ua}-\varphi_{j+1,\ua}+2(\theta_{j,\ua}+\theta_{j+1,\ua}+\theta_{j+1,\da})]$ \\
4 & 0 & 1 & 0 & -1 & 2 & 2 & 0 & 2 & $\exp i[\varphi_{j,\da}-\varphi_{j+1,\da}+2(\theta_{j,\da}+\theta_{j,\ua}+\theta_{j+1,\da})]$ \\
5 & 1 & 0 & -1 & 0 & 2 & 2 & 0 & 2 & $\exp i[\varphi_{j,\ua}-\varphi_{j+1,\ua}+2(\theta_{j,\ua}+\theta_{j,\da}+\theta_{j+1,\da})]$ \\
6 & 0 & 1 & 0 & -1 & 2 & 2 & 2 & 0 & $\exp i[\varphi_{j,\da}-\varphi_{j+1,\da}+2(\theta_{j,\ua}+\theta_{j,\da}+\theta_{j+1,\ua})]$ \\
7 & 1 & 0 & -1 & 0 & 0 & 2 & 2 & 2 & $\exp i[\varphi_{j,\ua}-\varphi_{j+1,\ua}+2(\theta_{j,\da}+\theta_{j+1,\ua}+\theta_{j+1,\da})]$ \\
8 & 0 & 1 & 0 & -1 & 2 & 0 & 2 & 2 & $\exp i[\varphi_{j,\da}-\varphi_{j+1,\da}+2(\theta_{j,\ua}+\theta_{j+1,\ua}+\theta_{j+1,\da})]$ \\
\end{tabular}
\end{ruledtabular}
\end{table*}
While their products are also allowed by symmetry and thus can be added to the Hamiltonian, the interactions shown in Table~\ref{tab:Int221} will be usually most relevant interactions at the fixed point of the SLL Hamiltonian \eqref{eq:SLLHam}. 
Furthermore, those interactions must commute with each other to simultaneously open a gap for different fields. 
A set of such interactions denoted by $\{ n_{p \sigma}^{(I)}, m_{p \sigma}^{(I)} \}$ must satisfy the Haldane's null vector condition \cite{Haldane95},
\begin{align}
\sum_{p=0,1} \sum_{\sigma=\ua,\da} \left( n_{p \sigma}^{(I)} m^{(J)}_{p+q, \sigma} +m_{p \sigma}^{(I)} n_{p+q, \sigma}^{(J)} \right) =0, 
\end{align}
for any integer $q$. 
From Table~\ref{tab:Int221}, the pairs of the interactions satisfying this conditions are given by $(I,J)=(1,2)$, $(1,3)$, $(2,4)$, and $(3,4)$. 
Among these pairs, either $(1,2)$ or $(3,4)$ leads to the Halperin (221) state, as we will see below. 
On the other hand, the other pairs $(1,3)$ or $(2,4)$ may give rise to a state in which one species of bosons forms the Laughlin $\nu=\frac{1}{2}$ state while the other species forms a charge-density-wave order. 

In the following, we pick up the following pair of the correlated hoppings $(I,J)=(1,2)$ from Table~\ref{tab:Int221}, 
\begin{align} \label{eq:Int221}
\begin{split}
\calO^t_{j,\ua} &= e^{i[ \varphi_{j,\ua} -\varphi_{j+1,\ua} +2(\theta_{j,\ua} +\theta_{j,\da} +\theta_{j+1,\ua}) ]}, \\
\calO^t_{j,\da} &= e^{i[ \varphi_{j,\da} -\varphi_{j+1,\da} +2(\theta_{j,\da} +\theta_{j+1,\ua} +\theta_{j+1,\da}) ]}, 
\end{split}
\end{align}
and consider the interaction Hamiltonian, 
\begin{align} \label{eq:IntHam221}
\calH_\textrm{int} = \sum_{j=1}^{N_w-1} \sum_{\sigma=\ua,\da} \int dx \left[ t_\sigma \calO^t_{j,\sigma}(x) +\textrm{H.c.} \right]. 
\end{align}
Then it is useful to introduce chiral fields, 
\begin{align} \label{eq:Chiral221}
\begin{split}
\tphi^R_{j,\ua} &= \varphi_{j,\ua} +2(\theta_{j,\ua}+\theta_{j,\da}), \\
\tphi^R_{j,\da} &= \varphi_{j,\da} +2\theta_{j,\da}, \\
\tphi^L_{j,\ua} &= \varphi_{j,\ua} -2\theta_{j,\ua}, \\
\tphi^L_{j,\da} &= \varphi_{j,\da} -2(\theta_{j,\ua}+\theta_{j,\da}). 
\end{split}
\end{align}
These fields obey the commutation relations, 
\begin{align}
[\partial_x \tphi^p_{j,\sigma}(x), \tphi^{p'}_{j',\sigma'}(x')] &= 2ip \pi \delta_{pp'} \delta_{jj'} K_{\sigma \sigma'} \delta (x-x'), 
\end{align}
with the $K$ matrix, 
\begin{align} \label{eq:Kmat221}
\bfK = \begin{pmatrix} 2 & 1 \\ 1 & 2 \end{pmatrix}, 
\end{align}
where $p=R,L$ also stand for $\pm 1$, respectively. 
In terms of these chiral fields, the interactions \eqref{eq:Int221} can be written as 
\begin{align} \label{eq:Int221a}
\calO^t_{j,\sigma} = e^{i\tphi^R_{j, \sigma} -i\tphi^L_{j+1, \sigma}} .
\end{align}
It is easy to see that these interactions open a bulk gap by introducing the link fields, 
\begin{align}
\begin{split}
\ttheta_{j+\frac{1}{2},\sigma} &= \frac{1}{2} (\tphi^R_{j,\sigma} -\tphi^L_{j+1,\sigma}), \\
\tvarphi_{j+\frac{1}{2},\sigma} &= \frac{1}{2} (\tphi^R_{j,\sigma} +\tphi^L_{j+1,\sigma}), 
\end{split}
\end{align}
which satisfy the commutation relations, 
\begin{align}
[\partial_x \ttheta_{\ell,\sigma}(x), \tvarphi_{\ell',\sigma'}(x')] = i\pi \delta_{\ell \ell'} K_{\sigma \sigma'} \delta(x-x'). 
\end{align}
Then the Hamiltonian is written as 
\begin{align} \label{eq:LinkSineGordon221}
\calH = \calH_\textrm{SLL} +\sum_{j=1}^{N_w-1} \sum_{\sigma=\ua,\da} \int dx \ 2t_\sigma \cos (2\ttheta_{j+\frac{1}{2},\sigma}). 
\end{align}
If the SLL Hamiltonian is appropriately tuned such that $t_\sigma$ become relevant, $t_\sigma$ flow to the strong-coupling limit under the renormalization group transformation. 
We can then simultaneously localize the link fields $\ttheta_{\ell,\sigma}$ to minima of the cosine potentials. 
This opens a gap for the link fields with $1 < \ell < N_w$, resulting in the bulk gap. 
However, at the leftmost $(j=1)$ and rightmost $(j=N_w)$ wires, the chiral fields cannot be paired into the link fields. 
These unpaired fields $\tphi^L_{1,\sigma}$ and $\tphi^R_{N_w,\sigma}$ behave as the edge states of the Halperin $(221)$ state. 

One may have noticed that the form of the interaction \eqref{eq:Int221} and the resulting Hamiltonian is not symmetric under the exchange of up and down spins. 
Although the exchange symmetry of two species of boson in the microscopic Hamiltonian is not necessarily required to stabilize the Halperin $(221)$ state, it is naturally assumed for several experimental setups such as rotating Bose gases \cite{Cooper08,Grass12,Furukawa12}. 
This asymmetric form of the interactions may be resolved when we also consider the correlated hopping \emph{inside} the wires. 
We will come back to this issue in Sec.~\ref{sec:CorrHopping}. 

\subsection{$SU(3)_1$ currents} \label{sec:SU31Currents}

To understand the underlying $SU(3)$-algebraic structure of the Halperin $(221)$ state from the coupled-wire Hamiltonian, we further introduce new linear combinations of the chiral fields \eqref{eq:Chiral221} by 
\begin{align} \label{eq:ChiralBoson2212}
\begin{split}
\tchi^p_{j,1} &= \frac{1}{\sqrt{2}} (\tphi^p_{j,\ua}-\tphi^p_{j,\da}), \\
\tchi^p_{j,2} &= \frac{1}{\sqrt{6}} (\tphi^p_{j,\ua}+\tphi^p_{j,\da}). 
\end{split}
\end{align}
These fields satisfy the commutation relations, 
\begin{align}
[\partial_x \tchi^p_{j,l}(x), \tchi^{p'}_{j',l'}(x')] = 2ip\pi \delta_{pp'} \delta_{jj'} \delta_{ll'} \delta (x-x'). 
\end{align}
The interactions \eqref{eq:Int221a} are rewritten as 
\begin{align} \label{eq:Int221b}
\begin{split}
\calO^t_{j,\ua} &= e^{i\bfalpha_\ua \cdot \tbfchi^R_j -i\bfalpha_\ua \cdot \tbfchi^L_{j+1}}, \\
\calO^t_{j,\da} &= e^{i\bfalpha_\da \cdot \tbfchi^R_j -i\bfalpha_\da \cdot \tbfchi^L_{j+1}}, 
\end{split}
\end{align}
where 
\begin{align} \label{eq:SU3Root}
\begin{split}
\bfalpha_\ua &= \begin{pmatrix} \frac{1}{\sqrt{2}}, & \sqrt{\frac{3}{2}} \end{pmatrix}, \\
\bfalpha_\da &= \begin{pmatrix} -\frac{1}{\sqrt{2}}, & \sqrt{\frac{3}{2}} \end{pmatrix}, 
\end{split}
\end{align}
and $\tbfchi^p_j = (\tchi^p_{j,1}, \tchi^p_{j,2})$. 
The vectors $\bfalpha_\sigma$ are roots of $SU(3)$, which comprise a set of primitive vectors of the $SU(3)$ root lattice. 

Let us suppose that the SLL Hamiltonian takes the diagonal form in each chiral field, 
\begin{align} \label{eq:SLLHamSU3}
\calH_\textrm{SLL} = \frac{v}{4\pi} \sum_{j=1}^{N_w} \int dx \left[ (\partial_x \tbfchi^R_j)^2 +(\partial_x \tbfchi^L_j)^2 \right]. 
\end{align}
For each wire, this Hamiltonian gives a bosonic description of the $SU(3)_1$ WZW CFT \cite{dFMS}. 
In the Sugawara construction, this Hamiltonian can be written in terms of the $SU(3)_1$ currents as 
\begin{align}
\calH_\textrm{SLL} = \frac{v}{16\pi} \sum_{j=1}^{N_w} \int dx \left[ :\mathrel{\bfJ_j \cdot \bfJ_j}: +:\mathrel{\bbfJ_j \cdot \bbfJ_j}: \right], 
\end{align}
where $:\mathrel{X}:$ denotes the normal-ordered product of an operator $X$, and $\bfJ_j=(J^1_j,\cdots,J^8_j)$ and $\bbfJ_j=(\bJ^1_j,\cdots,\bJ^8_j)$ stand for the right and left $SU(3)_1$ currents in the orthonormal basis, respectively. 
This equivalence can be expected from the fact that both a two-component free boson and the $SU(3)_1$ WZW CFT have the same central charge $c=2$. 
Introducing the complex coordinate $z=v\tau+ix$ and $\bz=v\tau-ix$, the currents $\bfJ_j(z)$ and $\bbfJ_j(\bz)$ respectively satisfy the $SU(3)_1$ current (or Kac-Moody) algebra, 
\begin{align}
\begin{split}
J^\alpha_j(z) J^\beta_{j'}(w) &\sim \delta_{jj'} \left[ \frac{\delta_{\alpha \beta}}{(z-w)^2} +\sum_\gamma \frac{if_{\alpha \beta \gamma} J^\gamma_j(w)}{z-w} \right], \\
\bJ^\alpha_j(\bz) \bJ^\beta_{j'}(\bw) &\sim \delta_{jj'} \left[ \frac{\delta_{\alpha \beta}}{(\bz-\bw)^2} +\sum_\gamma \frac{if_{\alpha \beta \gamma} \bJ^\gamma_j(\bw)}{\bz-\bw} \right], 
\end{split}
\end{align}
where $f_{\alpha \beta \gamma}$ is a structure constant. 
Here the symbol $\sim$ means an equivalence relation in the sense of operator-product expansion (OPE), namely the only singular terms are kept in the right-hand side. 
On the other hand, the right and left current algebras are independent of each other: 
\begin{align}
J^\alpha_j(z) \bJ^\beta_{j'}(\bw) \sim 0. 
\end{align}

Since the vectors $\bfalpha_\sigma$ are roots of $SU(3)$, the vertex operators appearing in Eq.~\eqref{eq:Int221b} can be regarded as $SU(3)_1$ currents. 
By taking the Cartan currents to be 
\begin{align}
J^3_j(x) = \partial_x \tchi^R_{j,1}(x), \hspace{10pt} J^8_j(x) = \partial_x \tchi^R_{j,2}(x), 
\end{align}
the explicit relations between the vertex operators and the $SU(3)_1$ currents are given by 
\begin{align} \label{eq:VertexRepSU3}
J^{2 \pm}_j(x) = \frac{\pm i}{x_c} e^{\pm i\bfalpha_\ua \cdot \tbfchi^R_j(x)}, \hspace{10pt} J^{3 \pm}_j(x) = \frac{\pm i}{x_c} e^{\pm i\bfalpha_\da \cdot \tbfchi^R_j(x)}, 
\end{align}
where
\begin{align}
J^{2 \pm}_j \equiv \frac{1}{\sqrt{2}} (J^4_j \pm iJ^5_j), \hspace{10pt} J^{3 \pm}_j \equiv \frac{1}{\sqrt{2}} (J^6_j \pm iJ^7_j). 
\end{align}
Here $x_c$ is a short-distance cutoff. 
The left currents $\bbfJ_j$ are obtained by replacing $\tbfchi^R_j$ with $\tbfchi^L_j$. 
In contrast to the $SU(2)_1$ current algebra, which appears in the construction of the Laughlin $\nu=\frac{1}{2}$ state \cite{Teo14}, we need a special care for the cocycle factor when we construct a faithful vertex representation of the current algebra associated with the higher-rank Lie algebra \cite{Frenkel80,Segal81,Goddard86,dFMS}. 
The phase factors appearing in Eq.~\eqref{eq:VertexRepSU3} originate from our convention of the cocycle factor.
In Appendix~\ref{app:VertexRep}, we explicitly show that the coupled-wire construction yields a faithful vertex representation of the $SU(N)_1$ current algebra. 
Then the interactions \eqref{eq:Int221b} can be expressed in terms of the $SU(3)_1$ currents as 
\begin{align}
\begin{split}
\calO^t_{j,\ua} &= x_c^2 J^{2+}_j \bJ^{2-}_{j+1}, \\
\calO^t_{j,\da} &= x_c^2 J^{3+}_j \bJ^{3-}_{j+1}. \\
\end{split}
\end{align}
Thus the interaction Hamiltonian \eqref{eq:IntHam221} is given by 
\begin{align}
\calH_\textrm{int} =& \ x_c^2 \sum_{j=1}^{N_w-1} \int dx \left[ t_\ua \left( J^4_j \bJ^4_{j+1} +J^5_j \bJ^5_{j+1} \right) \right. \nonumber \\
& \left. +t_\da \left( J^6_j \bJ^6_{j+1} +J^7_j \bJ^7_{j+1} \right) \right]. 
\end{align}

\subsection{Bulk quasiparticles and edge states} \label{sec:SU31Quasiparticles}

We now consider that the coupling constants $t_\sigma$ flow to the strong coupling limit. 
As discussed in Ref.~\cite{Teo14}, the quasiparticle excitations can be seen as the kinks of the link fields $\ttheta_{j,\sigma}$ in the Hamiltonian \eqref{eq:LinkSineGordon221}. 
Different minima of the cosine potentials are connected via the gauge transformation $\varphi_{j,\sigma} \to \varphi_{j,\sigma}+2\pi$. 
This is translated into the creation of kinks for the link fields, $\ttheta_{j \pm \frac{1}{2},\sigma} \to \ttheta_{j \pm \frac{1}{2},\sigma} \pm \pi$. 
This jump of $\ttheta_{j+\frac{1}{2},\sigma}$ by $\pi$ corresponds to the creation of a quasiparticle specified by $\sigma$ at the link $j+\frac{1}{2}$. 
From Eq.~\eqref{eq:ChainFieldComm}, the operators that transfer the quasiparticles at $x$ from the link $j+\frac{1}{2}$ to $j-\frac{1}{2}$ are the $2k_F$ backscattering operators, 
\begin{align}
B_{j,\sigma}(x) = e^{2i\theta_{j,\sigma}(x)}. 
\end{align}
These operators are written in terms of the chiral fields as 
\begin{align}
\begin{split}
B_{j,\ua}(x) &= e^{\frac{2i}{3} \tphi^R_{j,\ua}(x) -\frac{i}{3} \tphi^R_{j,\da}(x) -\frac{2i}{3} \tphi^L_{j,\ua}(x) +\frac{i}{3} \tphi^L_{j,\da}(x)}, \\
B_{j,\da}(x) &= e^{-\frac{i}{3} \tphi^R_{j,\ua}(x) +\frac{2i}{3} \tphi^R_{j,\da}(x) +\frac{i}{3} \tphi^L_{j,\ua}(x) -\frac{2i}{3} \tphi^L_{j,\da}(x)}. 
\end{split}
\end{align}
One may write these operators as  
\begin{align}
B_{j,\sigma}(x) = e^{-\frac{i\pi}{3}} \Psi_{\textrm{QP},j+\frac{1}{2},\sigma}^{R \dagger}(x) \Psi_{\textrm{QP},j-\frac{1}{2},\sigma}^L(x), 
\end{align}
through quasiparticle operators defined by 
\begin{align} \label{eq:QuasiParticles221}
\begin{split}
\Psi_{\textrm{QP},j+\frac{1}{2},\ua}^{R\dagger}(x) &= e^{\frac{2i}{3} \tphi^R_{j,\ua}(x) -\frac{i}{3} \tphi^R_{j,\da}(x)}, \\
\Psi_{\textrm{QP},j+\frac{1}{2},\da}^{R\dagger}(x) &= e^{-\frac{i}{3} \tphi^R_{j,\ua}(x) +\frac{2i}{3} \tphi^R_{j,\da}(x)}, \\
\Psi_{\textrm{QP},j+\frac{1}{2},\ua}^{L\dagger}(x) &= e^{\frac{2i}{3} \tphi^L_{j+1,\ua}(x) -\frac{i}{3} \tphi^L_{j+1,\da}(x)}, \\
\Psi_{\textrm{QP},j+\frac{1}{2},\da}^{L\dagger}(x) &= e^{-\frac{i}{3} \tphi^L_{j+1,\ua}(x) +\frac{2i}{3} \tphi^L_{j+1,\da}(x)}, 
\end{split}
\end{align}
which create the quasiparticles specified by $\sigma$ at $x$ on the link $j+\frac{1}{2}$. 
From this, we can read off the charges $Q^\sigma$ associated with $\sigma$-spin boson for the two quasiparticles of the Halperin (221) state, which are given by $(Q^\ua, Q^\da) = (\frac{2}{3}q,-\frac{1}{3}q)$ and $(-\frac{1}{3}q, \frac{2}{3}q)$ in the unit of $q$. 
Therefore, the two quasiparticles have the same total charge $Q^\ua+Q^\da=\frac{1}{3}q$ while the opposite spins $Q^\ua-Q^\da=\pm q$. 
These are consistent with the quasiparticle charges computed from the Chern-Simons theory with the $K$ matrix \eqref{eq:Kmat221}. 

If the bulk is gapped, there remain the unpaired gapless modes $\tphi^L_{1,\sigma}$ at $j=1$. 
The charge-$q$ particle operators at the edge are given by 
\begin{align}
\Psi_{q,\frac{1}{2},\sigma}^{L\dagger}(x) = e^{i\tphi^L_{1,\sigma}(x)}. 
\end{align}
These operators are nothing but the generators of the $SU(3)_1$ current algebra identified in Eq.~\eqref{eq:VertexRepSU3}. 
Corresponding primary fields must be local with respect to these operators, i.e. they must be single valued when taken around the particle operators \cite{Nayak08}. 
A set of such fields that are independent under arbitrary actions of the particle operators may be identified from the quasiparticle operators \eqref{eq:QuasiParticles221} as 1 (identity), $\Psi_{\textrm{QP},\frac{1}{2},\ua}^{L\dagger}$, and $\Psi_{\textrm{QP},\frac{1}{2},\ua}^{L}$. 
[To be precise, these operators should be the left-right product of the quasiparticle operators from the edges $j=1$ and $N_w$ in order to satisfy the faithful operator algebra of the $SU(3)_1$ WZW CFT (see also the discussion in Appendix~\ref{app:SUkNPara}).]
As we will discuss in the next section, these fields are associated with a weight in the trivial, fundamental, and conjugate representations of $SU(3)$, respectively. 
The number of the operators coincides with the number of primary fields of the $SU(3)_1$ current algebra, and hence the ground-state degeneracy on a torus, which is three, as computed from the determinant of the $K$ matrix \eqref{eq:Kmat221}. 

\section{Non-Abelian spin-singlet FQH state} \label{sec:NASSState}

We next consider the bosonic NASS states at $\nu_\textrm{tot}=2k/3$ \cite{Ardonne99,Ardonne01a}. 
These states are intimately related to the chiral $SU(3)_k$ WZW CFT. 
In the coupled-wire construction, they can be constructed by taking $k$ copies of the Halperin (221) state and by introducing interactions among them. 
This exactly follows the manner of Ref.~\cite{Teo14} for the construction of a bosonic Read-Rezayi state from $k$ copies of the Laughlin $\nu=\frac{1}{2}$ state. 
We here mainly discuss the simplest case of $k=2$. 

\subsection{Interactions at $\nu_\textrm{tot}=2k/3$} \label{sec:IntSU3k}

\begin{figure}
\includegraphics[clip,width=0.35\textwidth]{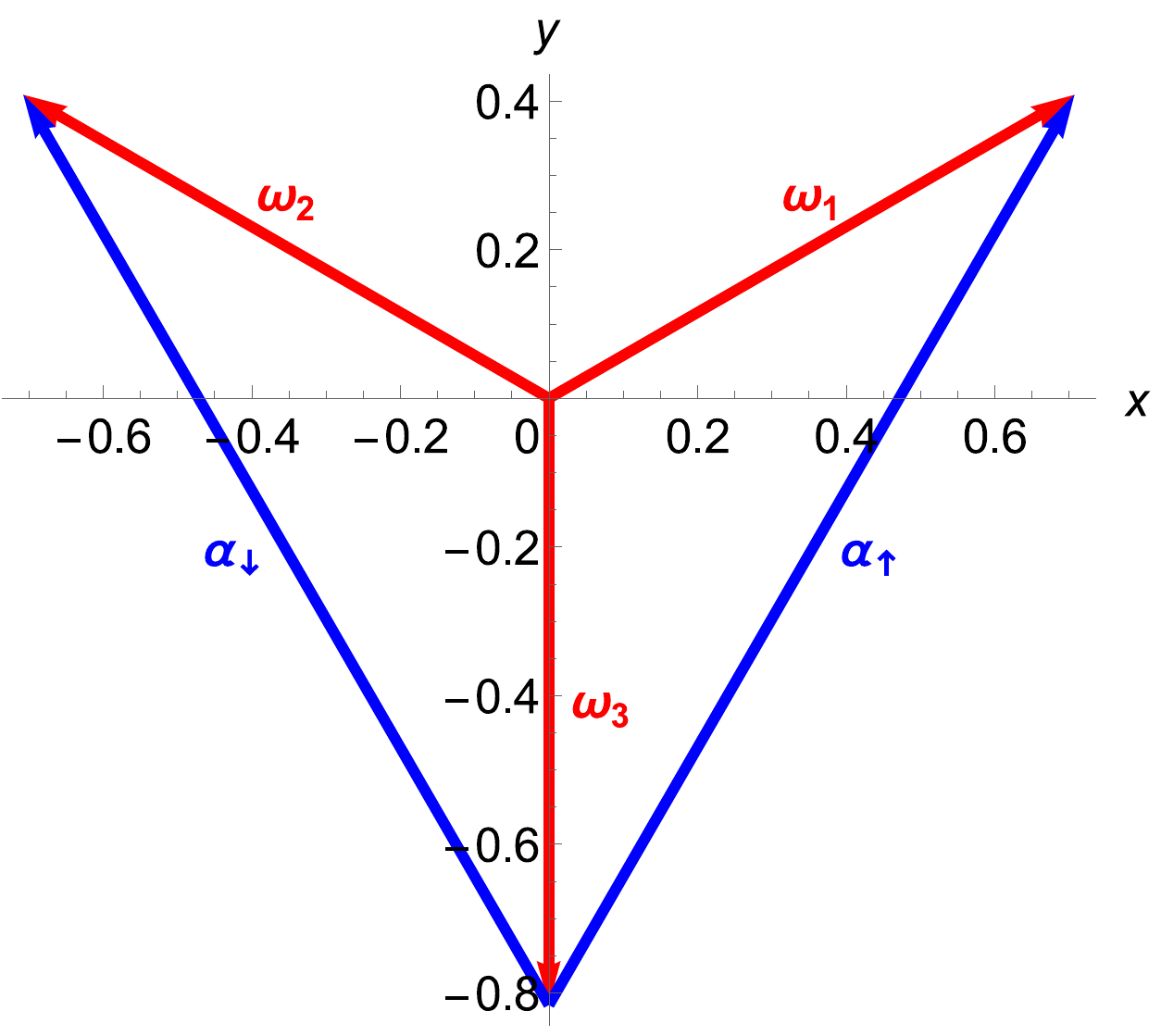}
\caption{Weight diagram of $SU(3)$. 
The red solid arrows represent the weight vectors $\bfomega_s$ in the fundamental representation. 
The blue solid arrows represent the root vectors $\bfalpha_\sigma$.}
\label{fig:SU3Weights}
\end{figure}

Now a single wire is composed of $k$ copies of the two-component bosons labeled by $\sigma = \ua,\da$. 
The decoupled-wire Hamiltonian is given by the Luttinger-liquid Hamiltonian in Eq.~\eqref{eq:DecoupledWires} with $a=1, \cdots, k$, and the bosonic fields $\theta_{j, \sigma, a}$ and $\varphi_{j, \sigma, a}$ obey the commutation relations \eqref{eq:ChainFieldComm}. 
We again consider the same forms of the interwire correlated hoppings as given in Eq.~\eqref{eq:Int221}, but now each component of the boson also hops between different copies: 
\begin{align}
\begin{split}
\calO^t_{j,\ua,ab} &= e^{i[\varphi_{j,\ua,a} -\varphi_{j+1,\ua,b} +2(\theta_{j,\ua,a} +\theta_{j,\da,a} +\theta_{j+1,\ua,b})]}, \\
\calO^t_{j,\da,ab} &= e^{i[\varphi_{j,\da,a} -\varphi_{j+1,\da,b} +2(\theta_{j,\da,a} +\theta_{j+1,\ua,b} +\theta_{j+1,\da,b})]}. 
\end{split}
\end{align}
We also consider the following intrawire interactions: 
\begin{align}
\begin{split}
\calO^{u,11}_{j,ab} &= \exp 2i(\theta_{j,\ua,a}-\theta_{j,\ua,b}), \\
\calO^{u,12}_{j,ab} &= \exp i[\varphi_{j,\ua,a}-\varphi_{j,\ua,b}-\varphi_{j,\da,a}+\varphi_{j,\da,b} \\
&\ \ \ +2(\theta_{j,\ua,a}-\theta_{j,\ua,b}+\theta_{j,\da,a}-\theta_{j,\da,b})], \\
\calO^{u,13}_{j,ab} &= \exp i(\varphi_{j,\ua,a}-\varphi_{j,\ua,b}), \\
\calO^{u,21}_{j,ab} &= \exp i(-\varphi_{j,\ua,a}+\varphi_{j,\ua,b}+\varphi_{j,\da,a}-\varphi_{j,\da,b}), \\
\calO^{u,22}_{j,ab} &= \exp 2i(\theta_{j,\da,a}-\theta_{j,\da,b}), \\
\calO^{u,23}_{j,ab} &= \exp i[\varphi_{j,\da,a}-\varphi_{j,\da,b}-2(\theta_{j,\ua,a}-\theta_{j,\ua,b})], \\
\calO^{u,31}_{j,ab} &= \exp i[-\varphi_{j,\ua,a}+\varphi_{j,\ua,b}-2(\theta_{j,\ua,a}-\theta_{j,\da,b})], \\
\calO^{u,32}_{j,ab} &= \exp i[-\varphi_{j,\da,a}+\varphi_{j,\da,b}], \\
\calO^{u,33}_{j,ab} &= \exp 2i(-\theta_{j,\ua,a}+\theta_{j,\ua,b}-\theta_{j,\da,a}+\theta_{j,\da,b}).
\end{split}
\end{align}
These interactions satisfy the particle and momentum conservations. 
Then the full interaction Hamiltonian is given by
\begin{align} \label{eq:IntHamSU3k}
\calH_\textrm{int} =& \int dx \left[ \sum_{j=1}^{N_w-1} \sum_{\sigma=\ua,\da} \sum_{a,b=1}^k t_{\sigma,ab} \calO^t_{j,\sigma,ab}(x) \right. \nonumber \\
& \left. +\frac{1}{2} \sum_{j=1}^{N_w} \sum_{a<b} \sum_{s,s'=1}^3 u^{ss'}_{ab} \calO^{u,ss'}_{j,ab}(x) +\textrm{H.c.} \right]. 
\end{align}
Similarly to the previous section, we introduce the chiral fields for each copy, 
\begin{align}
\begin{split}
\tchi^p_{j,1,a} &= \frac{1}{\sqrt{2}} (\tphi^p_{j,\ua,a} -\tphi^p_{j,\da,a}), \\
\tchi^p_{j,2,a} &= \frac{1}{\sqrt{6}} (\tphi^p_{j,\ua,a} +\tphi^p_{j,\da,a}), 
\end{split}
\end{align}
through 
\begin{align}
\begin{split}
\tphi^R_{j,\ua,a} &= \varphi_{j,\ua,a} +2(\theta_{j,\ua,a}+\theta_{j,\da,a}), \\
\tphi^R_{j,\da,a} &= \varphi_{j,\da,a} +2\theta_{j,\da,a}, \\
\tphi^L_{j,\ua,a} &= \varphi_{j,\ua,a} -2\theta_{j,\ua,a}, \\
\tphi^L_{j,\da,a} &= \varphi_{j,\da,a} -2(\theta_{j,\ua,a} +\theta_{j,\da,a}). 
\end{split}
\end{align}
They obey the commutation relations, 
\begin{align}
[\partial_x \tchi^p_{j,l,a}(x), \tchi^{p'}_{j',l',a'}(x')] = 2ip \pi \delta_{pp'} \delta_{jj'} \delta_{ll'} \delta_{aa'} \delta(x-x'). 
\end{align}
In terms of these bosonic fields, the interactions are expressed as 
\begin{align} \label{eq:InterIntSU3k}
\calO^t_{j,\sigma,ab} = e^{i \bfalpha_\sigma \cdot \tbfchi^R_{j,a} -i\bfalpha_\sigma \cdot \tbfchi^L_{j+1,b}}
\end{align}
and
\begin{align} \label{eq:IntraIntSU3k}
\calO^{u,ss'}_{j,ab} = e^{i \bfomega_s \cdot (\tbfchi^R_{j,a} -\tbfchi^R_{j,b}) -i\bfomega_{s'} \cdot (\tbfchi^L_{j,a} -\tbfchi^L_{j,b})}, 
\end{align}
where $\bfalpha_\sigma$ are the roots of $SU(3)$ given in Eq.~\eqref{eq:SU3Root}, $\bfomega_s$ are vectors given by
\begin{align}
\begin{split}
\bfomega_1 &= \begin{pmatrix} \frac{1}{\sqrt{2}}, & \frac{1}{\sqrt{6}} \end{pmatrix}, \\
\bfomega_2 &= \begin{pmatrix} -\frac{1}{\sqrt{2}}, & \frac{1}{\sqrt{6}} \end{pmatrix}, \\
\bfomega_3 &= \begin{pmatrix} 0, & -\sqrt{\frac{2}{3}} \end{pmatrix}, 
\end{split}
\end{align}
and $\tbfchi^p_{j,a} = (\tchi^p_{j,1,a}, \tchi^p_{j,2,a})$. 
The vectors $\bfomega_s$ form the fundamental representation of $SU(3)$,  as depicted on the weight diagram of $SU(3)$ in Fig.~\ref{fig:SU3Weights}. 

If one assumes that each copy of the coupled-wire system independently forms the Halperin (221) state, we have a simple physical understanding of these interactions in terms of excitations of the Halperin state. 
The interwire interactions \eqref{eq:InterIntSU3k} are now understood as tunnelings of the charge-$q$ particle excitations between the $k$ copies. 
On the other hand, the intrawire interactions \eqref{eq:IntraIntSU3k} are interactions among nontrivial quasiparticle excitations from different copies, as the quasiparticle operators in Sec.~\ref{sec:SU31Quasiparticles} are written as 
\begin{align}
\Psi^{R\dagger}_{QP,j+\frac{1}{2},\sigma} = e^{i\bfomega_\sigma \cdot \tbfchi^R_j}, \hspace{10pt}
\Psi^{L\dagger}_{QP,j-\frac{1}{2},\sigma} = e^{i\bfomega_\sigma \cdot \tbfchi^L_j}, 
\end{align}
where $\bfomega_\ua=\bfomega_1$ and $\bfomega_\da = \bfomega_2$. 
Equation \eqref{eq:IntraIntSU3k} gives the most natural interactions among the quasiparticles that manifestly preserve the particle and momentum conservations.

For a general form of the Hamiltonian $\calH = \calH_\textrm{SLL} +\calH_\textrm{int}$, it is practically hard to see how these interactions generate a gap. 
Therefore, in the following, we will concentrate on a special case in which we can translate the Hamiltonian in the language of certain CFTs. 
In that case, we can show that the resulting ground state has the same topological properties as those of the NASS state at $\nu_\textrm{tot}=2k/(2km+3)$ with $m=0$ \cite{Ardonne99,Ardonne01a}. 

\subsection{$SU(3)_k$ currents}
\label{sec:SU3kCurrents}

We first suppose that the SLL Hamiltonian takes the following form, 
\begin{align} \label{eq:SLLHamSU3kCopies}
\calH_\textrm{SLL} = \frac{v}{4\pi} \sum_{j=1}^{N_w} \sum_{a=1}^k \int dx \left[ (\partial_x \tbfchi^R_{j,a})^2 +(\partial_x \tbfchi^L_{j,a})^2 \right]. 
\end{align}
As in Sec.~\ref{sec:SU31Currents}, by introducing the right and left $SU(3)_1$ currents for each copy, the SLL Hamiltonian is written as 
\begin{align} \label{eq:kSU3WZWHam}
\calH_\textrm{SLL} = \frac{v}{16\pi} \sum_{j=1}^{N_w} \sum_{a=1}^k \int dx \left[ :\mathrel{\bfJ_{j,a} \cdot \bfJ_{j,a}}: +:\mathrel{\bbfJ_{j,a} \cdot \bbfJ_{j,a}}: \right], 
\end{align}
where $\bfJ_{j,a} = (J_{j,a}^1, \cdots, J_{j,a}^8)$ and $\bbfJ_{j,a} = (\bJ_{j,a}^1, \cdots, \bJ_{j,a}^8)$. 
They satisfy the $SU(3)_1$ current algebra, 
\begin{align}
\begin{split}
J^\alpha_{j,a}(z) J^\beta_{j',b}(w) &\sim \delta_{jj'} \delta_{ab} \left[ \frac{\delta_{\alpha \beta}}{(z-w)^2} +\sum_\gamma \frac{if_{\alpha \beta \gamma} J^\gamma_{j,a}(w)}{z-w} \right], \\
\bJ^\alpha_{j,a}(\bz) \bJ^\beta_{j',b}(\bw) &\sim \delta_{jj'} \delta_{ab} \left[ \frac{\delta_{\alpha \beta}}{(\bz-\bw)^2} +\sum_\gamma \frac{if_{\alpha \beta \gamma} \bJ^\gamma_{j,a}(\bw)}{\bz-\bw} \right]. \\
\end{split}
\end{align}
Equation \eqref{eq:kSU3WZWHam} nothing but represents $kN_w$ copies of the $SU(3)_1$ WZW CFT. 

\begin{figure}
\includegraphics[clip,width=0.46\textwidth]{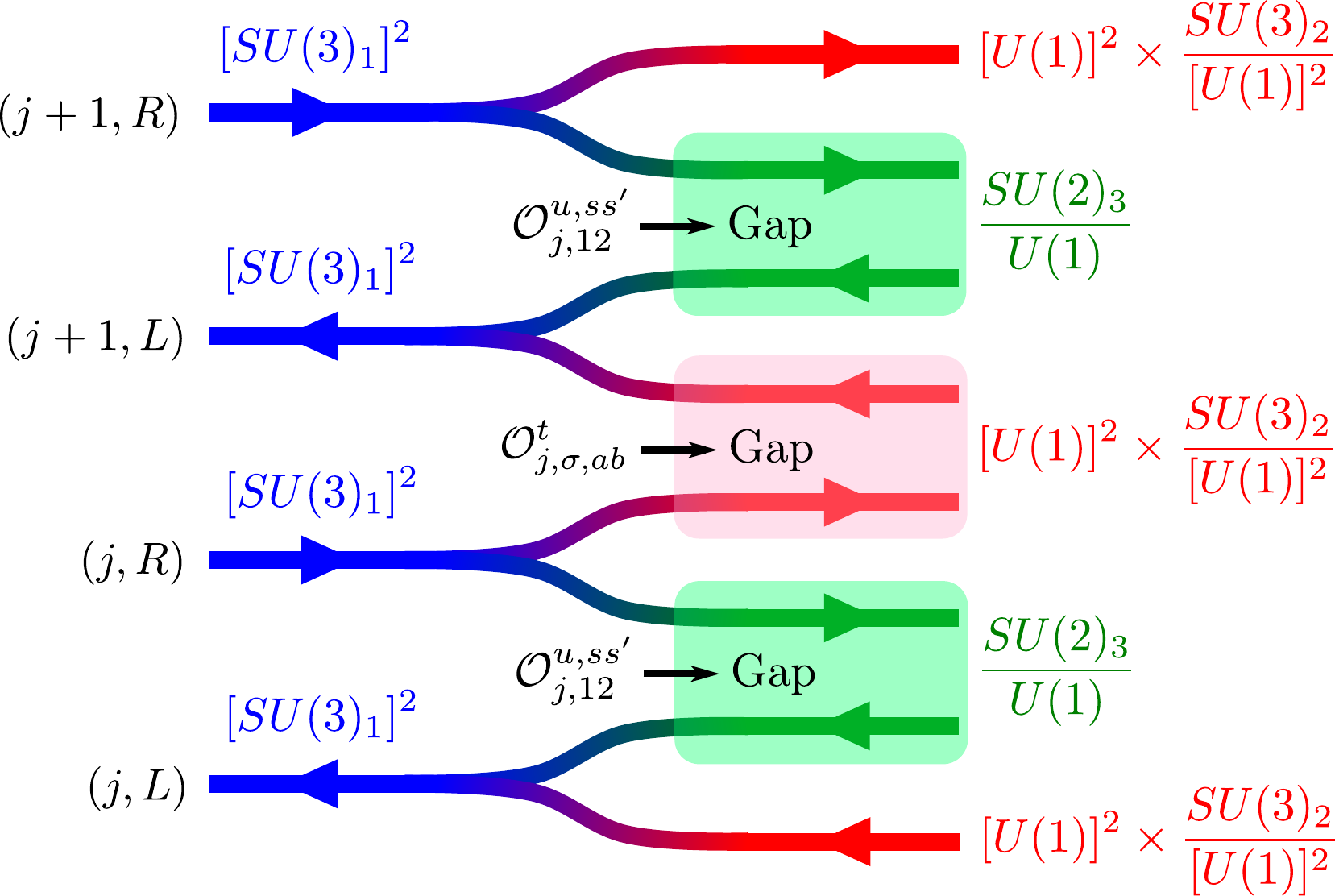}
\caption{Schematic view of the construction of a non-Abelian $SU(2)$-singlet FQH state for $k=2$.
The left and right $[SU(3)_1]^2$ CFTs from each wire are decomposed into the two sectors $[U(1)]^2 \times SU(3)_2/[U(1)]^2$ and $[SU(3)_1]^2/SU(3)_2 \sim SU(2)_3/U(1)$. 
The interwire interactions $\calO^t_{j,\sigma,ab}$ open a gap in the former sector, leaving unpaired chiral gapless modes at the edges, while the intrawire interactions $\calO^{u,ss'}_{j,12}$ open a gap in the latter sector.}
\label{fig:NonAbelianSU32}
\end{figure}

In a spirit of Ref.~\cite{Teo14}, we employ the coset construction \cite{dFMS} to decompose the $k$ copies of the $SU(3)_1$ WZW CFT $[SU(3)_1]^k$ in each wire as 
\begin{align}
[SU(3)_1]^k \sim SU(3)_k \times \frac{[SU(3)_1]^k}{SU(3)_k}. 
\end{align}
For the later purpose, we further write this as 
\begin{align} \label{eq:EmbeddingSU3k}
[SU(3)_1]^k \sim [U(1)]^2 \times \frac{SU(3)_k}{[U(1)]^2} \times \frac{SU(k)_3}{[U(1)]^{k-1}}. 
\end{align}
Here the $SU(3)_k$ WZW CFT is further decomposed into two CFTs: $[U(1)]^2$ and $SU(3)_k/[U(1)]^2$. 
The CFT $[U(1)]^2$ represents two free boson CFTs with $c=2$, which correspond to the two charge modes of the $m=0$ NASS state. 
The coset CFT $SU(3)_k/[U(1)]^2$ is a Gepner parafermion CFT \cite{Gepner87} with central charge, 
\begin{align}
c=\frac{6(k-1)}{k+3}. 
\end{align}
The residual coset CFT $[SU(3)_1]^k/SU(3)_k$ is equivalent to the Gepner parafermion CFT $SU(k)_3/[U(1)]^{k-1}$, whose central charge is 
\begin{align}
c=\frac{2k(k-1)}{k+3}. 
\end{align}
The conformal embedding \eqref{eq:EmbeddingSU3k} is demonstrated in Appendix~\ref{app:EMTensor} by directly decomposing the energy-momentum tensor corresponding to the SLL Hamiltonian \eqref{eq:SLLHamSU3kCopies} via the vertex representation. 
Although the last two coset CFTs in Eq.~\eqref{eq:EmbeddingSU3k} are both charge neutral, the only $SU(3)_k/[U(1)]^2$ CFT actually appears as the neutral mode of the $m=0$ NASS state. 
Then our task is to show that the $[U(1)]^2$ and $SU(3)_k/[U(1)]^2$ sectors are gapped by the \emph{interwire} interaction, while the $SU(k)_3/[U(1)]^{k-1}$ sector is gapped by the \emph{intrawire} interaction, as schematically shown in Fig.~\ref{fig:NonAbelianSU32} for $k=2$. 
These interactions leave gapless edge modes described by the chiral $SU(3)_k$ WZW CFTs at the outermost wires $j=1$ and $N_w$. 

The interwire interactions \eqref{eq:InterIntSU3k} are easily identified as products of the $SU(3)_k$ currents, when the coupling constants are identical for all possible pairs among the $k$ copies.
Indeed, if $t_{\sigma,ab} \equiv t_\sigma$, one can write 
\begin{align}
\begin{split}
t_\ua \sum_{a,b=1}^k \calO^t_{j,\ua,ab} &= t_\ua x_c^2 \sum_{a=1}^k J^{2+}_{j,a} \sum_{b=1}^k \bJ^{2-}_{j+1,b}, \\
t_\da \sum_{a,b=1}^k \calO^t_{j,\da,ab} &= t_\da x_c^2 \sum_{a=1}^k J^{3+}_{j,a} \sum_{b=1}^k \bJ^{3-}_{j+1,b},
\end{split}
\end{align}
where we have used Eq.~\eqref{eq:VertexRepSU3} for each copy. 
This can be rewritten by the $SU(3)_k$ currents, 
\begin{align} \label{eq:SU3kCurrent}
\calJ^\alpha_j = \sum_{a=1}^k J^\alpha_{j,a}, \hspace{10pt}
\bcalJ^\alpha_j = \sum_{a=1}^k \bJ^\alpha_{j,a}, 
\end{align}
which satisfy the $SU(3)_k$ current algebra, 
\begin{align}
\begin{split}
\calJ^\alpha_j(z) \calJ^\beta_{j'}(w) &\sim \delta_{jj'} \left[ \frac{k\delta_{\alpha \beta}}{(z-w)^2} +\sum_\gamma \frac{if_{\alpha \beta \gamma} \calJ^\gamma_j(w)}{z-w} \right], \\
\bcalJ^\alpha_j(\bz) \bcalJ^\beta_{j'}(\bw) &\sim \delta_{jj'} \left[ \frac{k\delta_{\alpha \beta}}{(\bz-\bw)^2} +\sum_\gamma \frac{if_{\alpha \beta \gamma} \bcalJ^\gamma_j(\bw)}{\bz-\bw} \right].
\end{split}
\end{align}
Then we have 
\begin{align} \label{eq:InterIntSU3k2}
\begin{split}
t_\ua \sum_{a,b=1}^k \calO^t_{j,\ua,ab} +\textrm{H.c.} &= t_\ua x_c^2 \left( \calJ^4_j \bcalJ^4_{j+1} +\calJ^5_j \bcalJ^5_{j+1} \right), \\
t_\da \sum_{a,b=1}^k \calO^t_{j,\da,ab} +\textrm{H.c.} &= t_\da x_c^2 \left( \calJ^6_j \bcalJ^6_{j+1} +\calJ^7_j \bcalJ^7_{j+1} \right). 
\end{split}
\end{align}
Thus these interactions obviously act on the nonchiral $SU(3)_k$ WZW CFTs consisting of neighboring wires. 

\subsection{Parafermions for $k=2$}
\label{sec:SU32Parafermion}

If the coupling constants are fine tuned, the intrawire interactions in Eq.~\eqref{eq:IntHamSU3k} can be identified as products of $SU(k)_3/[U(1)]^{k-1}$ parafermionic fields, or equivalently products of $SU(3)_1$ WZW primary fields, with conformal weight $2/3$. 
For later convenience, we also show that the interwire interactions in Eq.~\eqref{eq:InterIntSU3k2} are further written in terms of products of $[U(1)]^2$ vertex operators and $SU(3)_k/[U(1)]^2$ parafermionic fields, whose conformal weights are $1/k$ and $1-1/k$, respectively. 
This manifests the conformal embedding $SU(3)_k \sim [U(1)]^2 \times SU(3)_k/[U(1)]^2$. 
In the following, we focus on the case of $k=2$, for which we can apply a powerful result from integrable field theory to prove the existence of bulk gap. 
The case for $k \geq 3$ will be discussed in Sec.~\ref{sec:SUN} along with the $(N-1)$-component generalization of the NASS state. 

Let us introduce two charge fields and two neutral fields for each wire and each chiral sector by 
\begin{align}
\begin{split}
\tX^p_{j,l} = \frac{1}{\sqrt{2}} \left( \tchi^p_{j,l,1} +\tchi^p_{j,l,2} \right), \\
\tY^p_{j,l} = \frac{1}{\sqrt{2}} \left( \tchi^p_{j,l,1} -\tchi^p_{j,l,2} \right), 
\end{split}
\end{align}
which satisfy the commutation relations,
\begin{align}
\begin{split}
[\partial_x \tX^p_{j,l}(x), \tX^{p'}_{j',l'}(x')] &= 2ip\pi \delta_{pp'} \delta_{jj'} \delta_{ll'} \delta(x-x'), \\
[\partial_x \tY^p_{j,l}(x), \tY^{p'}_{j',l'}(x')] &= 2ip\pi \delta_{pp'} \delta_{jj'} \delta_{ll'} \delta(x-x'), \\
[\partial_x \tX^p_{j,l}(x), \tY^{p'}_{j',l'}(x')] &= 0. 
\end{split}
\end{align}
The SLL Hamiltonian \eqref{eq:SLLHamSU3kCopies} then becomes 
\begin{align}
\calH_\textrm{SLL} =& \ \frac{v}{4\pi} \sum_{j=1}^{N_w} \int dx \left[ (\partial_x \tbfX^R_j)^2 +(\partial_x \tbfX^L_j)^2 \right. \nonumber \\
&\left. +(\partial_x \tbfY^R_j)^2 +(\partial_x \tbfY^L_j)^2 \right], 
\end{align}
where $\tbfX^p_j = (\tX^p_{j,1}, \tX^p_{j,2})$ and $\tbfY^p_j = (\tY^p_{j,1}, \tY^p_{j,2})$. 
Using Eqs.~\eqref{eq:VertexRepSU3} and \eqref{eq:SU3kCurrent}, we find for the right $SU(3)_2$ currents, 
\begin{align} \label{eq:VertexRepSU32Cur}
\begin{split}
\calJ^{2 \pm}_j(x) &= \frac{\sqrt{2} e^{\pm \frac{i\pi}{4}}}{\sqrt{x_c}} e^{\pm \frac{i}{\sqrt{2}} \bfalpha_\ua \cdot \tbfX^R_j(x)} \Psi^{\pm \bfalpha_\ua, 1}_j(x), \\
\calJ^{3 \pm}_j(x) &= \frac{\sqrt{2} e^{\pm \frac{i\pi}{4}}}{\sqrt{x_c}} e^{\pm \frac{i}{\sqrt{2}}\bfalpha_\da \cdot \tbfX^R_j(x)} \Psi^{\pm \bfalpha_\da, 1}_j(x).
\end{split}
\end{align}
Here $\Psi^{\bfalpha,I}_j$ is the $SU(3)_2/[U(1)]^2$ parafermionic field associated with a root $\bfalpha$ of $SU(3)$ and the $I$-th antisymmetric representation of $SU(2)$.
$SU(2)$ only has a trivial antisymmetric representation with a single box ($I=1$) in Young tableau, in which there are only two weights $\pm \frac{1}{\sqrt{2}}$. 
The parafermionic field $\Psi^{\bfalpha,1}_j$ has conformal weight $\frac{1}{2}$ and has an obvious vertex representation \cite{Dunne89}, 
\begin{align} \label{eq:VertexRepSU32Para}
\Psi^{\pm \bfalpha_\sigma,1}_j(x) = \frac{\sqrt{2} e^{\pm \frac{i\pi}{4}}}{\sqrt{x_c}} \cos \left( \frac{\bfalpha_\sigma \cdot \tbfY^R_j(x)}{\sqrt{2}} \right). 
\end{align}
Since we still need some care about the ``parafermionic cocycles'', the details about this identification are given in Appendix~\ref{app:SUNkPara}. 
Similarly for the left $SU(3)_2$ currents, one can find 
\begin{align}
\begin{split}
\bcalJ^{2 \pm}_j(x) &= \frac{\sqrt{2} e^{\pm \frac{i\pi}{4}}}{\sqrt{x_c}} e^{\pm \frac{i}{\sqrt{2}} \bfalpha_\ua \cdot \tbfX^L_j(x)} \bPsi^{\pm \bfalpha_\ua,1}_j(x), \\
\bcalJ^{3 \pm}_j(x) &= \frac{\sqrt{2} e^{\pm \frac{i\pi}{4}}}{\sqrt{x_c}} e^{\pm \frac{i}{\sqrt{2}} \bfalpha_\da \cdot \tbfX^L_j(x)} \bPsi^{\pm \bfalpha_\da,1}_j(x),
\end{split}
\end{align}
with
\begin{align}
\bPsi^{\pm \bfalpha_\sigma,1}_j(x) = \frac{\sqrt{2} e^{\pm \frac{i\pi}{4}}}{\sqrt{x_c}} \cos \left( \frac{\bfalpha_\sigma \cdot \tbfY^L_j(x)}{\sqrt{2}} \right). 
\end{align}
Once the charge modes $\tbfX^p_j$ are gapped, we can focus only on the neutral sector. 
In this case, the interwire interaction \eqref{eq:InterIntSU3k2} may become 
\begin{align}
t_\sigma \sum_{a,b=1}^2 \calO^t_{j,\sigma,ab} \propto t_\sigma \Psi^{\bfalpha_\sigma,1 \dagger}_j \bPsi^{\bfalpha_\sigma,1}_{j+1}. 
\end{align}
This is a clear manifestation of the level-rank duality between the $SU(2)_3$ and $SU(3)_2$ WZW CFTs \cite{dFMS}, by which the $SU(3)_2/[U(1)]^2$ parafermionic fields in the intrawire interaction for the $k=3$ Read-Rezayi state \cite{Teo14} appear in the interwire interaction for the $k=2$ NASS state. 
As seen below, this duality in fact exchanges the neutral sectors of the interwire and intrawire interactions between these two non-Abelian FQH states. 
Hence the $SU(2)_3/U(1)$ parafermionic fields in the interwire interaction for the Read-Rezayi state now appear in the intrawire interaction for the NASS state. 

The intrawire interactions \eqref{eq:IntraIntSU3k} are now written as 
\begin{align} \label{eq:IntraIntSU32}
\calO^{u,ss'}_{j,12} = e^{i\sqrt{2} \bfomega_s \cdot \tbfY^R_j -i\sqrt{2} \bfomega_{s'} \cdot \tbfY^L_j}. 
\end{align}
As we have mentioned, the vectors $\bfomega_s$ are weights in the fundamental representation of $SU(3)$. 
If all the coupling constants are identical, that is $u^{ss'}_{12} \equiv u$, we can express the interaction as (see Appendix~\ref{app:SUkNPara})
\begin{align} \label{eq:IntraIntSU32Para}
\sum_{s,s'=1}^3 \calO^{u,ss'}_{j,12} = e^{-\frac{2i\pi}{3}} x_c^\frac{4}{3} \Xi^{\sqrt{2}}_j \bXi^{\sqrt{2} \dagger}_j, 
\end{align}
where $\Xi^{\sqrt{2}}_j$ is an $SU(2)_3/U(1)$ primary field in the holomorphic sector with conformal weight $\frac{2}{3}$ and $\bXi^{\sqrt{2}}_j$ is its antiholomorphic counterpart. 
It may appear that from their conformal weights, each chiral primary field coincides with the parafermionic field of the $SU(2)_3/U(1) \sim \mathbb{Z}_3$ CFT \cite{Zamolodchikov85}. 
However, as discussed in Appendix~\ref{app:SUkNPara}, only their nonchiral product $\Upsilon^{\sqrt{2},1}_j \equiv \Xi^{\sqrt{2}}_j \bXi^{\sqrt{2} \dagger}_j$ behaves as the $\mathbb{Z}_3$ parafermionic field generating the correct parafermionic algebra. 
The interaction \eqref{eq:IntraIntSU32Para} has scaling dimension $\frac{4}{3}$ and acts only on the residual nonchiral $\mathbb{Z}_3$ coset CFT in each wire. 

One may naively expect that since the interaction \eqref{eq:IntraIntSU32Para} is relevant, it immediately gaps out the $\mathbb{Z}_3$ sector in each wire. 
However, an important remark is in order. 
Since the interaction \eqref{eq:IntraIntSU32Para} is now identified as the $\mathbb{Z}_3$ parafermionic field, we can import the knowledge from an integrable deformation of the $\mathbb{Z}_3$ parafermion theory given by Fateev and Zamolodchikov \cite{Fateev91a,Fateev91b}. 
It is known that a non-perturbative mass gap is generated only when its coupling constant is \emph{negative}, while a positive coupling constant induces a massless flow to the tricritical Ising CFT $\mathcal{M}_4$.
Therefore, if $u^{ss'}_{12} = u<0$ and they become relevant, the interactions open a bulk gap but leave gapless edge modes described by the chiral $SU(3)_2$ WZW CFT. 
This is a strong signature of the NASS state with $k=2$ and $m=0$ \cite{Ardonne99,Ardonne01a}. 

\subsection{Bulk quasiparticles}
\label{sec:SU32Quasiparticles}

We continue to focus on the case for $k=2$.
We here consider the quasiparticle excitations of the $k=2$ NASS state. 
As suggested in Ref.~\cite{Teo14}, we consider the $2k_F$ backscattering operators given by 
\begin{align}
B_{j,\sigma,a}(x) = e^{2i\theta_{j,\sigma,a}(x)}. 
\end{align}
In terms of the chiral fields, they are written as 
\begin{align}
\begin{split}
B_{j,\sigma,1} &= e^{\frac{i}{\sqrt{2}} \bfomega_\sigma \cdot (\tbfX^R_j -\tbfX^L_j)} e^{\frac{i}{\sqrt{2}} \bfomega_\sigma \cdot (\tbfY^R_j -\tbfY^L_j)}, \\
B_{j,\sigma,2} &= e^{\frac{i}{\sqrt{2}} \bfomega_\sigma \cdot (\tbfX^R_j -\tbfX^L_j)} e^{-\frac{i}{\sqrt{2}} \bfomega_\sigma \cdot (\tbfY^R_j -\tbfY^L_j)}. 
\end{split}
\end{align}
Then we may define the quasiparticle operators as
\begin{align}
\begin{split}
\Psi_{\textrm{QP},j+\frac{1}{2},\sigma,1}^{R\dagger} &= e^{\frac{i}{\sqrt{2}} \bfomega_\sigma \cdot \tbfX^R_j} \Sigma^{R\dagger}_{j,\sigma}, \\
\Psi_{\textrm{QP},j+\frac{1}{2},\sigma,2}^{R\dagger} &= e^{\frac{i}{\sqrt{2}} \bfomega_\sigma \cdot \tbfX^R_j} \Sigma^{R}_{j,\sigma}, \\
\Psi_{\textrm{QP},j+\frac{1}{2},\sigma,1}^{L\dagger} &= e^{\frac{i}{\sqrt{2}} \bfomega_\sigma \cdot \tbfX^L_{j+1}} \Sigma^{L\dagger}_{j+1,\sigma}, \\
\Psi_{\textrm{QP},j+\frac{1}{2},\sigma,2}^{L\dagger} &= e^{\frac{i}{\sqrt{2}} \bfomega_\sigma \cdot \tbfX^L_{j+1}} \Sigma^L_{j+1,\sigma}, 
\end{split}
\end{align}
where
\begin{align}
\Sigma_{j,\sigma}^{R/L \dagger} &= e^{\frac{i}{\sqrt{2}} \bfomega_\sigma \cdot \tbfY^{R/L}_j}. 
\end{align}
They create quasiparticles with total charge $\frac{1}{3} q$ and spins $\pm q$, and thus have the same actions on the charge part as those of the quasiparticles of the Halperin (221) state (see Sec.~\ref{sec:SU31Quasiparticles}). 
However, their actions on the neutral part are highly nontrivial, since we have to consider the situation in which both interwire and intrawire interactions flow to the strong-coupling limit. 
In the following, we consider the neutral part of the quasiparticle operators $\Sigma^{R/L}_{j,\sigma}$ when the intrawire interaction opens a gap. 

\begin{figure}
\includegraphics[clip,width=0.35\textwidth]{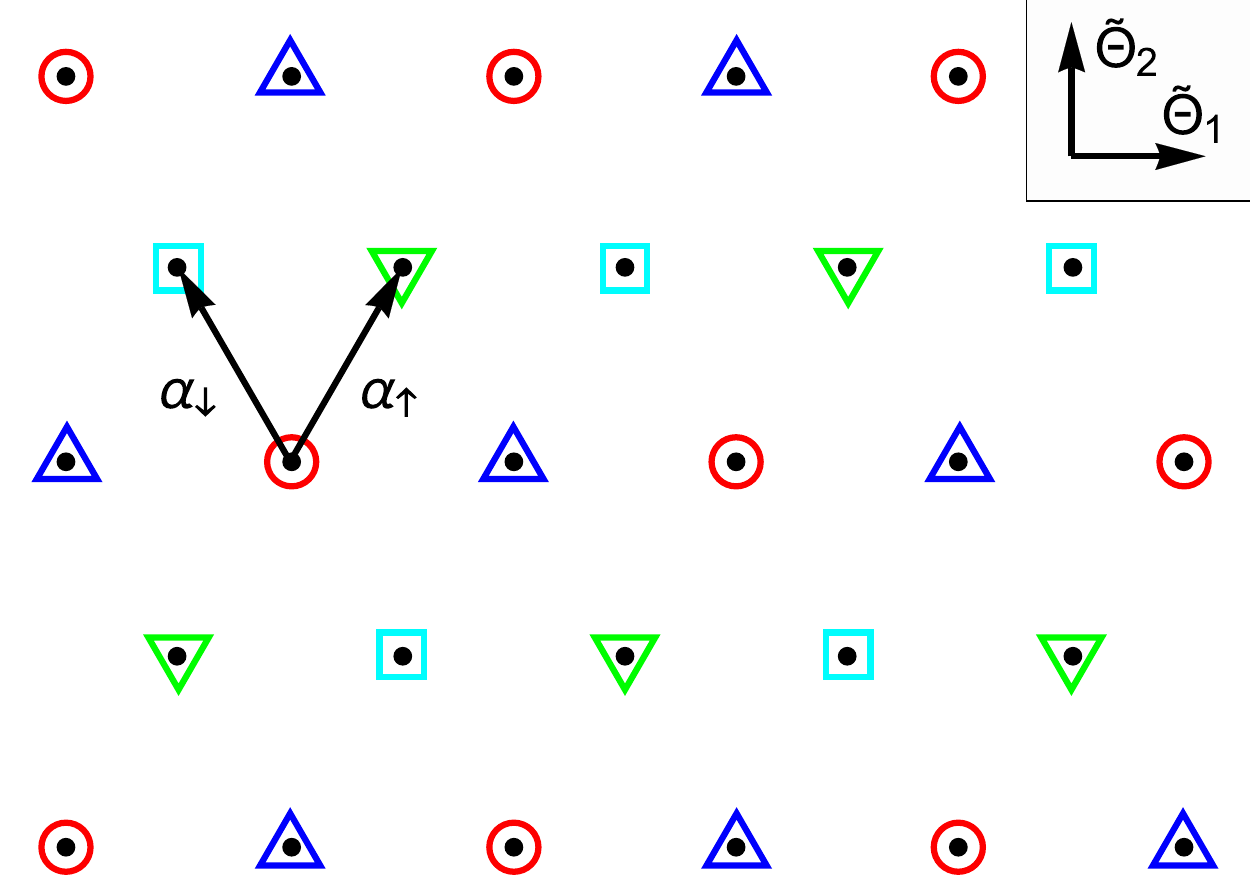}
\caption{Minima of the cosine potential \eqref{eq:CosinePotSU32b} are indicated by the black dots on the $(\tTheta_1,\tTheta_2)$ plane scaled by $\sqrt{2}/\pi$. 
Different symbols enclosing the dots represent four independent minima within the compactification radii of $\tbfTheta$.
The two vectors $\bfalpha_{\ua,\da}$ are roots of $SU(3)$ [see Eq.~\eqref{eq:SU3Root}]. }
\label{fig:MinimaSU3}
\end{figure}

Supposing that the charge modes are gapped, we focus only on the neutral part of a single-wire Hamiltonian
\begin{align}
\calH_j = \frac{v}{4\pi} \int dx \Bigl[ (\partial_x \tbfY^R_j)^2 +(\partial_x \tbfY^L_j)^2 \Bigr] +\int dx \ \calV [\tbfY^{R/L}_j], 
\end{align}
where the intrawire interaction is given by 
\begin{align} \label{eq:CosinePotSU32}
\calV[\tbfY^{R/L}_j] 
&= \frac{1}{2} \sum_{s,s'=1}^3 u^{ss'} \calO^{u,ss'}_{j,12} +\textrm{H.c.} \nonumber \\
&= \sum_{s,s'=1}^3 u^{ss'} \cos \bigl[ \sqrt{2} (\bfomega_s \cdot \tbfY^R_j -\bfomega_{s'} \cdot \tbfY^L_j) \bigr]. 
\end{align}
To investigate the ground state of this Hamiltonian, it is convenient to introduce the nonchiral fields, 
\begin{align}
\begin{split}
\tPhi_{l,j} &= \frac{1}{2} (\tY^R_{l,j} +\tY^L_{l,j}), \\
\tTheta_{l,j} &= \frac{1}{2} (\tY^R_{l,j} -\tY^L_{l,j}), 
\end{split}
\end{align}
which satisfy 
\begin{align}
[\partial_x \tTheta_{j,l}(x), \tPhi_{j',l'}(x')] = i\pi \delta_{jj'} \delta_{ll'} \delta(x-x'). 
\end{align}
Then the interaction \eqref{eq:CosinePotSU32} is written as 
\begin{align}
&\calV[\tbfPhi_j,\tbfTheta_j] \nonumber \\
&= \sum_{s=1}^3 u^{ss} \cos (2\sqrt{2} \bfomega_s \cdot \tbfTheta_j) \nonumber \\
&\ \ \ +\sum_{s \neq s'} u^{ss'} \cos \sqrt{2} \bigl[ (\bfomega_s -\bfomega_{s'}) \cdot \tbfPhi_j +(\bfomega_s +\bfomega_{s'}) \cdot \tbfTheta_j \bigr]. 
\end{align}
This Hamiltonian may be interpreted as a statistical mechanical model with $\mathbb{Z}_2 \times \mathbb{Z}_2$ symmetry as follows. 
Let us first assume that $u^{ss} \equiv u <0$ and $u^{ss'}=0$ for $s \neq s'$. 
Then the fields $\tbfTheta_j$ may be pinned at the minima of the cosine potential (dropping the wire index $j$), 
\begin{align} \label{eq:CosinePotSU32b}
-\cos \left(2\tTheta_1 +\frac{2\tTheta_2}{\sqrt{3}} \right) -\cos \left( 2\tTheta_1 -\frac{2\tTheta_2}{\sqrt{3}} \right)
-\cos \left( \frac{4\tTheta_2}{\sqrt{3}} \right). 
\end{align}
The minima form a triangular lattice with lattice constant $\pi$ as depicted in Fig.~\ref{fig:MinimaSU3}. 
Although the corresponding ground state seems to be infinitely degenerate, the degeneracy is actually lifted by the compactification conditions originally imposed for the bosonic fields $\tbfTheta_j$. 
Finite coupling constants $u^{ss'}$ for $s \neq s'$ dynamically resume the compactification of $\tbfTheta_j$, as the operator $e^{i\sqrt{2} \bfalpha \cdot \tbfPhi_j}$ creates the kink $\tbfTheta_j \to \tbfTheta_j +\pi \sqrt{2} \bfalpha$, where $\bfalpha = \bfomega_s -\bfomega_{s'}$ and $\bfalpha$ is a root of $SU(3)$. 
Therefore, the only minima within the unit cell of a larger triangular lattice with lattice constant $2\pi$ become independent, and the resulting ground-state degeneracy is four. 
The nonchiral products of the neutral parts of the quasiparticle operators take the form, 
\begin{align}
\Sigma^n_{j,\sigma} \equiv e^{-\frac{i\pi}{6}} \Sigma^{R\dagger}_{j,\sigma} \Sigma^L_{j,\sigma} = e^{i\sqrt{2} \bfomega_\sigma \cdot \tbfTheta_j}, 
\end{align}
and acquire finite expectation values in each of the four potential minima as shown in Table.~\ref{tab:ExpValueSU32}.
\begin{table}
\caption{Expectation values of $\Sigma^n_{j,\sigma}$ in the four independent minima of the cosine potential \eqref{eq:CosinePotSU32b}.}
\label{tab:ExpValueSU32}
\begin{tabular}{cc|cc}
\hline \hline
$\tTheta_{1,j}$ & $\tTheta_{2,j}$ & $\langle \Sigma^n_{j,\ua} \rangle$ & $\langle \Sigma^n_{j,\da} \rangle$ \\ \hline
0 & 0 & 1 & 1 \\
$\pi$ & 0 & $-1$ & $-1$ \\
$\frac{\pi}{2}$ & $\frac{\sqrt{3}\pi}{2}$ & $-1$ & 1 \\
$\frac{3\pi}{2}$ & $\frac{\sqrt{3} \pi}{2}$ & 1 & $-1$ \\ \hline \hline
\end{tabular}
\end{table}
These operators appear to behave as two Ising order parameters detecting a $\mathbb{Z}_2 \times \mathbb{Z}_2$ symmetry breaking. 
As discussed in detail in Sec.~\ref{sec:SUNkQuasiparticles}, the $\mathbb{Z}_2 \times \mathbb{Z}_2$ symmetry is actually the symmetry of the $SU(3)_2/[U(1)]^2$ parafermion CFT \cite{Dunne89}, and therefore the operators $\Sigma^n_{j,\sigma}$ will be regarded as spin fields of the corresponding nonchiral CFT. 
Thus the neutral parts of the backscattering operators $\Sigma^{R/L}_{j,\sigma}$ may be seen as spin fields of the chiral parafermion CFTs once the Hamiltonian is gapped by the intrawire interactions. 
Such spin fields, combined with the charge part, actually constitute quasihole operators inserted to the trial wave functions of the $m=0$ NASS states \cite{Ardonne01a}. 

\section{Abelian and non-Abelian $SU(N-1)$-singlet FQH states} \label{sec:SUN}

In this section, we give a general construction of Abelian and non-Abelian $SU(N-1)$-singlet FQH states. 
The Abelian $SU(N-1)$-singlet FQH state is described by the wave function \eqref{eq:MultiFQHS} with the $K$ matrix given in Eq.~\eqref{eq:KmatSUN}. 
It is nothing but an $(N-1)$-component analogue of the $\nu=\frac{1}{2}$ Laughlin state for $N=2$ and the Halperin (221) state \cite{Halperin83} for $N=3$ at filling fraction $\nu_\textrm{tot}=(N-1)/N$. 
Such a FQH state has gapless edge modes described by the chiral $SU(N)_1$ WZW CFT. 

The non-Abelian $SU(N-1)$-singlet FQH state that we construct here is obtained by symmetrizing $k$ copies of the above Abelian FQH state. 
This is a generalization of a bosonic Read-Rezayi state \cite{Read99} for $N=2$ and a bosonic NASS state \cite{Ardonne99,Ardonne01a} for $N=3$ to the $(N-1)$-component case. 
It is realized at filling fraction $\nu_\textrm{tot}=k(N-1)/N$ and has gapless edge modes described by the chiral $SU(N)_k$ WZW CFT. 

The following argument is in the same line as the previous two sections. 
However, we proceed in a more abstract fashion by largely relying on the $SU(N)$-algebraic structure of the $SU(N-1)$-singlet FQH states. 

\subsection{Abelian $SU(N-1)$-singlet FQH state} \label{sec:AbelianSUN}

\subsubsection{Interactions at $\nu_\textrm{tot}=(N-1)/N$}

We start from the array of wires with $(N-1)$-component bosons, each of which has a single channel and filling factor $\nu=1/N$. 
Hence, we drop the channel index $a$ for a moment. 
We then consider the interaction Hamiltonian, 
\begin{align} \label{eq:InterIntHamSUN}
\calH_\textrm{int} = \sum_{j=1}^{N_w-1} \sum_{\sigma=1}^{N-1} \int dx \left[ t_\sigma \calO^t_{j,\sigma}(x) +\textrm{H.c.} \right],
\end{align}
with the interwire correlated hoppings, 
\begin{align} \label{eq:InterIntSUN}
\calO^t_{j,\sigma} = e^{i\left[ \varphi_{j,\sigma} -\varphi_{j+1,\sigma} +\sum_{\sigma'} \left( M_{\sigma \sigma'} \theta_{j,\sigma'} +\bar{M}_{\sigma \sigma'} \theta_{j+1,\sigma'} \right) \right]}, 
\end{align}
where $\bfM$ and $\bar{\bfM}$ are $(N-1) \times (N-1)$ integer matrices. 
Since the interactions must be built from bosonic operators, we require that the entries of these matrices are even integers (this condition may be relaxed as discussed in Sec.~\ref{sec:CorrHopping}). 
From the momentum conservation \eqref{eq:MomConsv}, they must satisfy 
\begin{align} \label{eq:MtoFilling}
\sum_{\sigma'=1}^{N-1} \left( M_{\sigma \sigma'}+\bar{M}_{\sigma \sigma'} \right) = 2N, 
\end{align}
for any $\sigma$. 
By introducing the $2(N-1)$ chiral bosonic fields for each wire as 
\begin{align} \label{eq:ChiralField}
\begin{split}
\tphi^R_{j,\sigma} &= \varphi_{j,\sigma} +\sum_{\sigma'=1}^{N-1} M_{\sigma \sigma'} \theta_{j,\sigma'}, \\
\tphi^L_{j,\sigma} &= \varphi_{j,\sigma} -\sum_{\sigma'=1}^{N-1} \bar{M}_{\sigma \sigma'} \theta_{j,\sigma'}, 
\end{split}
\end{align}
the interactions \eqref{eq:InterIntSUN} are written as 
\begin{align} \label{eq:InterIntSUN2}
\calO^t_{j,\sigma} = e^{i\left( \tphi^R_{j,\sigma} -\tphi^L_{j+1,\sigma} \right)}. 
\end{align}
These chiral bosonic fields satisfy the commutation relations, 
\begin{align}
\begin{split}
[\partial_x \tphi^R_{j,\sigma}(x), \tphi^R_{j',\sigma'}(x')] &= 2i\pi \delta_{jj'} K_{\sigma \sigma'} \delta (x-x'), \\
[\partial_x \tphi^L_{j,\sigma}(x), \tphi^L_{j',\sigma'}(x')] &= -2i\pi \delta_{jj'} \bar{K}_{\sigma \sigma'} \delta (x-x'), 
\end{split}
\end{align}
where 
\begin{align}
\bfK = \frac{1}{2}(\bfM+\bfM^T), \hspace{10pt} \bar{\bfK} = \frac{1}{2}(\bar{\bfM} +\bar{\bfM}^T). 
\end{align}
The Haldane's null vector condition \cite{Haldane95} requires that 
\begin{align}
\bfM=\bar{\bfM}^T. 
\end{align}
This indicates that $\bfK=\bar{\bfK}$ and commutators vanish between different chiral fields, 
\begin{align}
[\partial_x \tphi^R_{j,\sigma}(x), \tphi^L_{j',\sigma'}(x')] =0. 
\end{align} 
We can find the matrix $\bfM$ satisfying Eq.~\eqref{eq:MtoFilling} and giving $\bfK=\bfK_{SU(N)}$ [$\bfK_{SU(N)}$ is defined in Eq.~\eqref{eq:KmatSUN}]; 
for example, the upper (or lower) triangular matrix whose allowed nonzero entries are two, 
\begin{align} \label{eq:Mmat}
\bfM = \begin{pmatrix}
2 & 2 & \cdots & 2 & 2 \\
0 & 2 & & 2 & 2 \\
\vdots & & \ddots & & \vdots \\
0 & 0 & & 2 & 2 \\
0 & 0 & \cdots & 0 & 2
\end{pmatrix}, 
\end{align}
is in the desired form. 
We hereafter call the matrix $\bfM$ as the \emph{interaction matrix}. 
Then we introduce dual fields associated with the link between $j$ and $j+1$ by 
\begin{align}
\begin{split}
\ttheta_{j+\frac{1}{2},\sigma} &= \frac{1}{2} (\tphi^R_{j,\sigma} -\tphi^L_{j+1,\sigma}), \\
\tvarphi_{j+\frac{1}{2},\sigma} &= \frac{1}{2} (\tphi^R_{j,\sigma} +\tphi^L_{j+1,\sigma}), 
\end{split}
\end{align}
which satisfy the commutation relations, 
\begin{align} \label{eq:LinkFieldComm}
\begin{split}
[\partial_x \ttheta_{\ell,\sigma}(x), \tvarphi_{\ell',\sigma'}(x')] &= i\pi K_{\sigma \sigma'} \delta_{\ell \ell'} \delta(x-x'), \\
[\partial_x \ttheta_{\ell,\sigma}(x), \ttheta_{\ell',\sigma'}(x')] &= [\partial_x \tvarphi_{\ell,\sigma}(x), \tvarphi_{\ell',\sigma'}(x')] =0. 
\end{split}
\end{align}
The interactions \eqref{eq:InterIntHamSUN} take the following form, 
\begin{align}
\calH_\textrm{int} = \sum_{j=1}^{N_w-1} \sum_{\sigma=1}^{N-1} \int dx \ 2 t_\sigma \cos (2\ttheta_{j+\frac{1}{2},\sigma}). 
\end{align}
If the coupling constants $t_\sigma$ flow to the strong-coupling limit under the renormalization group transformation, the system will acquire the bulk gap. 
Indeed, all the fields $\ttheta_{\ell,\sigma}$ can be simultaneously pinned at minima of the cosine potentials and individually acquire a gap, while unpaired chiral Luttinger liquids are left at the outermost wires. 

\subsubsection{$SU(N)_1$ currents} \label{sec:SUN1Currents}

In order to extend the Abelian $SU(N-1)$-singlet FQH states to non-Abelian FQH states and also apply them to chiral spin liquids (see Sec.~\ref{sec:SUNSpinSystems}), we here unveil their connection with the current algebra of the underlying $SU(N)_1$ WZW CFT. 
To this end, we further introduce a linear transformation of the chiral bosonic fields \eqref{eq:ChiralField} by 
\begin{align} \label{eq:ChiralBosonSUN2}
\tchi^p_{j,l} = \sum_{\sigma=1}^{N-1} \omega^l_\sigma \tphi^p_{j,\sigma}, 
\end{align}
such that they satisfy the commutation relations, 
\begin{align}
[\partial_x \tchi^p_{j,l}(x), \tchi^{p'}_{j',l'}(x')] = 2ip\pi \delta_{pp'} \delta_{jj'} \delta_{ll'} \delta (x-x'). 
\end{align}
Such a transformation is easily found by interpreting the matrix $\bfK_{SU(N)}$ as a Gram matrix constructed from the scalar products of $N-1$ basis vectors $\bfalpha_\sigma$ \cite{Read90}, 
\begin{align} \label{eq:KtoAlpha}
(\bfK_{SU(N)})_{\sigma \sigma'} = \sum_{l=1}^{N-1} \alpha^l_\sigma \alpha^l_{\sigma'}. 
\end{align}
In fact, the vectors $\bfalpha_\sigma$ are positive (or negative) roots of $SU(N)$, which are not simple roots but still form primitive vectors of the $SU(N)$ root lattice. 
The vectors $\bfomega_\sigma$ in Eq.~\eqref{eq:ChiralBosonSUN2} are chosen to be the vectors dual to $\bfalpha_\sigma$, 
\begin{align} \label{eq:DefOmega}
\omega^l_\sigma = \sum_{\sigma'=1}^{N-1} (\bfK_{SU(N)}^{-1})_{\sigma \sigma'} \alpha^l_{\sigma'}. 
\end{align}
These vectors are weights of the fundamental representation of $SU(N)$ and form primitive vectors of the $SU(N)$ weight lattice. 
The vectors $\bfalpha_\sigma$ and $\bfomega_\sigma$ satisfy the following relations, 
\begin{subequations} \label{eq:RelAlphaOmega}
\begin{align}
\sum_{l=1}^{N-1} \omega^l_\sigma \omega^l_{\sigma'} &= (\bfK_{SU(N)}^{-1})_{\sigma \sigma'}, \\
\label{eq:OrthAlphaOmega}
\sum_{l=1}^{N-1} \alpha^l_\sigma \omega^l_{\sigma'} &= \delta_{\sigma \sigma'}, \\
\sum_{\sigma=1}^{N-1} \alpha_\sigma^l \omega_\sigma^{l'} &= \delta_{ll'}. 
\end{align}
\end{subequations}
Using the transformation \eqref{eq:ChiralBosonSUN2}, the interwire interactions \eqref{eq:InterIntSUN2} are expressed as 
\begin{align} \label{eq:InterIntSUN3}
\calO^t_{j,\sigma} = e^{i\bfalpha_\sigma \cdot \tbfchi^R_j -i\bfalpha_\sigma \cdot \tbfchi^L_{j+1}}, 
\end{align}
where $\tbfchi^p_j = (\tchi^p_{j,1}, \cdots, \tchi^p_{j,N-1})$. 

\begin{figure}
\includegraphics[clip,width=0.35\textwidth]{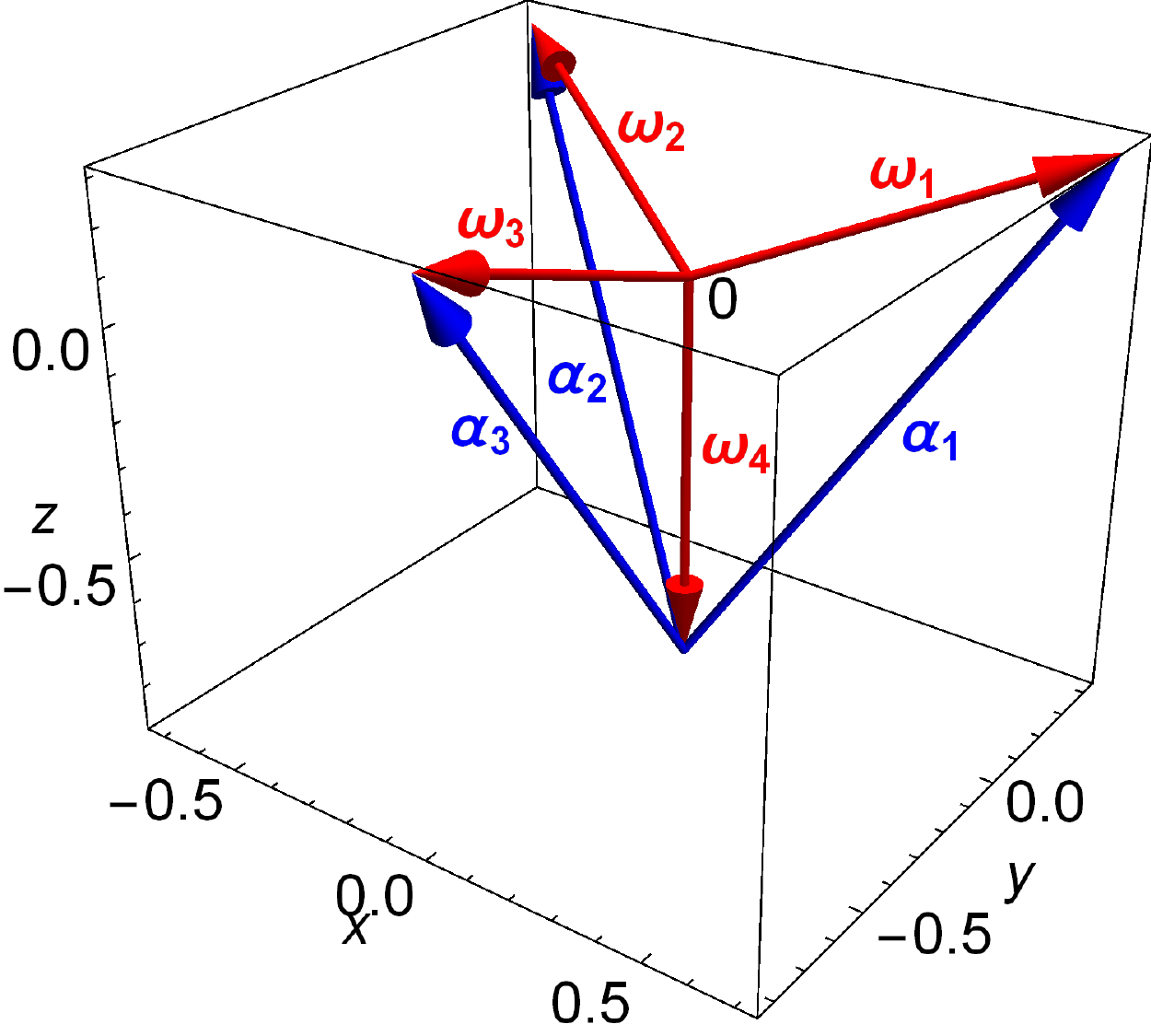}
\caption{Weight diagram of $SU(4)$. 
The red solid arrows represent the weight vectors $\bfomega_\sigma$. 
Combined with $\bfomega_4 = -\bfomega_1 -\bfomega_2 -\bfomega_3$, they form the weights of the fundamental representation of $SU(4)$. 
The blue solid arrows represent the root vectors $\bfalpha_\sigma$.}
\label{fig:SU4Weights}
\end{figure}

For instance, the vectors $\bfalpha_\sigma$ and $\bfomega_\sigma$ for $N=4$ are given by 
\begin{align} \label{eq:SU4Root}
\begin{split}
\bfalpha_1 &= \begin{pmatrix} \frac{1}{\sqrt{2}}, & \frac{1}{\sqrt{6}}, & \frac{2}{\sqrt{3}} \end{pmatrix}, \\
\bfalpha_2 &= \begin{pmatrix} -\frac{1}{\sqrt{2}}, & \frac{1}{\sqrt{6}}, & \frac{2}{\sqrt{3}} \end{pmatrix}, \\
\bfalpha_3 &= \begin{pmatrix} 0, & -\sqrt{\frac{2}{3}}, & \frac{2}{\sqrt{3}} \end{pmatrix}, 
\end{split}
\end{align}
and 
\begin{align}
\begin{split}
\bfomega_1 &= \begin{pmatrix} \frac{1}{\sqrt{2}}, & \frac{1}{\sqrt{6}}, & \frac{1}{2\sqrt{3}} \end{pmatrix}. \\
\bfomega_2 &= \begin{pmatrix} -\frac{1}{\sqrt{2}}, & \frac{1}{\sqrt{6}}, & \frac{1}{2\sqrt{3}} \end{pmatrix}, \\
\bfomega_3 &= \begin{pmatrix} 0, & -\sqrt{\frac{2}{3}}, & \frac{1}{2\sqrt{3}} \end{pmatrix}, 
\end{split}
\end{align}
These vectors are displayed on the weight diagram of $SU(4)$ in Fig.~\ref{fig:SU4Weights}. 
We note that the choice of $\{ \bfalpha_\sigma, \bfomega_\sigma \}$ is generally not unique. 
The above choice is made just for our convenience; if we assign the same charge $q$ for each component of boson, only the field $\tchi^p_{j,N-1}$ carries charge while the others only carry ``spins''---differences of the charges of different components. 
Details about our convention are provided in Appendix~\ref{app:RootWeightSUN}.

Suppose that the SLL Hamiltonian is appropriately tuned to be diagonal with the chiral bosonic fields: 
\begin{align} \label{eq:ChiralHam}
\calH_\textrm{SLL} = \sum_{j=1}^{N_w} \frac{v}{4\pi} \int dx \left[ (\partial_x \tbfchi^R_j)^2 +(\partial_x \tbfchi^L_j)^2 \right]. 
\end{align}
This free-boson Hamiltonian of each wire is equivalent to the Hamiltonian of the $SU(N)_1$ WZW CFT. 
Choosing the Cartan-Weyl basis, this can be written in the Sugawara form, 
\begin{align}
\calH_\textrm{SLL} 
=& \ \frac{v}{4\pi (N+1)} \sum_{j=1}^{N_w} \int dx \nonumber \\
&\times \Biggl[ \sum_{l=1}^{N-1} :\mathrel{H^l_j H^l_j}: +\sum_{\bfalpha \in \Delta_N} :\mathrel{E^\bfalpha_j E^{-\bfalpha}_j}: \nonumber \\
&+\sum_{l=1}^{N-1} :\mathrel{\bH^l_j \bH^l_j}: +\sum_{\bfalpha \in \Delta_N} :\mathrel{\bE^\bfalpha_j \bE^{-\bfalpha}_j}: \Biggr], 
\end{align}
where $H^l_j$ and $E^\bfalpha_j$ are the $SU(N)_1$ currents for the right-moving mode, $\bH^l_j$ and $\bE^\bfalpha_j$ are those for the left-moving mode, and $\Delta_N$ denotes the set of all roots of $SU(N)$. 
$H^l_j$ and $E^\bfalpha_j$ satisfy the $SU(N)_1$ current algebra, 
\begin{align}
\label{eq:SUNCurrentAlgebra1}
H^l_j(z) H^{l'}_{j'}(w) &\sim \frac{\delta_{jj'} \delta_{ll'}}{(z-w)^2}, \\
\label{eq:SUNCurrentAlgebra2}
H^l_j(z) E^\bfalpha_{j'}(w) &\sim \frac{\delta_{jj'} \alpha^l E^\bfalpha_j(w)}{z-w}, 
\end{align}
and
\begin{align} \label{eq:SUNCurrentAlgebra3}
&E^\bfalpha_j(z) E^\bfbeta_{j'}(w) \nonumber \\
&\sim 
\begin{cases}
\delta_{jj'} \left[ \dfrac{1}{(z-w)^2} +\dfrac{\sum_l \alpha^l H^l_j(w)}{z-w} \right] & (\bfalpha \cdot \bfbeta =-2) \\[10pt]
\dfrac{\delta_{jj'} \epsilon(\bfalpha,\bfbeta) E^{\bfalpha+\bfbeta}_j(w)}{z-w} & (\bfalpha \cdot \bfbeta =-1) \\
0 & \textrm{(otherwise)} 
\end{cases}
\end{align}
where $\epsilon(\bfalpha,\bfbeta)=\pm 1$. 
Similar relations hold for $\bH^l_j$ and $\bE^\bfalpha_j$, while the left and right currents are independent of each other. 
In terms of the chiral fields $\tbfchi^{R/L}_j$, the $SU(N)_1$ currents are given by (see Appendix~\ref{app:VertexRep})
\begin{align}
H^l_j(x) = \partial_x \tchi^R_{j,l}(x), \hspace{10pt} \bH^l_j(x) = \partial_x \tchi^L_{j,l}(x),
\end{align}
and
\begin{align}
E^{\pm \bfalpha}_j(x) = \frac{\pm i}{x_c} e^{\pm i\bfalpha \cdot \tbfchi^R_j(x)}, \hspace{10pt} \bE^{\pm \bfalpha}_j(x) = \frac{\pm i}{x_c} e^{\pm i\bfalpha \cdot \tbfchi^L(x)}, 
\end{align}
for positive roots $\bfalpha$. 
Thus the interwire interaction \eqref{eq:InterIntSUN} is further written in terms of the $SU(N)_1$ currents as 
\begin{align}
\calO^t_{j, \sigma} = x_c^2 E^{\bfalpha_\sigma}_j \bE^{-\bfalpha_\sigma}_{j+1}. 
\end{align}
Then the interaction Hamiltonian \eqref{eq:InterIntHamSUN} becomes 
\begin{align}
\calH_\textrm{int} = \sum_{j=1}^{N_w-1}\sum_{\sigma=1}^{N-1} \int dx \ t_\sigma x_c^2 \left( E^{\bfalpha_\sigma}_j \bE^{-\bfalpha_\sigma}_{j+1} +E^{-\bfalpha_\sigma}_j \bE^{\bfalpha_\sigma}_{j+1} \right). 
\end{align}
Since $\bfalpha_\sigma$ are primitive vectors of the $SU(N)$ root lattice, this interaction is sufficient to produce a bulk gap, while gapless edge states described by the chiral $SU(N)_1$ WZW CFT are left at the outermost wires $j=1$ and $N_w$. 

\subsection{Non-Abelian $SU(N-1)$-singlet FQH state} \label{sec:NonAbelianSUN}

\subsubsection{Interactions at $\nu_\textrm{tot}=k(N-1)/N$}

We next consider the coupled-wire construction of a bosonic non-Abelian $SU(N-1)$-singlet FQH state that has gapless edge states described by the chiral $SU(N)_k$ WZW CFT. 
Following a similar strategy to that for the $m=0$ NASS state in Sec.~\ref{sec:NASSState}, we introduce $k$ channels to each $(N-1)$-component bosonic wires and consider filling fraction $\nu_\textrm{tot}=k(N-1)/N$. 
The bosonic fields $\theta_{j,\sigma,a}$ and $\varphi_{j,\sigma,a}$ obey the commutation relations in Eq.~\eqref{eq:ChainFieldComm}. 
For the correlated hoppings between neighboring wires, we assume the following form: 
\begin{align} \label{eq:InterIntSUNk}
\calO^t_{j,\sigma,ab} =& \ \exp i\Biggl[ \varphi_{j,\sigma,a} -\varphi_{j+1,\sigma,b} \nonumber \\
&+\sum_{\sigma'=1}^{N-1} \left( M_{\sigma \sigma'} \theta_{j,\sigma,a} +M^T_{\sigma \sigma'} \theta_{j+1,\sigma',b} \right) \Biggr], 
\end{align}
where $\bfM$ is an even-integer interaction matrix satisfying $\frac{1}{2}(\bfM+\bfM^T)=\bfK_{SU(N)}$. 
By introducing the chiral fields, 
\begin{align}
\begin{split}
\tphi^R_{j,\sigma,a} &= \varphi_{j,\sigma,a} +\sum_{\sigma'=1}^{N-1} M_{\sigma \sigma'} \theta_{j,\sigma',a}, \\
\tphi^L_{j,\sigma,a} &= \varphi_{j,\sigma,a} -\sum_{\sigma'=1}^{N-1} M^T_{\sigma \sigma'} \theta_{j,\sigma',a}, 
\end{split}
\end{align}
and then defining, 
\begin{align} \label{eq:ChiralFieldSUNk}
\tchi^p_{j,l,a} = \sum_{\sigma=1}^{N-1} \omega^l_\sigma \tphi^p_{j,\sigma,a}, 
\end{align}
with the weight vectors $\bfomega_\sigma$ given in Eq.~\eqref{eq:DefOmega}, the interwire interactions \eqref{eq:InterIntSUNk} are written as 
\begin{align} \label{eq:InterIntSUNk2}
\calO^t_{j,\sigma,ab} = e^{i\bfalpha_\sigma \cdot \tbfchi^R_{j,a}-i\bfalpha_\sigma \cdot \tbfchi^L_{j+1,b}}, 
\end{align}
where the root vectors $\bfalpha_\sigma$ are again defined by Eq.~\eqref{eq:KtoAlpha} and $\tbfchi^p_{j,a}=(\tchi^p_{j,1,a}, \cdots, \tchi^p_{j,N-1,a})$. 
As discussed for $N=3$, these interactions are tunnelings of charge-$q$ particle excitations between the Abelian $SU(N-1)$-singlet states. 
The chiral fields \eqref{eq:ChiralFieldSUNk} satisfy the commutation relations, 
\begin{align}
[\partial_x \tchi^p_{j,l,a}(x), \tchi^{p'}_{j',l',a'}(x')] = 2ip \pi \delta_{pp'} \delta_{jj'} \delta_{ll'} \delta_{aa'} \delta(x-x'). 
\end{align}

\begin{figure}
\includegraphics[clip,width=0.48\textwidth]{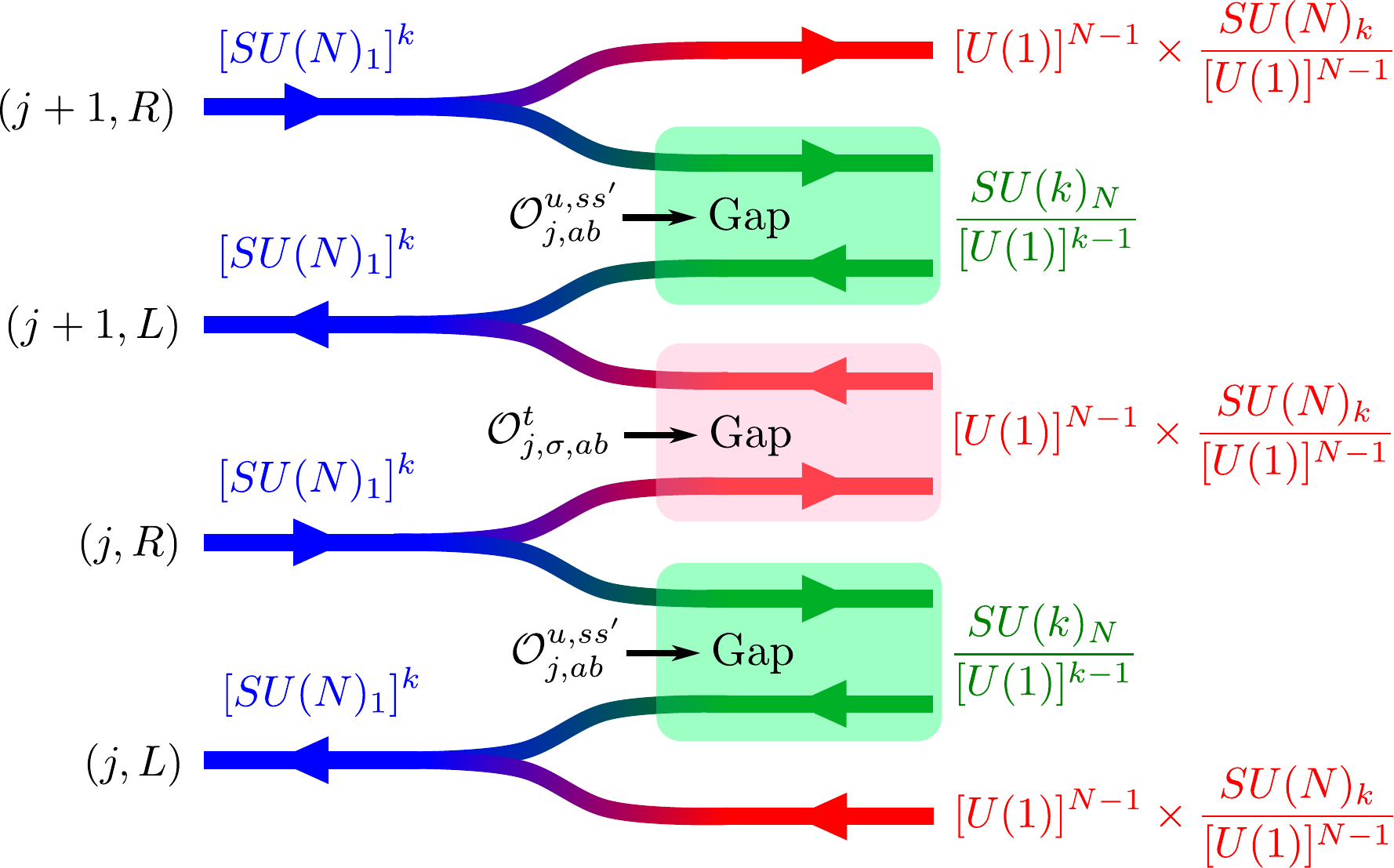}
\caption{Schematic view of the construction of a non-Abelian $SU(N-1)$-singlet FQH state.
The left and right $[SU(N)_1]^k$ CFTs from each wire are decomposed into the two sectors $[U(1)]^{N-1} \times SU(N)_k/[U(1)]^{N-1}$ and $SU(N)_k/[U(1)]^{k-1}$, which are respectively gapped by the interwire interactions $\calO^t_{j,\sigma,ab}$ and the intrawire interactions $\calO^{u,ss'}_{j,ab}$, leaving unpaired chiral gapless modes at the edges.}
\label{fig:NonAbelianSUNk}
\end{figure}

Let us suppose that the SLL Hamiltonian is of the form, 
\begin{align} \label{eq:SLLHamSUNk}
\calH_\textrm{SLL} = \sum_{j=1}^{N_w} \sum_{a=1}^k \frac{v}{4\pi} \int dx \left[ (\partial_x \tbfchi^R_{j,a})^2 +(\partial_x \tbfchi^L_{j,a})^2 \right]. 
\end{align}
As discussed in Sec.~\ref{sec:SUN1Currents}, this is the vertex representation of the $kN_w$ copies of the $SU(N)_1$ WZW CFT. 
For each wire, we adopt the following conformal embedding, 
\begin{align} \label{eq:EmbeddingSUNk}
[SU(N)_1]^k \sim [U(1)]^{N-1} \times \frac{SU(N)_k}{[U(1)]^{N-1}} \times \frac{SU(k)_N}{[U(1)]^{k-1}}, 
\end{align}
where we have used the equivalence of two coset CFTs $[SU(N)_1]^k/SU(N)_k$ and $SU(k)_N/[U(1)]^{k-1}$. 
This can be understood by checking that the energy-momentum tensor corresponding to $[SU(N)_1]^k$ is written as a sum of the energy-momentum tensors of the CFTs in r.h.s of Eq.~\eqref{eq:EmbeddingSUNk} (see Appendix~\ref{app:EMTensor}). 
The CFT $[U(1)]^{N-1}$ represents $N-1$ free bosons corresponding to the charge modes of the non-Abelian $SU(N-1)$-singlet FQH state. 
The coset CFT $SU(N)_k/[U(1)]^{N-1}$ is a Gepner parafermion CFT with central charge \cite{Gepner87}, 
\begin{align}
c=\frac{N(N-1)(k-1)}{k+N}, 
\end{align}
which corresponds to the neutral mode of the non-Abelian $SU(N-1)$-singlet state. 
The remaining coset CFT $SU(k)_N/[U(1)]^{k-1}$ is also a Gepner parafermion CFT with 
\begin{align}
c=\frac{(N-1)k(k-1)}{k+N}, 
\end{align}
but it does not enter the topological property of the resulting non-Abelian state. 
Then we consider the interaction Hamiltonian, 
\begin{align} \label{eq:IntHamSUNk}
\calH_\textrm{int} =& \int dx \left[ \sum_{j=1}^{N_w-1} \sum_{\sigma=1}^{N-1} \sum_{a,b=1}^k t_{\sigma,ab} \calO^t_{j,\sigma,ab}(x) \right. \nonumber \\
&\left. +\frac{1}{2} \sum_{j=1}^{N_w} \sum_{s,s'=1}^N \sum_{a<b} u^{ss'}_{ab} \calO^{u,ss'}_{j,ab}(x) +\textrm{H.c.} \right]. 
\end{align}
The interwire interaction $\calO^t_{j,\sigma,ab}$ is the correlated hopping given in Eq.~\eqref{eq:InterIntSUNk2} and is expected to open a gap in the sector $SU(N)_k \sim [U(1)]^{N-1} \times SU(N)_k/[U(1)]^{N-1}$. 
The intrawire interaction $\calO^{u,ss'}_{j,ab}$ is chosen to open a gap in the remaining sector $SU(k)_N/[U(1)]^{k-1}$. 
This coupled-wire system is schematically shown in Fig.~\ref{fig:NonAbelianSUNk}. 

As one might expect from Sec.~\ref{sec:IntSU3k}, for a large value of $N$, the desired intrawire interactions take quite complicated forms in terms of the original physical bosonic fields $\theta_{j, \sigma, a}$ and $\varphi_{j, \sigma, a}$. 
We here do not write down their explicit forms but only show that there exist such interactions respecting the symmetry constraints. 
We want interactions of the following form: 
\begin{align} \label{eq:IntraIntSUNk}
\calO^{u,ss'}_{j,ab} = e^{i\bfomega_s \cdot (\tbfchi^R_{j,a}-\tbfchi^R_{j,b}) -i\bfomega_{s'} \cdot (\tbfchi^L_{j,a}-\tbfchi^L_{j,b})}, 
\end{align}
where $s,s'=1,\cdots,N$. 
Here the vectors $\bfomega_1, \cdots, \bfomega_{N-1}$ are those given in Eq.~\eqref{eq:DefOmega}, while $\bfomega_N$ is given by 
\begin{align}
\bfomega_N = -\sum_{\sigma=1}^{N-1} \bfomega_\sigma. 
\end{align}
These $(N-1)$-dimensional vectors $\bfomega_s$ satisfy the relations, 
\begin{align}
\begin{split}
\sum_{s=1}^N \omega^l_s &= 0, \\
\sum_{s=1}^N \omega^l_s \omega^{l'}_s &= \delta_{ll'}, \\
\sum_{l=1}^{N-1} \omega^l_s \omega^l_{s'} &= \delta_{ss'} -\frac{1}{N}, 
\end{split}
\end{align}
and thus form the regular simplex in the $(N-1)$-dimensional space (this has been shown in Fig.~\ref{fig:SU4Weights} for $N=4$). 
Indeed, they are weight vectors in the fundamental representation of $SU(N)$. 
The interactions \eqref{eq:IntraIntSUNk} are made of the quasiparticle excitations of $k$ copies of the Abelian $SU(N-1)$-singlets state and obviously satisfy the (separate) charge and momentum conservations. 
In terms of the original bosonic fields $\theta_{j,\sigma,a}$ and $\varphi_{j,\sigma,a}$, it can be written as 
\begin{align}
\calO^{u,ss'}_{j,ab} =& \ \exp i\Biggl[ \sum_{\sigma=1}^{N-1} (\bfomega_s \cdot \bfomega_\sigma -\bfomega_{s'} \cdot \bfomega_\sigma) (\varphi_{j,\sigma,a} -\varphi_{j,\sigma,b}) \nonumber \\
&+ \sum_{\sigma,\sigma'=1}^{N-1} (\bfomega_s \cdot \bfomega_\sigma M_{\sigma \sigma'} +\bfomega_{s'} \cdot \bfomega_\sigma M^T_{\sigma \sigma'}) \nonumber \\
&\times (\theta_{j,\sigma',a}-\theta_{j,\sigma',b}) \Biggr]. 
\end{align}
For given $s$ and $s'$, the coefficients of $\varphi_{j,\sigma,a}$ and $\theta_{j,\sigma',a}$ are found to be
\begin{align}
\bfomega_s \cdot \bfomega_\sigma -\bfomega_{s'} \cdot \bfomega_\sigma = \delta_{s \sigma} -\delta_{s' \sigma}, 
\end{align}
and
\begin{align}
&\sum_{\sigma=1}^{N-1} (\bfomega_s \cdot \bfomega_\sigma M_{\sigma \sigma'} +\bfomega_{s'} \cdot \bfomega_\sigma M^T_{\sigma \sigma'}) \nonumber \\
&= (1-\delta_{sN}) M_{s \sigma'} +(1-\delta_{s' N}) M^T_{s' \sigma'} -2
\end{align}
Since the latter is even integer, the intrawire interactions can be constructed from bosonic operators. 

\subsubsection{$SU(N)_k$ currents and parafermions}
\label{sec:SUNkCurrent}

We now argue that the interwire interaction \eqref{eq:InterIntSUNk2} opens a gap in the $[U(1)]^{N-1} \times SU(N)_k/[U(1)]^{N-1}$ sector, while the intrawire interaction \eqref{eq:IntraIntSUNk} opens a gap in the $SU(k)_N/[U(1)]^{k-1}$ sector. 
This is most easily understood by expressing those interactions in terms of the field content of the corresponding CFTs. 

Let us first introduce the $SU(N)_k$ currents. 
Using $k$ copies of the $SU(N)_1$ currents, the $SU(N)_k$ currents associated with a root $\bfalpha$ of $SU(N)$ are given by
\begin{align} \label{eq:SUNkCurrentCartan}
\calE^\bfalpha_j = \sum_{a=1}^k E^\bfalpha_{j,a}, \hspace{10pt}
\bcalE^\bfalpha_j = \sum_{a=1}^k \bE^\bfalpha_{j,a}, 
\end{align}
while the $N-1$ Cartan currents are given by 
\begin{align} \label{eq:SUNkCurrentLadder}
\calH^l_j = \sum_{a=1}^k H^l_{j,a}, \hspace{10pt}
\bcalH^l_j = \sum_{a=1}^k \bH^l_{j,a}. 
\end{align}
They satisfy the $SU(N)_k$ current algebra, 
\begin{align}
\label{eq:SUNkCurrentAlgebra1}
&\calH^l_j(z) \calH^{l'}_{j'}(w) \sim \frac{k\delta_{jj'} \delta_{ll'}}{(z-w)^2}, \\
\label{eq:SUNkCurrentAlgebra2}
&\calH^l_j(z) \calE^\bfalpha_{j'}(w) \sim \frac{\delta_{jj'} \alpha^l \calE^\bfalpha_j(w)}{z-w}, \\
\label{eq:SUNkCurrentAlgebra3}
&\calE^\bfalpha_j(z) \calE^\bfbeta_{j'}(w) \nonumber \\
&\sim 
\begin{cases}
\delta_{jj'} \left[ \dfrac{k}{(z-w)^2} +\dfrac{\sum_l \alpha^l \calH^l_j(w)}{z-w} \right] & (\bfalpha \cdot \bfbeta =-2) \\[10pt]
\dfrac{\delta_{jj'} \epsilon(\bfalpha,\bfbeta) \calE^{\bfalpha+\bfbeta}_j(w)}{z-w} & (\bfalpha \cdot \bfbeta =-1) \\
0 & \textrm{(otherwise)} 
\end{cases}
\end{align}
for the right currents. 
The left currents $\bcalH^l_j$ and $\bcalE^\bfalpha_j$ also satisfy a similar algebra. 

Using the vertex representation of the $SU(N)_1$ current associated with a positive root $\bfalpha$ for each copy, 
\begin{align}
\begin{split}
E^{\pm \bfalpha}_{j,a}(x) &= \frac{\pm i}{x_c} e^{\pm i\bfalpha \cdot \tbfchi^R_{j,a}(x)}, \\
\bE^{\pm \bfalpha}_{j,a}(x) &= \frac{\pm i}{x_c} e^{\pm i\bfalpha \cdot \tbfchi^L_{j,a}(x)}, 
\end{split}
\end{align}
the interwire interaction \eqref{eq:InterIntSUNk2} can be written as 
\begin{align}
\calO^t_{j,\sigma,ab} = x_c^2 E^{\bfalpha_\sigma}_{j,a} \bE^{-\bfalpha_\sigma}_{j+1,b}. 
\end{align}
If the coupling constants are fine tuned such that $t_{\sigma,ab} \equiv t_\sigma$, this can be represented solely by the $SU(N)_k$ currents. 
Thus we find 
\begin{align} \label{eq:InterIntSUNk3}
\sum_{a,b=1}^k \calO^t_{j,\sigma,ab} = x_c^2 \calE^{\bfalpha_\sigma}_j \bcalE^{-\bfalpha_\sigma}_{j+1}. 
\end{align}
It obviously acts only on the nonchiral $SU(N)_k$ WZW CFT composed of the neighboring wires $j$ and $j+1$ and will produce a gap. 

In order to further translate them into the language of the bosonic CFT $[U(1)]^{N-1}$ and the Gepner parafermion CFT $SU(N)_k/[U(1)]^{N-1}$, we introduce $N-1$ charge modes and $(k-1)(N-1)$ neutral modes for each wire in terms of the chiral bosonic fields \eqref{eq:ChiralFieldSUNk}, 
\begin{align} \label{eq:ChargeNeutralFieldSUNk}
\begin{split}
\tX^p_{j,l} &= \frac{1}{\sqrt{k}} \sum_{a=1}^k \tchi^p_{j,l,a}, \\
\tY^{p,\mu}_{j,l} &= \sum_{a=1}^k W^\mu_a \tchi^p_{j,l,a}, 
\end{split}
\end{align}
where $\bfW_a$ are $k$ $(k-1)$-dimensional vectors satisfying
\begin{align} \label{eq:VectorW}
\begin{split}
\sum_{a=1}^k W^\mu_a &= 0, \\
\sum_{a=1}^k W^\mu_a W^\nu_a &= \delta_{\mu \nu}, \\
\sum_{\mu=1}^{k-1} W^\mu_a W^\mu_b &= \delta_{ab}-\frac{1}{k}. 
\end{split}
\end{align}
Specifically, such vectors can be chosen as, for example, 
\begin{align} \label{eq:SpecificVectorW}
W^\mu_a = \begin{cases} \frac{1}{\sqrt{\mu(\mu+1)}} & (\mu \geq a) \\ -\sqrt{\frac{\mu}{\mu+1}} & (\mu=a-1) \\ 0 & (\mu < a-1) \end{cases}.
\end{align}
These new fields satisfy the commutation relations, 
\begin{align} \label{eq:ChiralFieldSUNkComm}
\begin{split}
[\partial_x \tX^p_{j,l}(x), \tX^{p'}_{j',l'}(x')] &= 2ip\pi \delta_{pp'} \delta_{jj'} \delta_{ll'} \delta(x-x'), \\
[\partial_x \tY^{p,\mu}_{j,l}(x), \tY^{p',\mu'}_{j',l'}(x')] &= 2ip\pi \delta_{pp'} \delta_{jj'} \delta_{ll'} \delta_{\mu \mu'} \delta(x-x'), \\
[\partial_x \tX^p_{j,l}(x), \tY^{p',\mu}_{j',l'}(x')] &=0. 
\end{split}
\end{align}
Now the SLL Hamiltonian \eqref{eq:SLLHamSUNk} becomes 
\begin{align}
\calH_\textrm{SLL} =& \ \frac{v}{4\pi} \sum_{j=1}^{N_w} \int dx \biggl[ (\partial_x \tbfX^R_j)^2 +(\partial_x \tbfX^L_j)^2 \nonumber \\
&+\sum_{\mu=1}^{k-1} \left\{ (\partial_x \tbfY^{R,\mu}_j)^2 +(\partial_x \tbfY^{L,\mu}_j)^2 \right\} \biggr], 
\end{align}
where $\tbfX^p_j = (\tX^p_{j,1}, \cdots, \tX^p_{j,N-1})$ and $\tbfY^{p,\mu}_j = (\tY^{p,\mu}_{j,1}, \cdots, \tY^{p,\mu}_{j,N-1})$. 

Using these fields, the $SU(N)_k$ currents associated with a positive root $\bfalpha$ are written as 
\begin{align} \label{eq:VertexRepSUNkCur}
\begin{split}
\calE_j^{\pm \bfalpha}(x) &= \frac{\sqrt{k} e^{\pm i\pi/2k}}{x_c^{1/k}} e^{\pm \frac{i}{\sqrt{k}} \bfalpha \cdot \tbfX^R_j(x)} \Psi^{\pm \bfalpha,1}_j(x), \\
\bcalE_j^{\pm \bfalpha}(x) &=\frac{\sqrt{k} e^{\pm i\pi/2k}}{x_c^{1/k}} e^{\pm \frac{i}{\sqrt{k}} \bfalpha \cdot \tbfX^L_j(x)} \bPsi^{\pm \bfalpha,1}_j(x). 
\end{split}
\end{align}
Here $\Psi^{\bfalpha,1}_j$ ($\bPsi^{\bfalpha,1}_j$) is the right (left) $SU(N)_k/[U(1)]^{N-1}$ parafermionic field associated with a root $\bfalpha$ of $SU(N)$ and the fundamental representation of $SU(k)$, whose conformal weight is $1-1/k$. 
Their vertex representations involve only the neutral modes $\tbfY^{p,\mu}_j$ and are given by 
\begin{align} \label{eq:VertexRepSUNkPara}
\begin{split}
\Psi^{\pm \bfalpha,1}_j(x) &= \frac{e^{\pm i\pi(k-1)/2k}}{\sqrt{k} x_c^{1-1/k}} \sum_{a=1}^k e^{\pm i\bfalpha \bfW_a \cdot \tbfY^R_j(x)}, \\
\bPsi^{\pm \bfalpha,1}_j(x) &= \frac{e^{\pm i\pi(k-1)/2k}}{\sqrt{k} x_c^{1-1/k}} \sum_{a=1}^k e^{\pm i\bfalpha \bfW_a \cdot \tbfY^L_j(x)}, 
\end{split}
\end{align}
where we have introduced a shorthand notation, 
\begin{align} \label{eq:ShortNote}
\bfalpha \bfW \cdot \tbfY^p_j \equiv \sum_{l=1}^{N-1} \sum_{\mu=1}^{k-1} \alpha^l W^\mu \tY^{p,\mu}_{j,l}. 
\end{align}
In fact, the vectors $\bfW_a$ satisfying Eq.~\eqref{eq:VectorW} generate a set of weights in the fundamental representation of $SU(k)$. 
Thus Eq.~\eqref{eq:VertexRepSUNkPara} is consistent with the vertex representation of the $SU(N)_k/[U(1)]^{N-1}$ parafermionic field in Ref.~\cite{Dunne89} with an appropriate choice of parafermionic cocycle (see Appendix~\ref{app:SUNkPara}).
Then the interwire interaction \eqref{eq:InterIntSUNk3} is finally written as 
\begin{align}
\sum_{a,b=1}^k \calO^t_{j,\sigma,ab} = k x_c^{2-2/k} e^{i\bfalpha_\sigma \cdot (\tbfX^R_j-\tbfX^L_{j+1})/\sqrt{k}} \Psi^{\bfalpha_\sigma,1}_j \bPsi^{-\bfalpha_\sigma,1}_{j+1}. 
\end{align}
This indicates that the charge part involving $\tbfX^p_j$ acts only on the $[U(1)]^{N-1}$ bosonic CFT, while the neutral part involving $\tbfY^{p,\mu}_j$ acts only on the $SU(N)_k/[U(1)]^{N-1}$ parafermion CFT. 
We also note that since the Cartan generators of the $SU(N)_1$ currents are expressed as 
\begin{align}
H^l_{j,a}(x) = \partial_x \tchi^R_{j,l,a}(x), \hspace{10pt}
\bH^l_{j,a}(x) = \partial_x \tchi^L_{j,l,a}(x), 
\end{align}
those of the $SU(N)_k$ currents are given by 
\begin{align}
\calH^l_j(x) = \sqrt{k} \partial_x \tX^R_{j,l}(x), \hspace{10pt}
\bcalH^l_j(x) = \sqrt{k} \partial_x \tX^L_{j,l}(x). 
\end{align}
Hence they involve only the charge modes. 

We next see that the intrawire interaction \eqref{eq:IntraIntSUNk} can be written in terms of $SU(k)_N/[U(1)]^{k-1}$ parafermionic fields when $u^{ss'}_{ab} \equiv u_{ab}$. 
Using Eq.~\eqref{eq:ChargeNeutralFieldSUNk}, we can write the interaction only in terms of the neutral modes as 
\begin{align} \label{eq:IntraIntSUNk2}
\sum_{s,s'=1}^N \calO^{u,ss'}_{j,ab} = \sum_{s,s'=1}^N e^{i\bfomega_s (\bfW_a-\bfW_b) \cdot \tbfY^R_j -i\bfomega_{s'} (\bfW_a-\bfW_b) \cdot \tbfY^L_j}. 
\end{align}
Here the vectors $\bfomega_s$ are the weights in the fundamental representation of $SU(N)$, while the vectors $\bfW_a-\bfW_b$ for $a \neq b$ span all the roots of $SU(k)$ since $\bfW_a$ are weights in the fundamental representation of $SU(k)$. 
Then we can identify this interaction as the products of left and right $SU(k)_N/[U(1)]^{k-1}$ primary fields with scaling dimension $2(1-1/N)$ (see Appendix~\ref{app:SUkNPara}), 
\begin{align}
\sum_{s,s'=1}^N \calO^{u,ss'}_{j,ab} = e^{i\pi (1-N)/N} x_c^{2-2/N} \Xi^\bfA_j \bXi^{\bfA \dagger}_j, 
\end{align}
which is associated with a root of $SU(k)$, $\bfA = \bfW_a-\bfW_b$. 
The chiral primary fields are given by
\begin{align}
\begin{split}
\Xi^\bfA_j(x) &= \frac{e^{i\pi (N-1)/2N}}{x_c^{1-1/N}} \sum_{s=1}^N e^{-\frac{2i\pi s}{N}} e^{i\bfomega_s \bfA \cdot \tbfY^R_j(x)}, \\
\bXi^\bfA_j(x) &= \frac{e^{i\pi (N-1)/2N}}{x_c^{1-1/N}} \sum_{s=1}^N e^{-\frac{2i\pi s}{N}} e^{i\bfomega_s \bfA \cdot \tbfY^L_j(x)}.  
\end{split}
\end{align}
Then we can write 
\begin{align} \label{eq:IntraIntSUNk3}
&\sum_{a<b}^k \sum_{s,s'=1}^N u_{ab} \calO^{u,ss'}_{j,ab} \nonumber \\
&= e^{i\pi (1-N)/N} x_c^{2-2/N} \sum_{\bfA \in \Delta^+_k} \frac{u_\bfA}{2} \Xi^\bfA_j \bXi^{\bfA \dagger}_j, 
\end{align}
where $\Delta^+_k$ is the set of all positive roots of $SU(k)$ and we have written $u_{ab} \equiv u_\bfA$. 
We remark that, as shown in Appendix~\ref{app:SUkNPara}, $\Xi^\bfA_j$ and $\bXi^\bfA_j$ are primary fields of the chiral $SU(k)_N/[U(1)]^{k-1}$ CFTs with conformal weight $1-1/N$, but they \emph{cannot} generate the parafermionic algebra in the respective \emph{chiral} sectors. 
Instead, the nonchiral product $\Xi^\bfA_j \bXi^{\bfA \dagger}_j$ generates the parafermionic algebra of the \emph{nonchiral} $SU(k)_N/[U(1)]^{k-1}$ CFT and therefore serves as the (unnormalized) first parafermionic field associated with a root $\bfA$ of $SU(k)$ and the fundamental representation of $SU(N)$. 

This interaction acts only on the $SU(k)_N/[U(1)]^{k-1}$ parafermion CFT residing in each wire and is a strictly relevant perturbation with scaling dimension $2(1-1/N)$. 
For $k=2$, the interaction \eqref{eq:IntraIntSUNk3} is known to be an integrable deformation of the $\mathbb{Z}_N$ parafermion theory \cite{Fateev91a,Fateev91b}.
When $N$ is even, a mass gap is generally produced for any sign of the coupling constant \cite{Fateev91a}. 
On the other hand, for odd $N$, a mass is generated when the coupling constant is \emph{negative}, while there is a massless flow to the minimal unitary CFT $\mathcal{M}_{N+1}$ when the coupling constant is positive \cite{Fateev91a,Fateev91b}.
For $k \geq 3$, to the best of our knowledge, we are not aware of any investigation about the integrable deformation of the Gepner parafermion CFT, which generalizes results known for the $\mathbb{Z}_k$ parafermion theory. 
However, it can be shown that there exists a massive flow for $N=2$ and when $u_\bfA \equiv u <0$, by the analogy with the $\mathbb{Z}_k$ statistical-mechanical model \cite{Teo14} or more rigorously by the self-dual sine-Gordon model \cite{Lecheminant07,Lecheminant12}. 
For $N \geq 3$ and $k \geq 3$, no rigorous proof of the massive flow is available. 
Nevertheless, we expect that the relevant interaction \eqref{eq:IntraIntSUNk3} will produce a gap in the $SU(k)_N/[U(1)]^{k-1}$ parafermion sector for $u_\bfA \equiv u <0$ from the analogy with a $\mathbb{Z}_k^{N-1}$ statistical-mechanical model as discussed below in Sec.~\ref{sec:NalCriticalPoint}. 
Therefore, we conclude that if the coupling constants are fine tuned in such a way that the interaction Hamiltonian \eqref{eq:IntHamSUNk} takes the form, 
\begin{align}
\calH_\textrm{int} =& \ \int dx \Biggl[ \sum_{j=1}^{N_w-1} \sum_{\sigma=1}^{N-1} k x_c^{2-2/k} t_\sigma e^{i\bfalpha_\sigma \cdot (\tbfX^R_j-\tbfX^L_{j+1})/\sqrt{k}} \nonumber \\
&\times \Psi^{\bfalpha_\sigma,1}_j \bPsi^{-\bfalpha_\sigma,1}_{j+1} \nonumber \\
&+\frac{u}{2} e^{i\pi(1-N)/N} x_c^{2-2/N} \sum_{j=1}^{N_w} \sum_{\bfA \in \Delta^+_k} \Xi^\bfA_j \bXi^{\bfA \dagger}_j +\textrm{H.c.} \Biggr], 
\end{align}
it produce a bulk gap but leaves gapless edge modes described by the chiral $SU(N)_k$ WZW CFTs at $j=1$ and $N_w$. 

\subsection{Bulk quasiparticles and $\mathbb{Z}_k^{N-1}$ statistical mechanical model}
\label{sec:SUNkQuasiparticles}

\subsubsection{Quasiparticle operators}

We again consider the $2k_F$ backscattering operators in terms of the original bosonic fields,
\begin{align}
B_{j,\sigma,a}(x) = e^{2i\theta_{j,\sigma,a}(x)}, 
\end{align}
which may transfer quasiparticles at $x$ from the link $j+1/2$ to $j-1/2$. 
Using the chiral fields, we can rewrite them as 
\begin{align}
B_{j,\sigma,a} &= e^{i \sum_{\sigma'=1}^{N-1} (\bfK^{-1}_{SU(N)})_{\sigma \sigma'} (\tphi^R_{j,\sigma',a} -\tphi^L_{j,\sigma',a})} \nonumber \\
&= e^{i\bfomega_\sigma \cdot (\tbfchi^R_{j,a} -\tbfchi^L_{j,a})} \nonumber \\
&= e^{\frac{i}{\sqrt{k}} \bfomega_\sigma \cdot (\tbfX^R_j -\tbfX^L_j)} e^{i\bfomega_\sigma \bfW_a \cdot (\tbfY^R_j -\tbfY^L_j)}. 
\end{align}
Thus the quasiparticle operators may be defined as 
\begin{align} \label{eq:QuasiparticleOpSUNk}
\begin{split}
\Psi_{\textrm{QP},j+\frac{1}{2},\sigma,a}^{R\dagger} &= e^{\frac{i}{\sqrt{k}} \bfomega_\sigma \cdot \tbfX^R_j} \Sigma^{R\dagger}_{j,\sigma,a}, \\
\Psi_{\textrm{QP},j+\frac{1}{2},\sigma,a}^{L\dagger} &= e^{\frac{i}{\sqrt{k}} \bfomega_\sigma \cdot \tbfX^L_{j+1}} \Sigma^{L\dagger}_{j+1,\sigma,a}, 
\end{split}
\end{align}
where
\begin{align}
\Sigma^{R/L\dagger}_{j,\sigma,a} = e^{i\bfomega_\sigma \bfW_a \cdot \tbfY^{R/L}_j}. 
\end{align}

When $k=1$, and thus for the Abelian case, the quasiparticle operators are just reduced to 
\begin{align}
\Psi_{\textrm{QP},j+\frac{1}{2},\sigma}^{R\dagger} = e^{i\sum_{\sigma'=1}^{N-1} (\bfK^{-1}_{SU(N)})_{\sigma \sigma'} \tphi^R_{j,\sigma'}}, 
\end{align}
and similarly for the left part. 
The operator labeled by $\sigma$ creates a quasiparticle with charge $Q^\rho_\sigma = q(\delta_{\sigma \rho}-1/N)$, where $Q^\rho_\sigma$ is the quasiparticle charge associated with the $\rho$-th component of boson. 
If we define the total charge as the sum of charges over all the $N-1$ components, $Q^\textrm{tot}_\sigma = \sum_{\rho=1}^{N-1} Q^\rho_\sigma$, these operators create quasiparticles with the same total charge $Q^\textrm{tot}_\sigma = q/N$. 
This is consistent with the one calculated by the Chern-Simons theory, $Q^\textrm{tot}_\sigma = q\bft^T \bfK_{SU(N)}^{-1} \bfl_\sigma$ with the substitutions of the charge vector $(\bft)_\sigma=1$ and the quasiparticle vectors $(\bfl_\sigma)_{\sigma'} =\delta_{\sigma \sigma'}$. 
For $k \geq 2$, there are $k$ quasiparticle operators for each label $\sigma$. 
Their actions on the charge sector are, however, the same for any $k$ and equally create quasiparticles with total charge $q/N$. 

Now we turn to the neutral sector. 
As we have seen, the neutral modes $\tbfY^{p,\mu}_j$ are contained in the intrawire interaction as well as the interwire interaction, both of which are supposed to flow to the strong-coupling limit. 
Therefore, the actions of the quasiparticle operators on the neutral sector are nontrivial. 
We thus again follow the strategy of Ref.~\cite{Teo14}: we investigate how the operators $\Sigma^{R/L}_{j,\sigma,a}$ behave when the intrawire interaction opens a gap. 
In the following, we argue that the nonchiral products of these operators, 
\begin{align} \label{eq:OpSigma}
\Sigma^n_{j,\sigma,a} &\equiv e^{i\pi(N-1)(k-1)/Nk} \Sigma^{R\dagger}_{j,\sigma,a} \Sigma^L_{j,\sigma,a} \nonumber \\
&= e^{i\bfomega_\sigma \bfW_a \cdot (\tbfY^R_j -\tbfY^L_j)}, 
\end{align}
behave as order parameters detecting the $\mathbb{Z}_k^{N-1}$ symmetry breaking. 

This result implies that the chiral operators $\Sigma^{R/L}_{j,\sigma,a}$ contain the fundamental spin field of each chiral $SU(N)_k/[U(1)]^{N-1}$ CFT. 
If the Hamiltonian of each wire is fine-tuned to be the SLL Hamiltonian \eqref{eq:SLLHamSUNk}, these operators become primary fields of the $SU(N)_k/[U(1)]^{N-1} \times SU(k)_N/[U(1)]^{k-1}$ CFT with conformal weight $(N-1)(k-1)/2Nk$. 
We conjecture that these operators are the products of the fundamental spin field of the $SU(N)_k/[U(1)]^{N-1}$ parafermion CFT and that of the $SU(k)_N/[U(1)]^{k-1}$ parafermion CFT, whose conformal weights are given by $(N-1)(k-1)/2k(N+k)$ and $(N-1)(k-1)/2N(N+k)$, respectively.

\subsubsection{Symmetry of the Gepner parafermion} \label{sec:SymPara}

Following Ref.~\cite{Dunne89}, we can easily extract the symmetry of the Gepner parafermion CFT from its vertex representation. 
From the vertex representation of the energy-momentum tensor in Eq.~\eqref{eq:SUNkParaEMTensor}, the $SU(N)_k/[U(1)]^{N-1}$ parafermion CFT is invariant under
\begin{align} \label{eq:SymSUNkPara}
\tY^{R,\mu}_{j,l} \to \tY^{R,\mu}_{j,l} +2\pi \upsilon^l V^\mu, 
\end{align}
where $\bfupsilon$ and $\bfV$ are arbitrary vectors on the weight lattices of $SU(N)$ and $SU(k)$, respectively. 
A vector on the $SU(k)$ weight lattice can be written as 
\begin{align} \label{eq:SUNWeightParam}
\bfV = m \bfW_1 +\bfB, 
\end{align}
where $m = 0, \cdots, k-1$ and $\bfB$ is some vector on the $SU(k)$ root lattice. 
Substituting Eqs.~\eqref{eq:SymSUNkPara} and \eqref{eq:SUNWeightParam} into the vertex representation of the $SU(N)_k/[U(1)]^{N-1}$ parafermionic field in Eq.~\eqref{eq:VertexRepSUNkPara}, we find \cite{Dunne89}
\begin{align}
\Psi^{\bfalpha,1}_j \to e^{-\frac{2i \pi m}{k} (\bfalpha \cdot \bfupsilon)} \Psi^{\bfalpha,1}_j, 
\end{align}
where we have used $\bfW_1 \cdot \bfW_a = \delta_{a,1} -1/k$ and $\bfalpha \cdot \bfupsilon \in \mathbb{Z}$. 
Since $\bfalpha$ is a root of $SU(N)$ and the $SU(N)$ root lattice is spanned by $N-1$ primitive vectors, this indicates that the $SU(N)_k/[U(1)]^{N-1}$ parafermion CFT has a $\mathbb{Z}_k^{N-1}$ symmetry. 
For $N=2$, this reduces to the well-known $\mathbb{Z}_k$ symmetry of the $SU(2)_k/U(1)$ parafermion \cite{Zamolodchikov85}. 
A similar argument is also applied to the left-moving sector. 

In each chiral sector of the $SU(N)_k/[U(1)]^{k-1}$ parafermion CFT, there are in general $N$ spin fields that are associated with the weights in the fundamental representation of $SU(N)$ and have conformal weight $(N-1)(k-1)/2k(N+k)$. 
Two of these spin fields from the left- and right-moving sectors may be combined into a nonchiral spin field, which serves as a physical order parameter to partially detect the $\mathbb{Z}_k^{N-1}$ symmetry breaking. 
There will be $N(N-1)$ such spin fields associated with the roots of $SU(N)$ and may transform under the $\mathbb{Z}_k^{N-1}$ symmetry in the same way as the parafermions, 
\begin{align}
\sigma_{\rho,j} \to e^{-\frac{2i\pi m}{k} (\bfalpha_\rho \cdot \bfupsilon)} \sigma_{\rho,j}, 
\end{align}
where $\{ \bfalpha_\rho \}$ is the set of roots of $SU(N)$. 
Among such spin fields, only those related to primitive vectors of the $SU(N)$ root lattice are independent in the sense of order parameter. 
We expect that the operators in Eq.~\eqref{eq:OpSigma} are related to such ``primitive'' spin fields. 

\subsubsection{Identification of spin fields}

We consider a single-wire Hamiltonian with the intrawire interaction \eqref{eq:IntraIntSUNk}. 
Focusing on the neutral sector, the Hamiltonian is given by 
\begin{align} \label{eq:SingleWireHam}
\calH_j =& \ \frac{v}{4\pi} \int dx \sum_{\mu=1}^{k-1} \bigl[ (\partial_x \tbfY^{R,\mu}_j)^2 + (\partial_x \tbfY^{L,\mu}_j)^2 \bigr] \nonumber \\
&+\int dx \ \calV [\tbfY^{R/L,\mu}_j], 
\end{align}
where the intrawire interaction is given by 
\begin{align} \label{eq:IntraIntNeutral}
& \calV [\tbfY^{R/L,\mu}_j] \nonumber \\
&= \sum_{s ,s'=1}^N \sum_{\bfA \in \Delta^+_k} u^{ss'}_\bfA \cos (\bfomega_s \bfA \cdot \tbfY^R_j -\bfomega_{s'} \bfA \cdot \tbfY^L_j). 
\end{align}
In the following, we omit the wire index $j$ for brevity. 
We now introduce nonchiral fields by 
\begin{align}
\begin{split}
\tPhi^\mu_l &= \frac{1}{2} (\tY^{R,\mu}_l +\tY^{L,\mu}_l), \\
\tTheta^\mu_l &= \frac{1}{2} (\tY^{R,\mu}_l -\tY^{L,\mu}_l), 
\end{split}
\end{align}
which satisfy the commutation relations, 
\begin{align}
[\partial_x \tTheta^\mu_l(x), \tPhi^{\mu'}_{l'}(x')] = i\pi \delta_{\mu \mu'} \delta_{ll'} \delta (x-x'). 
\end{align}
In terms of these fields, the intrawire interaction \eqref{eq:IntraIntNeutral} is written as 
\begin{align} \label{eq:CosinePotSUNk}
\calV[\tbfPhi^\mu, \tbfTheta^\mu] =& \ \sum_{s=1}^N \sum_{\bfA \in \Delta^+_k} u^{ss}_\bfA \cos (2\bfomega_s \bfA \cdot \tbfTheta) \nonumber \\
& +\sum_{s \neq s'} \sum_{\bfA \in \Delta^+_k} u^{ss'}_\bfA \cos \Bigl[ (\bfomega_s -\bfomega_{s'}) \bfA \cdot \tbfPhi \nonumber \\
& + (\bfomega_s +\bfomega_{s'}) \bfA \cdot \tbfTheta \Bigr], 
\end{align}
where we have used the shorthand notations similar to Eq.~\eqref{eq:ShortNote}. 

\begin{figure}
\includegraphics[clip,width=0.4\textwidth]{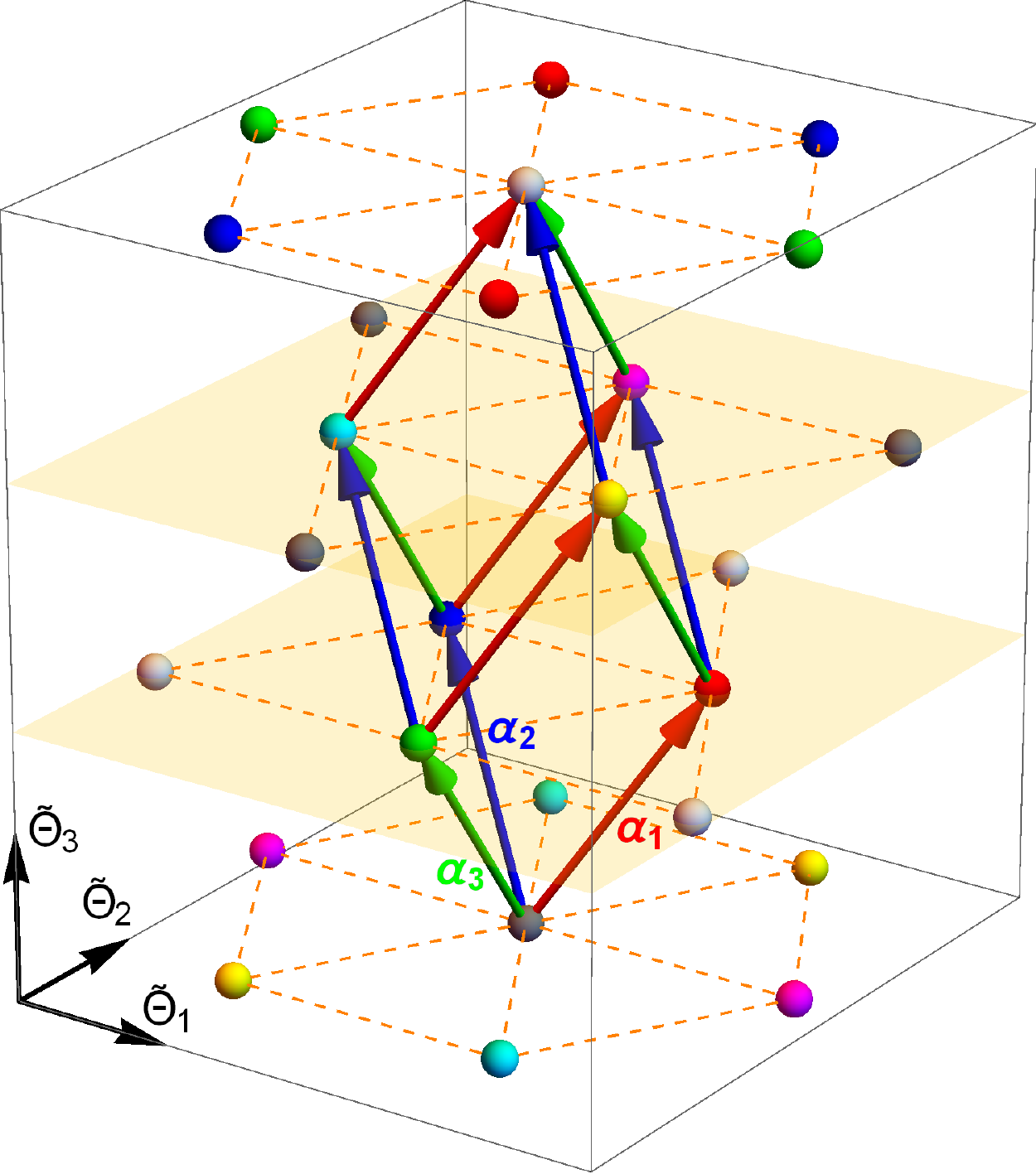}
\caption{Minima of the cosine potentials \eqref{eq:CosinePotSUNk} for $N=4$ and $k=2$, which are indicated by the spheres in the $(\tTheta_1,\tTheta_2,\tTheta_3)$ space scaled by $\sqrt{2}/\pi$. 
Different colors of the spheres represent eight independent minima within the compactification radii of $\tbfTheta$.
The three vectors $\bfalpha_{1,2,3}$ are roots of $SU(4)$ [see Eq.~\eqref{eq:SU4Root}]. }
\label{fig:MinimaSU4}
\end{figure}

Let us parametrize the coupling constants as $u^{ss}_\bfA \equiv u$ and $u^{ss'}_\bfA \equiv u'$ ($s \neq s'$) for simplicity. 
If $u <0$ and $u' = 0$, the fields $\tbfTheta^\mu$ may be pinned at the potential minima, 
\begin{align} \label{eq:MinimaOfV}
\langle \tTheta^\mu_l \rangle = \pi \beta^l V^\mu, 
\end{align}
with $\bfbeta$ and $\bfV$ being arbitrary vectors on the $SU(N)$ root lattice and the $SU(k)$ weight lattice, respectively. 
However, not all the minima of $\tbfTheta^\mu$ are independent because of the compactification of the fields $\tbfTheta^\mu$. 
One way to extract this compactification condition is to examine kinks of $\tbfTheta^\mu$ created by the intrawire interactions proportional to $u'$. 
Since the vectors $\bfomega_s -\bfomega_{s'}$ for $s \neq s'$ span all the roots of $SU(N)$, the compactification condition is read off as 
\begin{align}
\tTheta^\mu_l \sim \tTheta^\mu_l + \pi \alpha^l A^\mu, 
\end{align}
where $\bfalpha$ and $\bfA$ are roots of $SU(N)$ and $SU(k)$, respectively. 
Therefore, only the potential minima within these compactification radii are independent among those given in Eq.~\eqref{eq:MinimaOfV}. 
Since the ratio of the root lattice to the weight lattice for $SU(k)$ is $\mathbb{Z}_k$ and $\mathbb{Z}_k$ is associated with each primitive vector of the $SU(N)$ root lattice, the independent minima are labeled by $\mathbb{Z}_k^{N-1}$. 
The potential minima for $N=4$ and $k=2$ are shown in Fig.~\ref{fig:MinimaSU4} and only $2^3=8$ minima among them are independent in this case. 
Hence, we expect that in the phase realized for $u<0$ and $u'=0$, the $\mathbb{Z}_k^{N-1}$ symmetry of the corresponding parafermion CFT is broken so that the ground state is $k^{N-1}$-fold degenerate. 
This symmetry-breaking phase may be extended for $|u'| < |u|$. 

We are now ready to examine the expectation values of the operators $\Sigma^n_{\sigma,a}$ defined in Eq.~\eqref{eq:OpSigma}, which are now written as 
\begin{align}
\Sigma^n_{\sigma,a} = e^{2i\bfomega_\sigma \bfW_a \cdot \tbfTheta}.
\end{align}
We can label the independent potential minima by a set of $N-1$ integers $\{ m_1, \cdots, m_{N-1} \}$ with $m_\rho \in \mathbb{Z}_k$ as 
\begin{align}
\langle \tTheta^\mu_l \rangle =  \pi W_1^\mu \sum_{\rho=1}^{N-1} m_\rho \alpha^l_\rho, 
\end{align}
where $\bfalpha_\rho$ are primitive vectors of the $SU(N)$ root lattice. 
Correspondingly, the operators $\Sigma^n_{\sigma,a}$ acquire the expectation values, 
\begin{align}
\langle \Sigma^n_{\sigma,a} \rangle &= e^{-\frac{2i \pi}{k} \sum_{\rho=1}^{N-1} m_\rho (\bfalpha_\rho \cdot \bfomega_\sigma)} \nonumber \\
&= e^{-\frac{2i\pi m_\sigma}{k}}. 
\end{align}
Thus, irrespective of the index $a$, the set of $\langle \Sigma^n_{\sigma,a} \rangle$ specifies one of the degenerate ground states. 
This result is consistent with the expected behavior of the spin fields in the nonchiral $SU(N)_k/[U(1)]^{N-1}$ parafermion CFT, which are supposed to detect the $\mathbb{Z}_k^{N-1}$ symmetry breaking. 
Therefore, we conclude that once the $SU(k)_N/[U(1)]^{k-1}$ CFT is gapped, the neutral parts of the backscattering operators $\Sigma^n_{\sigma,a}$ behave as the spin fields of the nonchiral $SU(N)_k/[U(1)]^{N-1}$ CFT. 
Since $\Sigma^n_{\sigma,a}$ is a product of the chiral vertex operators $\Sigma^R_{\sigma,a}$ and $\Sigma^L_{\sigma,a}$, it is natural to expect that each chiral operator contains a fundamental spin field of the chiral $SU(N)_k/[U(1)]^{N-1}$ CFT, which may be labeled by a weight $\bfomega_\sigma$ in the fundamental representation of $SU(N)$. 
At the self-$N$-al critical point discussed below, where only the $SU(k)_N/[U(1)]^{k-1}$ CFT is gapped while the $SU(N)_k/[U(1)]^{N-1}$ CFT remains gapless, the quasiparticle operators \eqref{eq:QuasiparticleOpSUNk} may be seen as the products of the fundamental spin field and the charge vertex operator $e^{i\bfomega_\sigma \cdot \tbfX^{R/L}/\sqrt{k}}$, which are natural generalizations of the quasihole operators discussed in Ref.~\cite{Ardonne01a}. 

\subsubsection{$N$-ality of critical point}
\label{sec:NalCriticalPoint}

We here argue that the Hamiltonian \eqref{eq:SingleWireHam} has a massless flow to the $SU(N)_k/[U(1)]^{N-1}$ CFT at some special values of the coupling constants, where the Hamiltonian is invariant under $N$-ality transformations. 
This will reinforce our previous argument that the intrawire interaction \eqref{eq:IntraIntSUNk} gaps out the $SU(k)_N/[U(1)]^{k-1}$ sector in each wire, while it leaves the $SU(N)_k/[U(1)]^{N-1}$ sector gapless.

Let us first recall the case for $N=2$, which has been discussed in Ref.~\cite{Teo14}. 
Now $\tbfPhi^\mu$ and $\tbfTheta^\mu$ become $(k-1)$-component bosonic fields and the Hamiltonian is written as 
\begin{align} \label{eq:DualHam}
\calH =& \ \frac{v}{2\pi} \int dx \bigl[ (\partial_x \tbfPhi)^2 + (\partial_x \tbfTheta)^2 \bigr] \nonumber \\
&+ \int dx \sum_{\bfA \in \Delta^+_k} \bigl[ u_\bfA \cos (\sqrt{2} \bfA \cdot \tbfTheta) +u'_\bfA \cos (\sqrt{2} \bfA \cdot \tbfPhi) \bigr]. 
\end{align}
This Hamiltonian is invariant under the duality transformation $\tbfPhi \leftrightarrow \tbfTheta$ and $u_\bfA \leftrightarrow u'_\bfA$. 
This is nothing but the Kramers-Wannier duality, which interchanges the $\mathbb{Z}_k$ symmetry-breaking phase with the disordered phase in the corresponding statistical mechanical models. 
The phase transition between these two phases, which is described by the $\mathbb{Z}_k$ parafermion CFT \cite{Zamolodchikov85}, is expected to occur at a self-dual point in the parameter space. 
In fact, there exist several statistical mechanical models that exhibit both the $\mathbb{Z}_k$ symmetry-breaking phase and the $\mathbb{Z}_k$ criticality at which the models host the self-dual property \cite{Baxter89a,Baxter89b,Fendley14}. 
This self-dual point corresponds to $u_\bfA = u'_\bfA$ in the Hamiltonian \eqref{eq:DualHam}. 
In Refs.~\cite{Lecheminant07,Lecheminant12}, it has been shown that the Hamiltonian has a massless flow to the $\mathbb{Z}_k$ parafermion CFT when $u \equiv u_\bfA = u'_\bfA$ and $u<0$. 
One can alternatively say that the Hamiltonian has a gap only in the $SU(k)_2/[U(1)]^{k-1}$ sector. 

The situation becomes involved for $N \geq 3$. 
Now the Hamiltonian \eqref{eq:SingleWireHam} possesses the $N$-ality property as a generalization of the Kramers-Wannier duality. 
Explicitly, the $N$-ality transformation gives $N-1$ sets of new dual fields $\bfvarphi^{(n)}$ and $\bfvartheta^{(n)}$ ($n=1,\cdots,N-1$), which are defined by 
\begin{align} \label{eq:N-alTrans}
\begin{split}
\varphi^{(n) \mu}_l =& \ \frac{1}{2} (\tPhi^\mu_l +\tTheta^\mu_l) -\frac{1}{2} \sum_{s=1}^N \omega^l_s \sum_{m=1}^{N-1} \omega^m_{s-n} (\tPhi^\mu_m-\tTheta^\mu_m), \\
\vartheta^{(n) \mu}_l =& \ \frac{1}{2} (\tPhi^\mu_l +\tTheta^\mu_l) \\ 
&+\frac{1}{2} \sum_{s=1}^{N-1} \omega^l_s \sum_{m=1}^{N-1} (\omega^m_{s-n} -\omega^m_{N-n}) (\tPhi^\mu_m -\tTheta^\mu_m), 
\end{split}
\end{align}
and satisfy 
\begin{align}
[\partial_x \vartheta^{(n) \mu}_l(x), \varphi^{(n) \mu'}_{l'}(x')] = i\pi \delta_{\mu \mu'} \delta_{ll'} \delta(x-x'). 
\end{align}
Under the substitution of $\bfvarphi^{(n)}$ and $\bfvartheta^{(n)}$ and the replacement $u^{(ss')}_\bfA \to u^{(s,s'+n)}_\bfA$ ($s'+n$ is defined modulo $N$), the Hamiltonian \eqref{eq:SingleWireHam} still keeps the same form. 
[We note that the $N$-ality transformation \eqref{eq:N-alTrans} is defined up to $\mathbb{Z}_N$ transformations corresponding to $u^{ss'}_\bfA \to u^{(s+r,s'+r)}_\bfA$ with $r \in \mathbb{Z}_N$.] 
We expect that the $N$-ality transformation permutes the $\mathbb{Z}_k^{N-1}$ symmetry-breaking phase and the other $N-1$ phases, although the physical properties of the latter phases are not clear. 
We then conjecture that the self-$N$-al sine-Gordon Hamiltonian, 
\begin{align} \label{eq:SelfN-alHam}
\calH =& \ \frac{v}{2\pi} \int dx \bigl[ (\partial_x \tbfPhi)^2 +(\partial_x \tbfTheta)^2 \bigr] \nonumber \\
&+\int dx \sum_{s,s'=1}^N \sum_{\bfA \in \Delta^+_k} u_\bfA \cos \bigl[ (\bfomega_s-\bfomega_{s'}) \bfA \cdot  \tbfPhi \nonumber \\
&+(\bfomega_s +\bfomega_{s'}) \bfA \cdot \tbfTheta \bigr], 
\end{align}
has a massless flow to the $SU(N)_k/[U(1)]^{N-1}$ parafermion CFT. 
Again, one can alternatively say that the Hamiltonian will have a gap only in the $SU(k)_N/[U(1)]^{k-1}$ CFT. 
According to the discussion about the bulk gap from integrable field theory in Sec.~\ref{sec:SUNkCurrent}, this holds true at least for $k=2$ and $u_{\sqrt{2}}<0$. 
The other case may require further fine tuning of the coupling constants but seems plausible at least for $u \equiv u_\bfA <0$. 
Since the interactions in Eq.~\eqref{eq:SelfN-alHam} have scaling dimension $2(1-1/N)$ and thus are relevant, there is no obvious way to access the infrared properties of the Hamiltonian \eqref{eq:SelfN-alHam}. 
In connection with statistical mechanical models, there exists an exactly solvable model that may correspond to the $SU(N)_k/[U(1)]^{N-1}$ parafermion theory for $N \geq 3$ \cite{Bazhanov91,Kashaev91}. 
The model hosts the $\mathbb{Z}_k^{N-1}$ symmetry as expected. 
Although the critical exponents and the $N$-ality nature of the critical point are still not very conclusive, the model seems to have the same properties conjectured for our self-$N$-al sine-Gordon Hamiltonian.

\section{Generalization to other filling factors}
\label{sec:GeneralFilling}

We can generalize the above non-Abelian $SU(N-1)$-singlet FQH states to other filling factors, such that the resulting states have the same $SU(N)_k/[U(1)]^{N-1}$ parafermion CFT in the neutral sector of edge states but have different $U(1)$ CFTs in the charge sector. 
Most natural generalizations appear at filling factor, 
\begin{align}
\nu_\textrm{tot} = \frac{k(N-1)}{N+k(N-1)m}, 
\end{align}
with $k>1$ and $m \geq 0$. 
The states with $m=0$ correspond to the non-Abelian $SU(N-1)$-singlet FQH states constructed in the previous sections. 
This generalization includes the series of non-Abelian state proposed by Read and Rezayi ($N=2$) \cite{Read99} and by Ardonne and Schoutens ($N=3$) \cite{Ardonne99,Ardonne01a}.
The first step on constructing these non-Abelian states is to find Abelian FQH states described by the $k(N-1)$-dimensional $K$ matrix, 
\begin{align} \label{eq:GeneralKMat}
\bfK = \begin{pmatrix} \bfK_{SU(N)} & & & \\ & \bfK_{SU(N)} & & \\ & & \ddots & \\ & & & \bfK_{SU(N)} \end{pmatrix} + m\bfC_{k(N-1)}, 
\end{align}
where the first matrix is the block-diagonal matrix with $k$ blocks of $\bfK_{SU(N)}$ and $\bfC_{k(N-1)}$ is the $k(N-1)$-dimensional pseudo-identity matrix whose entries are all one. 
Since the diagonal entries of this $K$ matrix are all $2+m$, even $m$ corresponds to a bosonic FQH state while odd $m$ to a fermionic FQH state. 

These parent Abelian states will be realized when the tunnelings between different copies (channels) and the intrawire interactions are absent. 
Turning on these interactions, the coupled-wire system undergoes a transition to the non-Abelian FQH states. 
For instance, such a transition is expected to occur between the Halperin (331) state and Moore-Read state in bilayer or thick monolayer systems at $\nu=1/2$ (see Ref.~\cite{Read00} and references therein).

The resulting non-Abelian state inherits the same fractional charges from the parent Abelian state, while the neutral sector is still described by the $SU(N)_k/[U(1)]^{N-1}$ parafermion CFT. 
One way to recognize this fact is again to regard the $K$ matrix \eqref{eq:GeneralKMat} as a Gram matrix \cite{Read90}. 
Writing the $K$ matrix as 
\begin{align}
K_{\sigma a; \sigma' a'} = (\bfK_{SU(N)})_{\sigma \sigma'} \delta_{aa'} +m, 
\end{align}
a set of the basis vectors $\{ \bfb_{\sigma a} \}$ to be 
\begin{align}
K_{\sigma a; \sigma' a'} = \sum_{l=1}^{N-1} \sum_{\mu=1}^k b^{l\mu}_{\sigma a} b^{l\mu}_{\sigma' a'}, 
\end{align}
is given by 
\begin{align}
b^{l\mu}_{\sigma a} &= \begin{cases} \alpha^l_\sigma W^\mu_a & (1 \leq \mu < k) \\ \frac{1}{\sqrt{k}} \alpha^l_\sigma(m) & (\mu=k) \end{cases}. 
\end{align}
Here $\bfalpha_\sigma$ and $\bfW_a$ are defined from Eqs.~\eqref{eq:RelAlphaOmega} and \eqref{eq:VectorW}, respectively, and $\bfalpha_\sigma(m)$ are defined by 
\begin{align}
\begin{split}
\alpha^l_\sigma (m) &\equiv \sum_{l'=1}^{N-1} \Lambda^{-1}_{ll'}(m) \alpha^{l'}_\sigma, \\
{\bm \Lambda}^{-1}(m) &= \textrm{diag} \left( 1,1, \cdots, 1, \sqrt{\frac{N+k(N-1)m}{N}} \right). 
\end{split}
\end{align}
The $(N-1)(k-1)$-dimensional lattice spanned by $\alpha^l_\sigma W^\mu_a$ is exactly what the neutral sector of the non-Abelian $SU(N-1)$-singlet FQH state in Sec.~\ref{sec:NonAbelianSUN} lies on. 
According to the modified $SU(N)$ roots $\bfalpha_\sigma(m)$, the fractional charge in the $U(1)$ sector depends on the value of $m$. 

We also remark that the Read-Rezayi states have been obtained by a projection of the trial wave functions for the parent Abelian FQH states with the $K$ matrix \eqref{eq:GeneralKMat} for $N=2$ and general $k$ and $m$ \cite{Cappelli01,Regnault08}. 
This idea will be similarly generalized for $N \geq 3$. 
In fact, the same $K$ matrix as Eq.~\eqref{eq:GeneralKMat} has been proposed to study several properties of the non-Abelian states from the point of view of exclusion statistics of electron and quasihole excitations \cite{Ardonne00,Ardonne01b}. 
This may be related to the lattice structure of the $K$ matrix as we discussed above.

We can actually generate more sequences by taking the freedom to choose the basis vectors of the $SU(N)$ root lattice while fixing the assignment of charge to the multi-layer basis. 
This may result in choosing a new Gram matrix $\bfK'_{SU(N)}$ for the block-diagonal elements of Eq.~\eqref{eq:GeneralKMat}, in stead of $\bfK_{SU(N)}$ defined in Eq.~\eqref{eq:KmatSUN}. 
However, the $K$ matrix obtained in this way generally requires different filling factors for different components in the corresponding Abelian FQH state. 
If one can find a $K$ matrix that allows the same filling factor for different components, we expect that the resulting non-Abelian FQH states for given $N$ and $k$ have different $U(1)$ CFTs in the charge sector but the same parafermion CFT in the neutral sector. 
Examples of such $K$ matrices are
\begin{align}
\bfK'_{SU(5)} = \begin{pmatrix} 2 & -1 & -1 & 1 \\ -1 & 2 & 1 & -1 \\ -1 & 1 & 2 & -1 \\ 1 & -1 & -1 & 2 \end{pmatrix}, 
\end{align}
and $\bfK'_{SU(3)}$ discussed below. 

In the following, we specifically consider two different sequences for $N=3$ and $k=2$. 
The first one is that by Ardonne and Schoutens \cite{Ardonne99,Ardonne01a} at $\nu=4/(4m+3)$, which corresponds to the matrix $\bfK_{SU(3)}$ defined in Eq.~\eqref{eq:KmatSUN}. 
The second one corresponds to another choice of $\bfK_{SU(3)}$,
\begin{align} \label{eq:Kmat22-1}
\bfK'_{SU(3)} = \begin{pmatrix} 2 & -1 \\ -1 & 2 \end{pmatrix}
\end{align}
and is realized at $\nu=4/(4m+1)$. 
The latter sequence includes a bilayer non-Abelian FQH state at $\nu=4/5$, proposed by Barkeshli and Wen \cite{Barkeshli10}. 

\subsection{Ardonne-Schoutens series at $\nu_\textrm{tot}=4/(4m+3)$} \label{sec:ASSeries}

\begin{figure}
\includegraphics[clip,width=0.45\textwidth]{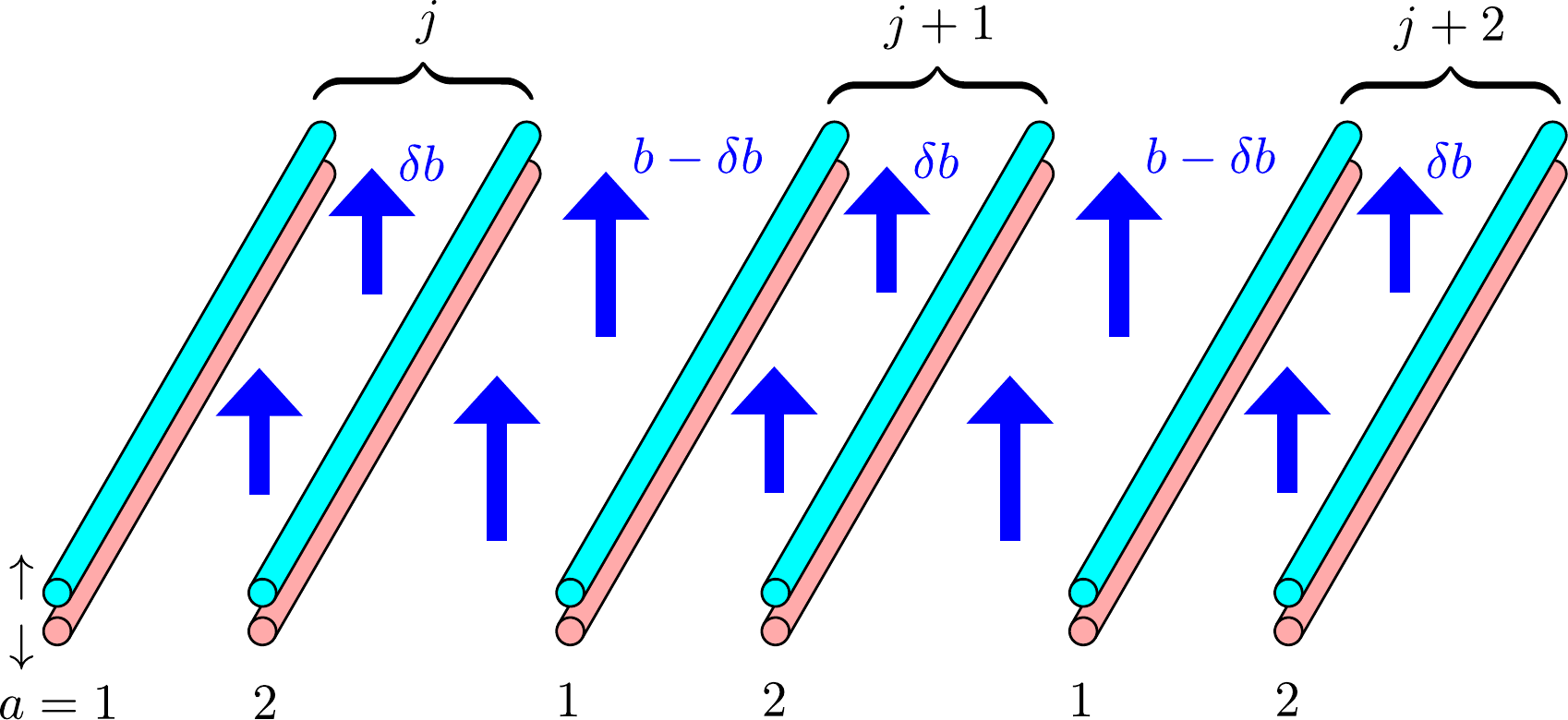}
\caption{Flux configuration required for the generalized non-Abelian FQH states. 
For each $j$, the particle hopping from the copy $1$ to $2$ feels the magnetic flux $\delta b$, while the average flux over copies is set to be a uniform value $b$.}
\label{fig:FermionicNASS}
\end{figure}

The sequence at $\nu_\textrm{tot}=4/(4m+3)$ corresponds to the $m$-generalization of the NASS state, which has been proposed by Ardonne and Schoutens \cite{Ardonne99}. 
In order to find appropriate interactions leading to these states, we have to spatially separate the two copies of wire, which are put on the uniform magnetic flux $b=(8m+6)k_F$, such that the particle hopping from $(j,\sigma,1)$ to $(j,\sigma,2)$ feels the magnetic flux $\delta b = 4mk_F$, as similarly done for the construction of the $q$-Pfaffian state \cite{Teo14}. 
Accordingly, the particle hopping from $(j,\sigma,1)$ to $(j+1,\sigma,2)$ feels the magnetic flux $b+\delta b = (12m+6)k_F$ while that from $(j,\sigma,2)$ to $(j+1,\sigma,1)$ feels $b-\delta b = (4m+6)k_F$. 
This is schematically shown in Fig.~\ref{fig:FermionicNASS}

We consider the interaction Hamiltonian of the form given in Eq.~\eqref{eq:IntHamSU3k}. 
The interwire interactions are now given by 
\begin{align} \label{eq:InterIntAS}
\calO^t_{j,\sigma,ab} =& \ \exp i\Biggl[ \varphi_{j,\sigma,a} -\varphi_{j+1,\sigma,b} \nonumber \\
&+\sum_{\sigma',c} \left( M_{\sigma a; \sigma' c} \theta_{j,\sigma',c} +M^T_{\sigma b; \sigma' c} \theta_{j+1,\sigma',c} \right) \Biggr], 
\end{align}
where $\bfM$ is the $4 \times 4$ matrix given by 
\begin{align}
\bfM = \begin{pmatrix} m+2 & 2(m+1) & 2m & 2m \\ 0 & m+2 & 2m & 2m \\ 0 & 0 & m+2 & 2(m+1) \\ 0 & 0 & 0 & m+2 \end{pmatrix}, 
\end{align}
in the basis of $(\sigma,a)=(\ua,1),(\da,1),(\ua,2),(\da,2)$. 
If these interactions act only within the same channel $a=b$, we will find an Abelian hierarchy FQH state given by the following $4 \times 4$ $K$ matrix, 
\begin{align} \label{eq:KMatAS}
\bfK = \begin{pmatrix} m+2 & m+1 & m & m \\ m+1 & m+2 & m & m \\ m & m & m+2 & m+1 \\ m & m & m+1 & m+2 \end{pmatrix}. 
\end{align}
This Abelian state has quasiparticles with the fractional charge $q/(4m+3)$ and the unfractionalized spin $\pm q$. 
We then introduce the chiral fields, 
\begin{align} \label{eq:ChiralFieldAS}
\begin{split}
\tphi^R_{j,\sigma,a} &= \varphi_{j,\sigma,a} +\sum_{\sigma'=\ua,\da} \sum_{a'=1,2} M_{\sigma a; \sigma' a'} \theta_{j,\sigma',a'}, \\
\tphi^L_{j,\sigma,a} &= \varphi_{j,\sigma,a} -\sum_{\sigma'=\ua,\da} \sum_{a'=1,2} M^T_{\sigma a; \sigma' a'} \theta_{j,\sigma',a'}, 
\end{split}
\end{align}
which satisfy the commutation relations, 
\begin{align} \label{eq:CommAS}
[\partial_x \tphi^p_{j,\sigma,a}(x), \tphi^{p'}_{j',\sigma',a'}(x')] = 2ip\pi \delta_{pp'} \delta_{jj'} K_{\sigma a; \sigma' a'} \delta(x-x'). 
\end{align}
We further introduce the chiral fields, 
\begin{align}
\begin{split}
\tX^p_{j,1} &= \frac{1}{2} (\tphi^p_{j,\ua,1} -\tphi^p_{j,\da,1} +\tphi^p_{j,\ua,2} -\tphi^p_{j,\da,2}), \\
\tX^p_{j,2} &= \frac{1}{2\sqrt{4m+3}} (\tphi^p_{j,\ua,1} +\tphi^p_{j,\da,1} +\tphi^p_{j,\ua,2} +\tphi^p_{j,\da,2}), \\
\tY^p_{j,1} &= \frac{1}{2} (\tphi^p_{j,\ua,1} -\tphi^p_{j,\da,1} -\tphi^p_{j,\ua,2} +\tphi^p_{j,\da,2}), \\
\tY^p_{j,2} &= \frac{1}{2\sqrt{3}} (\tphi^p_{j,\ua,1} +\tphi^p_{j,\da,1} -\tphi^p_{j,\ua,2} -\tphi^p_{j,\da,2}), 
\end{split}
\end{align}
which satisfy 
\begin{align} \label{eq:CommXYAS}
\begin{split}
[\partial_x \tX^p_{j,l}(x), \tX^{p'}_{j',l'}(x')] &= 2ip\pi \delta_{pp'} \delta_{jj'} \delta_{ll'} \delta(x-x'), \\
[\partial_x \tY^p_{j,l}(x), \tY^{p'}_{j',l'}(x')] &= 2ip\pi \delta_{pp'} \delta_{jj'} \delta_{ll'} \delta(x-x'), \\
[\partial_x \tX^p_{j,l}(x), \tY^{p'}_{j',l'}(x')] &=0. 
\end{split}
\end{align}
The interwire interactions \eqref{eq:InterIntAS} are then written as
\begin{align} \label{eq:InterIntAS2}
\calO^t_{j,\sigma,ab} = e^{\frac{i}{\sqrt{2}} \bfalpha_\sigma(m) \cdot (\tbfX^R_j -\tbfX^L_{j+1}) +i\bfalpha_\sigma \bfW_a \cdot \tbfY^R_j -i\bfalpha_\sigma \bfW_b \cdot \tbfY^L_{j+1}}, 
\end{align}
where 
\begin{align}
\begin{split}
\bfalpha_\ua (m) &= \begin{pmatrix} \frac{1}{\sqrt{2}}, & \sqrt{\frac{4m+3}{2}} \end{pmatrix}, \\
\bfalpha_\da (m) &= \begin{pmatrix} -\frac{1}{\sqrt{2}}, & \sqrt{\frac{4m+3}{2}} \end{pmatrix}, 
\end{split}
\end{align}
$\bfalpha_\sigma = \bfalpha_\sigma(0)$ and $\bfW_1 = -\bfW_2 = 1/\sqrt{2}$. 
We can then find the intrawire interactions being of the form, 
\begin{align} \label{eq:IntraIntAS}
\calO^{u,ss'}_{j,12} &= e^{i\sqrt{2} \bfomega_s \cdot \tbfY^R_j -i\sqrt{2} \bfomega_{s'} \cdot \tbfY^L_j},
\end{align}
whose explicit forms in terms of the original bosonic fields are given in Appendix~\ref{app:ExplicitIntraInt}. 
Once again, it is possible to interpret these interactions in terms of the excitations of the parent Abelian FQH state described by Eq.~\eqref{eq:KMatAS}. 
The interwire interactions \eqref{eq:InterIntAS2} are tunnelings of the charge-$q$ particle excitations, while the intrawire interactions \eqref{eq:IntraIntAS} are interactions among the quasipartcle excitations. 

As we have already expected from the underlying lattice structure of the $K$ matrix, the neutral modes appearing in the interactions take exactly the same forms as those for the $m=0$ NASS states in Sec.~\ref{sec:NASSState}. 
In Appendix~\ref{app:IntGenNASS}, we show that if the SLL Hamiltonian and the coupling constants are appropriately tuned, the interwire interactions \eqref{eq:InterIntAS2} are identified as the product of a $[U(1)]^2$ vertex operator and an $SU(3)_2/[U(1)]^2$ parafermionic field. 
The modification of the vertex operator in the charge part indicates that now quasiparticles carry the fractional charge $q/(4m+3)$. 
The intrawire interactions \eqref{eq:IntraIntAS} are identified as the nonchiral products of $SU(2)_3/U(1)$ parafermionic fields. 

\subsection{Barkeshli-Wen series at $\nu_\textrm{tot}=4/(4m+1)$}

The other series associated with the $K$ matrix \eqref{eq:Kmat22-1} is realized at $\nu=4/(4m+1)$. 
The $m=1$ state corresponds to the non-Abelian state proposed by Barkeshli and Wen for a bilayer quantum Hall system at $\nu=4/5$ \cite{Barkeshli10}. 
Similarly to the Ardonne-Schoutens series, we consider an array of the Luttinger liquids put on alternating magnetic fields with $b=(8m+2)k_F$ and $\delta b = 4mk_F$. 
This flux structure allows the interwire interactions \eqref{eq:InterIntAS} but with the following interaction matrix, 
\begin{align}
\bfM = \begin{pmatrix} m+2 & 2(m-1) & 2m & 2m \\ 0 & m+2 & 2m & 2m \\ 0 & 0 & m+2 & 2(m-1) \\ 0 & 0 & 0 & m+2 \end{pmatrix}. 
\end{align}
The corresponding chiral fields \eqref{eq:ChiralFieldAS} now satisfy the commutation relations \eqref{eq:CommAS} with the $K$ matrix, 
\begin{align}
\bfK = \begin{pmatrix} m+2 & m-1 & m & m \\ m-1 & m+2 & m & m \\ m & m & m+2 & m-1 \\ m & m & m-1 & m+2 \end{pmatrix}. 
\end{align}
The Abelian FQH state with this $K$ matrix will be realized when the interwire interactions within the same channel $a=b$ flow to the strong-coupling limit. 
This state has quasiparticles with the fractional charge $q/(4m+1)$ as well as the fractional spin $\pm q/3$. 
By further introducing the chiral fields, 
\begin{align}
\begin{split}
\tX^p_{j,1} &= \frac{1}{2\sqrt{4m+1}} (\tphi^p_{j,\ua,1} +\tphi^p_{j,\da,1} +\tphi^p_{j,\ua,2} +\tphi^p_{j,\da,2}), \\
\tX^p_{j,2} &= \frac{1}{2\sqrt{3}} (\tphi^p_{j,\ua,1} -\tphi^p_{j,\da,1} +\tphi^p_{j,\ua,2} -\tphi^p_{j,\da,2}), \\
\tY^p_{j,1} &= \frac{1}{2} (\tphi^p_{j,\ua,1} +\tphi^p_{j,\da,1} -\tphi^p_{j,\ua,2} -\tphi^p_{j,\da,2}), \\
\tY^p_{j,2} &= \frac{1}{2\sqrt{3}} (\tphi^p_{j,\ua,1} -\tphi^p_{j,\da,1} -\tphi^p_{j,\ua,2} +\tphi^p_{j,\da,2}), 
\end{split}
\end{align}
which satisfy the commutation relations \eqref{eq:CommXYAS}, we can write the interwire interactions as 
\begin{align}
\calO^t_{j,\sigma,ab} = e^{\frac{i}{\sqrt{2}} \bfbeta_\sigma(m) \cdot (\tbfX^R_j -\tbfX^L_{j+1}) +i\bfbeta_\sigma \bfW_a \cdot \tbfY^R_j -i\bfbeta_\sigma \bfW_b \cdot \tbfY^L_{j+1}}, 
\end{align}
where 
\begin{align}
\begin{split}
\bfbeta_\ua(m) &= \begin{pmatrix} \sqrt{\frac{4m+1}{2}}, & \sqrt{\frac{3}{2}} \end{pmatrix}, \\
\bfbeta_\da(m) &= \begin{pmatrix} \sqrt{\frac{4m+1}{2}}, & -\sqrt{\frac{3}{2}} \end{pmatrix},
\end{split}
\end{align}
and $\bfbeta_\sigma = \bfbeta_\sigma(0)$. 
We consider the same form of the intrawire interaction as in the previous case [see Eq.~\eqref{eq:IntraIntAS}], whose explicit expressions are given in Appendix~\ref{app:ExplicitIntraInt}. 
Apart from the modification of the charge part and the different choice of $SU(3)$ roots, the interactions take the same form as those for the NASS state. 
Thus the neutral sector of this non-Abelian FQH state is also given by the $SU(3)_2/[U(1)]^2$ parafermion CFT. 

\section{Application to lattice systems}
\label{sec:LatticeSystems}

In lattice systems, the above FQH states can be realized as fractional Chern insulators or chiral spin liquids. 
A crucial difference from the coupled-wire system is that the momenta along both two spatial directions are conserved only modulo $2\pi$. 
This allows the umklapp scattering and induces the numerous instabilities to long-range orders, such as charge- and spin-density waves and valence-bond solids, which usually overwhelm the FQH states. 
Thus the realization of the FQH states in the lattice system requires frustration suppressing the long-range orders as well as strong interaction. 
Here we provide several key ingredients towards their realization, especially in lattice bosonic systems and $SU(N)$ spin systems, from the perspective of coupled-wire construction.

\subsection{Bosons with correlated hopping}
\label{sec:CorrHopping}

The coupled-wire construction of the FQH state extensively uses correlated hoppings as its building blocks. 
The correlated hopping can be thought of as a virtual process due to the interaction strongly enhanced in the flat band. 
However, it can actually be engineered in cold-atom systems with periodically modulated on-site interactions \cite{Rapp12} and recently realized experimentally \cite{Meinert16}. 
Hence, it is interesting to seek the possibility of FQH states in lattice systems with correlated hoppings. 

We start from the array of Bose-Hubbard chains, 
\begin{align}
\calH_\textrm{BH} =& \ \sum_{j,\ell} \sum_{\sigma=1}^{N-1} \Bigl[ -t \bigl( b^\dagger_{j,\ell,\sigma} b_{j,\ell+1,\sigma} +\textrm{H.c.} \bigr) \nonumber \\
&+\frac{U}{2} n_{j,\ell,\sigma} (n_{j,\ell,\sigma}-1) -\mu n_{j,\ell,\sigma} \Bigr], 
\end{align}
where $b^\dagger_{j,\ell,\sigma}$ creates a boson with $\sigma$ species at site $(j,\ell)$ and $n_{j,\ell,\sigma} = b^\dagger_{j,\ell,\sigma} b_{j,\ell,\sigma}$. 
Taking the continuum limit with respect to the index $\ell$, the low-energy effective Hamiltonian is given by the array of Luttinger liquids \cite{Giamarchi}, 
\begin{align} \label{eq:LLArray}
\calH_0 = \sum_{j} \sum_{\sigma=1}^{N-1} \frac{v_F}{2\pi} \int dx \biggl[ \frac{1}{g} (\partial_x \theta_{j,\sigma})^2 +g(\partial_x \varphi_{j,\sigma})^2 \biggr], 
\end{align}
where $x=\ell a_0$ with $a_0$ being the lattice spacing and $v_F$ and $g$ depend on the parameters $t$, $U$, and $\mu$ in the original model. 
The lattice bosonic operators are now given by \cite{Haldane81}
\begin{align} \label{eq:ParticleOpsBoson}
\begin{split}
b^\dagger_{j,\ell,\sigma} &\sim \Bigl[ \brho -\frac{1}{\pi} \partial_x \theta_{j,\sigma}(x) \Bigr]^\frac{1}{2} e^{i\varphi_{j,\sigma}(x)} \sum_{n \in \mathbb{Z}} e^{2in (\pi \brho x +\theta_{j,\sigma}(x))}, \\
n_{j,\ell,\sigma} &\sim \Bigl( \brho -\frac{1}{\pi} \partial_x \theta_{j,\sigma}(x) \Bigr) \sum_{n \in \mathbb{Z}} e^{2in (\pi \brho x +\theta_{j,\sigma}(x))}. 
\end{split}
\end{align}
In order to realize the Abelian $SU(N-1)$-singlet FQH state with the $K$ matrix \eqref{eq:KmatSUN}, we need the interchain interaction of the form \eqref{eq:InterIntSUN} from the analysis in Sec.~\ref{sec:AbelianSUN}. 
Such interactions are obtained from the continuum expressions of the following correlated hopping, 
\begin{align}
\calO^{(\ell,\ell')}_{j,\sigma} = b^\dagger_{j,\ell,\sigma} b_{j+1,\ell',\sigma} \prod_{\sigma_1>\sigma} n_{j,\ell,\sigma_1} \prod_{\sigma_2<\sigma} n_{j+1,\ell',\sigma_2}, 
\end{align}
which now corresponds to the interaction matrix $\bfM$ given in Eq.~\eqref{eq:Mmat}. 
Here, $|\ell -\ell'|$ must be of the order of the lattice spacing. 
To be precise, these lattice interactions generate an infinite number of vertex operators in the continuum limit, while most of their combinations do not lead to the FQH state. 
Thus we need fine tuning of the (generally complex) coupling constants under an appropriate average density $\brho a_0 = \langle n_{j,\ell,\sigma} \rangle$ to suppress unwanted vertex operators. 

We do not go into further details about how to realize the FQH states from this setup. 
We here only state a general remark. 
Since we start from purely bosonic chains, the resulting interaction matrix $\bfM$ relevant for the Abelian $SU(N-1)$-singlet FQH state must be an even-integer matrix. 
This intrinsically disables us from finding a \emph{symmetric} matrix $\bfM$, which preserves the permutation symmetry among the components of boson.
Although this symmetry is not necessarily required to realize the FQH states, it can exactly or approximately appear in several physical setups such as rotating Bose gases \cite{Cooper08,Reijnders02,Reijnders04,Grass12,Furukawa12} or fractional Chern insulators \cite{Barkeshli12,Sterdyniak13}. 

An intriguing way to obtain a symmetric matrix $\bfM$ is again to consider correlated hoppings, but this time for the \emph{intrachain} couplings, 
\begin{align} \label{eq:CorrelatedHoppingChain}
\calH_\textrm{cor} =& \ \sum_{j,\ell} \sum_{\sigma=1}^{N-1} \biggl[ t \Bigl( b_{j,\ell,\sigma}^\dagger b_{j,\ell+1,\sigma} \prod_{\sigma' \neq \sigma} (2n_{j,\ell,\sigma}-1) +\textrm{H.c.} \Bigr) \nonumber \\
&+ \frac{U}{2} n_{j,\ell,\sigma} (n_{j,\ell,\sigma}-1) -\mu n_{j,\ell,\sigma} \biggr]. 
\end{align}
Due to the correlated hoppings, it is not clear that the low-energy description in terms of the Luttinger liquid is generally applicable to this Hamiltonian. 
However, in the limit of $U \to \infty$ where the hard-core treatment of bosons $b_{j,\ell,\sigma}^2=0$ is justified, we can introduce a variant of the Jordan-Wigner transformation, 
\begin{align} \label{eq:JordanWigner}
b_{j,\sigma,\ell} = \tb_{j,\ell,\sigma} \tK_{j,\ell,\sigma} 
\end{align}
where $\tK_{j,\ell,\sigma}$ is a string operator defined by 
\begin{align}
\tK_{j,\ell,\sigma} = \exp \left[ i\pi \sum_{\sigma' \neq \sigma} \left( \sum_{\ell'<\ell} \tn_{j,\ell',\sigma'} +\sum_{j'<j} \sum_{\ell'} \tn_{j',\ell',\sigma'} \right) \right] 
\end{align}
and $\tn_{j,\ell,\sigma} = n_{j,\ell,\sigma}$. 
The new particle operators $\tb_{j,\ell,\sigma}$ with the same component commute with each other, while those with different components anticommute except when they are at the same position. 
After this transformation, the correlated hoppings becomes the standard hoppings, 
\begin{align}
\calH_\textrm{cor} = \sum_{j,\ell} \sum_{\sigma=1}^{N-1} \Bigl[ -t \bigl( \tb^\dagger_{j,\ell,\sigma} \tb_{j,\ell+1,\sigma} +\textrm{H.c.} \bigr) -\mu \tn_{j,\sigma,\ell} \Bigr]. 
\end{align}
Thus the low-energy theory of the Hamiltonian \eqref{eq:CorrelatedHoppingChain} is still described by the array of Luttinger liquids in Eq.~\eqref{eq:LLArray} for a sufficiently large $U$. 
The lattice operators $\tb_{j,\ell,\sigma}$ and $\tn_{j,\ell,\sigma}$ are also bosonized as in Eq.~\eqref{eq:ParticleOpsBoson}. 
On the other hand, the continuum expressions of the original bosonic operators $b_{j,\sigma,\ell}$ are modified because of the string operators. 
Noting that the lattice string operator is Hermitian, it may be bosonized as 
\begin{align}
\tK_{j,\ell,\sigma} \sim \kappa_{j,\sigma} \cos \Biggl[ (N-2) \pi \brho x +\sum_{\sigma' \neq \sigma} \theta_{j,\sigma'}(x) \Biggr], 
\end{align}
where $\kappa_{j,\sigma}$ is the Klein factor, 
\begin{align}
\kappa_{j,\sigma} = \prod_{\sigma' \neq \sigma} \prod_{j'<j} e^{i\int_{-\infty}^\infty dx \partial_x \theta_{j',\sigma'}(x)}, 
\end{align}
which must be multiplied to keep the anticommuting property of $\tb_{j,\sigma,\ell}$ between different components. 
Since the string operator $\tK_{j,\ell,\sigma}$ contains \emph{odd} multiples of $\theta_{j,\ell,\sigma'}$ with $\sigma' \neq \sigma$, interchain hoppings (or generally correlated hoppings) 
\begin{align}
\calO^{(\ell,\ell')}_{j,\sigma} = b^\dagger_{j,\ell,\sigma} b_{j+1,\ell',\sigma}, 
\end{align}
can produce vertex operators of the form \eqref{eq:InterIntSUN} but with the interaction matrix $\bfM=\bfK_{SU(N)}$, up to the Klein factors. 
Indeed, the off-diagonal entries of $\bfM$ now become odd integers while the diagonal entries still remain even integers. 
This allows us to obtain a symmetric matrix $\bfM$ and therefore the $SU(N-1)$-singlet FQH states with manifestly keeping the symmetries among the components. 

The above idea for $N=3$ is explicitly demonstrated in Ref.~\cite{Fuji16} for a lattice bosonic model with correlated hoppings \cite{YCHe15} to obtain the Halperin (221) state as well as a bosonic integer quantum Hall state protected by $U(1)$ symmetry \cite{Senthil13}. 
The Jordan-Wigner transformation \eqref{eq:JordanWigner} is reminiscent of a mutual flux attachment \cite{Senthil13} in lattice systems. 
This suggests that the constraints imposed to the allowed interactions \eqref{eq:GeneralInt} may be relaxed for the coupled-wire system of multi-component bosons, if we think that the wires are initially composed of mutual composite bosons due to strong interaction. 

\subsection{$SU(N)$ spin systems}
\label{sec:SUNSpinSystems}

We have shown in Sec.~\ref{sec:SUN} that the interwire interactions for the Abelian and non-Abelian $SU(N-1)$-singlet FQH states can be written in terms of the $SU(N)_k$ currents. 
These currents naturally appear in $SU(N)$ spin chains rather than in bosonic or fermionic chains, as low-energy properties of the former are described by the $SU(N)_k$ WZW CFT \cite{Affleck86,Affleck88}. 
Thus we here consider the realizations of FQH states in $SU(N)$ spin systems, which are usually dubbed as the chiral spin liquids. 
Our basic idea follows a detailed study for the $SU(2)$ case by Gorohovsky, Pereira, and Sela \cite{Gorohovsky15}. 

We start from the array of $SU(N)$ Heisenberg chains, 
\begin{align}
\calH_\textrm{Heis} = J \sum_{j,\ell} \sum_{a=1}^{N^2-1} T^a_{j,\ell} T^a_{j,\ell+1}, 
\end{align}
where $T^a_{j,\ell}$ are the generators of $SU(N)$ in some representation. 
We here focus on the fundamental representation, and hence $T^a_{j,\ell}$ are $N \times N$ matrices satisfying 
\begin{align}
[T^a_{j,\ell}, T^b_{j',\ell'}] = i\delta_{jj'} \delta_{\ell \ell'} \sum_{c=1}^{N^2-1} f_{abc} T^c_{j,\ell}, 
\end{align}
where $f_{abc}$ is the structure constant. 
The low-energy effective Hamiltonian is given by the $SU(N)_1$ WZW CFT \cite{Affleck86,Affleck88}, 
\begin{align}
\calH_0 = \frac{v}{4\pi (N+1)} \int dx \sum_j \Bigl( :\mathrel{\bfJ_j \cdot \bfJ_j}: + :\mathrel{\bbfJ_j \cdot \bbfJ_j}: \Bigr), 
\end{align}
where $\bfJ_j(x)$ and $\bbfJ_j(x)$ are the right and left $SU(N)_1$ currents, respectively. 
The lattice spin operators are expressed as \cite{Affleck86,Affleck88}
\begin{align} \label{eq:SUNOp}
T^a_{j,\ell} \sim& \ a_0 \bigl[ J^a_j(x) +\bJ^a_j(x) \bigr] +e^{\frac{2i\pi x}{Na_0}} N^a_j(x) \nonumber \\
&+e^{-\frac{2i\pi x}{Na_0}} N^{a\dagger}_j(x) +\cdots,
\end{align}
where $a_0$ is the lattice spacing and the ellipsis stands for less relevant terms. 
Here $N^a_j(x)$ is related with the WZW primary field $\bfg(x)$, which is an $N \times N$ matrix field with scaling dimension $1-1/N$, through 
\begin{align}
N^a_j(x) = c \ \textrm{Tr} \bigl[ \bfg_j(x) T^a \bigr],
\end{align}
where $c$ is some nonuniversal constant. 

\begin{figure}
\includegraphics[clip,width=0.43\textwidth]{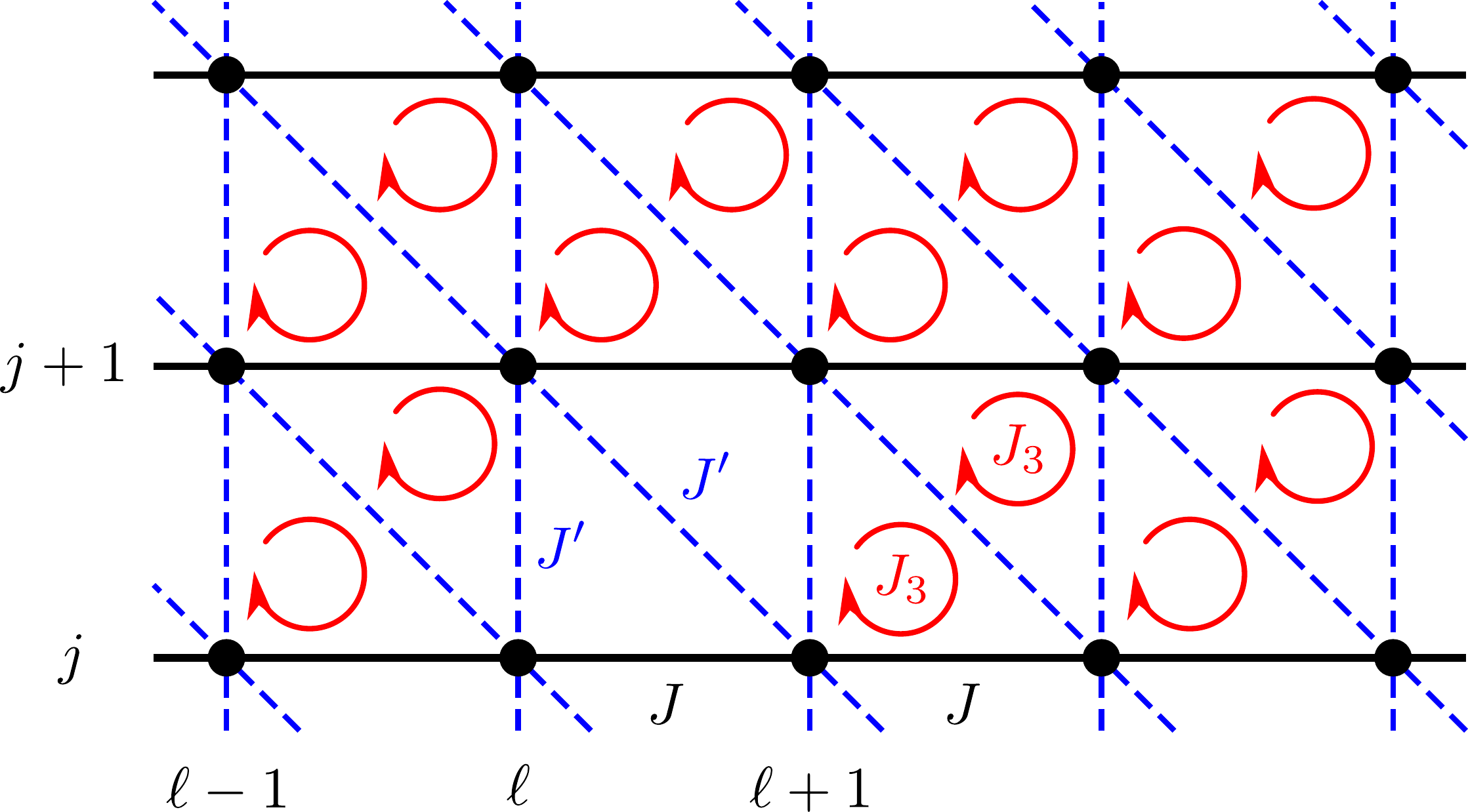}
\caption{$SU(N)$ spin model defined on an anisotropic triangular lattice.
The lattice sites are specified by the indices $j$ and $\ell$.
The solid and dashed lines represent the exchange couplings $J$ and $J'$, respectively. 
The circle arrows represent the three-body interactions $J_3$ on triangles.}
\label{fig:TriangularLattice}
\end{figure}

We consider the model on an anisotropic triangular lattice studied in Ref.~\cite{Gorohovsky15}, which is now generalized to $SU(N)$ spins as 
\begin{align}
\calH_1 =& \ \sum_{j,\ell} \Biggl[ J' \sum_{a=1}^{N^2-1} \bigl( T^a_{j,\ell} T^a_{j+1,\ell} +T^a_{j,\ell+1} T^a_{j+1,\ell} \bigr) \nonumber \\
&+J_3 \sum_{a,b,c} f_{abc} \bigl( T^a_{j,\ell} T^b_{j+1,\ell} T^c_{j,\ell+1} +T^a_{j,\ell} T^b_{j+1,\ell-1} T^c_{j+1,\ell} \bigr) \Biggr]. 
\end{align}
The first term is the standard exchange coupling between neighboring chains. 
The second term represents three-body interactions acting on triangles, which explicitly break both time-reversal and parity symmetries. 
Such interactions can be generated in the strong-coupling limit of $SU(N)$ Hubbard models at $1/N$ filling with a finite flux. 
The model is depicted in Fig.~\ref{fig:TriangularLattice}. 
Substituting Eq.~\eqref{eq:SUNOp}, we find 
\begin{align} \label{eq:SUNIntBoson}
\calH_1 \sim& \ \int dx \sum_j \Bigl[ 2J'a_0 \bigl( \bfJ_j \cdot \bbfJ_{j+1} +\bbfJ_j \cdot \bfJ_{j+1} \bigr) \nonumber \\
&+J'\bigl( (1+e^\frac{2i\pi}{N}) \bfN_j \cdot \bfN^\dagger_{j+1} +\textrm{H.c.} \bigr) \nonumber \\
&+ 4N J_3 a_0 \bigl( \bfJ_j \cdot \bbfJ_{j+1} -\bbfJ_j \cdot \bfJ_{j+1} \bigr) \Bigr] +\cdots, 
\end{align}
where we have used the OPEs, 
\begin{align}
\begin{split}
J^a_j(x) J^b_{j'}(y) &\sim \delta_{jj'} \biggl[ -\frac{\delta_{ab}}{(x-y)^2} +\sum_{c=1}^{N^2-1} \frac{f_{abc}}{x-y} J^c_j(y) \biggr], \\
\bJ^a_j(x) \bJ^b_{j'}(y) &\sim \delta_{jj'} \biggl[ -\frac{\delta_{ab}}{(x-y)^2} -\sum_{c=1}^{N^2-1} \frac{f_{abc}}{x-y} \bJ^c_j(y) \biggr],
\end{split}
\end{align}
and $\sum_{b,c} f_{abc} f_{dbc} = 2N\delta_{ad}$ to obtain the third term. 
In a similar manner to Ref.~\cite{Gorohovsky15}, if $J'$ and $J_3$ are appropriately tuned, we can find the marginal current-current interaction, only one of either $\bfJ_j \cdot \bbfJ_{j+1}$ or $\bbfJ_j \cdot \bfJ_{j+1}$, which potentially stabilizes the $SU(N)$ chiral spin liquid as we have seen. 
However, in contrast to the $N=2$ case \cite{Gorohovsky15} where $\bfN_j^\dagger=\bfN_j$, the backscattering terms $\bfN_j \cdot \bfN^\dagger_{j+1}$ do not cancel for general $N$.
Since these terms have scaling dimension $2(1-1/N)$ and thus are strictly relevant for a finite $N$, they will usually govern the low-energy physics and drive the system into some long-range ordered phase. 
We note that even for $N=2$, it is likely that the system develops some magnetic order by relevant perturbations generated at higher order \cite{Gorohovsky15}. 

In the limit $N \to \infty$, both current-current and backscattering interactions become marginal. 
Thus the resulting physics will be simply determined by the magnitudes of the coupling constants. 
As also discussed in several large-$N$ analyses \cite{Khveshchenko89,Khveshchenko90,Hermele09,Hermele11}, the chiral spin liquid can be more stable in this limit. 
It is interesting to note that since the contribution of the three-body interaction in the low-energy theory is of the order of $N$ [see Eq.~\eqref{eq:SUNIntBoson}], an infinitesimal $J_3 \sim O(1/N)$ may be enough to stabilize the chiral spin liquid. 
This indicates the possibility of a chiral spin liquid with spontaneous time-reversal symmetry breaking in the large $N$ limit. 
A parent Hamiltonian approach in Ref.~\cite{HHTu14} also suggests this possibility, while the Hamiltonian is long ranged in that case. 

Although a recent numerical study \cite{Nataf16} argues that the same model exhibits the $SU(N)$ chiral spin liquid for $J = J' \sim J_3$ and $3 \leq N \leq 9$, it may not be the case at least for the spatially anisotropic case studied above. 
Nevertheless, it is interesting to ask whether we can find a robust chiral spin liquid phase within the coupled-wire approach in other lattice models. 
Such models may require different lattices, representations of $SU(N)$, and/or interactions. 
We leave this question for the future study. 

\section{Conclusion} \label{sec:Conclusion}

We developed the coupled-wire construction of FQH states for the Abelian and non-Abelian FQH states related to $[U(1)]^{N-1} \times SU(N)_k/[U(1)]^{N-1}$ CFTs. 
These FQH states include the $SU(N-1)$-singlet generalization of the NASS state proposed in Refs.~\cite{Ardonne99,Ardonne01a}, which will be relevant for multi-component quantum Hall systems. 
We also proposed a procedure to generate different non-Abelian FQH states whose neutral sector is still described by the $SU(N)_k/[U(1)]^{N-1}$ parafermion CFT, and demonstrated it for a bilayer non-Abelian state at $\nu=4/5$ \cite{Barkeshli10} as the simplest example. 
By employing the underlying $SU(N)$-algebraic structure of the FQH states and the vertex representations of the associated CFTs, we explicitly constructed the microscopic interactions from coupled wires that will open a bulk gap while leave the desired chiral gapless excitations at the boundaries. 
This approach allows us to rigorously prove the existence of the bulk gap for $k=2$ by utilizing the results from an integrable deformation of the $\mathbb{Z}_k$ parafermion theory, in addition to the previously addressed $N=2$ case \cite{Teo14}. 
We also argued that from the analogue with a $\mathbb{Z}_k^{N-1}$ statistical mechanical model, bulk quasiparticles of those non-Abelian FQH states are related to spin fields of the $SU(N)_k/[U(1)]^{N-1}$ parafermion CFT. 

In the present construction, the non-Abelian FQH states are obtained by turning on interactions in certain parent Abelian FQH states. 
This construction may be extended to non-Abelian topological ordered states related to more general CFTs by similarly taking Abelian states with the $K$ matrices proposed in Ref.~\cite{Ardonne03}. 
We hope that our approach opens the door to understand the microscopic origins of more exotic topological orders.

We also discussed how our results are applied to lattice systems towards the realization of those FQH states. 
For a bosonic system with correlated hoppings, the Abelian $SU(N-1)$-singlet FQH state becomes a natural candidate of ground state due to an effect like the mutual flux attachment. 
This effect actually also forbids the simple boson condensation at least in the quasi-1D limit. 
It is thus interesting to ask whether we can engineer a realistic optical lattice system with correlated hoppings by using the Floquet theory for periodically modulated interactions. 

Another system that we worked out is the $SU(N)$ Heisenberg model, which may be realized as the Mott-insulating phase of multi-component Hubbard models. 
In the model with chiral three-body exchange interactions, we found that the Abelian $SU(N-1)$-singlet chiral spin state tends to be stable for large $N$. 
Although the present coupled-wire approach based on the $SU(N)_k$ WZW CFT most naturally fits into this system, the actual realization of the chiral spin state for a finite $N$ seems quite subtle due to competing long-range orders. 
Nevertheless, it is an intriguing future direction to seek the possibility of the chiral spin states by manipulating the microscopic interactions and especially the representation of $SU(N)$ which may lead to non-Abelian topological orders. 

Finally, we believe that the field-theoretical machinery developed in the present study facilitates the deep microscopic understanding of exotic topological phenomena as well as the investigation of novel quantum critical phenomena. 
One example is topological defects between distinct topologically ordered states or topologically distinct boundary states \cite{Barkeshli13,TLan15}. 
Interestingly, the $SU(N)_k/[U(1)]^{N-1}$ parafermion may emerge as a localized zero mode of such a topological defect, and the associated criticality as discussed in Sec.~\ref{sec:NalCriticalPoint} may be realized as the phase transition between certain topologically distinct gapped boundary states \cite{Barkeshli13}. 
From the perspective of quantum field theory, our analysis poses two intimately related questions remaining unanswered: whether there exists an integrable deformation of the $SU(N)_k/[U(1)]^{N-1}$ parafermion CFT and whether there exists a massless flow to the $SU(N)_k/[U(1)]^{N-1}$ CFT in the self-$N$-al sine-Gordon theory \eqref{eq:SelfN-alHam}. 
These questions will be worth addressing also in the relations with statistical mechanical models.

\acknowledgments

Y.F. is grateful to S. Furukawa for collaboration at initial stage. 
Y.F. would like to thank E. Ardonne, M. Hermanns, M. Oshikawa, K. Penc, and K. Schoutens for insightful discussions, Y.-C. He, S. Bhattacharjee, and F. Pollmann for related collaborations, and the hospitality of LPTM at Universit\'e de Cergy-Pontoise. 
P.L. would like to thank CNRS for financial support (PICS grant).

\appendix

\section{Roots and weights of $SU(N)$} \label{app:RootWeightSUN}

We here summarize our convention of roots and weights of $SU(N)$, which are used in the main text and the other appendices. 
With $(N-1)$-dimensional vectors $\bfw_\sigma=(w^1_\sigma, \cdots, w^{N-1}_\sigma)$, a full set of the weights in the fundamental representation of $SU(N)$ is generated by $N$ vectors satisfying 
\begin{subequations}
\begin{align}
\sum_{\sigma=1}^N w^l_\sigma &= 0, \\
\sum_{\sigma=1}^N w^l_\sigma w^{l'}_\sigma &= \delta_{ll'}, \\
\sum_{l=1}^{N-1} w^l_\sigma w^l_{\sigma'} &= \delta_{\sigma \sigma'} -\frac{1}{N}. 
\end{align}
\end{subequations}
By definition, their differences generate all the $N(N-1)$ roots of $SU(N)$: 
\begin{align}
\Delta_N = \{ \bfw_\sigma -\bfw_{\sigma'} \ | \ \sigma \neq \sigma', \ 1 \leq \sigma \leq N, \ 1 \leq \sigma' \leq N \}. 
\end{align}
Now any root of $SU(N)$, $\bfalpha \in \Delta_N$, is normalized as $|\bfalpha|^2=2$. 
Let us specifically consider the following choice of $\bfw_\sigma$, 
\begin{align} \label{eq:SpecificWeight}
\omega^l_\sigma = \begin{cases} \frac{1}{\sqrt{l(l+1)}} & (l \geq \sigma) \\ -\sqrt{\frac{l}{l+1}} & (l=\sigma-1) \\ 0 & (l<\sigma-1) \end{cases}. 
\end{align}
The first $N-1$ vectors form an orthogonal matrix diagonalizing the $K$ matrix \eqref{eq:KmatSUN} as
\begin{align}
\sum_{\sigma,\sigma'=1}^{N-1} \omega^l_\sigma (\bfK_{SU(N)})_{\sigma \sigma'} \omega^{l'}_{\sigma'} = \delta_{ll'}, 
\end{align}
and thus have been used in the linear transformations of the bosonic fields in Eqs.~\eqref{eq:ChiralBoson2212} and \eqref{eq:ChiralBosonSUN2}. 
They also span the weight lattice of $SU(N)$, 
\begin{align}
\Lambda^W_{SU(N)} = \Biggl\{ \sum_{\sigma=1}^{N-1} r_\sigma \bfomega_\sigma \ \Bigg| \ r_\sigma \in \mathbb{Z} \Biggr\}. 
\end{align}
Upon the choice of Eq.~\eqref{eq:SpecificWeight}, a set of positive roots of $SU(N)$ is generated by 
\begin{align}
\Delta^+_N = \{ \bfomega_\sigma-\bfomega_{\sigma'} \ | \ 1 \leq \sigma < \sigma' \leq N \}. 
\end{align}
It also generates a set of $N-1$ simple roots by 
\begin{align} \label{eq:SimpleRoots}
\tbfalpha_\sigma = \bfomega_\sigma -\bfomega_{\sigma+1}. 
\end{align}
The simple roots $\tbfalpha_\sigma$ constitute the Cartan matrix given in Eq.~\eqref{eq:CartanSUN} by 
\begin{align}
(\bfA_{SU(N)})_{\sigma \sigma'} = \sum_{l=1}^{N-1} \talpha^l_\sigma \talpha^l_{\sigma'}, 
\end{align}
and become primitive vectors of the root lattice of $SU(N)$,
\begin{align}
\Lambda^R_{SU(N)} = \Biggl\{ \sum_{\sigma=1}^{N-1} r_\sigma \tbfalpha_\sigma \ \Bigg| \ r_\sigma \in \mathbb{Z} \Biggr\}.
\end{align}

The transformation matrix $\bfG$ from the $K$ matrix $\bfK_{SU(N)}$ to the Cartan matrix $\bfA_{SU(N)}$ is given by the $(N-1)$-dimensional upper triangular matrix, 
\begin{align} \label{eq:MatG}
\bfG = \begin{pmatrix}
1 & 1 & \cdots & 1 & 1 \\
0 & 1 & & 1 & 1 \\
\vdots & & \ddots & & \vdots \\
0 & 0 & & 1 & 1 \\
0 & 0 & \cdots & 0 & 1
\end{pmatrix}, 
\end{align}
whose allowed nonzero entries are all one. 
Since $|\det \bfG|$ is obviously unity, this matrix belongs to $GL(N-1,\mathbb{Z})$ with determinant $\pm 1$. 
While the $GL(N-1,\mathbb{Z})$ transformation generally changes the primitive vectors of the root lattice $\Lambda^R_{SU(N)}$, the lattice structure itself is unchanged \cite{Read90}. 
The set of the simple roots $\tbfalpha_\sigma$ is transformed by $\bfG$ to the other set of roots $\bfalpha_\sigma$ with $\sigma=1,\cdots,N-1$, 
\begin{align}
\bfalpha_\sigma &= \sum_{\sigma'=1}^{N-1} G_{\sigma \sigma'} \tbfalpha_{\sigma'} \nonumber \\
&= \bfomega_\sigma-\bfomega_N, 
\end{align}
which constitute $\bfK_{SU(N)}$ [see Eq.~\eqref{eq:KtoAlpha}]. 
They obviously satisfy Eq.~\eqref{eq:RelAlphaOmega} and still form primitive vectors of the root lattice $\Lambda^R_{SU(N)}$. 
In Ref.~\cite{Frohlich91}, Fr\"ohlich and Zee have discussed that the $m$-dimensional $K$ matrix of the form $\bfK = {\bm 1} +p\bfC$ (${\bm 1}$ and $\bfC$ are the identity and pseudo-identity matrices, respectively) is related to the Cartan matrix of the $SU(m)$ algebra by the same transformation matrix $\bfG$. 
Our case corresponds to $p=1$ in their case which is, as we have shown, related to the Cartan matrix of the $SU(m+1)$ algebra. 

Taking the first $N-1$ vectors from Eq.~\eqref{eq:SpecificWeight}, the transformation $\bfG^T$ generates a set of the fundamental weights of $SU(N)$, 
\begin{align} \label{eq:FundamentalWeights}
\tbfomega_\sigma = \sum_{\sigma'=1}^{N-1} G^T_{\sigma \sigma'} \bfomega_{\sigma'}, 
\end{align}
which are dual to the simple roots, 
\begin{align}
\sum_{l=1}^{N-1} \tbfomega^l_\sigma \tbfalpha^l_{\sigma'} = \delta_{\sigma \sigma'}. 
\end{align}
The $I$-th antisymmetric tensor representation of $SU(N)$, which is represented by the Young tableau with a single column with $I$ boxes ($1 \leq I < N$), is uniquely characterized by the highest weight $\tbfomega_I$. 
Thus the $I$-th antisymmetric representation denoted by $[\tbfomega_I]$ consists of the weights, 
\begin{align}
[\tbfomega_I] = \{ \bfomega_{\sigma_1} +\cdots +\bfomega_{\sigma_I} \ | \ 1 \leq \sigma_1 < \cdots < \sigma_I \leq N \}. 
\end{align}
The number of weights in $[\tbfomega_I]$ is given by $N!/I!(N-I)!$ and $|\tbfomega_I|^2=I(N-I)/N$. 

\section{Vertex representation of $SU(N)_1$ currents} \label{app:VertexRep}

We here explicitly construct the $SU(N)_1$ current algebra from free bosons defined in Eq.~\eqref{eq:ChainFieldComm}. 
In contrast to the $SU(2)_1$ case, a special care on the cocycle factor is required to obtain a faithful representation for $SU(N)_1$ with $N \geq 3$. 

\subsection{Free bosons and their linear transformations}

Let us consider a single wire and the bosonic fields $(\theta_\sigma, \varphi_\sigma)$ with $\sigma=1, \cdots,N-1$, which obeys the commutation relations, 
\begin{align} \label{eq:DualComm}
[\theta_\sigma(x), \varphi_{\sigma'}(x')] = i\pi \delta_{\sigma \sigma'} \Theta(x-x'), 
\end{align}
where $\Theta(x)$ is a step function that takes $1$ for $x>0$ while $0$ for $x<0$. 
This makes the creation operator $\Phi_\sigma^\dagger (x) \propto e^{i\varphi_\sigma(x)}$ and the density operator $\rho^{(n)}_\sigma (x) \propto e^{2in[k_F x +\theta_\sigma(x)]}$ at different positions commutative. 
These fields are now compactified as 
\begin{align} \label{eq:CompactChainField}
\begin{split}
\varphi_\sigma &\sim \varphi_\sigma +2\pi n_\sigma, \\
\theta_\sigma &\sim \theta_\sigma +\pi m_\sigma, 
\end{split}
\end{align}
where $n_\sigma, m_\sigma \in \mathbb{Z}$. 
Using Eq.~\eqref{eq:DualComm}, the chiral fields \eqref{eq:ChiralBosonSUN2} obey the commutation relations, 
\begin{align}
[\tchi^R_l(x), \tchi^R_{l'}(x')] =& \ i\pi \delta_{ll'} \sgn(x-x') \nonumber \\ 
&+\frac{i\pi}{2} \sum_{\sigma,\sigma'} \omega^l_\sigma \omega^{l'}_{\sigma'} (M_{\sigma \sigma'}-M_{\sigma' \sigma}), \\
[\tchi^L_l(x), \tchi^L_{l'}(x')] =& -i\pi \delta_{ll'} \sgn(x-x') \nonumber \\ 
&+\frac{i\pi}{2} \sum_{\sigma,\sigma'} \omega^l_\sigma \omega^{l'}_{\sigma'} (M_{\sigma \sigma'}-M_{\sigma' \sigma}), \\
\label{eq:tchiLRComm}
[\tchi^R_l(x), \tchi^L_{l'}(x')] =& \ i\pi \sum_{\sigma,\sigma'} \omega^l_\sigma \omega^{l'}_{\sigma'} M_{\sigma \sigma'}. 
\end{align}
Thus these fields do not commute not only between different chiralities but also between different components. 
We note that this bosonic theory can be equivalently formulated in the toroidal compactification of a bosonic string action with a symmetric and antisymmetric constant background fields \cite{Griffin89}. 
In the notation of Ref.~\cite{Griffin89}, the antisymmetric field may be chosen to be $B_{\mu \nu} = \frac{1}{2} \sum_{\sigma,\sigma'} \omega^\mu_\sigma \omega^{\nu}_{\sigma'} (M-M^T)_{\sigma \sigma'}$. 
To obtain a commutative algebra, we introduce coordinate-free operators $P^{R/L}_l$ satisfying, 
\begin{align}
[\tchi^p_l(x), P^{p'}_{l'}] = \frac{i\delta_{pp'}}{2} \sum_{\sigma, \sigma'} \omega^l_\sigma \omega^{l'}_{\sigma'} M_{\sigma \sigma'}, 
\end{align}
and consider 
\begin{align} \label{eq:FVBoson}
\begin{split}
\tchi^R_l &= \chi^R_l +\pi P^R_l, \\
\tchi^L_l &= \chi^L_l +\pi P^L_l +2\pi P^R_l. 
\end{split}
\end{align}
The new chiral fields $\chi^p_l$ satisfy the standard commutation relations, 
\begin{align}
[\chi^p_l(x), \chi^{p'}_{l'}(x')] = i\pi p\delta_{pp'} \delta_{ll'} \sgn(x-x'). 
\end{align}
Below we show that the operators $P^{R/L}_l$ can be constructed from the zero-mode part of the mode expansion of $\chi^p_l$ and give the correct cocycle factor to a vertex representation of the $SU(N)_1$ current algebra. 

Since $P^{R/L}_l$ are assumed to be independent of the coordinate, the Hamiltonian \eqref{eq:ChiralHam} for each wire is equivalent to 
\begin{align}
\calH_\textrm{SLL} = \frac{v}{4\pi} \sum_{l=1}^{N-1} \int dx \left[ (\partial_x \chi^R_l)^2 +(\partial_x \chi^L_l)^2 \right]. 
\end{align}
On a cylinder of circumference $L$, the fields $\chi^p_l$ have mode expansions \cite{dFMS}:
\begin{align} \label{eq:ModeExpansions}
\begin{split}
\chi^R_l(z) &= q^R_l -ip^R_l \ln z +\sum_{n=1}^\infty \frac{1}{\sqrt{n}} \left( a^R_{ln} z^{-n} +a^{R \dagger}_{ln} z^n \right), \\
\chi^L_l(\bz) &= q^L_l -ip^L_l \ln \bz +\sum_{n=1}^\infty \frac{1}{\sqrt{n}} \left( a^L_{ln} \bz^{-n} +a^{L \dagger}_{ln} \bz^n \right), \\
\end{split}
\end{align}
where
\begin{align}
z = e^{2\pi(\tau+ix)/L}, \hspace{10pt} \bz = e^{2\pi (\tau-ix)/L}, 
\end{align}
and the mode operators satisfy the commutation relations, 
\begin{align}
[q^R_l, p^R_{l'}] &= [q^L_l,p^L_{l'}] = i\delta_{ll'}, \\
[a^R_{ln}, a^{R \dagger}_{l'n'}] &= [a^L_{ln}, a^{L \dagger}_{l'n'}] = \delta_{ll'} \delta_{nn'}, 
\end{align}
while the other commutators vanish. 
Utilizing the compactification condition \eqref{eq:CompactChainField}, $\chi^p_l$ are compactified as 
\begin{align} \label{eq:compactification}
\begin{split}
\chi^R_l &\sim \chi^R_l + 2\pi \sum_\sigma \omega^l_\sigma \left( n_\sigma +\frac{1}{2} \sum_\rho M_{\sigma \rho} m_\rho \right), \\
\chi^R_l &\sim \chi^R_l + 2\pi \sum_\sigma \omega^l_\sigma \left( n_\sigma -\frac{1}{2} \sum_\rho M_{\rho \sigma} m_\rho \right). 
\end{split}
\end{align}
This indicates that eigenvalues of the zero-mode operators $p^{R/L}_l$ lie in 
\begin{align}
\begin{split}
p^R_l &\in \sum_\sigma \omega^l_\sigma \left( n_\sigma +\frac{1}{2} \sum_\rho M_{\sigma \rho} m_\rho \right), \\
p^L_l &\in \sum_\sigma \omega^l_\sigma \left( n_\sigma -\frac{1}{2} \sum_\rho M_{\rho \sigma} m_\rho \right), 
\end{split}
\end{align}
since the mode expansion \eqref{eq:ModeExpansions} must reproduce the compactification condition \eqref{eq:compactification} under $x \to x+L$. 
Using the zero-mode operators $p^{R/L}_l$, we find that the operators $P^{R/L}_l$ are given by 
\begin{align} \label{eq:OperatorP}
P^{R/L}_l = \frac{1}{2} \sum_{\sigma, \sigma'} (\bfomega_\sigma \cdot \bfp^{R/L}) \omega^l_{\sigma'} M_{\sigma \sigma'}. 
\end{align}

\subsection{Operator product expansions}

The OPEs of the chiral fields $\bfchi^p$ are given by
\begin{align} \label{eq:FreeFieldOPE}
\begin{split}
\chi^R_l(z) \chi^R_{l'}(w) &\sim -\delta_{ll'} \ln (z-w), \\
\chi^L_l(\bz) \chi^L_{l'}(\bw) &\sim -\delta_{ll'} \ln (\bz-\bw). 
\end{split}
\end{align}
From these OPEs, we can derive several useful OPEs: 
\begin{subequations} \label{eq:FirstOPEs}
\begin{align}
\partial_z \chi^R_l(z) \partial_w \chi^R_m(w) &\sim -\frac{\delta_{lm}}{(z-w)^2}, \\
\partial_z \chi^R_l(z) :\mathrel{e^{i\bfa \cdot \bfchi^R(w)}}: &\sim \frac{-ia^l}{z-w} :\mathrel{e^{i\bfa \cdot \bfchi^R(w)}}:, \\
:\mathrel{e^{i\bfa \cdot \bfchi^R(z)}}: \partial_w \chi^R_l(w) &\sim \frac{ia^l}{z-w} :\mathrel{e^{i\bfa \cdot \bfchi^R(w)}}:, \\
\label{eq:VertexVertexOPE}
:\mathrel{e^{i\bfa \cdot \bfchi^R(z)}}: :\mathrel{e^{i\bfb \cdot \bfchi^R(w)}}: &\sim (z-w)^{\bfa \cdot \bfb} :\mathrel{e^{i\bfa \cdot \bfchi^R(z) +i\bfb \cdot \bfchi^R(w)}}:, 
\end{align}
\end{subequations}
where $\bfa$ and $\bfb$ are some $(N-1)$-dimensional vectors. 
If $\bfa \cdot \bfb <-1$ in Eq.~\eqref{eq:VertexVertexOPE}, further singular terms are generated by the Taylor expansion of $e^{i\bfa \cdot \bfchi^R(z)+i\bfb \cdot \bfchi^R(w)}$ about $w$. 
For example, if $\bfa=-\bfb$ and $|\bfa|^2=2$, we have 
\begin{align}
&:\mathrel{e^{i\bfa \cdot \bfchi^R(z)}}: :\mathrel{e^{-i\bfa \cdot \bfchi^R(w)}}: \nonumber \\
&\sim \frac{1}{(z-w)^2} +\frac{i\bfa \cdot \partial_w \bfchi^R(w)}{z-w} \nonumber \\
&\ \ \ -\frac{1}{2} :\mathrel{(\bfa \cdot \partial_w \bfchi^R(w))^2}: +\frac{i}{2} \bfa \cdot \partial_w^2 \bfchi^R(w), 
\end{align}
up to the zeroth order in $z-w$. 
Similar OPEs for $\bfchi^L$ can be obtained by simply replacing $z \to \bz$ and $w \to \bw$. 

\subsection{Vertex representation and cocycle}

We are now ready to consider a vertex representation of the $SU(N)_1$ current algebra \cite{Frenkel80,Segal81,Goddard86,dFMS}. 
In the Cartan-Weyl basis, the current corresponding to the Cartan subalgebra is given by
\begin{align} \label{eq:VertexRepCartan}
H^l(z) = i\partial_z \chi^R_l(z). 
\end{align}
To construct the other currents, let us consider the following form of vertex operator, 
\begin{align} \label{eq:VertexRepLadderFailed}
\tE^\bfalpha(z) \equiv \ :\mathrel{e^{i\bfalpha \cdot \bfchi^R(z)}}:, 
\end{align}
where $\bfalpha$ is a root of $SU(N)$. 
Such a vertex operator is compatible with the compactification condition (\ref{eq:compactification}) since $\bfalpha \cdot \bfomega_\sigma \in \mathbb{Z}$. 
Using Eq.~\eqref{eq:FirstOPEs}, those operators give the following OPEs: 
\begin{align}
&H^l(z) H^m(w) \sim \frac{\delta_{lm}}{(z-w)^2}, \\
&H^l(z) \tE^\bfalpha(w) \sim \frac{\alpha^l \tE^\bfalpha(w)}{z-w},
\end{align}
and
\begin{align}
&\tE^\bfalpha(z) \tE^\bfbeta(w) \nonumber \\
\label{eq:SUNCurrentOPE}
&\sim \begin{cases} 
\dfrac{1}{(z-w)^2} +\dfrac{\sum_l \alpha^l H^l(w)}{z-w} & (\bfalpha \cdot \bfbeta = -2) \\
\dfrac{\tE^{\bfalpha+\bfbeta}(w)}{z-w} & (\bfalpha \cdot \bfbeta = -1) \\
0 & (\textrm{otherwise}). 
\end{cases}
\end{align} 
Since $\bfalpha$ and $\bfbeta$ are roots, $\bfalpha \cdot \bfbeta = -2$ equivalently means $\bfalpha = -\bfbeta$, while $\bfalpha \cdot \bfbeta = -1$ means that $\bfalpha+\bfbeta$ is also a root of $SU(N)$. 

However, the vertex operators \eqref{eq:VertexRepLadderFailed} still do not give a faithful representation of the $SU(N)$ current algebra; 
we need to recover a correct sign factor in the second relation of Eq.~\eqref{eq:SUNCurrentOPE}. 
Following the prescription in Refs.~\cite{GSW1,dFMS}, we introduce the cocycle factor $c_\bfalpha(\bfp^R)$ dependent only upon the zero mode $\bfp^R$. 
Then we consider the modified vertex operator, 
\begin{align} \label{eq:VertexRepLadderPos}
E^\bfalpha(z) = c_\bfalpha(\bfp^R) \tE^\bfalpha(z),
\end{align}
for a positive root $\bfalpha$, while
\begin{align} \label{eq:VertexRepLadderNeg}
E^{-\bfalpha}(z) = \tE^{-\bfalpha}(z) c_{-\bfalpha}(\bfp^R),
\end{align}
for a negative root $-\bfalpha$. 
Since $c_\bfbeta(\bfp^R)$ depends only on $\bfp^R$, when it goes over $\tE^\bfalpha(z)$, we have 
\begin{align}
\tE^\bfalpha(z) c_\bfbeta(\bfp^R) = c_\bfbeta(\bfp^R-\bfalpha) \tE^\bfalpha(z). 
\end{align}
Hence, in order to obtain the correct algebra, we require that 
\begin{align} \label{eq:CocycleRequirement1}
c_\bfalpha(\bfp^R) c_\bfbeta(\bfp^R-\bfalpha) = (-1)^{\bfalpha \cdot \bfbeta} c_\bfbeta(\bfp^R) c_\bfalpha(\bfp^R-\bfbeta), 
\end{align}
and 
\begin{align} \label{eq:CocycleRequirement2}
c_\bfalpha(\bfp^R) c_\bfbeta(\bfp^R-\bfalpha) = \tepsilon(\bfalpha,\bfbeta) c_{\bfalpha+\bfbeta}(\bfp^R), 
\end{align}
with $\tepsilon(\bfalpha,\bfbeta) = \pm 1$. 
Such $c_\bfalpha(\bfp^R)$ can be chosen as 
\begin{align} \label{eq:CocycleFactor}
c_\bfalpha(\bfp^R) = e^{\frac{i\pi}{2} \sum_{\sigma,\sigma'} (\bfomega_\sigma \cdot \bfp^R) (\bfomega_{\sigma'} \cdot \bfalpha) M_{\sigma \sigma'}}, 
\end{align}
which only depends on the momentum part of the zero mode of $\bfchi^R$. 
Let us define the star product, 
\begin{align}
\bfalpha * \bfbeta \equiv \frac{1}{2} \sum_{\sigma,\sigma'} (\bfomega_\sigma \cdot \bfalpha) (\bfomega_{\sigma'} \cdot \bfbeta) M_{\sigma \sigma'}.  
\end{align}
Then we can write 
\begin{align}
c_\bfalpha(\bfp^R) = e^{i\pi \bfp^R * \bfalpha}. 
\end{align}
Since $\bfomega \cdot \bfalpha \in \mathbb{Z}$ for any two vectors $\bfalpha \in \Lambda^R_{SU(N)}$ and $\bfomega \in \Lambda^W_{SU(N)}$, $\bfalpha * \bfbeta \in \mathbb{Z}$ is obviously satisfied. 
The product also obeys the relation, 
\begin{align}\label{eq:Starproduct}
\bfalpha * \bfbeta 
&= \sum_{\sigma,\sigma'} (\bfomega_\sigma \cdot \bfalpha) (\bfomega_{\sigma'} \cdot \bfbeta) \left( \bfK_{SU(N)}-\frac{1}{2} \bfM^T \right)_{\sigma \sigma'} \nonumber \\
&= \bfalpha \cdot \bfbeta -\bfbeta * \bfalpha. 
\end{align}
Utilizing these facts, one can easily confirm that the cocycle factor defined in Eq.~\eqref{eq:CocycleFactor} satisfies the two requirements \eqref{eq:CocycleRequirement1} and \eqref{eq:CocycleRequirement2}. 
We can further find that $\tepsilon(\bfalpha,\bfbeta)=(-1)^{\bfalpha * \bfbeta}$. 
When $\bfalpha \cdot \bfbeta=-1$, the modified vertex operator $E^\bfalpha(z)$ satisfies the OPE, 
\begin{align}
E^\bfalpha(z) E^\bfbeta(w) \sim \frac{\epsilon(\bfalpha,\bfbeta) E^{\bfalpha+\bfbeta}(w)}{z-w},
\end{align}
where (denoting as $\bfgamma = \bfalpha+\bfbeta$)
\begin{align}
\epsilon(\bfalpha,\bfbeta) = \begin{cases} 
(-1)^{\bfalpha * \bfbeta} & (\bfalpha, \bfbeta, \bfgamma \in \Delta^+_N) \\
(-1)^{\bfbeta * \bfalpha} & (-\bfalpha, -\bfbeta, -\bfgamma \in \Delta^+_N) \\
(-1)^{\bfalpha * \bfbeta +1} & (\bfalpha, -\bfbeta, \bfgamma \in \Delta^+_N \\ & \ \textrm{or} \ -\bfalpha, \bfbeta, \bfgamma \in \Delta^+_N) \\
(-1)^{\bfbeta * \bfalpha +1} & (\bfalpha, -\bfbeta, -\bfgamma \in \Delta^+_N \\ & \ \textrm{or} \ -\bfalpha, \bfbeta, -\bfgamma \in \Delta^+_N)
\end{cases}
\end{align}
The interchange $z \leftrightarrow w$ with $\bfalpha \leftrightarrow \bfbeta$ pick up two minus signs from the single pole and $\epsilon(\bfalpha,\bfbeta)$, and thus they are canceled. 
Therefore, the ladder operators $E^\bfalpha(z)$, combined with the Cartan currents $H^l(z)$, give a faithful representation of the $SU(N)_1$ current algebra for the holomorphic (right-moving) part, as given in Eqs.~\eqref{eq:SUNCurrentAlgebra1}-\eqref{eq:SUNCurrentAlgebra3}. 

The exponent of the cocycle factor \eqref{eq:CocycleFactor} can be written as $i\pi \sum_l \alpha^l P^R_l$ with the operator $P^R_l$ defined in Eq.~\eqref{eq:OperatorP}.
Thus for a positive roots $\bfalpha$ of $SU(N)$, the currents $E^\bfalpha(z)$ are written in terms of the fields $\tbfchi^R(z)$ as
\begin{align} \label{eq:Vertexoperatoboso}
E^{\pm \bfalpha}(z) &= \frac{1}{x_c} e^{\pm i\pi \bfalpha * \bfalpha /2} e^{\pm i\bfalpha \cdot \tbfchi^R(z)} \nonumber \\
&=  \frac{\pm i}{x_c} e^{\pm i\bfalpha \cdot \tbfchi^R(z)},
\end{align} 
where $x_c$ is a microscopic cutoff and we have used the identity (\ref{eq:Starproduct}) and the normalization $|\bfalpha|^2 = 2$. 
Here the vertex operator in the right-hand side is not normal-ordered and thus is exactly what appeared in the interwire interaction \eqref{eq:InterIntSUN3}. 
Applying a similar argument to the vertex representation of the antiholomorphic (left-moving) current algebra, we find 
\begin{align} \label{eq:VertexRepSUN1CurrentL}
\begin{split}
\bH^l(\bz) &= i\partial_{\bz} \tchi^L_l(\bz), \\
\bE^{\pm \bfalpha}(\bz) &= \frac{\pm i}{x_c} e^{\pm i \bfalpha \cdot \tbfchi^L(\bz)}, 
\end{split}
\end{align}
for a positive root $\bfalpha$. 
We also remark that from the commutation relations \eqref{eq:tchiLRComm}, the left and right currents commute with each other since 
\begin{align}
e^{i\bfalpha \cdot \tbfchi^R(z)} e^{i\bfbeta \cdot \tbfchi^L(\bz)} = e^{-2i\pi \bfalpha * \bfbeta} e^{i\bfbeta \cdot \tbfchi^L(\bz)} e^{i\bfalpha \cdot \tbfchi^R(z)}, 
\end{align}
and $\bfalpha * \bfbeta \in \mathbb{Z}$. 

\begin{widetext}

\section{Vertex representations of energy-momentum tensors} \label{app:EMTensor}

We here prove the conformal embeddings given in Eqs.~\eqref{eq:EmbeddingSU3k} and \eqref{eq:EmbeddingSUNk} by explicit calculations of the corresponding energy-momentum tensors. 
In the following, we only focus on the right-moving sector, while the same results also hold for the left-moving sector. 

We first consider the energy-momentum tensor of the $SU(N)_1$ WZW CFT. 
In the Sugawara form, it is given by 
\begin{align}
T_{SU(N)_1}(z) = \frac{1}{2(N+1)} \Biggl[ \sum_{l=1}^{N-1} :\mathrel{H^l(z) H^l(z)}: 
+\sum_{\bfalpha \in \Delta_N} :\mathrel{E^\bfalpha(z) E^{-\bfalpha}(z)}: \Biggr], 
\end{align}
where $H^l(z)$ and $E^\bfalpha(z)$ are $SU(N)_1$ currents in the Cartan-Weyl basis. 
Their vertex representations are explicitly obtained in Eq.~\eqref{eq:VertexRepCartan}, \eqref{eq:VertexRepLadderPos}, and \eqref{eq:VertexRepLadderNeg}. 
Substituting those vertex representations and applying the OPE \eqref{eq:VertexVertexOPE}, we can easily find \cite{dFMS}
\begin{align} \label{eq:SUN1EMTensorVertex}
T_{SU(N)_1}(z) = -\frac{1}{2} :\mathrel{(\partial_z \bfchi^R(z))^2}:. 
\end{align}
The corresponding Hamiltonian is nothing but the right-moving sector of the SLL Hamiltonian given in Eq.~\eqref{eq:SLLHamSU3} for $N=3$ and that in Eq.~\eqref{eq:ChiralHam} for general $N$. 

We next consider the energy-momentum tensor of the $SU(N)_k$ WZW CFT. 
We follow the discussion by Dunne, Halliday, and Suranyi \cite{Dunne89} with a slight modification on the cocycles. 
In terms of the $SU(N)_k$ currents, it is given by 
\begin{align} \label{eq:SUNkEMTensorSugawara}
T_{SU(N)_k}(z) = \frac{1}{2(N+k)} \Biggl[ \sum_{l=1}^{N-1} :\mathrel{\calH^l(z) \calH^l(z)}:
+\sum_{\bfalpha \in \Delta_N} :\mathrel{\calE^\bfalpha(z) \calE^{-\bfalpha}(z)}: \Biggl]. 
\end{align}
Using Eqs.~\eqref{eq:SUNkCurrentCartan} and \eqref{eq:SUNkCurrentLadder} and the vertex representations of the $SU(N)_1$ currents, we obtain for a positive root $\bfalpha$, 
\begin{subequations} \label{eq:VertexSUNkCurrent}
\begin{align}
\calH^l(z) &= \sum_{a=1}^k i\partial_z \chi^R_{l,a}(z), \\
\calE^\bfalpha(z) &= \sum_{a=1}^k c_\bfalpha (\bfp^R_a) :\mathrel{e^{i\bfalpha \cdot \bfchi^R_a(z)}}:, \\
\calE^{-\bfalpha}(z) &= \sum_{a=1}^k :\mathrel{e^{-i\bfalpha \cdot \bfchi_a^R(z)}}: c_{-\bfalpha}(\bfp^R_a). 
\end{align}
\end{subequations}
Substituting these expressions into Eq.~\eqref{eq:SUNkEMTensorSugawara}, we find 
\begin{align}
T_{SU(N)_k} = \frac{1}{2(N+k)} \Biggl[ -\sum_{a,b=1}^k :\mathrel{\partial_z \bfchi^R_a \cdot \partial_z \bfchi^R_b}:
-\frac{1}{2} \sum_{a=1}^k \sum_{\bfalpha \in \Delta_N} :\mathrel{(\bfalpha \cdot \partial_z \bfchi^R_a)^2}:
+\sum_{a \neq b} \sum_{\bfalpha \in \Delta_N} c_\bfalpha(\bfp^R_a-\bfp^R_b-\bfalpha) :\mathrel{e^{i\bfalpha \cdot (\bfchi^R_a-\bfchi^R_b)}}: \Biggr]. 
\end{align}
Using the identity $\sum_{\bfalpha \in \Delta_N} \alpha^l \alpha^m = 2N\delta_{lm}$ and introducing the new fields [see Eq.~\eqref{eq:ChargeNeutralFieldSUNk}], 
\begin{align} \label{eq:ChargeNeutralChiral}
X^R_l = \frac{1}{\sqrt{k}} \sum_{a=1}^k \chi^R_{l,a}, \hspace{10pt} Y^{R,\mu}_l = \sum_{a=1}^k W^\mu_a \chi^R_{l,a}, 
\end{align}
we obtain 
\begin{align}
T_{SU(N)_k} = \frac{1}{2(N+k)} \Biggl[ -(k+N) :\mathrel{(\partial_z \bfX^R)^2}: -N \sum_{\mu=1}^{k-1} :\mathrel{(\partial_z \bfY^{R,\mu})^2}: 
+\sum_{a \neq b} \sum_{\bfalpha \in \Delta_N} c_\bfalpha(\bfp^R_a-\bfp^R_b-\bfalpha) :\mathrel{e^{i\bfalpha (\bfW_a-\bfW_b) \cdot \bfY^R}}: \Biggr]. 
\end{align}
By using the fact that the vectors $\bfA = \bfW_a-\bfW_b$ for $a \neq b$ form a set of roots of $SU(k)$, we finally obtain 
\begin{align} \label{eq:SUNkEMTensorVertex}
T_{SU(N)_k} = -\frac{1}{2} :\mathrel{(\partial_z \bfX^R)^2}: +\frac{1}{2(N+k)} \Biggl[ -N \sum_{\mu=1}^{k-1} :\mathrel{(\partial_z \bfY^{R,\mu})^2}:
-\sum_{\bfalpha \in \Delta_N} \sum_{\bfA \in \Delta_k} c_\bfalpha^\bfA (\bfp^R_Y) :\mathrel{e^{i\bfalpha \bfA \cdot \bfY^R}}: \Biggr], 
\end{align}
where the cocycle factor for $\bfA$ has been calculated as 
\begin{align}
c^\bfalpha(\bfp^R_a-\bfp^R_b-\bfalpha) = e^{-i\pi \bfalpha * \bfalpha} c_\bfalpha^\bfA (\bfp^R_Y), \hspace{10pt} 
c_\bfalpha^\bfA (\bfp) \equiv e^{i\pi \sum_\mu A^\mu \bfp^\mu * \bfalpha},
\end{align}
which depends only on the zero-mode momentum of $\bfY^{R,\mu}$ and thus commutes with $\bfX^R$. 
When the cocycle goes over a vertex operator involving $\bfY^{R,\mu}$, it gives a phase factor: 
\begin{align}
:\mathrel{e^{i\bfalpha \bfA \cdot \bfY^R(z)}}: c_\bfbeta^\bfB (\bfp^R_Y)
= e^{-i\pi (\bfA \cdot \bfB) (\bfalpha * \bfbeta)} c_\bfbeta^\bfB (\bfp^R_Y) :\mathrel{e^{i\bfalpha \bfA \cdot \bfY^R(z)}}:
\end{align}

Since the charge fields $\bfX^R$ and the neutral fields $\bfY^{R,\mu}$ are decoupled in Eq.~\eqref{eq:SUNkEMTensorVertex}, it can be naturally split into the energy-momentum tensors of the $[U(1)]^{N-1}$ and $SU(N)_k/[U(1)]^{N-1}$ CFTs as 
\begin{align}
T_{SU(N)_k} = T_{[U(1)]^{N-1}} +T_{SU(N)_k/[U(1)]^{N-1}}. 
\end{align}
Then we find
\begin{align}
\label{eq:U1EMTensorVertex}
T_{[U(1)]^{N-1}} &= -\frac{1}{2} :\mathrel{(\partial_z \bfX^R)^2}:, \\
\label{eq:SUNkParaEMTensor}
T_{SU(N)_k/[U(1)]^{N-1}} &= \frac{1}{2(N+k)} \Bigg[ -N \sum_{\mu=1}^{k-1} :\mathrel{(\partial_z \bfY^{R,\mu})^2}:
-\sum_{\bfalpha \in \Delta_N} \sum_{\bfA \in \Delta_k} c_\bfalpha^\bfA (\bfp^R_Y) :\mathrel{e^{i\bfalpha \bfA \cdot \bfY^R}}: \Biggr]. 
\end{align}
The conformal embedding \eqref{eq:EmbeddingSUNk} further implies that 
\begin{align}
T_{[SU(N)_1]^k} = T_{[U(1)]^{N-1}} +T_{SU(N)_k/[U(1)]^{N-1}}
+T_{SU(k)_N/[U(1)]^{k-1}}. 
\end{align}
Thus a vertex representation of the energy-momentum tensor for the $SU(k)_N/[U(1)]^{k-1}$ CFT may be obtained by subtracting the two tensors \eqref{eq:U1EMTensorVertex} and \eqref{eq:SUNkParaEMTensor} from that for the $[SU(N)_1]^k$ CFT. 
Using Eq.~\eqref{eq:SUN1EMTensorVertex}, the energy-momentum tensor for $k$ copies of the $SU(N)_1$ WZW CFT is given by 
\begin{align}
T_{[SU(N)_1]^k} = -\frac{1}{2} \sum_{a=1}^k :\mathrel{(\partial_z \bfchi^R_a)^2}:. 
\end{align}
It can be further written in terms of $\bfX^R$ and $\bfY^{R,\mu}$ as 
\begin{align}
T_{[SU(N)_1]^k} = -\frac{1}{2} :\mathrel{(\partial_z \bfX^R)^2}: -\frac{1}{2} \sum_{\mu=1}^{k-1} :\mathrel{(\partial_z \bfY^{R,\mu})^2}:. 
\end{align}
Hence we find 
\begin{align} \label{eq:SUkNParaEMTensor}
T_{SU(k)_N/[U(1)]^{k-1}} = \frac{1}{2(N+k)} \Biggl[ -k \sum_{\mu=1}^{k-1} :\mathrel{(\partial_z \bfY^{R,\mu})^2}:
+\sum_{\bfalpha \in \Delta_N} \sum_{\bfA \in \Delta_k} c_\bfalpha^\bfA (\bfp^R_Y) :\mathrel{e^{i\bfalpha \bfA \cdot \bfY^R}}: \Biggr]. 
\end{align}
Now the cocycle factor combined with a coordinate-free part of $:\mathrel{e^{i\bfalpha \bfA \cdot \bfY^R}}:$ [cf. Eq.~\eqref{eq:ModeExpansions}], 
\begin{align}
\hat{c}^\bfA_\bfalpha(\bfp^R_Y) \equiv -c^\bfA_\bfalpha(\bfp^R_Y) e^{i\bfalpha \bfA \cdot \bfq^R_Y}, 
\end{align}
satisfies the same algebraic properties as given in Eq.~(C.6) of Ref.~\cite{Dunne89}. 
Therefore, one can show by the same argument in Ref.~\cite{Dunne89} that the above vertex representations of the energy-momentum tensors fulfill the OPEs \cite{dFMS}, 
\begin{align} \label{eq:EMTensorOPE}
T(z) T(w) \sim \frac{c/2}{(z-w)^4} +\frac{2T(w)}{(z-w)^2} +\frac{\partial_w T(w)}{z-w}. 
\end{align}
The central charge of each CFT is given by 
\begin{subequations}
\begin{align}
c_{[SU(N)_1]^k} &= k(N-1), \\
c_{[U(1)]^{N-1}} &= N-1, \\
c_{SU(N)_k/[U(1)]^{N-1}} &= \frac{N(N-1)(k-1)}{N+k}, \\
c_{SU(k)_N/[U(1)]^{k-1}} &= \frac{k(k-1)(N-1)}{N+k}. 
\end{align}
\end{subequations}

\section{Vertex representation of $SU(N)_k/[U(1)]^{N-1}$ parafermionic fields} \label{app:SUNkPara}

We here explicitly construct the vertex representations of Gepner parafermionic fields of the $SU(N)_k/[U(1)]^{N-1}$ CFT \cite{Gepner87}. 
When the SLL Hamiltonian is fine-tuned to be $k$ copies of the $SU(N)_1$ WZW CFT in each wire, the interwire interactions can be identified as products of left- and right-moving parafermionic fields in neighboring wires. 

\subsection{Vertex representation}

In terms of the charge and neutral fields in Eq.~\eqref{eq:ChargeNeutralChiral}, the vertex representations of the $SU(N)_k$ currents \eqref{eq:VertexSUNkCurrent} are given by 
\begin{align}
\calE^\bfalpha(z) &= \sum_{a=1}^k c_{\bfalpha} \left( \frac{\bfp^R_X}{\sqrt{k}} \right) c_\bfalpha^{\bfW_a} (\bfp^R_Y) :\mathrel{e^{\frac{i}{\sqrt{k}} \bfalpha \cdot \bfX^R(z) +i\bfalpha \bfW_a \cdot \bfY^R(z)}}: \nonumber \\
&= c_\bfalpha \left( \frac{\bfp^R_X}{\sqrt{k}} \right) :\mathrel{e^{\frac{i}{\sqrt{k}} \bfalpha \cdot \bfX^R(z)}}: \tPsi^{\bfalpha,1}(z), 
\end{align}
for positive roots $\bfalpha \in \Delta^+_N$, where $\bfp^R_X$ is the zero-mode momentum of $\bfX^R$. 
Here we have defined the \emph{unnormalized} parafermionic operator, 
\begin{align}
\tPsi^{\bfalpha,1}(z) = \sum_{\bfW \in [\bfW_1]} c_\bfalpha^\bfW (\bfp^R_Y) :\mathrel{e^{i\bfalpha \bfW \cdot \bfY^R(z)}}:. 
\end{align}
We have also used the fact that the weights $\bfW_a$ belong to the first antisymmetric representation of $SU(k)$, which is characterized by the highest weight $\bfW_1$ under the specific choice of $\bfW_a$ in Eq.~\eqref{eq:SpecificVectorW}. 
We similarly define the parafermionic operator $\tPsi^{-\bfalpha,1}$ for a negative root $-\bfalpha$ as 
\begin{align}
\tPsi^{-\bfalpha,1}(z) = \sum_{\bfW \in [\bfW_1]} :\mathrel{e^{-i\bfalpha \bfW \cdot \bfY(z)}}: c_{-\bfalpha}^\bfW (\bfp_Y), 
\end{align}
through the $SU(N)_k$ current for $-\bfalpha$, 
\begin{align}
\calE^{-\bfalpha}(z) = \tPsi^{-\bfalpha,1}(z) :\mathrel{e^{-\frac{i}{\sqrt{k}} \bfalpha \cdot \bfX^R(z)}}: c_{-\bfalpha} \left( \frac{\bfp^R_X}{\sqrt{k}} \right). 
\end{align}
The above definition indicates that $\tPsi^{\bfalpha,1 \dagger} = \tPsi^{-\bfalpha,1}$. 

The normalization of $\tPsi^{\bfalpha,1}$ is fixed by demanding that the leading singularity of the OPE $\tPsi^{\bfalpha,1}(z) \tPsi^{-\bfalpha,1}(w)$ has a unity coefficient. 
This OPE is shown later along with the Gepner parafermionic algebra. 
As a result, the \emph{normalized} parafermionic operator is given by
\begin{align}
\Psi^{\bfalpha,1}(z) = \frac{1}{\sqrt{k}} \sum_{\bfW \in [\bfW_1]} c_\bfalpha^\bfW (\bfp^R_Y) :\mathrel{e^{i\bfalpha \bfW \cdot \bfY^R(z)}}:, 
\end{align}
for $\bfalpha \in \Delta^+_N$. 
In terms of the fields $\tbfY^{R,\mu}$ used in the main text, we can further write it as 
\begin{align}
\Psi^{\bfalpha,1}(z) = \frac{e^{\frac{i\pi}{2} (1-\frac{1}{k})}}{\sqrt{k} x_c^{1-\frac{1}{k}}} \sum_{\bfW \in [\bfW_1]} e^{i\bfalpha \bfW \cdot \tbfY^R(z)}. 
\end{align}
This gives the vertex representations in Eqs.~\eqref{eq:VertexRepSU32Para} and \eqref{eq:VertexRepSUNkPara}. 
Combining with 
\begin{align}
c_{\bfalpha} \left( \frac{\bfp^R_X}{\sqrt{k}} \right) :\mathrel{e^{\frac{i}{\sqrt{k}} \bfalpha \cdot \bfX^R(z)}}: \ = \frac{e^{i\pi/2k}}{x_c^{1/k}} e^{\frac{i}{\sqrt{k}} \bfalpha \cdot \tbfX^R(z)}, 
\end{align}
we find 
\begin{align}
\calE^\bfalpha(z) = \frac{\sqrt{k} e^{i\pi/2k}}{x_c^{1/k}} e^{\frac{i}{\sqrt{k}} \bfalpha \cdot \tbfX^R(z)} \Psi^{\bfalpha,1}(z). 
\end{align}
This gives Eqs.~\eqref{eq:VertexRepSU32Cur} and \eqref{eq:VertexRepSUNkCur}. 
Similar expressions for $\Psi^{-\bfalpha,1}$ are simply given by the Hermitian conjugates of those for $\Psi^{\bfalpha,1}$. 

We so far consider only the right-moving sector. 
From Eq.~\eqref{eq:FVBoson}, the parafermion in the left-moving sector must involve the parafermionic cocycle with the zero-mode momentum of the \emph{right-moving} field, which is of the form $c_\bfalpha^\bfW (2\bfp^R_Y)$. 
Despite this additional factor, the remaining discussion in this appendix similarly holds for the left-moving parafermionic fields, since the associated phase factor always appears in the form $e^{2i\pi (\bfA \cdot \bfW) (\bfalpha * \bfbeta)}=1$ and thus does not play any role. 
When we consider the interwire interaction, which is the left-right product of parafermionic fields, we do not need to care about this factor since the left and right parafermionic fields come from different wires.
On the other hand, we need a special care for the intrawire interaction which involves left and right fields from the same wire, as discussed in the next appendix.

\subsection{$\Psi$ is primary} \label{app:PsiIsPrimary}

Here we prove that $\tPsi^{\bfalpha,1}$ is a primary field of the $SU(N)_k/[U(1)]^{N-1}$ CFT. 
A similar procedure is also applied to $\tPsi^{-\bfalpha,1}$. 
Let us calculate the OPE with the energy-momentum tensor \eqref{eq:SUNkParaEMTensor}, 
\begin{align}
&T_{SU(N)_k/[U(1)]^{N-1}}(z) \tPsi^{\bfbeta,1}(w) \nonumber \\
&= \frac{1}{2(N+k)} \Biggl[ -N\sum_{\mu=1}^{k-1} :\mathrel{(\partial_z \bfY^{R,\mu}(z))^2}: \tPsi^{\bfbeta,1}(w)
-\sum_{\bfalpha \in \Delta_N} \sum_{\bfA \in \Delta_k} c_\bfalpha^\bfA (\bfp^R_Y) :\mathrel{e^{i\bfalpha \bfA \cdot \bfY^R(z)}}: \tPsi^{\bfbeta,1}(w) \Biggl] \nonumber \\
&\equiv \frac{1}{2(N+k)} \left[ -N \calO_{C \Psi}(z,w) -\calO_{L \Psi}(z,w) \right]. 
\end{align}
The following calculation takes a similar way to Ref.~\cite{Dunne89} for $N=2$ but we here explicitly keep the parafermionic cocycles for $N \geq 3$. 
Using
\begin{align}
\sum_{\mu=1}^{k-1} :\mathrel{(\partial_z \bfY^{R,\mu}(z))^2}: :\mathrel{e^{i\bfbeta \bfW \cdot \bfY^R(w)}}: \ \sim -\frac{|\bfbeta|^2 |\bfW|^2 :\mathrel{e^{i\bfbeta \bfW \cdot \bfY^R(w)}}:}{(z-w)^2} -\frac{2 :\mathrel{\partial_w e^{i\bfbeta \bfW \cdot \bfY^R(w)}}:}{z-w}, 
\end{align}
and $|\bfbeta|^2 |\bfW|^2 = 2(1-1/k)$, the OPE with the Cartan part, $\calO_{C\Psi}$, is given by 
\begin{align} \label{eq:SUNkCartanPara}
\calO_{C\Psi}(z,w) \sim \frac{2(1/k-1)}{(z-w)^2} \tPsi^{\bfbeta,1}(w) -\frac{2}{z-w} \partial_w \tPsi^{\bfbeta,1}(w). 
\end{align}
Using Eq.~\eqref{eq:VertexVertexOPE}, the OPE with the ladder part, $\calO_{L\Psi}$, is given by 
\begin{align} \label{eq:SUNkLadderPara}
\calO_{L\Psi}(z,w) \sim \sum_{\bfalpha \in \Delta_N} \sum_{\bfA \in \Delta_k} \sum_{\bfW \in [\bfW_1]} \frac{ e^{-i\pi (\bfA \cdot \bfW) (\bfalpha * \bfbeta)}}{(z-w)^{-(\bfalpha \cdot \bfbeta) (\bfA \cdot \bfW)}}
c_\bfalpha^\bfA (\bfp^R_Y) c_\bfbeta^\bfW (\bfp^R_Y):\mathrel{e^{i\bfalpha \bfA \cdot \bfY^R(z) +i\bfbeta \bfW \cdot \bfY^R(w)}}:. 
\end{align}
Thus the singular parts will appear when one of the following conditions is satisfied: (i) $\bfalpha \cdot \bfbeta=2$ and $\bfA \cdot \bfW=-1$, (ii) $\bfalpha \cdot \bfbeta=-2$ and $\bfA \cdot \bfW=1$, (iii) $\bfalpha \cdot \bfbeta=1$ and $\bfA \cdot \bfW=-1$, or (iv) $\bfalpha \cdot \bfbeta=-1$ and $\bfA \cdot \bfW=1$. 
For the case (i), since $\bfalpha=\bfbeta$, Eq.~\eqref{eq:SUNkLadderPara} can be written as 
\begin{align} \label{eq:SUNkLadderPara2}
\calO_{L\Psi}^\textrm{(i)}(z,w) &\sim \sum_{\bfA \in \Delta_k} \sum_{\substack{\bfW \in [\bfW_1] \\ \bfA \cdot \bfW=-1}} \frac{e^{i\pi \bfbeta * \bfbeta}}{(z-w)^2} c_\bfbeta^\bfA (\bfp^R_Y) c_\bfbeta^\bfW (\bfp^R_Y)
:\mathrel{e^{i\bfbeta \bfA \cdot \bfY^R(z)+i\bfbeta \bfW \cdot \bfY^R(w)}}: \nonumber \\
&\sim -\sum_{\bfA \in \Delta_k} \sum_{\substack{\bfW \in [\bfW_1] \\ \bfA \cdot \bfW=-1}} c_\bfbeta^{\bfA+\bfW} (\bfp^R_Y)
\biggl[ \frac{:\mathrel{e^{i\bfbeta (\bfA+\bfW) \cdot \bfY^R(w)}}:}{(z-w)^2}
+\frac{:\mathrel{i\bfbeta \bfA \cdot \partial_w \bfY^R(w) e^{i\bfbeta (\bfA+\bfW) \cdot \bfY^R(w)}}:}{z-w} \biggr]. 
\end{align}
Using the identities for an arbitrary scalar function $f$ and vector function $\bff$, 
\begin{align}
\begin{split}
\sum_{\bfA \in \Delta_k} \sum_{\substack{\bfW \in [\bfW_1] \\ \bfA \cdot \bfW=-1}} f(\bfA+\bfW) &= (k-1) \sum_{\bfW \in [\bfW_1]} f(\bfW), \\
\sum_{\bfA \in \Delta_k} \sum_{\substack{\bfW \in [\bfW_1] \\ \bfA \cdot \bfW=-1}} \bfA \cdot \bff(\bfA+\bfW) &= k \sum_{\bfW \in [\bfW_1]} \bfW \cdot \bff(\bfW), 
\end{split}
\end{align}
Eq.~\eqref{eq:SUNkLadderPara2} is finally written as 
\begin{align} \label{eq:SUNkLadderPara3}
\calO^\textrm{(i)}_{L\Psi}(z,w) \sim \frac{1-k}{(z-w)^2} \tPsi^{\bfbeta,1}(w) -\frac{k}{z-w} \partial_w \tPsi^{\bfbeta,1}(w). 
\end{align}
Since the case (ii) is obtained by $\bfalpha \to -\bfalpha$ and $\bfA \to -\bfA$ in the case (i), it yields the same contribution as Eq.~\eqref{eq:SUNkLadderPara3}. 
For the case (iv), Eq.~\eqref{eq:SUNkLadderPara} is written as 
\begin{align} \label{eq:SUNkLadderPara4}
\calO^\textrm{(iv)}_{L\Psi}(z,w) \sim \sum_{\substack{\bfalpha \in \Delta_N \\ \bfalpha \cdot \bfbeta=-1}} \sum_{\bfA \in \Delta_k} \sum_{\substack{\bfW \in [\bfW_1] \\ \bfA \cdot \bfW =1}} \frac{e^{-i\pi \bfalpha * \bfbeta}}{z-w} c_\bfalpha^\bfA (\bfp^R_Y)
c_\bfbeta^\bfW (\bfp^R_Y) :\mathrel{e^{i(\bfalpha \bfA +\bfbeta \bfW) \cdot \bfY^R(w)}}:
\end{align}
In fact, thanks to the parafermionic cocycle, the contributions from the cases (iii) and (iv) identically vanish. 
One way to recognize this is to notice that the condition (iv) is equivalently written as $\bfalpha +\bfbeta \in \Delta_N$ and $\bfW -\bfA \in [\bfW_1]$. 
Thus we can change the variables as $\bfalpha \to \bfalpha-\bfbeta$ and $\bfW \to \bfA +\bfW$. 
This change yields the same condition as (iii) and hence Eq.~\eqref{eq:SUNkLadderPara4} becomes 
\begin{align} \label{eq:SUNkLadderPara5}
\calO_{L\Psi}^\textrm{(iv)}(z,w) &\sim \sum_{\substack{\bfalpha \in \Delta_N \\ \bfalpha \cdot \bfbeta=1}} \sum_{\bfA \in \Delta_k} \sum_{\substack{\bfW \in [\bfW_1] \\ \bfA \cdot \bfW =-1}} \frac{e^{-i\pi (\bfalpha-\bfbeta) * \bfbeta}}{z-w} c_{\bfalpha-\bfbeta}^\bfA (\bfp^R_Y)
c_\bfbeta^{\bfA+\bfW} (\bfp^R_Y) :\mathrel{e^{i[(\bfalpha-\bfbeta) \bfA +\bfbeta (\bfA+\bfW)] \cdot \bfY^R(w)}}: \nonumber \\
&\sim \sum_{\substack{\bfalpha \in \Delta_N \\ \bfalpha \cdot \bfbeta=1}} \sum_{\bfA \in \Delta_k} \sum_{\substack{\bfW \in [\bfW_1] \\ \bfA \cdot \bfW =-1}} \frac{-e^{-i\pi \bfalpha * \bfbeta}}{z-w} c_\bfalpha^\bfA (\bfp^R_Y)
c_\bfbeta^\bfW (\bfp^R_Y) :\mathrel{e^{i(\bfalpha \bfA +i\bfbeta \bfW) \cdot \bfY^R(w)}}:. 
\end{align}
Since the parafermionic cocycle picks up an extra minus sign, Eq.~\eqref{eq:SUNkLadderPara5} is exactly canceled with the contribution from the case (iii). 
Combining with Eqs.~\eqref{eq:SUNkCartanPara} and \eqref{eq:SUNkLadderPara3}, we finally obtain 
\begin{align}
T_{SU(N)_k/[U(1)]^{N-1}}(z) \tPsi^{\bfbeta,1}(w)
\sim \frac{1-1/k}{(z-w)^2} \tPsi^{\bfbeta,1}(w) +\frac{1}{z-w} \tPsi^{\bfbeta,1}(w). 
\end{align}
This indicates that $\tPsi^{\bfbeta,1}$ is a primary field of the $SU(N)_k/[U(1)]^{N-1}$ CFT with conformal weight $1-1/k$. 
Following the above procedure, it is also straightforward to show that 
\begin{align}
T_{SU(k)_N/[U(1)]^{k-1}}(z) \tPsi^{\bfbeta,1}(w) \sim 0. 
\end{align}
Thus the parafermion $\tPsi^{\bfalpha,1}$ is independent of the $SU(k)_N/[U(1)]^{k-1}$ CFT. 

\subsection{Gepner parafermionic algebra}

The OPE of $\tPsi^{\bfalpha,1}$ generates the parafermionic algebra generalized to $SU(N)$ by Gepner \cite{Gepner87}. 
To obtain the closed form, we further define the parafermions associated with the $I$-th antisymmetric representation of $SU(k)$, 
\begin{align}
\tPsi^{\bfalpha,I}(z) = \sum_{\bfW \in [\tbfW_I]} c_\bfalpha^\bfW (\bfp^R_Y) :\mathrel{e^{i\bfalpha \bfW \cdot \bfY^R(z)}}:, 
\end{align}
where the corresponding highest weight is given by 
\begin{align}
\tbfW_I = \sum_{a=1}^I \bfW_a, 
\end{align}
upon the choice in Eq.~\eqref{eq:SpecificVectorW}. 
In our convention, we multiply the parafermionic cocycle from the left side of the vertex operator when $\bfalpha$ is a positive root of $SU(N)$ after we fix the representation of $SU(k)$, while the cocycle is multiplied from the right side when $\bfalpha$ is a negative root. 
By a straightforward application of the above analysis, one can show that $\tPsi^{\bfalpha,I}$ is a primary field of the $SU(N)_k/[U(1)]^{N-1}$ CFT with conformal weight $I(k-I)/k$. 
The parafermion associated with a positive root $\bfalpha$ and $[\tbfW_I]$ is related to that with the negative root $-\bfalpha$ and the conjugate representation $[\tbfW_{k-I}]$ by 
\begin{align}
\tPsi^{\pm \bfalpha,I}(z) = e^{\pm i\pi I(k-I)/k} \tPsi^{\mp \bfalpha,k-I}(z), 
\end{align}
since $[\tbfW_I] = -[\tbfW_{k-I}]$. 
Keeping only the most singular part, the OPE system is given as follows:
The OPE of $\Psi^{\bfalpha,I}$ with its conjugate is given by 
\begin{align}
\tPsi^{\bfalpha,I}(z) \tPsi^{-\bfalpha,I}(w) \sim \frac{k!/I!(k-I)!}{(z-w)^{2I(k-I)/k}}. 
\end{align}
This fixes the appropriate normalization of $\tPsi^{\bfalpha,I}$, which is given by $\sqrt{I!(k-I)!/k!}$. 
For $I+J<k$ and $\bfalpha \in \Delta^+_N$,
\begin{align}
\tPsi^{\pm \bfalpha,I}(z) \tPsi^{\pm \bfalpha,J}(w) \sim \frac{(I+J)!}{I!J!} \frac{e^{\pm i\pi IJ/k}}{(z-w)^{2IJ/k}} \tPsi^{\pm \bfalpha,I+J}(w). 
\end{align}
For $I>J$ and $\bfalpha \in \Delta^+_N$, 
\begin{align}
\tPsi^{\pm \bfalpha,I}(z) \tPsi^{\mp \bfalpha,J}(w)
\sim \frac{(k-I+J)!}{(k-I)!J!} \frac{e^{\pm i\pi J(J-I)/k}}{(z-w)^{2(k-I)J/k}} \tPsi^{\pm \bfalpha,I-J}(w). 
\end{align}
For $\bfalpha$ and $\bfbeta$ such that $\bfgamma = \bfalpha+\bfbeta \in \Delta_N$, 
\begin{align}
\tPsi^{\bfalpha,I}(z) \tPsi^{\bfbeta,I}(w) \sim \frac{e^{i\pi C(\bfalpha, \bfbeta) I(k-I)/k}}{(z-w)^{I(k-I)/k}} \tPsi^{\bfgamma,I}(w), 
\end{align}
where
\begin{align}
C(\bfalpha,\bfbeta) = \begin{cases}
-\bfalpha * \bfbeta & (\bfalpha,\bfbeta,\bfgamma \in \Delta^+_N) \\
\bfbeta * \bfalpha & (-\bfalpha,-\bfbeta,-\bfgamma \in \Delta^+_N) \\
-\bfalpha * \bfbeta -1 & (\bfalpha, -\bfbeta, \bfgamma \in \Delta^+_N \ \textrm{or} \ -\bfalpha, \bfbeta, \bfgamma \in \Delta^+_N ) \\
\bfbeta * \bfalpha +1 & (\bfalpha, -\bfbeta, -\bfgamma \in \Delta^+_N \ \textrm{or} \ -\bfalpha, \bfbeta, -\bfgamma \in \Delta^+_N)
\end{cases}
\end{align}
These OPEs coincide with those obtained in Ref.~\cite{Dunne89} up to the normalization and the phase factors appearing from the parafermionic cocycle. 

\section{Vertex representation of $SU(k)_N/[U(1)]^{k-1}$ parafermionic fields} \label{app:SUkNPara}

Similarly to Appendix~\ref{app:SUNkPara}, we construct the vertex representations of Gepner parafermionic fields of the $SU(k)_N/[U(1)]^{k-1}$ CFTs \cite{Gepner87}. 
Now the intrawire interactions are identified as those parafermionic fields when the SLL Hamiltonian is fine-tuned. 
In contrast to the previous case, there are no decompositions of the intrawire interactions into left- and right-moving parafermionic fields for $N \geq 3$, while we can decompose them into chiral primary fields. 
Nevertheless, we show that the intrawire interactions themselves behave as nonchiral products of left- and right-moving parafermionic fields. 

\subsection{Vertex representation in the chiral sector}

For each \emph{chiral} sector of the $SU(k)_N/[U(1)]^{k-1}$ CFT, we find that the following vertex representation becomes a primary field with conformal weight $1-1/N$: 
\begin{align} \label{eq:VertexRepXi}
\Xi^\bfA(z) = \sum_{\sigma=1}^N e^{-ik_\sigma} c_{\bfomega_\sigma}^\bfA (\bfp^R_Y) :\mathrel{e^{i\bfomega_\sigma \bfA \cdot \bfY^R(z)}}:, 
\end{align}
where $\bfA$ is a root of $SU(k)$ and $\bfomega_\sigma$ constitutes the fundamental representation of $SU(N)$. 
The phases $k_\sigma$ are determined later. 
For $N=2$, the parafermionic cocycle is redundant so that this operator simply reduces to the operator defined in Eq.~(4.21) of Ref.~\cite{Teo14} up to the phase factor, which is the $SU(k)_2/[U(1)]^{k-1}$ parafermionic field associated with a root of $SU(k)$. 
For $N \geq 3$, the operator $\Xi^\bfA$ has the same conformal weight $1-1/N$ as those of the Gepner parafermionic fields associated with the first or $(N-1)$-th antisymmetric representations. 
However, it appears that the operator does \emph{not} generate the parafermionic algebra for $N \geq 3$; operators generated by the OPE of $\Xi^\bfA$ with itself is not primary. 

\subsection{Chiral field $\Xi$ is primary}

As before, we examine the OPE with the energy-momentum tensor \eqref{eq:SUkNParaEMTensor}, 
\begin{align}
T_{SU(k)_N/[U(1)]^{k-1}}(z) \Xi^\bfB(w)
&= \frac{1}{2(N+k)} \Biggl[ -k \sum_{\mu=1}^{k-1} :\mathrel{(\partial_z \bfY^{R,\mu}(z))^2}: \Xi^\bfB(w)
+\sum_{\bfalpha \in \Delta_N} \sum_{\bfA \in \Delta_k} c_\bfalpha^\bfA (\bfp^R_Y) :\mathrel{e^{i\bfalpha \bfA \cdot \bfY^R(z)}}: \Xi^\bfB(w) \Biggr] \nonumber \\
&= \frac{1}{2(N+k)} \left[ -k \calO_{C\Xi}(z,w) +\calO_{L\Xi}(z,w) \right].
\end{align}
We again separately consider the OPEs of $\Xi^\bfB$ with the Cartan part and the ladder part of the energy-momentum tensor. 
The Cartan part $\calO_{C\Xi}$ can be easily found to be 
\begin{align}
\calO_{C\Xi}(z,w) \sim \frac{2(1/N-1)}{(z-w)^2} \Xi^\bfB(w) -\frac{2}{z-w} \partial_w \Xi^\bfB(w), 
\end{align}
where we have used $|\bfomega_\sigma|^2 |\bfB|^2 = 2(1-1/N)$. 
Using Eq.~\eqref{eq:VertexVertexOPE}, the ladder part $\calO_{L\Xi}$ is written as 
\begin{align}
\calO_{L\Xi}(z,w) \sim \sum_{\bfalpha \in \Delta_N} \sum_{\bfA \in \Delta_k} \sum_{\sigma=1}^N \frac{e^{-ik_\sigma -i\pi (\bfA \cdot \bfB) (\bfalpha * \bfomega_\sigma)}}{(z-w)^{-(\bfalpha \cdot \bfomega_\sigma) (\bfA \cdot \bfB)}}
c_\bfalpha^\bfA (\bfp^R_Y) c_{\bfomega_\sigma}^\bfB (\bfp^R_Y) :\mathrel{e^{i\bfalpha \bfA \cdot \bfY^R(z) +i\bfomega_\sigma \bfB \cdot \bfY^R(w)}}:. 
\end{align}
Then we separately treat four cases that give rise to singular terms: (i) $\bfalpha \cdot \bfomega_\sigma=-1$ and $\bfA \cdot \bfB=2$, (ii) $\bfalpha \cdot \bfomega_\sigma=1$ and $\bfA \cdot \bfB=-2$, (iii) $\bfalpha \cdot \bfomega_\sigma=-1$ and $\bfA \cdot \bfB=1$, or (iv) $\bfalpha \cdot \bfomega_\sigma=1$ and $\bfA \cdot \bfB=-1$. 
For the case (i), we can equivalently write $\bfalpha = \bfomega_\rho -\bfomega_\sigma$ with $\rho \neq \sigma$ and $\bfA=\bfB$. 
Hence we have 
\begin{align}
\calO^\textrm{(i)}_{L\Xi}(z,w) &\sim \sum_{\rho=1}^N \sum_{\sigma \neq \rho} e^{-ik_\sigma -2i\pi (\bfomega_\rho -\bfomega_\sigma) * \bfomega_\sigma} c_{\bfomega_\rho}^\bfB (\bfp^R_Y) 
\biggl[ \frac{:\mathrel{e^{i\bfomega_\rho \bfB \cdot \bfY^R(w)}}:}{(z-w)^2}
+\frac{:\mathrel{i(\bfomega_\rho-\bfomega_\sigma) \bfB \cdot \partial_w \bfY^R(w) e^{i\bfomega_\rho \bfB \cdot \bfY^R(w)}}:}{z-w} \biggr]. 
\end{align}
Thus if the phase $k_\sigma$ satisfies
\begin{align} \label{eq:PhasekEq}
k_\rho -k_\sigma = 2\pi (\bfomega_\rho -\bfomega_\sigma) * \bfomega_\sigma \mod 2\pi, 
\end{align}
$\calO^\textrm{(i)}_{L\Xi}$ can be written in terms of $\Xi^\bfB$. 

We here show the solution of Eq.~\eqref{eq:PhasekEq} for the specific choice of the weights $\bfomega_\sigma$ in Eq.~\eqref{eq:SpecificWeight}. 
Let us define
\begin{align}
\Omega_{\rho \sigma} \equiv (\bfomega_\rho -\bfomega_\sigma) * \bfomega_\sigma. 
\end{align}
Obviously $\Omega_{\rho \rho}=0$. 
We first consider the upper-triangular elements of the $N$-dimensional matrix $\bfOmega$. 
Denoting $\sigma=\rho+n$ ($1 \leq n \leq N-\rho$), we can write 
\begin{align}
\bfomega_\rho-\bfomega_\sigma = \bfomega_\rho -\bfomega_{\rho+n} = \sum_{m=0}^{n-1} \tbfalpha_{\rho+m}, 
\end{align}
where $\tbfalpha_\sigma$ ($1 \leq \sigma \leq N-1$) are simple roots of $SU(N)$ given in Eq.~\eqref{eq:SimpleRoots}. 
Then we find 
\begin{align}
\Omega_{\rho,\rho+n} &= \sum_{m=0}^{n-1} \tbfalpha_{\sigma+m} * \bfomega_{\rho+n} \nonumber \\
&= \frac{1}{2} \sum_{\lambda,\lambda'=1}^{N-1} \sum_{m=0}^{n-1} (\bfomega_\lambda \cdot \tbfalpha_{\rho+m}) (\bfomega_{\lambda'} \cdot \bfomega_{\rho+n}) M_{\lambda \lambda'}. 
\end{align}
Since we can choose half of the interaction matrix $\bfM$ to be the upper unitriangular matrix $\bfG$ in Eq.~\eqref{eq:MatG}, the above equation can be expressed in terms of the fundamental weights of $SU(N)$, $\tbfomega_\sigma$, defined in Eq.~\eqref{eq:FundamentalWeights}. 
Thus we have 
\begin{align} \label{eq:Omega1}
\Omega_{\rho,\rho+n} &= \sum_{\lambda'=1}^{N-1} \sum_{m=0}^{n-1} (\tbfomega_{\lambda'} \cdot \tbfalpha_{\rho+m}) (\bfomega_{\lambda'} \cdot \bfomega_{\rho+n}) \nonumber \\
&= \sum_{m=0}^{n-1} \bfomega_{\rho+m} \cdot \bfomega_{\rho+n} \nonumber \\
&= -\frac{n}{N}, 
\end{align}
where we have used the orthogonality between $\tbfalpha_\sigma$ and $\tbfomega_\sigma$. 
Similarly, the lower-triangular elements of $\bfOmega$ are given by 
\begin{align} \label{eq:Omega2}
\Omega_{\sigma+n,\sigma} = \frac{n}{N}-1 \hspace{10pt} (1 \leq n \leq N-\sigma). 
\end{align}
Finally Eq.~\eqref{eq:PhasekEq} is reduced to 
\begin{align}
k_\rho -k_\sigma = \frac{2\pi (\rho-\sigma)}{N} \mod 2\pi, 
\end{align}
and therefore we find 
\begin{align}
k_\sigma = \frac{2\pi \sigma}{N} \mod 2\pi. 
\end{align}
We expect that there always exists a solution of Eq.~\eqref{eq:PhasekEq} for any choice of $\bfomega_\sigma$ and $\bfM$. 
By taking the solution of Eq.~\eqref{eq:PhasekEq}, we reach the final expression of $\calO^\textrm{(i)}_{L\Xi}$: 
\begin{align}
\calO^\textrm{(i)}_{L\Xi}(z,w) \sim \frac{N-1}{(z-w)^2} \Xi^\bfB(w) +\frac{N}{z-w} \partial_w \Xi^\bfB(w). 
\end{align}
The case (ii) precisely produces the same expression. 

Under the condition (iii), $\calO_{L\Xi}$ is expressed as 
\begin{align}
\calO^\textrm{(iii)}_{L\Xi}(z,w) \sim \sum_{\sigma=1}^N \sum_{\rho \neq \sigma} \sum_{\substack{\bfA \in \Delta_k \\ \bfA \cdot \bfB =1}} \frac{e^{-ik_\sigma -i\pi (\bfomega_\rho-\bfomega_\sigma) * \bfomega_\sigma}}{z-w}
c_{\bfomega_\rho-\bfomega_\sigma}^\bfA (\bfp^R_Y) c_{\bfomega_\sigma}^\bfB (\bfp^R_Y)
:\mathrel{e^{i[(\bfomega_\rho-\bfomega_\sigma) \bfA +\bfomega_\sigma \bfB] \cdot \bfY^R(w)}}:, 
\end{align}
where we have rewritten the root $\bfalpha$ as $\bfalpha = \bfomega_\rho-\bfomega_\sigma$ with $\rho \neq \sigma$. 
Using Eq.~\eqref{eq:PhasekEq}, the phase factor is written as 
\begin{align}
e^{-ik_\sigma -i\pi (\bfomega_\rho-\bfomega_\sigma) * \bfomega_\sigma} = e^{-ik_\rho +i\pi (\bfomega_\rho -\bfomega_\sigma) * \bfomega_\sigma}. 
\end{align}
Furthermore, the condition $\bfalpha \cdot \bfomega_\sigma=-1$ equivalently means that $\bfalpha \cdot \bfomega_\rho =1$. 
Thus we can write
\begin{align}
\calO^\textrm{(iii)}_{L\Xi}(z,w) \sim \sum_{\rho=1}^N \sum_{\substack{\bfalpha \in \Delta_N \\ \bfalpha \cdot \bfomega_\rho =1}} \sum_{\substack{\bfA \in \Delta_k \\ \bfA \cdot \bfB =1}} \frac{e^{-ik_\rho +i\pi \bfalpha * (\bfomega_\rho -\bfalpha)}}{z-w} c_{\bfalpha}^\bfA (\bfp^R_Y)
c_{\bfomega_\rho-\bfalpha}^\bfB (\bfp^R_Y) :\mathrel{e^{i[\bfalpha \bfA +(\bfomega_\rho-\bfalpha) \bfB] \cdot \bfY^R(w)}}:. 
\end{align}
Since $\bfA \cdot \bfB=1$ indicates that $\bfA-\bfB \in \Delta_k$, we can change the summation variable as $\bfA \to \bfA+\bfB$ while the constraint becomes $\bfA \cdot \bfB =-1$. 
Then we finally obtain 
\begin{align}
\calO^\textrm{(iii)}_{L\Xi}(z,w) \sim \sum_{\rho=1}^N \sum_{\substack{\bfalpha \in \Delta_N \\ \bfalpha \cdot \bfomega_\rho=1}} \sum_{\substack{\bfA \in \Delta_k \\ \bfA \cdot \bfB=-1}} \frac{e^{-ik_\rho +i\pi \bfalpha * \bfomega_\rho -i\pi \bfalpha * \bfalpha}}{z-w}
c_\bfalpha^\bfA (\bfp^R_Y) c_{\bfomega_\rho}^\bfB (\bfp^R_Y) :\mathrel{e^{i(\bfalpha \bfA +\bfomega_\rho \bfB) \cdot \bfY^R(w)}}:. 
\end{align}
Except for a minus sign $e^{-i\pi \bfalpha * \bfalpha}=-1$, this expression coincides with the contribution from $\calO_{L\Xi}$ under the condition (iv). 
Hence they are canceled. 
Then we finally obtain 
\begin{align}
T_{SU(k)_N/[U(1)]^{k-1}}(z) \Xi^\bfB(w)
\sim \frac{1-1/N}{(z-w)^2} \Xi^\bfB(w) +\frac{1}{z-w} \partial_w \Xi^\bfB(w). 
\end{align}
Therefore, $\Xi^\bfB$ gives the vertex representation for a primary field of the $SU(k)_N/[U(1)]^{k-1}$ CFT with conformal weight $1-1/N$. 
We can also show that 
\begin{align}
T_{SU(N)_k/[U(1)]^{N-1}}(z) \Xi^\bfB(w) \sim 0. 
\end{align}
Thus the operator $\Xi^\bfB$ is independent of the $SU(N)_k/[U(1)]^{N-1}$ CFT. 

We can also generate an operator associated with the second antisymmetric representation of $SU(N)$ for $N \geq 3$ by using the OPE of $\Xi^\bfA$ with itself. 
While such an operator is naively expected to be the second parafermionic field of the $SU(k)_N/[U(1)]^{k-1}$ CFT, the operator is in fact not the primary field. 
A suitable linear combination of $\Xi^\bfA$ and its conjugate may give a proper vertex representation for the $SU(k)_N/[U(1)]^{k-1}$ first parafermionic field, which generates the full parafermionic algebra, while we were not aware of such possibility. 

Presumably, we cannot construct the faithful vertex representations for the \emph{chiral} $SU(k)_N/[U(1)]^{k-1}$ parafermions for $N \geq 3$ by some intrinsic reasons. 
For example, a subtlety on the construction of the $SU(2)_N/U(1)$ parafermion from the $[SU(N)_1]^2/SU(N)_2$ CFT has already been implied from the mismatch between the compactifications of field and lattice \cite{Griffin89}.
This problem may also be related to the fact that we cannot construct the faithful vertex representations of chiral $SU(N)_1$ primary fields unless supplementing with extra zero modes \cite{Chu95}. 
The addition of the zero modes requires an extension of the Hilbert space beyond the physical one and thus does not suit to our analysis. 
We below instead construct the \emph{nonchiral} vertex representations of those primary fields, which indeed generate the parafermionic algebra. 

\subsection{Vertex representation in the nonchiral sector}

We here consider the left-right product of the primary fields of the $SU(k)_N/[U(1)]^{k-1}$ CFT. 
The primary field in the right-moving sector is defined in Eq.~\eqref{eq:VertexRepXi}, while that in the left-moving sector is given by
\begin{align}
\bXi^\bfA(\bz) = \sum_{\sigma=1}^N e^{-ik_\sigma} c_{\bfomega_\sigma}^\bfA (\bfp^L_Y +2\bfp^R_Y) :\mathrel{e^{i\bfomega_\sigma \bfA \cdot \bfY^L(\bz)}}:. 
\end{align}
Below we show that the intrawire interactions given in Eqs.~\eqref{eq:IntraIntSU32} and \eqref{eq:IntraIntSUNk2} are precisely written in terms of the product $\Xi^\bfA(z) \bXi^{\bfA \dagger}(\bz)$. 
Since both left and right primary fields belong to the same wire, we need a care about the parafermionic cocycle with $\bfp^{R,\mu}_Y$ involved in $\bXi^\bfA$ when we split the exponentials. 
We then find 
\begin{align} \label{eq:Oss}
\sum_{s,s'=1}^N \calO^{u,ss'}_\bfA(z,\bz)
&= \sum_{s,s'=1}^N :\mathrel{e^{i\bfomega_s \bfA \cdot \tbfY^R(z) -i\bfomega_{s'} \bfA \cdot \tbfY^L(\bz)}}: \nonumber \\
&= \sum_{s,s'=1}^N e^{-i\pi (\bfomega_s * \bfomega_s -\bfomega_{s'} * \bfomega_{s'} +2\bfomega_s * \bfomega_{s'})} c_{\bfomega_s}^\bfA (\bfp^R_Y)
:\mathrel{e^{i\bfomega_s \bfA \cdot \bfY^R(z)}}: :\mathrel{e^{-i\bfomega_{s'} \bfA \cdot \bfY^L(\bz)}}: c_{-\bfomega_{s'}}^\bfA (\bfp^L_Y +2\bfp^R_Y). 
\end{align}
In order to see that the phase factor $e^{ik_\sigma}$ is recovered due to the parafermionic cocycle, let us examine $\bfomega_s * \bfomega_s$ by explicit calculations. 
For $1 \leq s \leq N-1$, we obtain 
\begin{align} \label{eq:ww1}
\bfomega_s * \bfomega_s &= \sum_{\sigma,\sigma'=1}^{N-1} (\bfomega_\sigma \cdot \bfomega_s) (\bfomega_{\sigma'} \cdot \bfomega_s) \frac{M_{\sigma \sigma'}}{2} \nonumber \\
&= \sum_{\sigma,\sigma'=1}^{N-1} \left( \delta_{\sigma s} -\frac{1}{N} \right) \left( \delta_{\sigma' s} -\frac{1}{N} \right) \frac{M_{\sigma \sigma'}}{2} \nonumber \\
&= \frac{N-1}{2N}. 
\end{align}
For $s=N$, we obtain 
\begin{align} \label{eq:ww2}
\bfomega_N * \bfomega_N &= \frac{1}{2N^2} \sum_{\sigma,\sigma'=1}^{N-1} M_{\sigma \sigma'}
= \frac{N-1}{2N}. 
\end{align}
Thus $\bfomega_s * \bfomega_s$ is independent of $s$. 
Substituting this into the solution of $k_s$ in Eq.~\eqref{eq:PhasekEq}, we find 
\begin{align}
k_s -k_{s'} = 2\pi \bfomega_s * \bfomega_{s'} -\frac{\pi (N-1)}{N} \mod 2\pi. 
\end{align}
Equation \eqref{eq:Oss} is finally written as 
\begin{align}
\sum_{s,s'=1}^N \calO^{u,ss'}_\bfA (z,\bz) &= e^{i\pi(1-N)/N} \Xi^\bfA(z) \bXi^{\bfA \dagger}(\bz). 
\end{align}
Therefore the intrawire interaction is an $SU(k)_N/[U(1)]^{k-1}$ primary field with scaling dimension $2(1-1/N)$. 
In fact, the interaction is now a vertex representation of the nonchiral $SU(k)_N/[U(1)]^{k-1}$ parafermion as we will demonstrate below. 

\subsection{Nonchiral product $\Upsilon$ is parafermion}

We now consider the left-right product of the operator $\Xi^\bfA$ and its conjugate, 
\begin{align} \label{eq:DefSigma1}
\Upsilon^{\bfA,1}(z,\bz) &\equiv \Xi^\bfA(z) \bXi^{\bfA \dagger}(\bz) \nonumber \\
&= \sum_{\sigma,\rho=1}^N e^{-ik_\sigma +ik_\rho} c_{\bfomega_\sigma}^\bfA (\bfp^R_Y) :\mathrel{e^{i\bfomega_\sigma \bfA \cdot \bfY^R(z)}}:
:\mathrel{e^{-i\bfomega_\rho \bfA \cdot \bfY^L(\bz)}}: c_{-\bfomega_\rho}^\bfA (\bfp^L_Y +2\bfp^R_Y). 
\end{align}
Here we show that this operator is the (unnormalized) first parafermion generating a nonchiral version of the Gepner parafermionic algebra. 
To do so, we first consider the OPE of $\Upsilon^{\bfA,1}$ with itself to confirm that it generates the second parafermion. 
Keeping only the most singular part, we find 
\begin{align} \label{eq:OPESigma1Sigma1}
\Upsilon^{\bfA,1}(z,\bz) \Upsilon^{\bfA,1}(w,\bw)
&\sim \frac{1}{|z-w|^{4/N}} \sum_{\sigma_1 \neq \sigma_2} \sum_{\rho_1 \neq \rho_2} e^{-i(k_{\sigma_1} +k_{\sigma_2} -k_{\rho_1} -k_{\rho_2}) -2i\pi (\bfomega_{\sigma_1} * \bfomega_{\sigma_2} -\bfomega_{\rho_2} * \bfomega_{\rho_1} +2\bfomega_{\sigma_2} * \bfomega_{\rho_1})} c_{\bfomega_{\sigma_1}+\bfomega_{\sigma_2}}^\bfA (\bfp^R_Y) \nonumber \\
&\ \ \ \times :\mathrel{e^{i(\bfomega_{\sigma_1}+\bfomega_{\sigma_2}) \bfA \cdot \bfY^R(w)}}: :\mathrel{e^{-i(\bfomega_{\rho_1}+\bfomega_{\rho_2}) \bfA \cdot \bfY^L(\bw)}}:
c_{-\bfomega_{\rho_1}-\bfomega_{\rho_2}}^\bfA (\bfp^L_Y +2\bfp^R_Y). 
\end{align}
A crucial observation at this point is that the phase factor arising from the parafermionic cocycle makes the whole expression symmetric under $\sigma_1 \leftrightarrow \sigma_2$ or $\rho_1 \leftrightarrow \rho_2$. 
This can be understood by substituting the explicit expressions of $\bfomega_\sigma * \bfomega_\rho$ [see Eqs.~\eqref{eq:Omega1}, \eqref{eq:Omega2}, \eqref{eq:ww1}, and \eqref{eq:ww2}];
\begin{align}
\bfomega_{\sigma_1} * \bfomega_{\sigma_2} -\bfomega_{\rho_2} * \bfomega_{\rho_1} +2\bfomega_{\sigma_2} * \bfomega_{\rho_1}
= \frac{1}{N} (\sigma_1 +\sigma_2 -\rho_1 -\rho_2 +N -1). 
\end{align}
This symmetric property does not hold for the operator generated by the OPE $\Xi^\bfA(z) \Xi^\bfA(w)$. 
Therefore we can write Eq.~\eqref{eq:OPESigma1Sigma1} as
\begin{align}
\Upsilon^{\bfA,1}(z,\bz) \Upsilon^{\bfA,1}(w,\bw) \sim \frac{4 e^{2i\pi/N}}{|z-w|^{4/N}} \Upsilon^{\bfA,2}(w,\bw), 
\end{align}
where 
\begin{align}
\Upsilon^{\bfA,2}(z,\bz) = \sum_{\bfomega, \bfomega' \in [\tbfomega_2]} e^{i\kappa^{(2)}_{\bfomega,\bfomega'}} c_\bfomega^\bfA (\bfp^R_Y) :\mathrel{e^{i\bfomega \bfA \cdot \bfY^R(z)}}:
:\mathrel{e^{-i\bfomega' \bfA \cdot \bfY^L(\bz)}}: c_{-\bfomega'}^\bfA (\bfp^L_Y +2\bfp^R_Y), 
\end{align}
where the phase $\kappa$ is explicitly given by 
\begin{align}
\kappa^{(2)}_{\bfomega,\bfomega'} = -\frac{4\pi}{N} (\sigma_1 +\sigma_2 -\rho_1 -\rho_2), 
\end{align}
for $\bfomega=\bfomega_{\sigma_1}+\bfomega_{\sigma_2}$ and $\bfomega' = \bfomega_{\rho_1}+\bfomega_{\rho_2}$. 

We next show that the operator $\Upsilon^{\bfA,2}$ is a primary field of the $SU(k)_N/[U(1)]^{k-1}$ CFT. 
We write the OPE of $\Upsilon^{\bfA,2}$ with the energy-momentum tensor in the right-moving sector as 
\begin{align}
T_{SU(k)_N/[U(1)]^{k-1}}(z) \Upsilon^{\bfB,2}(w,\bw)
= \frac{1}{2(N+k)} \left[ -k\calO_{C\Upsilon}(z,w,\bw) +\calO_{L\Upsilon}(z,w,\bw) \right]. 
\end{align}
The Cartan part is evaluated as 
\begin{align}
\calO_{C\Upsilon}(z,w,\bw) \sim -\frac{4(N-2)/N}{(z-w)^2} \Upsilon^{\bfB,2}(w,\bw) -\frac{2\partial_w \Upsilon^{\bfB,2}(w,\bw)}{z-w}, 
\end{align}
where we have used $|\bfomega|^2 = I(N-I)/N$ for $\bfomega \in [\tbfomega_I]$. 
The ladder part is written as 
\begin{align} \label{eq:TSigmaOPELad}
\calO_{L\Upsilon}(z,w,\bw) &\sim \sum_{\bfalpha \in \Delta_N} \sum_{\bfA \in \Delta_k} \sum_{\bfomega,\bfomega' \in [\tbfomega_2]} \frac{e^{i\kappa^{(2)}_{\bfomega,\bfomega'}-i\pi(\bfA \cdot \bfB)(\bfalpha * \bfomega)}}{(z-w)^{-(\bfA \cdot \bfB) (\bfalpha \cdot \bfomega)}}
c_\bfalpha^\bfA (\bfp^R_Y) c_\bfomega^\bfB (\bfp^R_Y) :\mathrel{e^{i\bfalpha \bfA \cdot \bfY^R(z) +i\bfomega \bfB \cdot \bfY^R(w)}}: \nonumber \\
&\ \ \ \times :\mathrel{e^{-i\bfomega' \bfB \cdot \bfY^L(\bw)}}: c_{-\bfomega'}^\bfB (\bfp^L_Y +2\bfp^R_Y). 
\end{align}
Singular contributions to the ladder part only come from the two cases (i) $\bfA \cdot \bfB=2$ and $\bfalpha \cdot \bfomega=-1$ or (ii) $\bfA \cdot \bfB=-2$ and $\bfalpha \cdot \bfomega =1$. 
Other contributions vanish by the same logic as discussed before. 
The case (i) equivalently means that $\bfA=\bfB$ and $\bfalpha +\bfomega \in [\tbfomega_2]$. 
The most important step is to check that the phase factor in Eq.~\eqref{eq:TSigmaOPELad} exactly produces $\kappa^{(2)}_{\bfalpha+\bfomega,\bfomega'}$. 
If we write $\bfomega = \bfomega_{\sigma_1}+\bfomega_{\sigma_2}$ and $\bfomega'=\bfomega_{\rho_1}+\bfomega_{\rho_2}$ and choose $\bfalpha=\bfomega_\lambda-\bfomega_{\sigma_1}$ ($\lambda \neq \sigma_1,\sigma_2$), the corresponding phase is explicitly calculated as 
\begin{align}
\kappa^{(2)}_{\bfomega,\bfomega'} -2\pi \bfalpha * \bfomega
&= -\frac{4\pi}{N}(\sigma_1+\sigma_2-\rho_1-\rho_2) -\frac{2\pi}{N} (2\lambda -2\sigma_1) \nonumber \\
&= -\frac{4\pi}{N}(\lambda+\sigma_2-\rho_1-\rho_2) \nonumber \\
&= \kappa^{(2)}_{\bfalpha+\bfomega,\bfomega'}. 
\end{align}
Thus Eq.~\eqref{eq:TSigmaOPELad} is reduced to 
\begin{align}
\calO_{L\Upsilon}^\textrm{(i)}(z,w,\bw) &\sim \sum_{\bfomega,\bfomega' \in [\tbfomega_2]} \sum_{\substack{\bfalpha \in \Delta_N \\ \bfalpha \cdot \bfomega =-1}} e^{i\kappa^{(2)}_{\bfalpha+\bfomega,\bfomega'}} c_{\bfalpha+\bfomega}^\bfB (\bfp^R_Y)
\biggl[ \frac{:\mathrel{e^{i(\bfalpha+\bfomega) \bfB \cdot \bfY^R(w)}}:}{(z-w)^2}
+\frac{:\mathrel{i\bfalpha \bfB \cdot \partial_w \bfY^R(w) e^{i(\bfalpha+\bfomega) \bfB \cdot \bfY^R(w)}}:}{z-w} \biggr] \nonumber \\
&\ \ \ \times :\mathrel{e^{-i\bfomega' \bfB \cdot \bfY^L(\bw)}}: c_{-\bfomega'}^\bfB (\bfp^L_Y +2\bfp^R_Y). 
\end{align}
Using the combinatorial identities (see also Ref.~\cite{Dunne89}), 
\begin{align}
\begin{split}
\sum_{\bfomega \in [\tbfomega_I]} \sum_{\substack{\bfalpha \in \Delta_N \\ \bfalpha \cdot \bfomega = -1}} f(\bfalpha+\bfomega) &= I(N-I) \sum_{\bfomega \in [\tbfomega_I]} f(\bfomega), \\
\sum_{\bfomega \in [\tbfomega_I]} \sum_{\substack{\bfalpha \in \Delta_N \\ \bfalpha \cdot \bfomega = -1}} \bfalpha \cdot \bff(\bfalpha+\bfomega) &= N \sum_{\bfomega \in [\tbfomega_I]} \bfomega \cdot \bff(\bfomega), 
\end{split}
\end{align}
we find 
\begin{align}
\calO^\textrm{(i)}_{L\Upsilon}(z,w,\bw) \sim \frac{2(N-2)}{(z-w)^2} \Upsilon^{\bfB,2}(w,\bw) +\frac{N}{z-w} \partial_w \Upsilon^{\bfB,2}(w,\bw). 
\end{align}
The case (ii) also produces the same contribution. 
We finally obtain 
\begin{align}
T_{SU(k)_N/[U(1)]^{k-1}}(z) \Upsilon^{\bfB,2}(w,\bw)
\sim \frac{2(N-2)/N}{(z-w)^2} \Upsilon^{\bfB,2}(w,\bw) +\frac{1}{z-w} \partial_w \Upsilon^{\bfB,2}(w,\bw). 
\end{align}
The OPE of $\Upsilon^{\bfB,2}$ with the energy-momentum tensor in the left-moving sector, 
\begin{align}
\bT_{SU(k)_N/[U(1)]^{k-1}}(\bz)
= \frac{1}{2(N+k)} \biggl[ -k\sum_{\mu=1}^{N-1} :\mathrel{(\partial_{\bz} \bfY^{R,\mu}(\bz))^2}:
+\sum_{\bfalpha \in \Delta_N} \sum_{\bfA \in \Delta_k} c_\bfalpha^\bfA (\bfp^L_Y +2\bfp^R_Y) :\mathrel{e^{i\bfalpha \bfA \cdot \bfY^L(\bz)}}: \biggr], 
\end{align}
can also be calculated in a similar way aside from a slightly different form of the phase factor. 
As a result, we find 
\begin{align}
\bT_{SU(k)_N/[U(1)]^{k-1}}(\bz) \Upsilon^{\bfB,2}(w,\bw)
\sim \frac{2(N-2)/N}{(\bz-\bw)^2} \Upsilon^{\bfB,2}(w,\bw) +\frac{1}{\bz-\bw} \partial_{\bw} \Upsilon^{\bfB,2}(w,\bw). 
\end{align}
Therefore the operator $\Upsilon^{\bfA,1}$ is a primary field of the $SU(k)_N/[U(1)]^{k-1}$ CFT with conformal weight $2(N-2)/N$ in both right- and left-moving sectors. 
This operator will serve as a vertex representation of a nonchiral product of the $SU(k)_N/[U(1)]^{k-1}$ parafermions associated with the second antisymmetric representation of $SU(N)$ \cite{Gepner87}. 

\subsection{Gepner parafermionic algebra}

We can generally define the nonchiral parafermion with the $I$-th antisymmetric representation of $SU(N)$ as 
\begin{align} \label{eq:SUkNNonchiralPara}
\Upsilon^{\bfA,I}(z,\bz) = \sum_{\bfomega, \bfomega' \in [\tbfomega_I]} e^{i\kappa^{(I)}_{\bfomega,\bfomega'}} c_\bfomega^\bfA (\bfp^R_Y) :\mathrel{e^{i\bfomega \bfA \cdot \bfY^R(z)}}:
:\mathrel{e^{-i\bfomega' \bfA \cdot \bfY^L(\bz)}}: c_{-\bfomega'}^\bfA (\bfp^L_Y +2\bfp^R_Y), 
\end{align}
where the phase is given by 
\begin{align}
\kappa^{(I)}_{\bfomega,\bfomega'} = -\frac{2\pi I}{N} \sum_{n=1}^I (\sigma_n-\rho_n), 
\end{align}
for $\bfomega=\bfomega_{\sigma_1}+\cdots+\bfomega_{\sigma_I}$ and $\bfomega'=\bfomega_{\rho_1}+\cdots+\bfomega_{\rho_I}$. 
Following the above proof, we can show that $\Upsilon^{\bfA,I}$ is a primary field of the $SU(k)_N/[U(1)]^{k-1}$ CFT with right and left conformal weights $I(N-I)/N$. 
The Hermitian conjugate of Eq.~\eqref{eq:SUkNNonchiralPara} corresponds to the conjugate representation of $SU(N)$: 
\begin{align}
\Upsilon^{\bfA,I \dagger}(z,\bz) = e^{-2i\pi I(N-I)/N} \Upsilon^{\bfA,N-I}(z,\bz). 
\end{align}
These parafermionic fields may constitute a nonchiral version of the Gepner parafermionic algebra, which is given as follows. 
The OPE of $\Upsilon^{\bfA,I}$ with its conjugate is given by 
\begin{align}
\Upsilon^{\bfA,I}(z,\bz) \Upsilon^{\bfA,I \dagger}(w,\bw)
\sim \left[ \frac{N!}{I!(N-I)!} \right]^2 \frac{1}{|z-w|^{4I(N-I)/N}}. 
\end{align}
This gives a proper normalization constant for $\Upsilon^{\bfA,I}$, which is read off as $I!(N-I)!/N!$. 
For $I+J<N$, 
\begin{align}
\Upsilon^{\bfA,I}(z,\bz) \Upsilon^{\bfA,J}(w,\bw)
\sim \left[ \frac{(I+J)!}{I!J!} \right]^2 \frac{e^{2i\pi IJ/N}}{|z-w|^{4IJ/N}} \Upsilon^{\bfA,I+J}(w,\bw). 
\end{align}
For $I>J$, 
\begin{align}
\Upsilon^{\bfA,I}(z,\bz) \Upsilon^{\bfA,J \dagger}(w,\bw)
\sim \left[ \frac{(N-I+J)!}{(N-I)!J!} \right]^2 \frac{e^{2i\pi J(J-I)/N}}{|z-w|^{4(N-I)J/N}} \Upsilon^{\bfA,I-J}(w,\bw). 
\end{align}
For $\bfA +\bfB \in \Delta_k$, 
\begin{align}
\Upsilon^{\bfA,I}(z,\bz) \Upsilon^{\bfB,I}(w,\bw) \sim \frac{e^{i\pi I^2 (N-1)/N}}{|z-w|^{2I(N-I)/N}} \Upsilon^{\bfA+\bfB,I}(w,\bw). 
\end{align}
Therefore we find vertex representations of the nonchiral products of the Gepner parafermions in the $SU(k)_N/[U(1)]^{k-1}$ CFT. 

\subsection{Decomposition into $SU(N)_1$ primary fields} \label{app:SUNPrimaries}

For a given weight of $SU(k)$, $\bfA = \bfW_a -\bfW_b$, we can express the unnormalized nonchiral parafermionic field \eqref{eq:DefSigma1} as 
\begin{align}
\Upsilon^{\bfA,1}(z,\bz) = e^{i\pi (N-1)/N} \textrm{Tr} \left[ \bfg_a(z,\bz) \bfg_b^\dagger(z,\bz) \right], 
\end{align}
where the $N \times N$ matrix field $\bfg_a$ corresponds to the primary field of the $SU(N)_1$ WZW CFT associated with the copy $a$ and has scaling dimension $1-1/N$. 
The explicit vertex representation of $\bfg_a$ is given by 
\begin{align}
g_{a,ss'}(z,\bz) 
&= e^{i\bfomega_s \cdot \tbfchi^R_a(z) -i\bfomega_{s'} \cdot \tbfchi^L_a(\bz)} \nonumber \\
&= e^{-i\pi \bfomega_s * \bfomega_{s'}} c_{\bfomega_s}(\bfp^R_a) :\mathrel{e^{i\bfomega_s \cdot \bfchi^R_a(z)}}:
:\mathrel{e^{-i\bfomega_{s'} \cdot \bfchi^L_a(\bz)}}: c_{-\bfomega_{s'}}(\bfp^L_a+2\bfp^R_a). 
\end{align}
One can show that these fields satisfy the following OPEs with the right-moving $SU(N)_1$ currents associated with the copy $a$ [see Eqs.~\eqref{eq:VertexRepCartan}, \eqref{eq:VertexRepLadderPos}, and \eqref{eq:VertexRepLadderNeg}], 
\begin{align}
\begin{split}
H^l_a(z) g_{a,ss'}(w,\bw) &\sim \frac{\omega^l_s}{z-w} g_{a,ss'}(w,\bw), \\
E^\bfalpha_a(z) g_{a,ss'}(w,\bw)
&\sim \begin{cases} \dfrac{1}{z-w} g_{a,rs'}(w,\bw) & \textrm{if} \ \bfalpha = \bfomega_r -\bfomega_s \ (r<s) \\ 0 & \textrm{otherwise} \end{cases} \\
E^{-\bfalpha}_a(z) g_{a,ss'}(w,\bw)
&\sim \begin{cases} \dfrac{-1}{z-w} g_{a,rs'}(w,\bw) & \textrm{if} \ \bfalpha = \bfomega_s -\bfomega_r \ (r>s) \\ 0 & \textrm{otherwise} \end{cases}
\end{split}
\end{align}
Therefore $\bfg_a$ transforms in the fundamental representation of $SU(N)$ under the action of the right-moving $SU(N)_1$ currents. 
Since the OPEs with the left-moving currents are similarly obtained except for an overall minus sign, $\bfg_a$ transforms in the conjugate representation of $SU(N)$ under the action of the left-moving currents. 
This in turn indicates that $\bfg_a$ is the primary field of the $SU(N)_1$ WZW CFT \cite{dFMS}. 
We also mention that the intrawire interaction \eqref{eq:IntraIntSUNk2} is simply expressed as 
\begin{align}
\calO^{u,ss'}_{\bfA}(z,\bz) = (\bfg_a)_{ss'}(z,\bz) (\bfg^\dagger_b)_{s's} (z,\bz). 
\end{align}

\section{Interactions for the generalized NASS states}
\label{app:IntGenNASS}

We show several technical details on the construction of the non-Abelian FQH states at $\nu=k(N-1)/[N+k(N-1)m]$ discussed in Sec.~\ref{sec:GeneralFilling}. 

\subsection{Explicit forms of intrawire interactions} \label{app:ExplicitIntraInt}

The intrawire interactions $\calO^{u,ss'}_{j,12}$ for the NASS state at $\nu=4/(4m+3)$ are explicitly given by 
\begin{align}
\begin{split}
\calO^{u,11}_{j,12} &= \exp 2i(\theta_{j,\ua,1} -\theta_{j,\ua,2}), \\
\calO^{u,12}_{j,12} &= \exp i\bigl[ \varphi_{j,\ua,1} -\varphi_{j,\da,1} -\varphi_{j,\ua,2} +\varphi_{j,\da,2} +(m+2) (\theta_{j,\ua,1} +\theta_{j,\da,1} -\theta_{j,\ua,2} -\theta_{j,\da,2}) \bigr], \\
\calO^{u,13}_{j,12} &= \exp i\bigl[ \varphi_{j,\ua,1} -\varphi_{j,\ua,2} +m(\theta_{j,\ua,1} +\theta_{j,\ua,2}) +2m \theta_{j,\da,1} \bigr], \\
\calO^{u,21}_{j,12} &= \exp i\bigl[ -\varphi_{j,\ua,1} +\varphi_{j,\da,1} +\varphi_{j,\ua,2} -\varphi_{j,\da,2} +m(-\theta_{j,\ua,1} -\theta_{j,\da,1} +\theta_{j,\ua,2} +\theta_{j,\da,2}) \bigr], \\
\calO^{u,22}_{j,12} &= \exp 2i(\theta_{j,\da,1} -\theta_{j,\da,2}), \\
\calO^{u,23}_{j,12} &= \exp i\bigl[ \varphi_{j,\da,1} -\varphi_{j,\da,2} -2\theta_{j,\ua,1} +m(\theta_{j,\da,1} +\theta_{j,\da,2}) +2(m+1) \theta_{j,\ua,2} \bigr], \\
\calO^{u,31}_{j,12} &= \exp i\bigl[ -\varphi_{j,\ua,1} +\varphi_{j,\ua,2} -m(\theta_{j,\ua,1} +\theta_{j,\ua,2}) -2(m+1) \theta_{j,\da,1} +2\theta_{j,\da,2} \bigr], \\
\calO^{u,32}_{j,12} &= \exp i\bigl[ -\varphi_{j,\da,1} +\varphi_{j,\da,2} -2m \theta_{j,\ua,2} -m(\theta_{j,\da,1} +\theta_{j,\da,2}) \bigr], \\
\calO^{u,33}_{j,12} &= \exp 2i(-\theta_{j,\ua,1} -\theta_{j,\da,1} +\theta_{j,\ua,2} +\theta_{j,\da,2}). 
\end{split}
\end{align}
For the bilayer state at $\nu=4/(4m+1)$, they are given by 
\begin{align}
\begin{split}
\calO^{u,11}_{j,12} &= \exp 2i(\theta_{j,\ua,1} -\theta_{j,\ua,2}), \\
\calO^{u,12}_{j,12} &= \exp i\bigl[ \varphi_{j,\ua,1} +\varphi_{j,\da,1} -\varphi_{j,\ua,2} -\varphi_{j,\da,2} +(m+2) \theta_{j,\ua,1} +(3m-2) (\theta_{j,\da,1} +\theta_{j,\ua,2}) +2\theta_{j,\da,2} \bigr], \\
\calO^{u,13}_{j,12} &= \exp i\bigl[ \varphi_{j,\ua,1} -\varphi_{j,\ua,2} +m(\theta_{j,\ua,1} +\theta_{j,\ua,2}) +2m \theta_{j,\da,1} \bigr], \\
\calO^{u,21}_{j,12} &= \exp i\bigl[ -\varphi_{j,\ua,1} -\varphi_{j,\da,1} +\varphi_{j,\ua,2} +\varphi_{j,\da,2} -m(-\theta_{j,\ua,1} -\theta_{j,\da,2}) -3m(\theta_{j,\da,1} +\theta_{j,\ua,2}) \bigr], \\
\calO^{u,22}_{j,12} &= \exp 2i(-\theta_{j,\da,1} +\theta_{j,\da,2}), \\
\calO^{u,23}_{j,12} &= \exp i\bigl[ -\varphi_{j,\da,1} +\varphi_{j,\da,2} -2\theta_{j,\ua,1} -m(\theta_{j,\da,1} +\theta_{j,\da,2}) -2(m-1) \theta_{j,\ua,2} \bigr], \\
\calO^{u,31}_{j,12} &= \exp i\bigl[ -\varphi_{j,\ua,1} +\varphi_{j,\ua,2} -m(\theta_{j,\ua,1} +\theta_{j,\ua,2}) -2(m-1) \theta_{j,\da,1} -2\theta_{j,\da,2} \bigr], \\
\calO^{u,32}_{j,12} &= \exp i\bigl[ \varphi_{j,\da,1} -\varphi_{j,\da,2} +2m \theta_{j,\ua,2} +m(\theta_{j,\da,1} +\theta_{j,\da,2}) \bigr], \\
\calO^{u,33}_{j,12} &= \exp 2i(-\theta_{j,\ua,1} +\theta_{j,\da,1} +\theta_{j,\ua,2} -\theta_{j,\da,2}). 
\end{split}
\end{align}
These interactions can be constructed from bosonic operators for even $m$ and from fermionic operators for odd $m$ under the given filling factors and flux structure shown in Fig.~\ref{fig:FermionicNASS}. 

\subsection{Vertex representations for even $m$}

We here sketch how to find the vertex representations of fields in the parafermion CFTs for general values of $N$, $k$, and $m$. 
As a special case, this leads to the identification of the interactions in Sec.~\ref{sec:ASSeries} as primary fields of the parafermion CFTs.
We first consider the case of bosonic FQH states with even $m$. 
Corresponding to the $K$ matrix \eqref{eq:GeneralKMat}, we choose the interaction matrix $\bfM$ to be 
\begin{align} \label{eq:IntMatGen}
M_{\sigma a; \sigma' a'}(m) = 2(m+1) \Theta(\sigma'-\sigma) +(m+2) \delta_{\sigma \sigma'} \delta_{aa'} +2m\Theta (a'-a). 
\end{align}
We then introduce the sequences of chiral fields, 
\begin{align}
\begin{split}
\tphi^R_{j,\sigma,a} &= \varphi_{j,\sigma,a} +\sum_{\sigma'=1}^{N-1} \sum_{a'=1}^k M_{\sigma a; \sigma' a'}(m) \theta_{j,\sigma',a'}, \\
\tphi^L_{j,\sigma,a} &= \varphi_{j,\sigma,a} -\sum_{\sigma'=1}^{N-1} \sum_{a'=1}^k M^T_{\sigma a; \sigma' a'}(m) \theta_{j,\sigma',a'}, 
\end{split}
\end{align}
and
\begin{align} \label{eq:ChiralFieldXY}
\tX^p_{j,l} = \frac{1}{\sqrt{k}} \sum_{\sigma=1}^{N-1} \sum_{a=1}^k \omega^l_\sigma(m) \tphi^p_{j,\sigma,a}, \hspace{10pt}
\tY^{p,\mu}_{j,l} = \sum_{\sigma=1}^{N-1} \sum_{a=1}^k \omega^l_\sigma W^\mu_a \tphi^p_{j,\sigma,a}, 
\end{align}
where $\bfomega_\sigma$ and $\bfW_a$ are given in Eqs.~\eqref{eq:SpecificWeight} and \eqref{eq:SpecificVectorW}, respectively, and $\bfomega_\sigma(m)$ are defined by
\begin{align}
\omega^l_\sigma(m) \equiv \sum_{l'=1}^{N-1} \Lambda_{ll'}(m) \omega^{l'}_\sigma, \hspace{10pt} {\bm \Lambda}(m) = \textrm{diag} \left( 1,1,\cdots,1,\sqrt{\frac{N}{N+k(N-1)m}} \right).
\end{align}
The fields \eqref{eq:ChiralFieldXY} satisfy the commutation relations, 
\begin{align} \label{eq:CommXYGen}
\begin{split}
[\tX^p_{j,l}(x), \tX^{p'}_{j',l'}(x')] 
&= \pi \delta_{jj'} \delta_{ll'} \Big[ ip \delta_{pp'} \sgn(x-x') -\sigma^y_{pp'} \Big] +\frac{i\pi}{2k} \delta_{jj'} \sum_{\sigma,\sigma'} \sum_{a,a'} \omega^l_\sigma(m) \omega^{l'}_{\sigma'}(m) \bigl[\bfM(m)-\bfM^T(m) \bigr]_{\sigma a; \sigma' a'}, \\
[\tY^{p,\mu}_{j,l}(x), \tY^{p',\mu'}_{j',l'}(x')] 
&= \pi \delta_{jj'} \delta_{ll'} \delta_{\mu \mu'} \Bigl[ ip \delta_{pp'} \sgn(x-x') -\sigma^y_{pp'} \Bigr] +\frac{i\pi}{2} \delta_{jj'} \sum_{\sigma,\sigma'} \sum_{a,a'} \omega^l_\sigma \omega^{l'}_{\sigma'} W^\mu_a W^{\mu'}_{a'} \bigl[ \bfM(m)-\bfM^T(m) \bigr]_{\sigma a; \sigma' a'}, \\
[\tX^p_{j,l}(x), \tY^{p',\mu'}_{j',l'}(x')] 
&= \frac{i\pi}{\sqrt{k}} \delta_{jj'} \sum_{\sigma, \sigma'} \sum_{a,a'} \omega^l_\sigma(m) \omega^{l'}_{\sigma'} W^{\mu'}_{a'} \bigl[ \bfM(m) -\bfM^T(m) \bigr]_{\sigma a; \sigma' a'}. 
\end{split}
\end{align}
As we have done in Appendix~\ref{app:VertexRep}, we now split the chiral fields $\tX^p_{j,l}$ and $\tY^{p,\mu}_{j,l}$ into the bosonic fields $X^p_{j,l}$ and $Y^{p,\mu}_{j,l}$ satisfying the standard commutation relations, 
\begin{align}
\begin{split}
[X^p_{j,l}(x), X^{p'}_{j',l'}(x')] &= i\pi p \delta_{pp'} \delta_{jj'} \delta_{ll'} \sgn(x-x'), \\
[Y^{p,\mu}_{j,l}(x), Y^{p',\mu'}_{j',l'}(x')] &= i\pi p \delta_{pp'} \delta_{jj'} \delta_{ll'} \delta_{\mu \mu'} \sgn(x-x'), \\
[X^p_{j,l}(x), Y^{p',\mu'}_{j',l'}(x')] &= 0, 
\end{split}
\end{align}
and their zero-mode parts $p^p_{X,j,l}$ and $p^{p,\mu}_{Y,j,l}$ [see Eq.~\eqref{eq:ModeExpansions}]. 
While such a way of splitting is not unique, we proceed to the further discussion with the following choice, 
\begin{align}
\begin{split}
\tX^R_{j,l}(x) &= X^R_{j,l}(x) +\sum_{\sigma,\sigma'} \sum_{a,a'} \biggl( \frac{\pi}{2k} \bigl[ \bfomega_\sigma(m) \cdot \bfp^R_{X,j} \bigr] \omega^l_{\sigma'}(m) -\frac{\pi}{\sqrt{k}} \omega^l_\sigma(m) \bigl[ \bfomega_{\sigma'} \bfW_{a'} \cdot (\bfp^R_{Y,j} +\bfp^L_{Y,j}) \bigr] \biggr) M_{\sigma a; \sigma' a'}(m), \\
\tX^L_{j,l}(x) &= X^L_{j,l}(x) +\sum_{\sigma,\sigma'} \sum_{a,a'} \biggl( \frac{\pi}{2k} \bigl[ \bfomega_\sigma(m) \cdot (\bfp^L_{X,j} +2\bfp^R_{X,j}) \bigr] \omega^l_{\sigma'}(m) -\frac{\pi}{\sqrt{k}} \omega^l_\sigma(m) \bigl[ \bfomega_{\sigma'} \bfW_{a'} \cdot (\bfp^R_{Y,j} +\bfp^L_{Y,j}) \bigr] \biggr) M_{\sigma a; \sigma' a'}(m), \\
\tY^{R,\mu}_{j,l}(x) &= Y^{R,\mu}_{j,l}(x) +\frac{\pi}{2} \sum_{\sigma, \sigma'} \sum_{a,a'} \bigl( \bfomega_\sigma \bfW_a \cdot \bfp^R_{Y,j} \bigr) \omega^l_{\sigma'} W^\mu_{a'} M_{\sigma a; \sigma' a'}(m), \\
\tY^{L,\mu}_{j,l}(x) &= Y^{L,\mu}_{j,l}(x) +\frac{\pi}{2} \sum_{\sigma, \sigma'} \sum_{a,a'} \bigl[ \bfomega_\sigma \bfW_a \cdot (\bfp^L_{Y,j} +2\bfp^R_{Y,j}) \bigr] \omega^l_{\sigma'} W^\mu_{a'} M_{\sigma a; \sigma' a'}(m). 
\end{split}
\end{align}

Let us focus on the right-moving sector in a single wire and thus omit the wire index $j$ for brevity.
We find that the energy-momentum tensors for the $SU(N)_k/[U(1)]^{N-1}$ and $SU(k)_N/[U(1)]^{k-1}$ CFTs have the vertex representations, 
\begin{align}
\label{eq:SUNkParaEMTensorGen}
T_{SU(N)_k/[U(1)]^{N-1}}(z) &= \frac{1}{2(N+k)} \Biggl[ -N \sum_{\mu=1}^{k-1} :\mathrel{(\partial_z \bfY^{R,\mu})^2}: +\sum_{\bfalpha \in \Delta_N} \sum_{a \neq b} d_\bfA^{\bfW_a,\bfW_b}(\bfp^R_Y) :\mathrel{e^{i\bfalpha (\bfW_a -\bfW_b) \cdot \bfY^R(z)}}: \Biggl], \\
\label{eq:SUkNParaEMTensorGen}
T_{SU(k)_N/[U(1)]^{k-1}}(z) &= \frac{1}{2(N+k)} \Biggl[ -k \sum_{\mu=1}^{k-1} :\mathrel{(\partial_z \bfY^{R,\mu})^2}: -\sum_{\bfalpha \in \Delta_N} \sum_{a \neq b} d_\bfA^{\bfW_a,\bfW_b}(\bfp^R_Y) :\mathrel{e^{i\bfalpha (\bfW_a -\bfW_b) \cdot \bfY^R(z)}}: \Biggl], 
\end{align}
where the cocycle factors are given by 
\begin{align}
d^{\bfW_a,\bfW_b}_\bfalpha (\bfp) = e^{\frac{i\pi}{\sqrt{k}} \bfalpha (\bfW_a-\bfW_b) * \bfalpha(m) -\frac{i\pi}{2} [\bfalpha \bfW_a * \bfalpha (\bfW_a-\bfW_b) -\bfalpha (\bfW_a-\bfW_b) * \bfalpha \bfW_b] +i\pi \bfp * \bfalpha (\bfW_a -\bfW_b)}. 
\end{align}
Here we have also defined the normalized root, 
\begin{align}
\alpha^l(m) \equiv \sum_{l'=1}^{N-1} \Lambda^{-1}_{ll'}(m) \alpha^{l'}, \hspace{10pt}
{\bm \Lambda}^{-1}(m) = \textrm{diag} \left( 1,1, \cdots, 1, \sqrt{\frac{N+k(N-1)m}{N}} \right), 
\end{align}
and star products, 
\begin{align}
\begin{split}
\bfalpha * \bfbeta &\equiv \frac{1}{2k} \sum_{\sigma,\sigma'} \sum_{a,a'} \bigl[ \bfomega_\sigma(m) \cdot \bfalpha \bigr] \bigl[ \bfomega_{\sigma'}(m) \cdot \bfbeta \bigr] M_{\sigma a; \sigma' a'}(m), \\
\bfalpha * \bfbeta \bfB &\equiv \frac{1}{2\sqrt{k}} \sum_{\sigma,\sigma'} \sum_{a,a'} \bigl[ \bfomega_\sigma(m) \cdot \bfalpha \bigr] (\bfomega_{\sigma'} \cdot \bfbeta) (\bfW_{a'} \cdot \bfB) M_{\sigma a; \sigma' a'}(m), \\
\bfalpha \bfA * \bfbeta &\equiv \frac{1}{2\sqrt{k}} \sum_{\sigma,\sigma'} \sum_{a,a'} (\bfomega_\sigma \cdot \bfalpha) (\bfW_a \cdot \bfA) \bigl[ \bfomega_{\sigma'}(m) \cdot \bfbeta \bigr] M_{\sigma a; \sigma' a'}(m), \\
\bfalpha \bfA * \bfbeta \bfB &\equiv \frac{1}{2} \sum_{\sigma, \sigma'} \sum_{a,a'} (\bfomega_\sigma \cdot \bfalpha) (\bfW_a \cdot \bfA) (\bfomega_{\sigma'} \cdot \bfbeta) (\bfW_{a'} \cdot \bfB) M_{\sigma a; \sigma' a'}(m).
\end{split}
\end{align}
The energy-momentum tensor for the $SU(N)_k/[U(1)]^{N-1}$ CFT in Eq.~\eqref{eq:SUNkParaEMTensorGen} is obtained as follows. 
We start from the energy-momentum tensor, 
\begin{align} \label{eq:U1EMTensorGen}
T_{[U(1)]^{k(N-1)}}(z) = -\frac{1}{2} \sum_{l'=1}^{N-1} \Lambda^{-2}_{ll'}(m) :\mathrel{(\partial_z X^R_l)^2}: -\frac{1}{2} \sum_{\mu=1}^{k-1} :\mathrel{(\partial_z \bfY^{R,\mu})^2}:, 
\end{align}
which normalize the vertex representations of current operators, 
\begin{align}
\calH^l(z) &= i\sqrt{k} \sum_{l'=1}^{N-1} \Lambda^{-1}_{ll'}(m) \partial_z X^R_{l'}(z), \\
\label{eq:SUNkCurrentGen}
\calE^\bfalpha(z) &= \sum_{a=1}^k :\mathrel{e^{\frac{i}{\sqrt{k}} \bfalpha(m) \cdot \tbfX^R(z) +i\bfalpha \bfW_a \cdot \tbfY^R(z)}}:,
\end{align}
to have conformal weights $1$. 
By substituting these expressions into the Sugawara energy-momentum tensor \eqref{eq:SUNkEMTensorSugawara} and subtracting the $U(1)$ part containing $\bfX^R$, we obtain Eq.~\eqref{eq:SUNkParaEMTensorGen}. 
The energy-momentum tensor for the $SU(k)_N/[U(1)]^{k-1}$ CFT in Eq.~\eqref{eq:SUkNParaEMTensorGen} is obtained by subtracting those for the $SU(N)_k/[U(1)]^{N-1}$ and $U(1)$ CFTs from Eq.~\eqref{eq:U1EMTensorGen}. 

By factorizing Eq.~\eqref{eq:SUNkCurrentGen} as
\begin{align}
\calE^\bfalpha(z) = \ :\mathrel{e^{\frac{i}{\sqrt{k}}\bfalpha(m) \cdot \bfX^R(z) +\frac{i\pi}{\sqrt{k}} \bfp^R_X * \bfalpha(m) +\frac{2i\pi}{\sqrt{k}} \bfp^L_Y *\bfalpha(m)}}: \tPsi^{\bfalpha,1}(z), 
\end{align}
one can find the vertex representation of the $SU(N)_k/[U(1)]^{N-1}$ parafermionic field associated with an $SU(N)$ root $\bfalpha$ and the fundamental representation of $SU(k)$, 
\begin{align}
\tPsi^{\bfalpha,1}(z) = \sum_{a=1}^k d^{\bfW_a}_\bfalpha (\bfp^R_Y) :\mathrel{e^{i\bfalpha \bfW_a \cdot \bfY^R(z)}}:, 
\end{align}
up to the normalization, where the parafermionic cocycle factor is given by 
\begin{align}
d^{\bfW_a}_\bfalpha (\bfp) = e^{-\frac{i\pi}{\sqrt{k}} \bfalpha \bfW_a * \bfalpha(m) -\frac{i\pi}{2} \bfalpha \bfW_a * \bfalpha \bfW_a +\frac{2i\pi}{\sqrt{k}} \bfp * \bfalpha(m) +i\pi \bfp * \bfalpha \bfW_a}. 
\end{align}

In the left-moving sector, we similarly find the energy-momentum tensors, 
\begin{align}
\bT_{SU(N)_k/[U(1)]^{N-1}}(\bz) &= \frac{1}{2(N+k)} \Biggl[ -N \sum_{\mu=1}^{k-1} :\mathrel{(\partial_{\bz} \bfY^{L,\mu})^2}: +\sum_{\bfalpha \in \Delta_N} \sum_{a \neq b} d_\bfA^{\bfW_a,\bfW_b}(\bfp^L_Y+2\bfp^R_Y) :\mathrel{e^{i\bfalpha (\bfW_a -\bfW_b) \cdot \bfY^L(\bz)}}: \Biggl], \\
\bT_{SU(k)_N/[U(1)]^{k-1}}(\bz) &= \frac{1}{2(N+k)} \Biggl[ -k \sum_{\mu=1}^{k-1} :\mathrel{(\partial_{\bz} \bfY^{L,\mu})^2}: -\sum_{\bfalpha \in \Delta_N} \sum_{a \neq b} d_\bfA^{\bfW_a,\bfW_b}(\bfp^L_Y+2\bfp^R_Y) :\mathrel{e^{i\bfalpha (\bfW_a -\bfW_b) \cdot \bfY^L(\bz)}}: \Biggl],
\end{align}
and the parafermionic fields, 
\begin{align}
\bcalE^\bfalpha(\bz) &= \ :\mathrel{e^{\frac{i}{\sqrt{k}}\bfalpha(m) \cdot \bfX^L(\bz) +\frac{i\pi}{\sqrt{k}} (\bfp^L_X+2\bfp^R_X) * \bfalpha(m) -\frac{2i\pi}{\sqrt{k}} \bfp^R_Y *\bfalpha(m)}}: \bar{\tPsi}^{\bfalpha,1}(\bz), \\
\bar{\tPsi}^{\bfalpha,1}(\bz) &= \sum_{a=1}^k d^{\bfW_a}_\bfalpha (\bfp^L_Y+2\bfp^R_Y) :\mathrel{e^{i\bfalpha \bfW_a \cdot \bfY^L(\bz)}}:. 
\end{align}
Therefore, the interwire interactions 
\begin{align}
\sum_{a,b=1}^k \calO^t_{j,\sigma,ab} = \sum_{a,b=1}^k :\mathrel{e^{\frac{i}{\sqrt{k}} \bfalpha_\sigma(m) \cdot (\tbfX^R_j -\tbfX^L_{j+1}) +i\bfalpha_\sigma \bfW_a \cdot \tbfY^R_j -i\bfalpha_\sigma \bfW_b \cdot \tbfY^L_{j+1}}}:
\end{align}
can be identified as the products of a $[U(1)]^{N-1}$ vertex operator from the charge sector and the $SU(N)_k/[U(1)]^{N-1}$ parafermionic field from the neutral sector. 
On the other hand, the intrawire interactions 
\begin{align}
\sum_{s,s'=1}^N e^{i\Gamma_m(s,s')} \calO^{u,ss'}_{j,ab} &= \sum_{s,s'=1}^N :\mathrel{e^{i\bfomega_s (\bfW_a-\bfW_b) \cdot \tbfY^R_j -i\bfomega_{s'} (\bfW_a-\bfW_b) \cdot \tbfY^L_j +i\Gamma_m(s,s')}}:,
\end{align}
are identified as the nonchiral products of the $SU(k)_N/[U(1)]^{k-1}$ parafermionic fields with an appropriate choice of the phases $e^{i\Gamma_m(s,s')}$. 
Upon the assumption that the matrix $\bfM$ is given by Eq.~\eqref{eq:IntMatGen}, the phase is chosen to be 
\begin{align}
\Gamma_m(s,s') &= \frac{\pi m}{2} (\delta_{sN} +\delta_{s'N}). 
\end{align}

Because of the complicated form of the parafermionic cocycle, we could not find the algebraically simple way to prove that these vertex representations satisfy the required OPEs for a general choice of the interaction matrix $\bfM$. 
However, for a specific choice given in Eq.~\eqref{eq:IntMatGen} with even integer $m$, we have confirmed them by the explicit computation. 

\subsection{Vertex representations for odd $m$}

For odd integer $m$, FQH states are constructed from the array of fermionic wires. 
Thus we now introduce the fermionic operators in each wire, 
\begin{align}
\begin{split}
\psi^R_{j,\sigma,a}(x) &= \kappa_{j,\sigma,a} e^{-i\varphi_{j,\sigma,a}(x) -i\theta_{j,\sigma,a}(x) -ik_F x}, \\
\psi^L_{j,\sigma,a}(x) &= \kappa_{j,\sigma,a} e^{-i\varphi_{j,\sigma,a}(x) +i\theta_{j,\sigma,a}(x) +ik_F x}, 
\end{split}
\end{align}
where $\kappa_{j,\sigma,a}$ are the Klein factors satisfying $\kappa_{j,\sigma,a}^\dagger = \kappa_{j,\sigma,a}$ and
\begin{align}
\{ \kappa_{j,\sigma,a}, \kappa_{j',\sigma',a'} \} = 2\delta_{jj'} \delta_{\sigma \sigma'} \delta_{aa'}, 
\end{align}
and hence ensuring the anticommutation relations of fermions between different wires, components, or channels. 
The anticommutation relations of the fermion with itself and those with different chiralities are automatically ensured by the commutation relations of the bosonic fields in Eq.~\eqref{eq:BosonComm}. 
Then the interwire interactions are constructed as 
\begin{align}
\calO^t_{j,\sigma,ab} &\propto \psi^{R \dagger}_{j,\sigma,a} \psi^L_{j+1,\sigma,b} \prod_{\sigma'=1}^{N-1} \prod_{a'=1}^k \bigl( \psi^{R\dagger}_{j,\sigma',a'} \psi^L_{j,\sigma',a'} \bigr)^{\frac{1}{2}[\bfM(m) -\mathbf{1}]_{\sigma a; \sigma' a'}} \bigl( \psi^{R\dagger}_{j+1,\sigma',a'} \psi^L_{j+1,\sigma',a'} \bigr)^{\frac{1}{2}[\bfM^T(m) -\mathbf{1}]_{\sigma b; \sigma' a'}} \nonumber \\
&\propto \kappa_{j,\sigma,a} e^{i\varphi_{j,\sigma,a} +i\sum_{\sigma',a'} M_{\sigma a; \sigma' a'}(m) \theta_{j,\sigma',a'}} \kappa_{j+1,\sigma,b} e^{-i\varphi_{j+1,\sigma,b} +i\sum_{\sigma',a'} M^T_{\sigma b; \sigma' a'}(m) \theta_{j+1,\sigma',a'}}, 
\end{align}
where $\frac{1}{2}[\bfM(m)-\mathbf{1}]$ is an integer matrix and the Klein factors involved in the products $\psi^{R\dagger}_{j,\sigma',a'} \psi^L_{j,\sigma',a'}$ just become unity. 
If we write 
\begin{align}
\begin{split}
e^{i\tphi^R_{j,\sigma,a}} &= \kappa_{j,\sigma,a} e^{i\varphi_{j,\sigma,a} +i\sum_{\sigma',a'} M_{\sigma a; \sigma' a'}(m) \theta_{j,\sigma',a'}}, \\
e^{i\tphi^L_{j,\sigma,a}} &= \kappa_{j,\sigma,a} e^{i\varphi_{j,\sigma,a} -i\sum_{\sigma',a'} M^T_{\sigma a; \sigma' a'}(m) \theta_{j,\sigma',a'}},
\end{split}
\end{align}
the anticommuting property of the Klein factor can be incorporated into the commutation relations of the chiral bosonic fields, 
\begin{align}
\begin{split}
[\tphi^p_{j,\sigma,a}(x), \tphi^p_{j',\sigma',a'}(x')] &= i\pi p \delta_{jj'} K_{\sigma a; \sigma' a'}(m) \sgn(x-x') +\frac{i\pi}{2} \delta_{jj'} [\bfM(m)-\bfM^T(m)]_{\sigma a; \sigma' a'} \\
&\ \ \ +i\pi \delta_{jj'} \delta_{aa'} \sgn(\sigma'-\sigma) +i\pi \delta_{jj'} \sgn(a'-a) +i\pi \sgn(j'-j), \\
[\tphi^R_{j,\sigma,a}(x), \tphi^L_{j',\sigma',a'}(x')] &= i\pi \delta_{jj'} M_{\sigma a; \sigma' a'}(m) +i\pi \delta_{jj'} \delta_{aa'} \sgn(\sigma'-\sigma) +i\pi \delta_{jj'} \sgn(a'-a) +i\pi \sgn(j'-j),
\end{split}
\end{align}
where $\bfK(m) = \frac{1}{2}[\bfM(m)+\bfM^T(m)]$. 
Upon the choice of the interaction matrix $\bfM(m)$ in Eq.~\eqref{eq:IntMatGen}, these commutation relations are rewritten as 
\begin{align}
&[\tphi^p_{j,\sigma,a}(x), \tphi^{p'}_{j',\sigma',a'}(x')] \nonumber \\
&= \pi \delta_{jj'} K_{\sigma a; \sigma' a'}(m) \Bigl[ ip \delta_{pp'} \sgn(x-x') -\sigma^y_{pp'} \Bigr] +\frac{i\pi}{2} \delta_{jj'} \bigl[ \bfM(m+1) -\bfM^T(m+1) \bigr]_{\sigma a; \sigma' a'} +i\pi \sgn(j'-j). 
\end{align}
After introducing the chiral fields \eqref{eq:ChiralFieldXY}, this commutation relation produces Eq.~\eqref{eq:CommXYGen} within the same wire except that $\bfM(m)$ is replaced by $\bfM(m+1)$. 
This replacement means that the parafermionic cocycles appearing in the vertex representations of fields for odd $m$ nothing but follow the property for even $m$. 
Therefore, we can construct the vertex representations for parafermionic fields and energy-momentum tensors for the fermionic case simply along with the line of argument for the bosonic case given above. 

\end{widetext}

\bibliography{Refs_NASS}

\begin{thebibliography}{112}%
\makeatletter
\providecommand \@ifxundefined [1]{%
 \@ifx{#1\undefined}
}%
\providecommand \@ifnum [1]{%
 \ifnum #1\expandafter \@firstoftwo
 \else \expandafter \@secondoftwo
 \fi
}%
\providecommand \@ifx [1]{%
 \ifx #1\expandafter \@firstoftwo
 \else \expandafter \@secondoftwo
 \fi
}%
\providecommand \natexlab [1]{#1}%
\providecommand \enquote  [1]{``#1''}%
\providecommand \bibnamefont  [1]{#1}%
\providecommand \bibfnamefont [1]{#1}%
\providecommand \citenamefont [1]{#1}%
\providecommand \href@noop [0]{\@secondoftwo}%
\providecommand \href [0]{\begingroup \@sanitize@url \@href}%
\providecommand \@href[1]{\@@startlink{#1}\@@href}%
\providecommand \@@href[1]{\endgroup#1\@@endlink}%
\providecommand \@sanitize@url [0]{\catcode `\\12\catcode `\$12\catcode
  `\&12\catcode `\#12\catcode `\^12\catcode `\_12\catcode `\%12\relax}%
\providecommand \@@startlink[1]{}%
\providecommand \@@endlink[0]{}%
\providecommand \url  [0]{\begingroup\@sanitize@url \@url }%
\providecommand \@url [1]{\endgroup\@href {#1}{\urlprefix }}%
\providecommand \urlprefix  [0]{URL }%
\providecommand \Eprint [0]{\href }%
\providecommand \doibase [0]{http://dx.doi.org/}%
\providecommand \selectlanguage [0]{\@gobble}%
\providecommand \bibinfo  [0]{\@secondoftwo}%
\providecommand \bibfield  [0]{\@secondoftwo}%
\providecommand \translation [1]{[#1]}%
\providecommand \BibitemOpen [0]{}%
\providecommand \bibitemStop [0]{}%
\providecommand \bibitemNoStop [0]{.\EOS\space}%
\providecommand \EOS [0]{\spacefactor3000\relax}%
\providecommand \BibitemShut  [1]{\csname bibitem#1\endcsname}%
\let\auto@bib@innerbib\@empty
\bibitem [{\citenamefont {Nayak}\ \emph {et~al.}(2008)\citenamefont {Nayak},
  \citenamefont {Simon}, \citenamefont {Stern}, \citenamefont {Freedman},\ and\
  \citenamefont {Das~Sarma}}]{Nayak08}%
  \BibitemOpen
  \bibfield  {author} {\bibinfo {author} {\bibfnamefont {C.}~\bibnamefont
  {Nayak}}, \bibinfo {author} {\bibfnamefont {S.~H.}\ \bibnamefont {Simon}},
  \bibinfo {author} {\bibfnamefont {A.}~\bibnamefont {Stern}}, \bibinfo
  {author} {\bibfnamefont {M.}~\bibnamefont {Freedman}}, \ and\ \bibinfo
  {author} {\bibfnamefont {S.}~\bibnamefont {Das~Sarma}},\ }\href {\doibase
  10.1103/RevModPhys.80.1083} {\bibfield  {journal} {\bibinfo  {journal} {Rev.
  Mod. Phys.}\ }\textbf {\bibinfo {volume} {80}},\ \bibinfo {pages} {1083}
  (\bibinfo {year} {2008})}\BibitemShut {NoStop}%
\bibitem [{\citenamefont {Zeng}\ \emph {et~al.}()\citenamefont {Zeng},
  \citenamefont {Chen}, \citenamefont {Zhou},\ and\ \citenamefont
  {Wen}}]{BZeng15}%
  \BibitemOpen
  \bibfield  {author} {\bibinfo {author} {\bibfnamefont {B.}~\bibnamefont
  {Zeng}}, \bibinfo {author} {\bibfnamefont {X.}~\bibnamefont {Chen}}, \bibinfo
  {author} {\bibfnamefont {D.-L.}\ \bibnamefont {Zhou}}, \ and\ \bibinfo
  {author} {\bibfnamefont {X.-G.}\ \bibnamefont {Wen}},\ }\href
  {http://arxiv.org/abs/1508.02595} {\enquote {\bibinfo {title} {Quantum
  information meets quantum matter -- from quantum entanglement to topological
  phase in many-body systems},}\ }\Eprint
  {http://arxiv.org/abs/arXiv:1508.02595} {arXiv:1508.02595} \BibitemShut
  {NoStop}%
\bibitem [{\citenamefont {Wen}(2016)}]{Wen16}%
  \BibitemOpen
  \bibfield  {author} {\bibinfo {author} {\bibfnamefont {X.-G.}\ \bibnamefont
  {Wen}},\ }\href {\doibase 10.1093/nsr/nwv077} {\bibfield  {journal} {\bibinfo
   {journal} {Nat. Sci. Rev.}\ }\textbf {\bibinfo {volume} {3}},\ \bibinfo
  {pages} {68} (\bibinfo {year} {2016})}\BibitemShut {NoStop}%
\bibitem [{\citenamefont {Schoutens}\ and\ \citenamefont
  {Wen}(2016)}]{Schoutens16}%
  \BibitemOpen
  \bibfield  {author} {\bibinfo {author} {\bibfnamefont {K.}~\bibnamefont
  {Schoutens}}\ and\ \bibinfo {author} {\bibfnamefont {X.-G.}\ \bibnamefont
  {Wen}},\ }\href {\doibase 10.1103/PhysRevB.93.045109} {\bibfield  {journal}
  {\bibinfo  {journal} {Phys. Rev. B}\ }\textbf {\bibinfo {volume} {93}},\
  \bibinfo {pages} {045109} (\bibinfo {year} {2016})}\BibitemShut {NoStop}%
\bibitem [{\citenamefont {Lan}\ \emph {et~al.}(2016)\citenamefont {Lan},
  \citenamefont {Kong},\ and\ \citenamefont {Wen}}]{TLan16}%
  \BibitemOpen
  \bibfield  {author} {\bibinfo {author} {\bibfnamefont {T.}~\bibnamefont
  {Lan}}, \bibinfo {author} {\bibfnamefont {L.}~\bibnamefont {Kong}}, \ and\
  \bibinfo {author} {\bibfnamefont {X.-G.}\ \bibnamefont {Wen}},\ }\href
  {\doibase 10.1103/PhysRevB.94.155113} {\bibfield  {journal} {\bibinfo
  {journal} {Phys. Rev. B}\ }\textbf {\bibinfo {volume} {94}},\ \bibinfo
  {pages} {155113} (\bibinfo {year} {2016})}\BibitemShut {NoStop}%
\bibitem [{\citenamefont {Kane}\ \emph {et~al.}(2002)\citenamefont {Kane},
  \citenamefont {Mukhopadhyay},\ and\ \citenamefont {Lubensky}}]{Kane02}%
  \BibitemOpen
  \bibfield  {author} {\bibinfo {author} {\bibfnamefont {C.~L.}\ \bibnamefont
  {Kane}}, \bibinfo {author} {\bibfnamefont {R.}~\bibnamefont {Mukhopadhyay}},
  \ and\ \bibinfo {author} {\bibfnamefont {T.~C.}\ \bibnamefont {Lubensky}},\
  }\href {\doibase 10.1103/PhysRevLett.88.036401} {\bibfield  {journal}
  {\bibinfo  {journal} {Phys. Rev. Lett.}\ }\textbf {\bibinfo {volume} {88}},\
  \bibinfo {pages} {036401} (\bibinfo {year} {2002})}\BibitemShut {NoStop}%
\bibitem [{\citenamefont {Teo}\ and\ \citenamefont {Kane}(2014)}]{Teo14}%
  \BibitemOpen
  \bibfield  {author} {\bibinfo {author} {\bibfnamefont {J.~C.~Y.}\
  \bibnamefont {Teo}}\ and\ \bibinfo {author} {\bibfnamefont {C.~L.}\
  \bibnamefont {Kane}},\ }\href {\doibase 10.1103/PhysRevB.89.085101}
  {\bibfield  {journal} {\bibinfo  {journal} {Phys. Rev. B}\ }\textbf {\bibinfo
  {volume} {89}},\ \bibinfo {pages} {085101} (\bibinfo {year}
  {2014})}\BibitemShut {NoStop}%
\bibitem [{\citenamefont {Klinovaja}\ and\ \citenamefont
  {Loss}(2014)}]{Klinovaja14a}%
  \BibitemOpen
  \bibfield  {author} {\bibinfo {author} {\bibfnamefont {J.}~\bibnamefont
  {Klinovaja}}\ and\ \bibinfo {author} {\bibfnamefont {D.}~\bibnamefont
  {Loss}},\ }\href {\doibase 10.1140/epjb/e2014-50395-6} {\bibfield  {journal}
  {\bibinfo  {journal} {Eur. Phys. J. B}\ }\textbf {\bibinfo {volume} {87}},\
  \bibinfo {pages} {171} (\bibinfo {year} {2014})}\BibitemShut {NoStop}%
\bibitem [{\citenamefont {Meng}\ \emph {et~al.}(2014)\citenamefont {Meng},
  \citenamefont {Stano}, \citenamefont {Klinovaja},\ and\ \citenamefont
  {Loss}}]{Meng14a}%
  \BibitemOpen
  \bibfield  {author} {\bibinfo {author} {\bibfnamefont {T.}~\bibnamefont
  {Meng}}, \bibinfo {author} {\bibfnamefont {P.}~\bibnamefont {Stano}},
  \bibinfo {author} {\bibfnamefont {J.}~\bibnamefont {Klinovaja}}, \ and\
  \bibinfo {author} {\bibfnamefont {D.}~\bibnamefont {Loss}},\ }\href {\doibase
  10.1140/epjb/e2014-50445-1} {\bibfield  {journal} {\bibinfo  {journal} {Eur.
  Phys. J. B}\ }\textbf {\bibinfo {volume} {87}},\ \bibinfo {pages} {203}
  (\bibinfo {year} {2014})}\BibitemShut {NoStop}%
\bibitem [{\citenamefont {Meng}\ and\ \citenamefont {Sela}(2014)}]{Meng14b}%
  \BibitemOpen
  \bibfield  {author} {\bibinfo {author} {\bibfnamefont {T.}~\bibnamefont
  {Meng}}\ and\ \bibinfo {author} {\bibfnamefont {E.}~\bibnamefont {Sela}},\
  }\href {\doibase 10.1103/PhysRevB.90.235425} {\bibfield  {journal} {\bibinfo
  {journal} {Phys. Rev. B}\ }\textbf {\bibinfo {volume} {90}},\ \bibinfo
  {pages} {235425} (\bibinfo {year} {2014})}\BibitemShut {NoStop}%
\bibitem [{\citenamefont {Cano}\ \emph {et~al.}(2015)\citenamefont {Cano},
  \citenamefont {Hughes},\ and\ \citenamefont {Mulligan}}]{Cano15}%
  \BibitemOpen
  \bibfield  {author} {\bibinfo {author} {\bibfnamefont {J.}~\bibnamefont
  {Cano}}, \bibinfo {author} {\bibfnamefont {T.~L.}\ \bibnamefont {Hughes}}, \
  and\ \bibinfo {author} {\bibfnamefont {M.}~\bibnamefont {Mulligan}},\ }\href
  {\doibase 10.1103/PhysRevB.92.075104} {\bibfield  {journal} {\bibinfo
  {journal} {Phys. Rev. B}\ }\textbf {\bibinfo {volume} {92}},\ \bibinfo
  {pages} {075104} (\bibinfo {year} {2015})}\BibitemShut {NoStop}%
\bibitem [{\citenamefont {Sagi}\ \emph {et~al.}(2015)\citenamefont {Sagi},
  \citenamefont {Oreg}, \citenamefont {Stern},\ and\ \citenamefont
  {Halperin}}]{Sagi15a}%
  \BibitemOpen
  \bibfield  {author} {\bibinfo {author} {\bibfnamefont {E.}~\bibnamefont
  {Sagi}}, \bibinfo {author} {\bibfnamefont {Y.}~\bibnamefont {Oreg}}, \bibinfo
  {author} {\bibfnamefont {A.}~\bibnamefont {Stern}}, \ and\ \bibinfo {author}
  {\bibfnamefont {B.~I.}\ \bibnamefont {Halperin}},\ }\href {\doibase
  10.1103/PhysRevB.91.245144} {\bibfield  {journal} {\bibinfo  {journal} {Phys.
  Rev. B}\ }\textbf {\bibinfo {volume} {91}},\ \bibinfo {pages} {245144}
  (\bibinfo {year} {2015})}\BibitemShut {NoStop}%
\bibitem [{\citenamefont {Meng}\ \emph {et~al.}(2015)\citenamefont {Meng},
  \citenamefont {Neupert}, \citenamefont {Greiter},\ and\ \citenamefont
  {Thomale}}]{Meng15a}%
  \BibitemOpen
  \bibfield  {author} {\bibinfo {author} {\bibfnamefont {T.}~\bibnamefont
  {Meng}}, \bibinfo {author} {\bibfnamefont {T.}~\bibnamefont {Neupert}},
  \bibinfo {author} {\bibfnamefont {M.}~\bibnamefont {Greiter}}, \ and\
  \bibinfo {author} {\bibfnamefont {R.}~\bibnamefont {Thomale}},\ }\href
  {\doibase 10.1103/PhysRevB.91.241106} {\bibfield  {journal} {\bibinfo
  {journal} {Phys. Rev. B}\ }\textbf {\bibinfo {volume} {91}},\ \bibinfo
  {pages} {241106} (\bibinfo {year} {2015})}\BibitemShut {NoStop}%
\bibitem [{\citenamefont {Gorohovsky}\ \emph {et~al.}(2015)\citenamefont
  {Gorohovsky}, \citenamefont {Pereira},\ and\ \citenamefont
  {Sela}}]{Gorohovsky15}%
  \BibitemOpen
  \bibfield  {author} {\bibinfo {author} {\bibfnamefont {G.}~\bibnamefont
  {Gorohovsky}}, \bibinfo {author} {\bibfnamefont {R.~G.}\ \bibnamefont
  {Pereira}}, \ and\ \bibinfo {author} {\bibfnamefont {E.}~\bibnamefont
  {Sela}},\ }\href {\doibase 10.1103/PhysRevB.91.245139} {\bibfield  {journal}
  {\bibinfo  {journal} {Phys. Rev. B}\ }\textbf {\bibinfo {volume} {91}},\
  \bibinfo {pages} {245139} (\bibinfo {year} {2015})}\BibitemShut {NoStop}%
\bibitem [{\citenamefont {Huang}\ \emph {et~al.}(2016)\citenamefont {Huang},
  \citenamefont {Chen}, \citenamefont {Gomes}, \citenamefont {Neupert},
  \citenamefont {Chamon},\ and\ \citenamefont {Mudry}}]{PHHuang16}%
  \BibitemOpen
  \bibfield  {author} {\bibinfo {author} {\bibfnamefont {P.-H.}\ \bibnamefont
  {Huang}}, \bibinfo {author} {\bibfnamefont {J.-H.}\ \bibnamefont {Chen}},
  \bibinfo {author} {\bibfnamefont {P.~R.~S.}\ \bibnamefont {Gomes}}, \bibinfo
  {author} {\bibfnamefont {T.}~\bibnamefont {Neupert}}, \bibinfo {author}
  {\bibfnamefont {C.}~\bibnamefont {Chamon}}, \ and\ \bibinfo {author}
  {\bibfnamefont {C.}~\bibnamefont {Mudry}},\ }\href {\doibase
  10.1103/PhysRevB.93.205123} {\bibfield  {journal} {\bibinfo  {journal} {Phys.
  Rev. B}\ }\textbf {\bibinfo {volume} {93}},\ \bibinfo {pages} {205123}
  (\bibinfo {year} {2016})}\BibitemShut {NoStop}%
\bibitem [{\citenamefont {Lecheminant}\ and\ \citenamefont
  {Tsvelik}()}]{Lecheminant16}%
  \BibitemOpen
  \bibfield  {author} {\bibinfo {author} {\bibfnamefont {P.}~\bibnamefont
  {Lecheminant}}\ and\ \bibinfo {author} {\bibfnamefont {A.~M.}\ \bibnamefont
  {Tsvelik}},\ }\href {http://arxiv.org/abs/1608.05977} {\enquote {\bibinfo
  {title} {Lattice spin models for non-abelian chiral spin liquids},}\ }\Eprint
  {http://arxiv.org/abs/arXiv:1608.05977} {arXiv:1608.05977} \BibitemShut
  {NoStop}%
\bibitem [{\citenamefont {Mong}\ \emph {et~al.}(2014)\citenamefont {Mong},
  \citenamefont {Clarke}, \citenamefont {Alicea}, \citenamefont {Lindner},
  \citenamefont {Fendley}, \citenamefont {Nayak}, \citenamefont {Oreg},
  \citenamefont {Stern}, \citenamefont {Berg}, \citenamefont {Shtengel},\ and\
  \citenamefont {Fisher}}]{Mong14}%
  \BibitemOpen
  \bibfield  {author} {\bibinfo {author} {\bibfnamefont {R.~S.~K.}\
  \bibnamefont {Mong}}, \bibinfo {author} {\bibfnamefont {D.~J.}\ \bibnamefont
  {Clarke}}, \bibinfo {author} {\bibfnamefont {J.}~\bibnamefont {Alicea}},
  \bibinfo {author} {\bibfnamefont {N.~H.}\ \bibnamefont {Lindner}}, \bibinfo
  {author} {\bibfnamefont {P.}~\bibnamefont {Fendley}}, \bibinfo {author}
  {\bibfnamefont {C.}~\bibnamefont {Nayak}}, \bibinfo {author} {\bibfnamefont
  {Y.}~\bibnamefont {Oreg}}, \bibinfo {author} {\bibfnamefont {A.}~\bibnamefont
  {Stern}}, \bibinfo {author} {\bibfnamefont {E.}~\bibnamefont {Berg}},
  \bibinfo {author} {\bibfnamefont {K.}~\bibnamefont {Shtengel}}, \ and\
  \bibinfo {author} {\bibfnamefont {M.~P.~A.}\ \bibnamefont {Fisher}},\ }\href
  {\doibase 10.1103/PhysRevX.4.011036} {\bibfield  {journal} {\bibinfo
  {journal} {Phys. Rev. X}\ }\textbf {\bibinfo {volume} {4}},\ \bibinfo {pages}
  {011036} (\bibinfo {year} {2014})}\BibitemShut {NoStop}%
\bibitem [{\citenamefont {Klinovaja}\ and\ \citenamefont
  {Tserkovnyak}(2014)}]{Klinovaja14b}%
  \BibitemOpen
  \bibfield  {author} {\bibinfo {author} {\bibfnamefont {J.}~\bibnamefont
  {Klinovaja}}\ and\ \bibinfo {author} {\bibfnamefont {Y.}~\bibnamefont
  {Tserkovnyak}},\ }\href {\doibase 10.1103/PhysRevB.90.115426} {\bibfield
  {journal} {\bibinfo  {journal} {Phys. Rev. B}\ }\textbf {\bibinfo {volume}
  {90}},\ \bibinfo {pages} {115426} (\bibinfo {year} {2014})}\BibitemShut
  {NoStop}%
\bibitem [{\citenamefont {Neupert}\ \emph {et~al.}(2014)\citenamefont
  {Neupert}, \citenamefont {Chamon}, \citenamefont {Mudry},\ and\ \citenamefont
  {Thomale}}]{Neupert14}%
  \BibitemOpen
  \bibfield  {author} {\bibinfo {author} {\bibfnamefont {T.}~\bibnamefont
  {Neupert}}, \bibinfo {author} {\bibfnamefont {C.}~\bibnamefont {Chamon}},
  \bibinfo {author} {\bibfnamefont {C.}~\bibnamefont {Mudry}}, \ and\ \bibinfo
  {author} {\bibfnamefont {R.}~\bibnamefont {Thomale}},\ }\href {\doibase
  10.1103/PhysRevB.90.205101} {\bibfield  {journal} {\bibinfo  {journal} {Phys.
  Rev. B}\ }\textbf {\bibinfo {volume} {90}},\ \bibinfo {pages} {205101}
  (\bibinfo {year} {2014})}\BibitemShut {NoStop}%
\bibitem [{\citenamefont {Oreg}\ \emph {et~al.}(2014)\citenamefont {Oreg},
  \citenamefont {Sela},\ and\ \citenamefont {Stern}}]{Oreg14}%
  \BibitemOpen
  \bibfield  {author} {\bibinfo {author} {\bibfnamefont {Y.}~\bibnamefont
  {Oreg}}, \bibinfo {author} {\bibfnamefont {E.}~\bibnamefont {Sela}}, \ and\
  \bibinfo {author} {\bibfnamefont {A.}~\bibnamefont {Stern}},\ }\href
  {\doibase 10.1103/PhysRevB.89.115402} {\bibfield  {journal} {\bibinfo
  {journal} {Phys. Rev. B}\ }\textbf {\bibinfo {volume} {89}},\ \bibinfo
  {pages} {115402} (\bibinfo {year} {2014})}\BibitemShut {NoStop}%
\bibitem [{\citenamefont {Sagi}\ and\ \citenamefont {Oreg}(2014)}]{Sagi14}%
  \BibitemOpen
  \bibfield  {author} {\bibinfo {author} {\bibfnamefont {E.}~\bibnamefont
  {Sagi}}\ and\ \bibinfo {author} {\bibfnamefont {Y.}~\bibnamefont {Oreg}},\
  }\href {\doibase 10.1103/PhysRevB.90.201102} {\bibfield  {journal} {\bibinfo
  {journal} {Phys. Rev. B}\ }\textbf {\bibinfo {volume} {90}},\ \bibinfo
  {pages} {201102} (\bibinfo {year} {2014})}\BibitemShut {NoStop}%
\bibitem [{\citenamefont {Seroussi}\ \emph {et~al.}(2014)\citenamefont
  {Seroussi}, \citenamefont {Berg},\ and\ \citenamefont {Oreg}}]{Seroussi14}%
  \BibitemOpen
  \bibfield  {author} {\bibinfo {author} {\bibfnamefont {I.}~\bibnamefont
  {Seroussi}}, \bibinfo {author} {\bibfnamefont {E.}~\bibnamefont {Berg}}, \
  and\ \bibinfo {author} {\bibfnamefont {Y.}~\bibnamefont {Oreg}},\ }\href
  {\doibase 10.1103/PhysRevB.89.104523} {\bibfield  {journal} {\bibinfo
  {journal} {Phys. Rev. B}\ }\textbf {\bibinfo {volume} {89}},\ \bibinfo
  {pages} {104523} (\bibinfo {year} {2014})}\BibitemShut {NoStop}%
\bibitem [{\citenamefont {Vaezi}(2014)}]{Vaezi14a}%
  \BibitemOpen
  \bibfield  {author} {\bibinfo {author} {\bibfnamefont {A.}~\bibnamefont
  {Vaezi}},\ }\href {\doibase 10.1103/PhysRevX.4.031009} {\bibfield  {journal}
  {\bibinfo  {journal} {Phys. Rev. X}\ }\textbf {\bibinfo {volume} {4}},\
  \bibinfo {pages} {031009} (\bibinfo {year} {2014})}\BibitemShut {NoStop}%
\bibitem [{\citenamefont {Vaezi}\ and\ \citenamefont
  {Barkeshli}(2014)}]{Vaezi14b}%
  \BibitemOpen
  \bibfield  {author} {\bibinfo {author} {\bibfnamefont {A.}~\bibnamefont
  {Vaezi}}\ and\ \bibinfo {author} {\bibfnamefont {M.}~\bibnamefont
  {Barkeshli}},\ }\href {\doibase 10.1103/PhysRevLett.113.236804} {\bibfield
  {journal} {\bibinfo  {journal} {Phys. Rev. Lett.}\ }\textbf {\bibinfo
  {volume} {113}},\ \bibinfo {pages} {236804} (\bibinfo {year}
  {2014})}\BibitemShut {NoStop}%
\bibitem [{\citenamefont {Santos}\ \emph {et~al.}(2015)\citenamefont {Santos},
  \citenamefont {Huang}, \citenamefont {Gefen},\ and\ \citenamefont
  {Gutman}}]{Santos15}%
  \BibitemOpen
  \bibfield  {author} {\bibinfo {author} {\bibfnamefont {R.~A.}\ \bibnamefont
  {Santos}}, \bibinfo {author} {\bibfnamefont {C.-W.}\ \bibnamefont {Huang}},
  \bibinfo {author} {\bibfnamefont {Y.}~\bibnamefont {Gefen}}, \ and\ \bibinfo
  {author} {\bibfnamefont {D.~B.}\ \bibnamefont {Gutman}},\ }\href {\doibase
  10.1103/PhysRevB.91.205141} {\bibfield  {journal} {\bibinfo  {journal} {Phys.
  Rev. B}\ }\textbf {\bibinfo {volume} {91}},\ \bibinfo {pages} {205141}
  (\bibinfo {year} {2015})}\BibitemShut {NoStop}%
\bibitem [{\citenamefont {Lu}\ and\ \citenamefont {Vishwanath}(2012)}]{YMLu12}%
  \BibitemOpen
  \bibfield  {author} {\bibinfo {author} {\bibfnamefont {Y.-M.}\ \bibnamefont
  {Lu}}\ and\ \bibinfo {author} {\bibfnamefont {A.}~\bibnamefont
  {Vishwanath}},\ }\href {\doibase 10.1103/PhysRevB.86.125119} {\bibfield
  {journal} {\bibinfo  {journal} {Phys. Rev. B}\ }\textbf {\bibinfo {volume}
  {86}},\ \bibinfo {pages} {125119} (\bibinfo {year} {2012})}\BibitemShut
  {NoStop}%
\bibitem [{\citenamefont {Fuji}\ \emph {et~al.}(2016)\citenamefont {Fuji},
  \citenamefont {He}, \citenamefont {Bhattacharjee},\ and\ \citenamefont
  {Pollmann}}]{Fuji16}%
  \BibitemOpen
  \bibfield  {author} {\bibinfo {author} {\bibfnamefont {Y.}~\bibnamefont
  {Fuji}}, \bibinfo {author} {\bibfnamefont {Y.-C.}\ \bibnamefont {He}},
  \bibinfo {author} {\bibfnamefont {S.}~\bibnamefont {Bhattacharjee}}, \ and\
  \bibinfo {author} {\bibfnamefont {F.}~\bibnamefont {Pollmann}},\ }\href
  {\doibase 10.1103/PhysRevB.93.195143} {\bibfield  {journal} {\bibinfo
  {journal} {Phys. Rev. B}\ }\textbf {\bibinfo {volume} {93}},\ \bibinfo
  {pages} {195143} (\bibinfo {year} {2016})}\BibitemShut {NoStop}%
\bibitem [{\citenamefont {Sagi}\ and\ \citenamefont {Oreg}(2015)}]{Sagi15b}%
  \BibitemOpen
  \bibfield  {author} {\bibinfo {author} {\bibfnamefont {E.}~\bibnamefont
  {Sagi}}\ and\ \bibinfo {author} {\bibfnamefont {Y.}~\bibnamefont {Oreg}},\
  }\href {\doibase 10.1103/PhysRevB.92.195137} {\bibfield  {journal} {\bibinfo
  {journal} {Phys. Rev. B}\ }\textbf {\bibinfo {volume} {92}},\ \bibinfo
  {pages} {195137} (\bibinfo {year} {2015})}\BibitemShut {NoStop}%
\bibitem [{\citenamefont {Meng}(2015)}]{Meng15b}%
  \BibitemOpen
  \bibfield  {author} {\bibinfo {author} {\bibfnamefont {T.}~\bibnamefont
  {Meng}},\ }\href {\doibase 10.1103/PhysRevB.92.115152} {\bibfield  {journal}
  {\bibinfo  {journal} {Phys. Rev. B}\ }\textbf {\bibinfo {volume} {92}},\
  \bibinfo {pages} {115152} (\bibinfo {year} {2015})}\BibitemShut {NoStop}%
\bibitem [{\citenamefont {Meng}\ \emph {et~al.}(2016)\citenamefont {Meng},
  \citenamefont {Grushin}, \citenamefont {Shtengel},\ and\ \citenamefont
  {Bardarson}}]{Meng16}%
  \BibitemOpen
  \bibfield  {author} {\bibinfo {author} {\bibfnamefont {T.}~\bibnamefont
  {Meng}}, \bibinfo {author} {\bibfnamefont {A.~G.}\ \bibnamefont {Grushin}},
  \bibinfo {author} {\bibfnamefont {K.}~\bibnamefont {Shtengel}}, \ and\
  \bibinfo {author} {\bibfnamefont {J.~H.}\ \bibnamefont {Bardarson}},\ }\href
  {\doibase 10.1103/PhysRevB.94.155136} {\bibfield  {journal} {\bibinfo
  {journal} {Phys. Rev. B}\ }\textbf {\bibinfo {volume} {94}},\ \bibinfo
  {pages} {155136} (\bibinfo {year} {2016})}\BibitemShut {NoStop}%
\bibitem [{\citenamefont {Iadecola}\ \emph {et~al.}(2016)\citenamefont
  {Iadecola}, \citenamefont {Neupert}, \citenamefont {Chamon},\ and\
  \citenamefont {Mudry}}]{Iadecola16}%
  \BibitemOpen
  \bibfield  {author} {\bibinfo {author} {\bibfnamefont {T.}~\bibnamefont
  {Iadecola}}, \bibinfo {author} {\bibfnamefont {T.}~\bibnamefont {Neupert}},
  \bibinfo {author} {\bibfnamefont {C.}~\bibnamefont {Chamon}}, \ and\ \bibinfo
  {author} {\bibfnamefont {C.}~\bibnamefont {Mudry}},\ }\href {\doibase
  10.1103/PhysRevB.93.195136} {\bibfield  {journal} {\bibinfo  {journal} {Phys.
  Rev. B}\ }\textbf {\bibinfo {volume} {93}},\ \bibinfo {pages} {195136}
  (\bibinfo {year} {2016})}\BibitemShut {NoStop}%
\bibitem [{\citenamefont {Mross}\ \emph {et~al.}(2015)\citenamefont {Mross},
  \citenamefont {Essin},\ and\ \citenamefont {Alicea}}]{Mross15}%
  \BibitemOpen
  \bibfield  {author} {\bibinfo {author} {\bibfnamefont {D.~F.}\ \bibnamefont
  {Mross}}, \bibinfo {author} {\bibfnamefont {A.}~\bibnamefont {Essin}}, \ and\
  \bibinfo {author} {\bibfnamefont {J.}~\bibnamefont {Alicea}},\ }\href
  {\doibase 10.1103/PhysRevX.5.011011} {\bibfield  {journal} {\bibinfo
  {journal} {Phys. Rev. X}\ }\textbf {\bibinfo {volume} {5}},\ \bibinfo {pages}
  {011011} (\bibinfo {year} {2015})}\BibitemShut {NoStop}%
\bibitem [{\citenamefont {Mross}\ \emph {et~al.}(2016)\citenamefont {Mross},
  \citenamefont {Alicea},\ and\ \citenamefont {Motrunich}}]{Mross16}%
  \BibitemOpen
  \bibfield  {author} {\bibinfo {author} {\bibfnamefont {D.~F.}\ \bibnamefont
  {Mross}}, \bibinfo {author} {\bibfnamefont {J.}~\bibnamefont {Alicea}}, \
  and\ \bibinfo {author} {\bibfnamefont {O.~I.}\ \bibnamefont {Motrunich}},\
  }\href {\doibase 10.1103/PhysRevLett.117.016802} {\bibfield  {journal}
  {\bibinfo  {journal} {Phys. Rev. Lett.}\ }\textbf {\bibinfo {volume} {117}},\
  \bibinfo {pages} {016802} (\bibinfo {year} {2016})}\BibitemShut {NoStop}%
\bibitem [{\citenamefont {Sahoo}\ \emph {et~al.}(2016)\citenamefont {Sahoo},
  \citenamefont {Zhang},\ and\ \citenamefont {Teo}}]{Sahoo16}%
  \BibitemOpen
  \bibfield  {author} {\bibinfo {author} {\bibfnamefont {S.}~\bibnamefont
  {Sahoo}}, \bibinfo {author} {\bibfnamefont {Z.}~\bibnamefont {Zhang}}, \ and\
  \bibinfo {author} {\bibfnamefont {J.~C.~Y.}\ \bibnamefont {Teo}},\ }\href
  {\doibase 10.1103/PhysRevB.94.165142} {\bibfield  {journal} {\bibinfo
  {journal} {Phys. Rev. B}\ }\textbf {\bibinfo {volume} {94}},\ \bibinfo
  {pages} {165142} (\bibinfo {year} {2016})}\BibitemShut {NoStop}%
\bibitem [{\citenamefont {Gogolin}\ \emph {et~al.}(1998)\citenamefont
  {Gogolin}, \citenamefont {Nersesyan},\ and\ \citenamefont
  {Tsvelik}}]{Gogolin}%
  \BibitemOpen
  \bibfield  {author} {\bibinfo {author} {\bibfnamefont {A.~O.}\ \bibnamefont
  {Gogolin}}, \bibinfo {author} {\bibfnamefont {A.~A.}\ \bibnamefont
  {Nersesyan}}, \ and\ \bibinfo {author} {\bibfnamefont {A.~M.}\ \bibnamefont
  {Tsvelik}},\ }\href@noop {} {\emph {\bibinfo {title} {Bosonization and
  Strongly Correlated Systems}}}\ (\bibinfo  {publisher} {Cambridge University
  Press},\ \bibinfo {address} {Cambridge},\ \bibinfo {year} {1998})\BibitemShut
  {NoStop}%
\bibitem [{\citenamefont {Giamarchi}(2003)}]{Giamarchi}%
  \BibitemOpen
  \bibfield  {author} {\bibinfo {author} {\bibfnamefont {T.}~\bibnamefont
  {Giamarchi}},\ }\href@noop {} {\emph {\bibinfo {title} {Quantum Physics in
  One Dimension}}}\ (\bibinfo  {publisher} {Oxford University Press},\ \bibinfo
  {address} {New York},\ \bibinfo {year} {2003})\BibitemShut {NoStop}%
\bibitem [{\citenamefont {Moore}\ and\ \citenamefont {Read}(1991)}]{Moore91}%
  \BibitemOpen
  \bibfield  {author} {\bibinfo {author} {\bibfnamefont {G.}~\bibnamefont
  {Moore}}\ and\ \bibinfo {author} {\bibfnamefont {N.}~\bibnamefont {Read}},\
  }\href {\doibase 10.1016/0550-3213(91)90407-O} {\bibfield  {journal}
  {\bibinfo  {journal} {Nucl. Phys. B}\ }\textbf {\bibinfo {volume} {360}},\
  \bibinfo {pages} {362 } (\bibinfo {year} {1991})}\BibitemShut {NoStop}%
\bibitem [{\citenamefont {Read}\ and\ \citenamefont {Rezayi}(1999)}]{Read99}%
  \BibitemOpen
  \bibfield  {author} {\bibinfo {author} {\bibfnamefont {N.}~\bibnamefont
  {Read}}\ and\ \bibinfo {author} {\bibfnamefont {E.}~\bibnamefont {Rezayi}},\
  }\href {\doibase 10.1103/PhysRevB.59.8084} {\bibfield  {journal} {\bibinfo
  {journal} {Phys. Rev. B}\ }\textbf {\bibinfo {volume} {59}},\ \bibinfo
  {pages} {8084} (\bibinfo {year} {1999})}\BibitemShut {NoStop}%
\bibitem [{\citenamefont {Zamolodchikov}\ and\ \citenamefont
  {Fateev}(1985)}]{Zamolodchikov85}%
  \BibitemOpen
  \bibfield  {author} {\bibinfo {author} {\bibfnamefont {A.~B.}\ \bibnamefont
  {Zamolodchikov}}\ and\ \bibinfo {author} {\bibfnamefont {V.~A.}\ \bibnamefont
  {Fateev}},\ }\href {http://www.jetp.ac.ru/cgi-bin/e/index/e/62/2/p215?a=list}
  {\bibfield  {journal} {\bibinfo  {journal} {Sov. Phys. JETP}\ }\textbf
  {\bibinfo {volume} {62}},\ \bibinfo {pages} {215} (\bibinfo {year}
  {1985})}\BibitemShut {NoStop}%
\bibitem [{\citenamefont {Halperin}(1983)}]{Halperin83}%
  \BibitemOpen
  \bibfield  {author} {\bibinfo {author} {\bibfnamefont {B.~I.}\ \bibnamefont
  {Halperin}},\ }\href {\doibase 10.5169/seals-115362} {\bibfield  {journal}
  {\bibinfo  {journal} {Helv. Phys. Acta.}\ }\textbf {\bibinfo {volume} {56}},\
  \bibinfo {pages} {75} (\bibinfo {year} {1983})}\BibitemShut {NoStop}%
\bibitem [{\citenamefont {{Das Sarma}}\ and\ \citenamefont
  {Pinczuk}(1997)}]{DasSarma}%
  \BibitemOpen
  \bibfield  {author} {\bibinfo {author} {\bibfnamefont {S.}~\bibnamefont {{Das
  Sarma}}}\ and\ \bibinfo {author} {\bibfnamefont {A.}~\bibnamefont
  {Pinczuk}},\ }\href@noop {} {\emph {\bibinfo {title} {Perspectives in Quantum
  Hall Effects}}}\ (\bibinfo  {publisher} {Wiley},\ \bibinfo {address} {New
  York},\ \bibinfo {year} {1997})\BibitemShut {NoStop}%
\bibitem [{\citenamefont {Arovas}\ \emph {et~al.}(1999)\citenamefont {Arovas},
  \citenamefont {Karlhede},\ and\ \citenamefont {Lillieh\"o\"ok}}]{Arovas99}%
  \BibitemOpen
  \bibfield  {author} {\bibinfo {author} {\bibfnamefont {D.~P.}\ \bibnamefont
  {Arovas}}, \bibinfo {author} {\bibfnamefont {A.}~\bibnamefont {Karlhede}}, \
  and\ \bibinfo {author} {\bibfnamefont {D.}~\bibnamefont {Lillieh\"o\"ok}},\
  }\href {\doibase 10.1103/PhysRevB.59.13147} {\bibfield  {journal} {\bibinfo
  {journal} {Phys. Rev. B}\ }\textbf {\bibinfo {volume} {59}},\ \bibinfo
  {pages} {13147} (\bibinfo {year} {1999})}\BibitemShut {NoStop}%
\bibitem [{\citenamefont {Nomura}\ and\ \citenamefont
  {MacDonald}(2006)}]{Nomura06}%
  \BibitemOpen
  \bibfield  {author} {\bibinfo {author} {\bibfnamefont {K.}~\bibnamefont
  {Nomura}}\ and\ \bibinfo {author} {\bibfnamefont {A.~H.}\ \bibnamefont
  {MacDonald}},\ }\href {\doibase 10.1103/PhysRevLett.96.256602} {\bibfield
  {journal} {\bibinfo  {journal} {Phys. Rev. Lett.}\ }\textbf {\bibinfo
  {volume} {96}},\ \bibinfo {pages} {256602} (\bibinfo {year}
  {2006})}\BibitemShut {NoStop}%
\bibitem [{\citenamefont {Goerbig}\ and\ \citenamefont
  {Regnault}(2007)}]{Goerbig07}%
  \BibitemOpen
  \bibfield  {author} {\bibinfo {author} {\bibfnamefont {M.~O.}\ \bibnamefont
  {Goerbig}}\ and\ \bibinfo {author} {\bibfnamefont {N.}~\bibnamefont
  {Regnault}},\ }\href {\doibase 10.1103/PhysRevB.75.241405} {\bibfield
  {journal} {\bibinfo  {journal} {Phys. Rev. B}\ }\textbf {\bibinfo {volume}
  {75}},\ \bibinfo {pages} {241405} (\bibinfo {year} {2007})}\BibitemShut
  {NoStop}%
\bibitem [{\citenamefont {Dean}\ \emph {et~al.}(2011)\citenamefont {Dean},
  \citenamefont {Young}, \citenamefont {Cadden-Zimansky}, \citenamefont {Wang},
  \citenamefont {Ren}, \citenamefont {Watanabe}, \citenamefont {Taniguchi},
  \citenamefont {Kim}, \citenamefont {Hone},\ and\ \citenamefont
  {Shepard}}]{Dean11}%
  \BibitemOpen
  \bibfield  {author} {\bibinfo {author} {\bibfnamefont {C.~R.}\ \bibnamefont
  {Dean}}, \bibinfo {author} {\bibfnamefont {A.~F.}\ \bibnamefont {Young}},
  \bibinfo {author} {\bibfnamefont {P.}~\bibnamefont {Cadden-Zimansky}},
  \bibinfo {author} {\bibfnamefont {L.}~\bibnamefont {Wang}}, \bibinfo {author}
  {\bibfnamefont {H.}~\bibnamefont {Ren}}, \bibinfo {author} {\bibfnamefont
  {K.}~\bibnamefont {Watanabe}}, \bibinfo {author} {\bibfnamefont
  {T.}~\bibnamefont {Taniguchi}}, \bibinfo {author} {\bibfnamefont
  {P.}~\bibnamefont {Kim}}, \bibinfo {author} {\bibfnamefont {J.}~\bibnamefont
  {Hone}}, \ and\ \bibinfo {author} {\bibfnamefont {K.~L.}\ \bibnamefont
  {Shepard}},\ }\href {\doibase 10.1038/nphys2007} {\bibfield  {journal}
  {\bibinfo  {journal} {Nat. Phys.}\ }\textbf {\bibinfo {volume} {7}},\
  \bibinfo {pages} {693} (\bibinfo {year} {2011})}\BibitemShut {NoStop}%
\bibitem [{\citenamefont {Francesco}\ \emph {et~al.}(1997)\citenamefont
  {Francesco}, \citenamefont {Mathieu},\ and\ \citenamefont
  {S\'en\'echal}}]{dFMS}%
  \BibitemOpen
  \bibfield  {author} {\bibinfo {author} {\bibfnamefont {P.~D.}\ \bibnamefont
  {Francesco}}, \bibinfo {author} {\bibfnamefont {P.}~\bibnamefont {Mathieu}},
  \ and\ \bibinfo {author} {\bibfnamefont {D.}~\bibnamefont {S\'en\'echal}},\
  }\href@noop {} {\emph {\bibinfo {title} {Conformal Field Theory}}}\ (\bibinfo
   {publisher} {Splinger-Verlag},\ \bibinfo {address} {New York},\ \bibinfo
  {year} {1997})\BibitemShut {NoStop}%
\bibitem [{\citenamefont {Ardonne}\ and\ \citenamefont
  {Schoutens}(1999)}]{Ardonne99}%
  \BibitemOpen
  \bibfield  {author} {\bibinfo {author} {\bibfnamefont {E.}~\bibnamefont
  {Ardonne}}\ and\ \bibinfo {author} {\bibfnamefont {K.}~\bibnamefont
  {Schoutens}},\ }\href {\doibase 10.1103/PhysRevLett.82.5096} {\bibfield
  {journal} {\bibinfo  {journal} {Phys. Rev. Lett.}\ }\textbf {\bibinfo
  {volume} {82}},\ \bibinfo {pages} {5096} (\bibinfo {year}
  {1999})}\BibitemShut {NoStop}%
\bibitem [{\citenamefont {Ardonne}\ \emph
  {et~al.}(2001{\natexlab{a}})\citenamefont {Ardonne}, \citenamefont {Read},
  \citenamefont {Rezayi},\ and\ \citenamefont {Schoutens}}]{Ardonne01a}%
  \BibitemOpen
  \bibfield  {author} {\bibinfo {author} {\bibfnamefont {E.}~\bibnamefont
  {Ardonne}}, \bibinfo {author} {\bibfnamefont {N.}~\bibnamefont {Read}},
  \bibinfo {author} {\bibfnamefont {E.}~\bibnamefont {Rezayi}}, \ and\ \bibinfo
  {author} {\bibfnamefont {K.}~\bibnamefont {Schoutens}},\ }\href {\doibase
  10.1016/S0550-3213(01)00224-3} {\bibfield  {journal} {\bibinfo  {journal}
  {Nucl. Phys. B}\ }\textbf {\bibinfo {volume} {607}},\ \bibinfo {pages} {549 }
  (\bibinfo {year} {2001}{\natexlab{a}})}\BibitemShut {NoStop}%
\bibitem [{\citenamefont {Reijnders}\ \emph {et~al.}(2002)\citenamefont
  {Reijnders}, \citenamefont {van Lankvelt}, \citenamefont {Schoutens},\ and\
  \citenamefont {Read}}]{Reijnders02}%
  \BibitemOpen
  \bibfield  {author} {\bibinfo {author} {\bibfnamefont {J.~W.}\ \bibnamefont
  {Reijnders}}, \bibinfo {author} {\bibfnamefont {F.~J.~M.}\ \bibnamefont {van
  Lankvelt}}, \bibinfo {author} {\bibfnamefont {K.}~\bibnamefont {Schoutens}},
  \ and\ \bibinfo {author} {\bibfnamefont {N.}~\bibnamefont {Read}},\ }\href
  {\doibase 10.1103/PhysRevLett.89.120401} {\bibfield  {journal} {\bibinfo
  {journal} {Phys. Rev. Lett.}\ }\textbf {\bibinfo {volume} {89}},\ \bibinfo
  {pages} {120401} (\bibinfo {year} {2002})}\BibitemShut {NoStop}%
\bibitem [{\citenamefont {Reijnders}\ \emph {et~al.}(2004)\citenamefont
  {Reijnders}, \citenamefont {van Lankvelt}, \citenamefont {Schoutens},\ and\
  \citenamefont {Read}}]{Reijnders04}%
  \BibitemOpen
  \bibfield  {author} {\bibinfo {author} {\bibfnamefont {J.~W.}\ \bibnamefont
  {Reijnders}}, \bibinfo {author} {\bibfnamefont {F.~J.~M.}\ \bibnamefont {van
  Lankvelt}}, \bibinfo {author} {\bibfnamefont {K.}~\bibnamefont {Schoutens}},
  \ and\ \bibinfo {author} {\bibfnamefont {N.}~\bibnamefont {Read}},\ }\href
  {\doibase 10.1103/PhysRevA.69.023612} {\bibfield  {journal} {\bibinfo
  {journal} {Phys. Rev. A}\ }\textbf {\bibinfo {volume} {69}},\ \bibinfo
  {pages} {023612} (\bibinfo {year} {2004})}\BibitemShut {NoStop}%
\bibitem [{\citenamefont {Sterdyniak}\ \emph {et~al.}(2013)\citenamefont
  {Sterdyniak}, \citenamefont {Repellin}, \citenamefont {Bernevig},\ and\
  \citenamefont {Regnault}}]{Sterdyniak13}%
  \BibitemOpen
  \bibfield  {author} {\bibinfo {author} {\bibfnamefont {A.}~\bibnamefont
  {Sterdyniak}}, \bibinfo {author} {\bibfnamefont {C.}~\bibnamefont
  {Repellin}}, \bibinfo {author} {\bibfnamefont {B.~A.}\ \bibnamefont
  {Bernevig}}, \ and\ \bibinfo {author} {\bibfnamefont {N.}~\bibnamefont
  {Regnault}},\ }\href {\doibase 10.1103/PhysRevB.87.205137} {\bibfield
  {journal} {\bibinfo  {journal} {Phys. Rev. B}\ }\textbf {\bibinfo {volume}
  {87}},\ \bibinfo {pages} {205137} (\bibinfo {year} {2013})}\BibitemShut
  {NoStop}%
\bibitem [{\citenamefont {Gepner}(1987)}]{Gepner87}%
  \BibitemOpen
  \bibfield  {author} {\bibinfo {author} {\bibfnamefont {D.}~\bibnamefont
  {Gepner}},\ }\href {\doibase 10.1016/0550-3213(87)90176-3} {\bibfield
  {journal} {\bibinfo  {journal} {Nucl. Phys. B}\ }\textbf {\bibinfo {volume}
  {290}},\ \bibinfo {pages} {10 } (\bibinfo {year} {1987})}\BibitemShut
  {NoStop}%
\bibitem [{\citenamefont {Ardonne}\ \emph {et~al.}(2000)\citenamefont
  {Ardonne}, \citenamefont {Bouwknegt}, \citenamefont {Guruswamy},\ and\
  \citenamefont {Schoutens}}]{Ardonne00}%
  \BibitemOpen
  \bibfield  {author} {\bibinfo {author} {\bibfnamefont {E.}~\bibnamefont
  {Ardonne}}, \bibinfo {author} {\bibfnamefont {P.}~\bibnamefont {Bouwknegt}},
  \bibinfo {author} {\bibfnamefont {S.}~\bibnamefont {Guruswamy}}, \ and\
  \bibinfo {author} {\bibfnamefont {K.}~\bibnamefont {Schoutens}},\ }\href
  {\doibase 10.1103/PhysRevB.61.10298} {\bibfield  {journal} {\bibinfo
  {journal} {Phys. Rev. B}\ }\textbf {\bibinfo {volume} {61}},\ \bibinfo
  {pages} {10298} (\bibinfo {year} {2000})}\BibitemShut {NoStop}%
\bibitem [{\citenamefont {Ardonne}\ \emph
  {et~al.}(2001{\natexlab{b}})\citenamefont {Ardonne}, \citenamefont
  {Bouwknegt},\ and\ \citenamefont {Schoutens}}]{Ardonne01b}%
  \BibitemOpen
  \bibfield  {author} {\bibinfo {author} {\bibfnamefont {E.}~\bibnamefont
  {Ardonne}}, \bibinfo {author} {\bibfnamefont {P.}~\bibnamefont {Bouwknegt}},
  \ and\ \bibinfo {author} {\bibfnamefont {K.}~\bibnamefont {Schoutens}},\
  }\href {\doibase 10.1023/A:1004878231034} {\bibfield  {journal} {\bibinfo
  {journal} {J. Stat. Phys.}\ }\textbf {\bibinfo {volume} {102}},\ \bibinfo
  {pages} {421} (\bibinfo {year} {2001}{\natexlab{b}})}\BibitemShut {NoStop}%
\bibitem [{\citenamefont {Senthil}\ and\ \citenamefont
  {Levin}(2013)}]{Senthil13}%
  \BibitemOpen
  \bibfield  {author} {\bibinfo {author} {\bibfnamefont {T.}~\bibnamefont
  {Senthil}}\ and\ \bibinfo {author} {\bibfnamefont {M.}~\bibnamefont
  {Levin}},\ }\href {\doibase 10.1103/PhysRevLett.110.046801} {\bibfield
  {journal} {\bibinfo  {journal} {Phys. Rev. Lett.}\ }\textbf {\bibinfo
  {volume} {110}},\ \bibinfo {pages} {046801} (\bibinfo {year}
  {2013})}\BibitemShut {NoStop}%
\bibitem [{\citenamefont {Rapp}\ \emph {et~al.}(2012)\citenamefont {Rapp},
  \citenamefont {Deng},\ and\ \citenamefont {Santos}}]{Rapp12}%
  \BibitemOpen
  \bibfield  {author} {\bibinfo {author} {\bibfnamefont {A.}~\bibnamefont
  {Rapp}}, \bibinfo {author} {\bibfnamefont {X.}~\bibnamefont {Deng}}, \ and\
  \bibinfo {author} {\bibfnamefont {L.}~\bibnamefont {Santos}},\ }\href
  {\doibase 10.1103/PhysRevLett.109.203005} {\bibfield  {journal} {\bibinfo
  {journal} {Phys. Rev. Lett.}\ }\textbf {\bibinfo {volume} {109}},\ \bibinfo
  {pages} {203005} (\bibinfo {year} {2012})}\BibitemShut {NoStop}%
\bibitem [{\citenamefont {Meinert}\ \emph {et~al.}(2016)\citenamefont
  {Meinert}, \citenamefont {Mark}, \citenamefont {Lauber}, \citenamefont
  {Daley},\ and\ \citenamefont {N\"agerl}}]{Meinert16}%
  \BibitemOpen
  \bibfield  {author} {\bibinfo {author} {\bibfnamefont {F.}~\bibnamefont
  {Meinert}}, \bibinfo {author} {\bibfnamefont {M.~J.}\ \bibnamefont {Mark}},
  \bibinfo {author} {\bibfnamefont {K.}~\bibnamefont {Lauber}}, \bibinfo
  {author} {\bibfnamefont {A.~J.}\ \bibnamefont {Daley}}, \ and\ \bibinfo
  {author} {\bibfnamefont {H.-C.}\ \bibnamefont {N\"agerl}},\ }\href {\doibase
  10.1103/PhysRevLett.116.205301} {\bibfield  {journal} {\bibinfo  {journal}
  {Phys. Rev. Lett.}\ }\textbf {\bibinfo {volume} {116}},\ \bibinfo {pages}
  {205301} (\bibinfo {year} {2016})}\BibitemShut {NoStop}%
\bibitem [{\citenamefont {Fukuhara}\ \emph {et~al.}(2007)\citenamefont
  {Fukuhara}, \citenamefont {Takasu}, \citenamefont {Kumakura},\ and\
  \citenamefont {Takahashi}}]{Fukuhara07}%
  \BibitemOpen
  \bibfield  {author} {\bibinfo {author} {\bibfnamefont {T.}~\bibnamefont
  {Fukuhara}}, \bibinfo {author} {\bibfnamefont {Y.}~\bibnamefont {Takasu}},
  \bibinfo {author} {\bibfnamefont {M.}~\bibnamefont {Kumakura}}, \ and\
  \bibinfo {author} {\bibfnamefont {Y.}~\bibnamefont {Takahashi}},\ }\href
  {\doibase 10.1103/PhysRevLett.98.030401} {\bibfield  {journal} {\bibinfo
  {journal} {Phys. Rev. Lett.}\ }\textbf {\bibinfo {volume} {98}},\ \bibinfo
  {pages} {030401} (\bibinfo {year} {2007})}\BibitemShut {NoStop}%
\bibitem [{\citenamefont {DeSalvo}\ \emph {et~al.}(2010)\citenamefont
  {DeSalvo}, \citenamefont {Yan}, \citenamefont {Mickelson}, \citenamefont
  {Martinez~de Escobar},\ and\ \citenamefont {Killian}}]{DeSalvo10}%
  \BibitemOpen
  \bibfield  {author} {\bibinfo {author} {\bibfnamefont {B.~J.}\ \bibnamefont
  {DeSalvo}}, \bibinfo {author} {\bibfnamefont {M.}~\bibnamefont {Yan}},
  \bibinfo {author} {\bibfnamefont {P.~G.}\ \bibnamefont {Mickelson}}, \bibinfo
  {author} {\bibfnamefont {Y.~N.}\ \bibnamefont {Martinez~de Escobar}}, \ and\
  \bibinfo {author} {\bibfnamefont {T.~C.}\ \bibnamefont {Killian}},\ }\href
  {\doibase 10.1103/PhysRevLett.105.030402} {\bibfield  {journal} {\bibinfo
  {journal} {Phys. Rev. Lett.}\ }\textbf {\bibinfo {volume} {105}},\ \bibinfo
  {pages} {030402} (\bibinfo {year} {2010})}\BibitemShut {NoStop}%
\bibitem [{\citenamefont {Taie}\ \emph {et~al.}(2012)\citenamefont {Taie},
  \citenamefont {Yamazaki}, \citenamefont {Sugawa},\ and\ \citenamefont
  {Takahashi}}]{Taie12}%
  \BibitemOpen
  \bibfield  {author} {\bibinfo {author} {\bibfnamefont {S.}~\bibnamefont
  {Taie}}, \bibinfo {author} {\bibfnamefont {R.}~\bibnamefont {Yamazaki}},
  \bibinfo {author} {\bibfnamefont {S.}~\bibnamefont {Sugawa}}, \ and\ \bibinfo
  {author} {\bibfnamefont {Y.}~\bibnamefont {Takahashi}},\ }\href {\doibase
  10.1038/nphys2430} {\bibfield  {journal} {\bibinfo  {journal} {Nat. Phys.}\
  }\textbf {\bibinfo {volume} {8}},\ \bibinfo {pages} {825} (\bibinfo {year}
  {2012})}\BibitemShut {NoStop}%
\bibitem [{\citenamefont {Hofrichter}\ \emph {et~al.}(2016)\citenamefont
  {Hofrichter}, \citenamefont {Riegger}, \citenamefont {Scazza}, \citenamefont
  {H\"ofer}, \citenamefont {Fernandes}, \citenamefont {Bloch},\ and\
  \citenamefont {F\"olling}}]{Hofrichter16}%
  \BibitemOpen
  \bibfield  {author} {\bibinfo {author} {\bibfnamefont {C.}~\bibnamefont
  {Hofrichter}}, \bibinfo {author} {\bibfnamefont {L.}~\bibnamefont {Riegger}},
  \bibinfo {author} {\bibfnamefont {F.}~\bibnamefont {Scazza}}, \bibinfo
  {author} {\bibfnamefont {M.}~\bibnamefont {H\"ofer}}, \bibinfo {author}
  {\bibfnamefont {D.~R.}\ \bibnamefont {Fernandes}}, \bibinfo {author}
  {\bibfnamefont {I.}~\bibnamefont {Bloch}}, \ and\ \bibinfo {author}
  {\bibfnamefont {S.}~\bibnamefont {F\"olling}},\ }\href {\doibase
  10.1103/PhysRevX.6.021030} {\bibfield  {journal} {\bibinfo  {journal} {Phys.
  Rev. X}\ }\textbf {\bibinfo {volume} {6}},\ \bibinfo {pages} {021030}
  (\bibinfo {year} {2016})}\BibitemShut {NoStop}%
\bibitem [{\citenamefont {Khveshchenko}\ and\ \citenamefont
  {Wiegmann}(1989)}]{Khveshchenko89}%
  \BibitemOpen
  \bibfield  {author} {\bibinfo {author} {\bibfnamefont {D.~V.}\ \bibnamefont
  {Khveshchenko}}\ and\ \bibinfo {author} {\bibfnamefont {P.~B.}\ \bibnamefont
  {Wiegmann}},\ }\href {\doibase 10.1142/S0217984989002089} {\bibfield
  {journal} {\bibinfo  {journal} {Mod. Phys. Lett. B}\ }\textbf {\bibinfo
  {volume} {03}},\ \bibinfo {pages} {1383} (\bibinfo {year}
  {1989})}\BibitemShut {NoStop}%
\bibitem [{\citenamefont {Khveshchenko}\ and\ \citenamefont
  {Wiegmann}(1990)}]{Khveshchenko90}%
  \BibitemOpen
  \bibfield  {author} {\bibinfo {author} {\bibfnamefont {D.~V.}\ \bibnamefont
  {Khveshchenko}}\ and\ \bibinfo {author} {\bibfnamefont {P.~B.}\ \bibnamefont
  {Wiegmann}},\ }\href {\doibase 10.1142/S0217984990000040} {\bibfield
  {journal} {\bibinfo  {journal} {Mod. Phys. Lett. B}\ }\textbf {\bibinfo
  {volume} {04}},\ \bibinfo {pages} {17} (\bibinfo {year} {1990})}\BibitemShut
  {NoStop}%
\bibitem [{\citenamefont {Hermele}\ \emph {et~al.}(2009)\citenamefont
  {Hermele}, \citenamefont {Gurarie},\ and\ \citenamefont {Rey}}]{Hermele09}%
  \BibitemOpen
  \bibfield  {author} {\bibinfo {author} {\bibfnamefont {M.}~\bibnamefont
  {Hermele}}, \bibinfo {author} {\bibfnamefont {V.}~\bibnamefont {Gurarie}}, \
  and\ \bibinfo {author} {\bibfnamefont {A.~M.}\ \bibnamefont {Rey}},\ }\href
  {\doibase 10.1103/PhysRevLett.103.135301} {\bibfield  {journal} {\bibinfo
  {journal} {Phys. Rev. Lett.}\ }\textbf {\bibinfo {volume} {103}},\ \bibinfo
  {pages} {135301} (\bibinfo {year} {2009})}\BibitemShut {NoStop}%
\bibitem [{\citenamefont {Hermele}\ and\ \citenamefont
  {Gurarie}(2011)}]{Hermele11}%
  \BibitemOpen
  \bibfield  {author} {\bibinfo {author} {\bibfnamefont {M.}~\bibnamefont
  {Hermele}}\ and\ \bibinfo {author} {\bibfnamefont {V.}~\bibnamefont
  {Gurarie}},\ }\href {\doibase 10.1103/PhysRevB.84.174441} {\bibfield
  {journal} {\bibinfo  {journal} {Phys. Rev. B}\ }\textbf {\bibinfo {volume}
  {84}},\ \bibinfo {pages} {174441} (\bibinfo {year} {2011})}\BibitemShut
  {NoStop}%
\bibitem [{\citenamefont {Chen}\ \emph {et~al.}(2016)\citenamefont {Chen},
  \citenamefont {Hazzard}, \citenamefont {Rey},\ and\ \citenamefont
  {Hermele}}]{GChen16}%
  \BibitemOpen
  \bibfield  {author} {\bibinfo {author} {\bibfnamefont {G.}~\bibnamefont
  {Chen}}, \bibinfo {author} {\bibfnamefont {K.~R.~A.}\ \bibnamefont
  {Hazzard}}, \bibinfo {author} {\bibfnamefont {A.~M.}\ \bibnamefont {Rey}}, \
  and\ \bibinfo {author} {\bibfnamefont {M.}~\bibnamefont {Hermele}},\ }\href
  {\doibase 10.1103/PhysRevA.93.061601} {\bibfield  {journal} {\bibinfo
  {journal} {Phys. Rev. A}\ }\textbf {\bibinfo {volume} {93}},\ \bibinfo
  {pages} {061601} (\bibinfo {year} {2016})}\BibitemShut {NoStop}%
\bibitem [{\citenamefont {Barkeshli}\ and\ \citenamefont
  {Wen}(2010)}]{Barkeshli10}%
  \BibitemOpen
  \bibfield  {author} {\bibinfo {author} {\bibfnamefont {M.}~\bibnamefont
  {Barkeshli}}\ and\ \bibinfo {author} {\bibfnamefont {X.-G.}\ \bibnamefont
  {Wen}},\ }\href {\doibase 10.1103/PhysRevB.82.245301} {\bibfield  {journal}
  {\bibinfo  {journal} {Phys. Rev. B}\ }\textbf {\bibinfo {volume} {82}},\
  \bibinfo {pages} {245301} (\bibinfo {year} {2010})}\BibitemShut {NoStop}%
\bibitem [{\citenamefont {Wen}\ and\ \citenamefont {Zee}(1992)}]{Wen92}%
  \BibitemOpen
  \bibfield  {author} {\bibinfo {author} {\bibfnamefont {X.~G.}\ \bibnamefont
  {Wen}}\ and\ \bibinfo {author} {\bibfnamefont {A.}~\bibnamefont {Zee}},\
  }\href {\doibase 10.1103/PhysRevB.46.2290} {\bibfield  {journal} {\bibinfo
  {journal} {Phys. Rev. B}\ }\textbf {\bibinfo {volume} {46}},\ \bibinfo
  {pages} {2290} (\bibinfo {year} {1992})}\BibitemShut {NoStop}%
\bibitem [{\citenamefont {Wen}(1995)}]{Wen95}%
  \BibitemOpen
  \bibfield  {author} {\bibinfo {author} {\bibfnamefont {X.-G.}\ \bibnamefont
  {Wen}},\ }\href {\doibase 10.1080/00018739500101566} {\bibfield  {journal}
  {\bibinfo  {journal} {Adv. Phys.}\ }\textbf {\bibinfo {volume} {44}},\
  \bibinfo {pages} {405} (\bibinfo {year} {1995})}\BibitemShut {NoStop}%
\bibitem [{\citenamefont {Fradkin}\ \emph {et~al.}(1999)\citenamefont
  {Fradkin}, \citenamefont {Nayak},\ and\ \citenamefont
  {Schoutens}}]{Fradkin99}%
  \BibitemOpen
  \bibfield  {author} {\bibinfo {author} {\bibfnamefont {E.}~\bibnamefont
  {Fradkin}}, \bibinfo {author} {\bibfnamefont {C.}~\bibnamefont {Nayak}}, \
  and\ \bibinfo {author} {\bibfnamefont {K.}~\bibnamefont {Schoutens}},\ }\href
  {\doibase 10.1016/S0550-3213(99)00039-5} {\bibfield  {journal} {\bibinfo
  {journal} {Nucl. Phys. B}\ }\textbf {\bibinfo {volume} {546}},\ \bibinfo
  {pages} {711 } (\bibinfo {year} {1999})}\BibitemShut {NoStop}%
\bibitem [{\citenamefont {Read}(1990)}]{Read90}%
  \BibitemOpen
  \bibfield  {author} {\bibinfo {author} {\bibfnamefont {N.}~\bibnamefont
  {Read}},\ }\href {\doibase 10.1103/PhysRevLett.65.1502} {\bibfield  {journal}
  {\bibinfo  {journal} {Phys. Rev. Lett.}\ }\textbf {\bibinfo {volume} {65}},\
  \bibinfo {pages} {1502} (\bibinfo {year} {1990})}\BibitemShut {NoStop}%
\bibitem [{\citenamefont {Haldane}(1983)}]{Haldane83}%
  \BibitemOpen
  \bibfield  {author} {\bibinfo {author} {\bibfnamefont {F.~D.~M.}\
  \bibnamefont {Haldane}},\ }\href {\doibase 10.1103/PhysRevLett.51.605}
  {\bibfield  {journal} {\bibinfo  {journal} {Phys. Rev. Lett.}\ }\textbf
  {\bibinfo {volume} {51}},\ \bibinfo {pages} {605} (\bibinfo {year}
  {1983})}\BibitemShut {NoStop}%
\bibitem [{\citenamefont {Halperin}(1984)}]{Halperin84}%
  \BibitemOpen
  \bibfield  {author} {\bibinfo {author} {\bibfnamefont {B.~I.}\ \bibnamefont
  {Halperin}},\ }\href {\doibase 10.1103/PhysRevLett.52.1583} {\bibfield
  {journal} {\bibinfo  {journal} {Phys. Rev. Lett.}\ }\textbf {\bibinfo
  {volume} {52}},\ \bibinfo {pages} {1583} (\bibinfo {year}
  {1984})}\BibitemShut {NoStop}%
\bibitem [{\citenamefont {Cappelli}\ \emph {et~al.}(2001)\citenamefont
  {Cappelli}, \citenamefont {Georgiev},\ and\ \citenamefont
  {Todorov}}]{Cappelli01}%
  \BibitemOpen
  \bibfield  {author} {\bibinfo {author} {\bibfnamefont {A.}~\bibnamefont
  {Cappelli}}, \bibinfo {author} {\bibfnamefont {L.~S.}\ \bibnamefont
  {Georgiev}}, \ and\ \bibinfo {author} {\bibfnamefont {I.~T.}\ \bibnamefont
  {Todorov}},\ }\href {\doibase 10.1016/S0550-3213(00)00774-4} {\bibfield
  {journal} {\bibinfo  {journal} {Nucl. Phys. B}\ }\textbf {\bibinfo {volume}
  {599}},\ \bibinfo {pages} {499 } (\bibinfo {year} {2001})}\BibitemShut
  {NoStop}%
\bibitem [{\citenamefont {Haldane}(1985)}]{Haldane85}%
  \BibitemOpen
  \bibfield  {author} {\bibinfo {author} {\bibfnamefont {F.~D.~M.}\
  \bibnamefont {Haldane}},\ }\href {\doibase 10.1103/PhysRevLett.55.2095}
  {\bibfield  {journal} {\bibinfo  {journal} {Phys. Rev. Lett.}\ }\textbf
  {\bibinfo {volume} {55}},\ \bibinfo {pages} {2095} (\bibinfo {year}
  {1985})}\BibitemShut {NoStop}%
\bibitem [{\citenamefont {Emery}\ \emph {et~al.}(2000)\citenamefont {Emery},
  \citenamefont {Fradkin}, \citenamefont {Kivelson},\ and\ \citenamefont
  {Lubensky}}]{Emery00}%
  \BibitemOpen
  \bibfield  {author} {\bibinfo {author} {\bibfnamefont {V.~J.}\ \bibnamefont
  {Emery}}, \bibinfo {author} {\bibfnamefont {E.}~\bibnamefont {Fradkin}},
  \bibinfo {author} {\bibfnamefont {S.~A.}\ \bibnamefont {Kivelson}}, \ and\
  \bibinfo {author} {\bibfnamefont {T.~C.}\ \bibnamefont {Lubensky}},\ }\href
  {\doibase 10.1103/PhysRevLett.85.2160} {\bibfield  {journal} {\bibinfo
  {journal} {Phys. Rev. Lett.}\ }\textbf {\bibinfo {volume} {85}},\ \bibinfo
  {pages} {2160} (\bibinfo {year} {2000})}\BibitemShut {NoStop}%
\bibitem [{\citenamefont {Vishwanath}\ and\ \citenamefont
  {Carpentier}(2001)}]{Vishwanath01}%
  \BibitemOpen
  \bibfield  {author} {\bibinfo {author} {\bibfnamefont {A.}~\bibnamefont
  {Vishwanath}}\ and\ \bibinfo {author} {\bibfnamefont {D.}~\bibnamefont
  {Carpentier}},\ }\href {\doibase 10.1103/PhysRevLett.86.676} {\bibfield
  {journal} {\bibinfo  {journal} {Phys. Rev. Lett.}\ }\textbf {\bibinfo
  {volume} {86}},\ \bibinfo {pages} {676} (\bibinfo {year} {2001})}\BibitemShut
  {NoStop}%
\bibitem [{\citenamefont {Sondhi}\ and\ \citenamefont {Yang}(2001)}]{Sondhi01}%
  \BibitemOpen
  \bibfield  {author} {\bibinfo {author} {\bibfnamefont {S.~L.}\ \bibnamefont
  {Sondhi}}\ and\ \bibinfo {author} {\bibfnamefont {K.}~\bibnamefont {Yang}},\
  }\href {\doibase 10.1103/PhysRevB.63.054430} {\bibfield  {journal} {\bibinfo
  {journal} {Phys. Rev. B}\ }\textbf {\bibinfo {volume} {63}},\ \bibinfo
  {pages} {054430} (\bibinfo {year} {2001})}\BibitemShut {NoStop}%
\bibitem [{\citenamefont {Mukhopadhyay}\ \emph {et~al.}(2001)\citenamefont
  {Mukhopadhyay}, \citenamefont {Kane},\ and\ \citenamefont
  {Lubensky}}]{Mukhopadhyay01}%
  \BibitemOpen
  \bibfield  {author} {\bibinfo {author} {\bibfnamefont {R.}~\bibnamefont
  {Mukhopadhyay}}, \bibinfo {author} {\bibfnamefont {C.~L.}\ \bibnamefont
  {Kane}}, \ and\ \bibinfo {author} {\bibfnamefont {T.~C.}\ \bibnamefont
  {Lubensky}},\ }\href {\doibase 10.1103/PhysRevB.64.045120} {\bibfield
  {journal} {\bibinfo  {journal} {Phys. Rev. B}\ }\textbf {\bibinfo {volume}
  {64}},\ \bibinfo {pages} {045120} (\bibinfo {year} {2001})}\BibitemShut
  {NoStop}%
\bibitem [{\citenamefont {Haldane}(1995)}]{Haldane95}%
  \BibitemOpen
  \bibfield  {author} {\bibinfo {author} {\bibfnamefont {F.~D.~M.}\
  \bibnamefont {Haldane}},\ }\href {\doibase 10.1103/PhysRevLett.74.2090}
  {\bibfield  {journal} {\bibinfo  {journal} {Phys. Rev. Lett.}\ }\textbf
  {\bibinfo {volume} {74}},\ \bibinfo {pages} {2090} (\bibinfo {year}
  {1995})}\BibitemShut {NoStop}%
\bibitem [{\citenamefont {Cooper}(2008)}]{Cooper08}%
  \BibitemOpen
  \bibfield  {author} {\bibinfo {author} {\bibfnamefont {N.}~\bibnamefont
  {Cooper}},\ }\href {\doibase 10.1080/00018730802564122} {\bibfield  {journal}
  {\bibinfo  {journal} {Adv. Phys.}\ }\textbf {\bibinfo {volume} {57}},\
  \bibinfo {pages} {539} (\bibinfo {year} {2008})}\BibitemShut {NoStop}%
\bibitem [{\citenamefont {Gra\ss{}}\ \emph {et~al.}(2012)\citenamefont
  {Gra\ss{}}, \citenamefont {Juli\'a-D\'iaz}, \citenamefont {Barber\'an},\ and\
  \citenamefont {Lewenstein}}]{Grass12}%
  \BibitemOpen
  \bibfield  {author} {\bibinfo {author} {\bibfnamefont {T.}~\bibnamefont
  {Gra\ss{}}}, \bibinfo {author} {\bibfnamefont {B.}~\bibnamefont
  {Juli\'a-D\'iaz}}, \bibinfo {author} {\bibfnamefont {N.}~\bibnamefont
  {Barber\'an}}, \ and\ \bibinfo {author} {\bibfnamefont {M.}~\bibnamefont
  {Lewenstein}},\ }\href {\doibase 10.1103/PhysRevA.86.021603} {\bibfield
  {journal} {\bibinfo  {journal} {Phys. Rev. A}\ }\textbf {\bibinfo {volume}
  {86}},\ \bibinfo {pages} {021603} (\bibinfo {year} {2012})}\BibitemShut
  {NoStop}%
\bibitem [{\citenamefont {Furukawa}\ and\ \citenamefont
  {Ueda}(2012)}]{Furukawa12}%
  \BibitemOpen
  \bibfield  {author} {\bibinfo {author} {\bibfnamefont {S.}~\bibnamefont
  {Furukawa}}\ and\ \bibinfo {author} {\bibfnamefont {M.}~\bibnamefont
  {Ueda}},\ }\href {\doibase 10.1103/PhysRevA.86.031604} {\bibfield  {journal}
  {\bibinfo  {journal} {Phys. Rev. A}\ }\textbf {\bibinfo {volume} {86}},\
  \bibinfo {pages} {031604} (\bibinfo {year} {2012})}\BibitemShut {NoStop}%
\bibitem [{\citenamefont {Frenkel}\ and\ \citenamefont
  {Kac}(1980)}]{Frenkel80}%
  \BibitemOpen
  \bibfield  {author} {\bibinfo {author} {\bibfnamefont {I.~B.}\ \bibnamefont
  {Frenkel}}\ and\ \bibinfo {author} {\bibfnamefont {V.~G.}\ \bibnamefont
  {Kac}},\ }\href {\doibase 10.1007/BF01391662} {\bibfield  {journal} {\bibinfo
   {journal} {Inv. Math.}\ }\textbf {\bibinfo {volume} {62}},\ \bibinfo {pages}
  {23} (\bibinfo {year} {1980})}\BibitemShut {NoStop}%
\bibitem [{\citenamefont {Segal}(1981)}]{Segal81}%
  \BibitemOpen
  \bibfield  {author} {\bibinfo {author} {\bibfnamefont {G.}~\bibnamefont
  {Segal}},\ }\href {\doibase 10.1007/BF01208274} {\bibfield  {journal}
  {\bibinfo  {journal} {Commun. Math. Phys}\ }\textbf {\bibinfo {volume}
  {80}},\ \bibinfo {pages} {301 } (\bibinfo {year} {1981})}\BibitemShut
  {NoStop}%
\bibitem [{\citenamefont {Goddard}\ and\ \citenamefont
  {Olive}(1986)}]{Goddard86}%
  \BibitemOpen
  \bibfield  {author} {\bibinfo {author} {\bibfnamefont {P.}~\bibnamefont
  {Goddard}}\ and\ \bibinfo {author} {\bibfnamefont {D.}~\bibnamefont
  {Olive}},\ }\href {\doibase 10.1142/S0217751X86000149} {\bibfield  {journal}
  {\bibinfo  {journal} {Int. J. Mod. Phys. A}\ }\textbf {\bibinfo {volume}
  {01}},\ \bibinfo {pages} {303} (\bibinfo {year} {1986})}\BibitemShut
  {NoStop}%
\bibitem [{\citenamefont {Dunne}\ \emph {et~al.}(1989)\citenamefont {Dunne},
  \citenamefont {Halliday},\ and\ \citenamefont {Suranyi}}]{Dunne89}%
  \BibitemOpen
  \bibfield  {author} {\bibinfo {author} {\bibfnamefont {G.}~\bibnamefont
  {Dunne}}, \bibinfo {author} {\bibfnamefont {I.}~\bibnamefont {Halliday}}, \
  and\ \bibinfo {author} {\bibfnamefont {P.}~\bibnamefont {Suranyi}},\ }\href
  {\doibase 10.1016/0550-3213(89)90465-3} {\bibfield  {journal} {\bibinfo
  {journal} {Nucl. Phys. B}\ }\textbf {\bibinfo {volume} {325}},\ \bibinfo
  {pages} {526 } (\bibinfo {year} {1989})}\BibitemShut {NoStop}%
\bibitem [{\citenamefont {Fateev}(1991)}]{Fateev91a}%
  \BibitemOpen
  \bibfield  {author} {\bibinfo {author} {\bibfnamefont {V.~A.}\ \bibnamefont
  {Fateev}},\ }\href {\doibase 10.1142/S0217751X91001052} {\bibfield  {journal}
  {\bibinfo  {journal} {Int. J. Mod. Phys. A}\ }\textbf {\bibinfo {volume}
  {06}},\ \bibinfo {pages} {2109} (\bibinfo {year} {1991})}\BibitemShut
  {NoStop}%
\bibitem [{\citenamefont {Fateev}\ and\ \citenamefont
  {Zamolodchikov}(1991)}]{Fateev91b}%
  \BibitemOpen
  \bibfield  {author} {\bibinfo {author} {\bibfnamefont {V.~A.}\ \bibnamefont
  {Fateev}}\ and\ \bibinfo {author} {\bibfnamefont {A.~B.}\ \bibnamefont
  {Zamolodchikov}},\ }\href {\doibase 10.1016/0370-2693(91)91283-2} {\bibfield
  {journal} {\bibinfo  {journal} {Phys. Lett. B}\ }\textbf {\bibinfo {volume}
  {271}},\ \bibinfo {pages} {91 } (\bibinfo {year} {1991})}\BibitemShut
  {NoStop}%
\bibitem [{\citenamefont {Lecheminant}(2007)}]{Lecheminant07}%
  \BibitemOpen
  \bibfield  {author} {\bibinfo {author} {\bibfnamefont {P.}~\bibnamefont
  {Lecheminant}},\ }\href {\doibase 10.1016/j.physletb.2006.12.079} {\bibfield
  {journal} {\bibinfo  {journal} {Phys. Lett. B}\ }\textbf {\bibinfo {volume}
  {648}},\ \bibinfo {pages} {323 } (\bibinfo {year} {2007})}\BibitemShut
  {NoStop}%
\bibitem [{\citenamefont {Lecheminant}\ and\ \citenamefont
  {Nonne}(2012)}]{Lecheminant12}%
  \BibitemOpen
  \bibfield  {author} {\bibinfo {author} {\bibfnamefont {P.}~\bibnamefont
  {Lecheminant}}\ and\ \bibinfo {author} {\bibfnamefont {H.}~\bibnamefont
  {Nonne}},\ }\href {\doibase 10.1103/PhysRevB.85.195121} {\bibfield  {journal}
  {\bibinfo  {journal} {Phys. Rev. B}\ }\textbf {\bibinfo {volume} {85}},\
  \bibinfo {pages} {195121} (\bibinfo {year} {2012})}\BibitemShut {NoStop}%
\bibitem [{\citenamefont {Baxter}(1989{\natexlab{a}})}]{Baxter89a}%
  \BibitemOpen
  \bibfield  {author} {\bibinfo {author} {\bibfnamefont {R.~J.}\ \bibnamefont
  {Baxter}},\ }\href {\doibase 10.1016/0375-9601(89)90884-0} {\bibfield
  {journal} {\bibinfo  {journal} {Phys. Lett. A}\ }\textbf {\bibinfo {volume}
  {140}},\ \bibinfo {pages} {155 } (\bibinfo {year}
  {1989}{\natexlab{a}})}\BibitemShut {NoStop}%
\bibitem [{\citenamefont {Baxter}(1989{\natexlab{b}})}]{Baxter89b}%
  \BibitemOpen
  \bibfield  {author} {\bibinfo {author} {\bibfnamefont {R.~J.}\ \bibnamefont
  {Baxter}},\ }\href {\doibase 10.1007/BF01023632} {\bibfield  {journal}
  {\bibinfo  {journal} {J. Stat. Phys.}\ }\textbf {\bibinfo {volume} {57}},\
  \bibinfo {pages} {1} (\bibinfo {year} {1989}{\natexlab{b}})}\BibitemShut
  {NoStop}%
\bibitem [{\citenamefont {Fendley}(2014)}]{Fendley14}%
  \BibitemOpen
  \bibfield  {author} {\bibinfo {author} {\bibfnamefont {P.}~\bibnamefont
  {Fendley}},\ }\href {\doibase 10.1088/1751-8113/47/7/075001} {\bibfield
  {journal} {\bibinfo  {journal} {J. Phys. A}\ }\textbf {\bibinfo {volume}
  {47}},\ \bibinfo {pages} {075001} (\bibinfo {year} {2014})}\BibitemShut
  {NoStop}%
\bibitem [{\citenamefont {Bazhanov}\ \emph {et~al.}(1991)\citenamefont
  {Bazhanov}, \citenamefont {Kashaev}, \citenamefont {Mangazeev},\ and\
  \citenamefont {Stroganov}}]{Bazhanov91}%
  \BibitemOpen
  \bibfield  {author} {\bibinfo {author} {\bibfnamefont {V.~V.}\ \bibnamefont
  {Bazhanov}}, \bibinfo {author} {\bibfnamefont {R.~M.}\ \bibnamefont
  {Kashaev}}, \bibinfo {author} {\bibfnamefont {V.~V.}\ \bibnamefont
  {Mangazeev}}, \ and\ \bibinfo {author} {\bibfnamefont {Y.~G.}\ \bibnamefont
  {Stroganov}},\ }\href {\doibase 10.1007/BF02099497} {\bibfield  {journal}
  {\bibinfo  {journal} {Commun. Math. Phys.}\ }\textbf {\bibinfo {volume}
  {138}},\ \bibinfo {pages} {393} (\bibinfo {year} {1991})}\BibitemShut
  {NoStop}%
\bibitem [{\citenamefont {Kashaev}\ \emph {et~al.}(1991)\citenamefont
  {Kashaev}, \citenamefont {Mangazeev},\ and\ \citenamefont
  {Nakanishi}}]{Kashaev91}%
  \BibitemOpen
  \bibfield  {author} {\bibinfo {author} {\bibfnamefont {R.}~\bibnamefont
  {Kashaev}}, \bibinfo {author} {\bibfnamefont {V.}~\bibnamefont {Mangazeev}},
  \ and\ \bibinfo {author} {\bibfnamefont {T.}~\bibnamefont {Nakanishi}},\
  }\href {\doibase 10.1016/0550-3213(91)90542-6} {\bibfield  {journal}
  {\bibinfo  {journal} {Nucl. Phys. B}\ }\textbf {\bibinfo {volume} {362}},\
  \bibinfo {pages} {563 } (\bibinfo {year} {1991})}\BibitemShut {NoStop}%
\bibitem [{\citenamefont {Read}\ and\ \citenamefont {Green}(2000)}]{Read00}%
  \BibitemOpen
  \bibfield  {author} {\bibinfo {author} {\bibfnamefont {N.}~\bibnamefont
  {Read}}\ and\ \bibinfo {author} {\bibfnamefont {D.}~\bibnamefont {Green}},\
  }\href {\doibase 10.1103/PhysRevB.61.10267} {\bibfield  {journal} {\bibinfo
  {journal} {Phys. Rev. B}\ }\textbf {\bibinfo {volume} {61}},\ \bibinfo
  {pages} {10267} (\bibinfo {year} {2000})}\BibitemShut {NoStop}%
\bibitem [{\citenamefont {Regnault}\ \emph {et~al.}(2008)\citenamefont
  {Regnault}, \citenamefont {Goerbig},\ and\ \citenamefont
  {Jolicoeur}}]{Regnault08}%
  \BibitemOpen
  \bibfield  {author} {\bibinfo {author} {\bibfnamefont {N.}~\bibnamefont
  {Regnault}}, \bibinfo {author} {\bibfnamefont {M.~O.}\ \bibnamefont
  {Goerbig}}, \ and\ \bibinfo {author} {\bibfnamefont {T.}~\bibnamefont
  {Jolicoeur}},\ }\href {\doibase 10.1103/PhysRevLett.101.066803} {\bibfield
  {journal} {\bibinfo  {journal} {Phys. Rev. Lett.}\ }\textbf {\bibinfo
  {volume} {101}},\ \bibinfo {pages} {066803} (\bibinfo {year}
  {2008})}\BibitemShut {NoStop}%
\bibitem [{\citenamefont {Haldane}(1981)}]{Haldane81}%
  \BibitemOpen
  \bibfield  {author} {\bibinfo {author} {\bibfnamefont {F.~D.~M.}\
  \bibnamefont {Haldane}},\ }\href {\doibase 10.1103/PhysRevLett.47.1840}
  {\bibfield  {journal} {\bibinfo  {journal} {Phys. Rev. Lett.}\ }\textbf
  {\bibinfo {volume} {47}},\ \bibinfo {pages} {1840} (\bibinfo {year}
  {1981})}\BibitemShut {NoStop}%
\bibitem [{\citenamefont {Barkeshli}\ and\ \citenamefont
  {Qi}(2012)}]{Barkeshli12}%
  \BibitemOpen
  \bibfield  {author} {\bibinfo {author} {\bibfnamefont {M.}~\bibnamefont
  {Barkeshli}}\ and\ \bibinfo {author} {\bibfnamefont {X.-L.}\ \bibnamefont
  {Qi}},\ }\href {\doibase 10.1103/PhysRevX.2.031013} {\bibfield  {journal}
  {\bibinfo  {journal} {Phys. Rev. X}\ }\textbf {\bibinfo {volume} {2}},\
  \bibinfo {pages} {031013} (\bibinfo {year} {2012})}\BibitemShut {NoStop}%
\bibitem [{\citenamefont {He}\ \emph {et~al.}(2015)\citenamefont {He},
  \citenamefont {Bhattacharjee}, \citenamefont {Moessner},\ and\ \citenamefont
  {Pollmann}}]{YCHe15}%
  \BibitemOpen
  \bibfield  {author} {\bibinfo {author} {\bibfnamefont {Y.-C.}\ \bibnamefont
  {He}}, \bibinfo {author} {\bibfnamefont {S.}~\bibnamefont {Bhattacharjee}},
  \bibinfo {author} {\bibfnamefont {R.}~\bibnamefont {Moessner}}, \ and\
  \bibinfo {author} {\bibfnamefont {F.}~\bibnamefont {Pollmann}},\ }\href
  {\doibase 10.1103/PhysRevLett.115.116803} {\bibfield  {journal} {\bibinfo
  {journal} {Phys. Rev. Lett.}\ }\textbf {\bibinfo {volume} {115}},\ \bibinfo
  {pages} {116803} (\bibinfo {year} {2015})}\BibitemShut {NoStop}%
\bibitem [{\citenamefont {Affleck}(1986)}]{Affleck86}%
  \BibitemOpen
  \bibfield  {author} {\bibinfo {author} {\bibfnamefont {I.}~\bibnamefont
  {Affleck}},\ }\href {\doibase 10.1016/0550-3213(86)90167-7} {\bibfield
  {journal} {\bibinfo  {journal} {Nucl. Phys. B}\ }\textbf {\bibinfo {volume}
  {265}},\ \bibinfo {pages} {409 } (\bibinfo {year} {1986})}\BibitemShut
  {NoStop}%
\bibitem [{\citenamefont {Affleck}(1988)}]{Affleck88}%
  \BibitemOpen
  \bibfield  {author} {\bibinfo {author} {\bibfnamefont {I.}~\bibnamefont
  {Affleck}},\ }\href {\doibase 10.1016/0550-3213(88)90117-4} {\bibfield
  {journal} {\bibinfo  {journal} {Nucl. Phys. B}\ }\textbf {\bibinfo {volume}
  {305}},\ \bibinfo {pages} {582 } (\bibinfo {year} {1988})}\BibitemShut
  {NoStop}%
\bibitem [{\citenamefont {Tu}\ \emph {et~al.}(2014)\citenamefont {Tu},
  \citenamefont {Nielsen},\ and\ \citenamefont {Sierra}}]{HHTu14}%
  \BibitemOpen
  \bibfield  {author} {\bibinfo {author} {\bibfnamefont {H.-H.}\ \bibnamefont
  {Tu}}, \bibinfo {author} {\bibfnamefont {A.~E.~B.}\ \bibnamefont {Nielsen}},
  \ and\ \bibinfo {author} {\bibfnamefont {G.}~\bibnamefont {Sierra}},\ }\href
  {\doibase 10.1016/j.nuclphysb.2014.06.027} {\bibfield  {journal} {\bibinfo
  {journal} {Nucl. Phys. B}\ }\textbf {\bibinfo {volume} {886}},\ \bibinfo
  {pages} {328 } (\bibinfo {year} {2014})}\BibitemShut {NoStop}%
\bibitem [{\citenamefont {Nataf}\ \emph {et~al.}(2016)\citenamefont {Nataf},
  \citenamefont {Lajk\'o}, \citenamefont {Wietek}, \citenamefont {Penc},
  \citenamefont {Mila},\ and\ \citenamefont {L\"auchli}}]{Nataf16}%
  \BibitemOpen
  \bibfield  {author} {\bibinfo {author} {\bibfnamefont {P.}~\bibnamefont
  {Nataf}}, \bibinfo {author} {\bibfnamefont {M.}~\bibnamefont {Lajk\'o}},
  \bibinfo {author} {\bibfnamefont {A.}~\bibnamefont {Wietek}}, \bibinfo
  {author} {\bibfnamefont {K.}~\bibnamefont {Penc}}, \bibinfo {author}
  {\bibfnamefont {F.}~\bibnamefont {Mila}}, \ and\ \bibinfo {author}
  {\bibfnamefont {A.~M.}\ \bibnamefont {L\"auchli}},\ }\href {\doibase
  10.1103/PhysRevLett.117.167202} {\bibfield  {journal} {\bibinfo  {journal}
  {Phys. Rev. Lett.}\ }\textbf {\bibinfo {volume} {117}},\ \bibinfo {pages}
  {167202} (\bibinfo {year} {2016})}\BibitemShut {NoStop}%
\bibitem [{\citenamefont {Ardonne}\ \emph {et~al.}(2003)\citenamefont
  {Ardonne}, \citenamefont {Bouwknegt},\ and\ \citenamefont
  {Dawson}}]{Ardonne03}%
  \BibitemOpen
  \bibfield  {author} {\bibinfo {author} {\bibfnamefont {E.}~\bibnamefont
  {Ardonne}}, \bibinfo {author} {\bibfnamefont {P.}~\bibnamefont {Bouwknegt}},
  \ and\ \bibinfo {author} {\bibfnamefont {P.}~\bibnamefont {Dawson}},\ }\href
  {\doibase 10.1016/S0550-3213(03)00223-2} {\bibfield  {journal} {\bibinfo
  {journal} {Nucl. Phys. B}\ }\textbf {\bibinfo {volume} {660}},\ \bibinfo
  {pages} {473 } (\bibinfo {year} {2003})}\BibitemShut {NoStop}%
\bibitem [{\citenamefont {Barkeshli}\ \emph {et~al.}(2013)\citenamefont
  {Barkeshli}, \citenamefont {Jian},\ and\ \citenamefont {Qi}}]{Barkeshli13}%
  \BibitemOpen
  \bibfield  {author} {\bibinfo {author} {\bibfnamefont {M.}~\bibnamefont
  {Barkeshli}}, \bibinfo {author} {\bibfnamefont {C.-M.}\ \bibnamefont {Jian}},
  \ and\ \bibinfo {author} {\bibfnamefont {X.-L.}\ \bibnamefont {Qi}},\ }\href
  {\doibase 10.1103/PhysRevB.88.235103} {\bibfield  {journal} {\bibinfo
  {journal} {Phys. Rev. B}\ }\textbf {\bibinfo {volume} {88}},\ \bibinfo
  {pages} {235103} (\bibinfo {year} {2013})}\BibitemShut {NoStop}%
\bibitem [{\citenamefont {Lan}\ \emph {et~al.}(2015)\citenamefont {Lan},
  \citenamefont {Wang},\ and\ \citenamefont {Wen}}]{TLan15}%
  \BibitemOpen
  \bibfield  {author} {\bibinfo {author} {\bibfnamefont {T.}~\bibnamefont
  {Lan}}, \bibinfo {author} {\bibfnamefont {J.~C.}\ \bibnamefont {Wang}}, \
  and\ \bibinfo {author} {\bibfnamefont {X.-G.}\ \bibnamefont {Wen}},\ }\href
  {\doibase 10.1103/PhysRevLett.114.076402} {\bibfield  {journal} {\bibinfo
  {journal} {Phys. Rev. Lett.}\ }\textbf {\bibinfo {volume} {114}},\ \bibinfo
  {pages} {076402} (\bibinfo {year} {2015})}\BibitemShut {NoStop}%
\bibitem [{\citenamefont {Fr\"ohlich}\ and\ \citenamefont
  {Zee}(1991)}]{Frohlich91}%
  \BibitemOpen
  \bibfield  {author} {\bibinfo {author} {\bibfnamefont {J.}~\bibnamefont
  {Fr\"ohlich}}\ and\ \bibinfo {author} {\bibfnamefont {A.}~\bibnamefont
  {Zee}},\ }\href {\doibase 10.1016/0550-3213(91)90275-3} {\bibfield  {journal}
  {\bibinfo  {journal} {Nucl. Phys. B}\ }\textbf {\bibinfo {volume} {364}},\
  \bibinfo {pages} {517 } (\bibinfo {year} {1991})}\BibitemShut {NoStop}%
\bibitem [{\citenamefont {Griffin}\ and\ \citenamefont
  {Nemeschansky}(1989)}]{Griffin89}%
  \BibitemOpen
  \bibfield  {author} {\bibinfo {author} {\bibfnamefont {P.}~\bibnamefont
  {Griffin}}\ and\ \bibinfo {author} {\bibfnamefont {D.}~\bibnamefont
  {Nemeschansky}},\ }\href {\doibase 10.1016/0550-3213(89)90123-5} {\bibfield
  {journal} {\bibinfo  {journal} {Nucl. Phys. B}\ }\textbf {\bibinfo {volume}
  {323}},\ \bibinfo {pages} {545 } (\bibinfo {year} {1989})}\BibitemShut
  {NoStop}%
\bibitem [{\citenamefont {Green}\ \emph {et~al.}(1987)\citenamefont {Green},
  \citenamefont {Schwarz},\ and\ \citenamefont {Witten}}]{GSW1}%
  \BibitemOpen
  \bibfield  {author} {\bibinfo {author} {\bibfnamefont {M.~B.}\ \bibnamefont
  {Green}}, \bibinfo {author} {\bibfnamefont {J.~H.}\ \bibnamefont {Schwarz}},
  \ and\ \bibinfo {author} {\bibfnamefont {E.}~\bibnamefont {Witten}},\
  }\href@noop {} {\emph {\bibinfo {title} {Superstring Theory}}},\
  Vol.~\bibinfo {volume} {1}\ (\bibinfo  {publisher} {Cambridge University
  Press},\ \bibinfo {address} {Cambridge},\ \bibinfo {year} {1987})\BibitemShut
  {NoStop}%
\bibitem [{\citenamefont {Chu}\ and\ \citenamefont {Goddard}(1995)}]{Chu95}%
  \BibitemOpen
  \bibfield  {author} {\bibinfo {author} {\bibfnamefont {M.}~\bibnamefont
  {Chu}}\ and\ \bibinfo {author} {\bibfnamefont {P.}~\bibnamefont {Goddard}},\
  }\href {\doibase 10.1016/0550-3213(95)00160-T} {\bibfield  {journal}
  {\bibinfo  {journal} {Nucl. Phys. B}\ }\textbf {\bibinfo {volume} {445}},\
  \bibinfo {pages} {145 } (\bibinfo {year} {1995})}\BibitemShut {NoStop}%
\end{thebibliography}%

\end{document}